\documentclass[traditabstract]{aa}

\usepackage[pdftex,pdfpagemode={UseOutlines},bookmarks,bookmarksopen,colorlinks,linkcolor={blue},citecolor={blue},urlcolor={red}]{hyperref}
\usepackage{txfonts}
\usepackage[table]{xcolor}
\usepackage{longtable}
\usepackage{pdflscape}
\usepackage[maxfloats=110]{morefloats}
\usepackage{graphicx}
\usepackage{hyperref}
\usepackage{subcaption}

\usepackage{amssymb}% http://ctan.org/pkg/amssymb
\usepackage{pifont}% http://ctan.org/pkg/pifont

\usepackage{natbib}
\bibpunct{(}{)}{;}{a}{}{,} % to follow the A&A style
\usepackage{epsfig}
\usepackage{footmisc}
\DefineFNsymbols{mySymbols}{{\ensuremath\dagger}{\ensuremath\ddagger}\S\P
   *{**}{\ensuremath{\dagger\dagger}}{\ensuremath{\ddagger\ddagger}}}
\setfnsymbol{mySymbols}

\everymath{\displaystyle}

\usepackage{etoolbox}
\makeatletter
\patchcmd\@combinedblfloats{\box\@outputbox}{\unvbox\@outputbox}{}{%
  \errmessage{\noexpand\@combinedblfloats could not be patched}%
}%
\makeatother

\begin{document}
\rowcolors{2}{gray!25}{white}
\title{The BEBOP radial-velocity survey for circumbinary planets\thanks{Based on photometric observations with the SuperWASP and SuperWASP-South instruments and radial velocity measurement from the CORALIE spectrograph, mounted on the Swiss 1.2\,m \textit{Euler} Telescope, located at ESO, La Silla, Chile. The data is publicly available at the \textit{CDS} Strasbourg and on demand to the main author.}
\\ \vspace{5mm}{\Large I. Eight years of CORALIE observations of 47 single-line eclipsing binaries \\
and abundance constraints on the masses of circumbinary planets}%\\ {\Large WASP-30b and J1219-39b}
}
\author{David V. Martin\inst{1,2}\thanks{Fellow of the Swiss National Science Foundation.}
\and Amaury H.M.J. Triaud\inst{3}
\and St\'ephane Udry \inst{1}
\and Maxime Marmier\inst{1}
\and Pierre F.L. Maxted\inst{4}
\and Andrew Collier Cameron\inst{5}
\and Coel Hellier\inst{4}
\and Francesco Pepe\inst{1}
\and Don Pollacco\inst{6}
\and Damien S\'egransan \inst{1}
\and Richard West\inst{6}
}

\offprints{david.martin@unige.ch}

\institute{Observatoire Astronomique de l'Universit\'e de Gen\`eve, Chemin des Maillettes 51, CH-1290 Sauverny, Switzerland
\and Department of Astronomy and Astrophysics, University of Chicago, 5640 South Ellis Avenue, Chicago, IL 60637, USA
\and School of Physics \& Astronomy, University of Birmingham, Edgbaston, Birmingham, B15 2TT, UK
\and Astrophysics Group, Keele University, Staffordshire, ST5 5BG, UK
\and SUPA, School of Physics \& Astronomy, University of St Andrews, North Haugh, KY16 9SS, St Andrews, Fife, Scotland, UK
\and Department of Physics, University of Warwick, Coventry CV4 7AL, UK
}

\date{Received date / accepted date}
\authorrunning{Martin et al.}
\titlerunning{BEBOP Circumbinary RV Survey}

\abstract{We introduce the BEBOP radial velocity survey for circumbinary planets. We initiated this survey using the CORALIE spectrograph on the  Swiss Euler Telescope at La Silla, Chile. An intensive four year observing campaign commenced in 2013, targeting 47 single-lined eclipsing binaries drawn from the EBLM survey for low mass eclipsing binaries. Our specific use of binaries with faint M dwarf companions avoids spectral contamination, providing observing conditions akin to single stars. By combining new BEBOP observations with existing ones from the EBLM programme, we report on the results of 1519 radial velocity measurements over timespans as long as eight years. For the best targets we are sensitive to planets down to $0.1M_{\rm Jup}$,  and our median sensitivity is $0.4M_{\rm Jup}$. In this initial survey we do not detect any planetary mass companions. Nonetheless, we present the first constraints on the abundance of circumbinary companions, as a function of mass and period. A comparison of our results to {\it Kepler}'s detections indicates a dispersion of planetary orbital inclinations less than $\sim 10^{\circ}$. 
%This programme, which has been recently extended to the HARPS spectrograph, aims to complement the founding sample of 11 transiting circumbinary planets discovered by the {\it Kepler} mission. 

 }

\maketitle

\section{Introduction}

The progression of the field of exoplanets has led to more and more diverse discoveries. A particular example is planets orbniting around both stars of a tight binary, known as circumbinary planets. These planets provide insight into planet formation in different, perturbative environments \citep{Meschiari:2012uq,Paardekooper:2012kx,Rafikov:2013xy,Lines:2014ko}  and of planetary migration in a protoplanetary disc that is sculpted by the inner binary \citep{Artymowicz:1994ty,pierens:2013kx,Martin:2013ab,Kley:2014rt,Pierens:2018fe}. Studying the abundance of circumbinary planets ellicits comparisons to that around single stars \citep{Martin:2014lr,Armstrong:2014yq,Bonavita:2016gf,Klagyivik:2017iz}, and the presence or even absence of planets sheds light on the formation of their host binary \citep{Munoz:2015uq,Martin:2015iu,Hamers:2016er,Xu:2016qw}. Furthermore, circumbinary planets have increased transit probabilities \citep{Borucki:1984fh,Schneider:1994lr,Martin:2015rf,Li:2016ng,Martin:2017qf}, making them often found in the habitable zone \citep{Kane:2013gf,Haghighipour:2013cu} and potentially aiding the use of transmission and emission spectroscopy \citep{Deming:2017jf}.

Roughly two dozen circumbinary planets have been discovered to date, with the majority coming from only two techniques: observing the planet transit in front of one or both of its host stars (e.g. Kepler 16, \citealt{Doyle:2011vn}) or inferring its existence by the measurement of eclipse timing variations (ETVs) of the binary (e.g. NN Serpentis, \citealt{Qian:2009qw,Beuermann:2010fk}). Only a handful of discoveries have come from other methods: gravitational microlensing (OGLE-2007-BLG-349, \citealt{Bennett:2016nr}), direct imaging (e.g. HD106906, \citealt{Bailey:2014we,Lagrange:2016ih}) and pulsar timing (PSR B1620-26, \citealt{Backer:1993ko,Thorsett:1993ol,Sigurdsson:2003lp}). Furthermore, out of the two dominant techniques only the transit discoveries are completely reliable, as the validity of the ETV planets is debated, particularly for post-common envelope binaries \citep{Zorotovic:2013jd,Bear:2014wf,Hardy:2016hb,Bours:2016to}.

Our knowledge of circumbinary planets is largely based on a sample of transiting planets that is both small in number, 11, and impacted by observational biases. Preliminary insights into their formation and distribution have been obtained (reviewed in \citealt{Welsh:2017tr,Martin:2018ya}) but a more comprehensive understanding demands not only more discoveries, but ones made with complementary observing techniques with different sensitivities.

Radial velocities (RVs) led to the first discovered exoplanet around a Sun-like star (51 Peg, \citealt{Mayor:1995uq}) and  hundreds more since. Radial velocities have also yielded dozens of circumstellar planets  orbiting one component of a wider binary, for instance 16 Cygni Bb \citep{Cochran:1997mn}, and WASP-94Ab \& Bb \citep{Neveu-VanMalle:2014rf}. There is not, however, a bonafide circumbinary planet discovered by RVs. This is in spite of attempts stretching back many years, for example the TATOOINE survey of double-lined spectroscopic binaries (SB2s, \citealt{Konacki:2009lr,Konacki:2010re,Heminiak:2012fk}). There was a potential RV discovery of a circumbinary planet in the HD 202206 system by \citet{Correia:2005lr}, but astrometric data has since revealed it to be a $17.9^{+2.9}_{-1.8}M_{\rm Jup}$ circumbinary brown dwarf \citep{Benedict:2017eg}.

In spite of these past difficulties, RVs still have the ability to expand our knowledge of circumbinaries. The sensitivity of RVs partially overlaps with that of transits. This means that we may use transit discoveries as a guide and motivation, but radial velocities push into new parameter spaces,  with a weaker dependence on period and inclination and also having detections that are mass-dependent, rather than radius-dependent. 

%While, both techniques favour short period planets, the Doppler method extends to longer separations. Both favour large planets, although one in terms of mass and one radius. Radial velocities are not as strictly sensitive to the inclination alignment of the planetary orbit.

In 2013 we intiated a radial velocity survey named Binaries Escorted By Orbiting Planets, henceforth (BEBOP). The programme was initiated on the 1.2 metre Swiss Euler Telescope using the CORALIE spectrograph. We targeted 47 known single-lined eclipsing binaries (SB1s) drawn from the EBLM programme (\citealt{Triaud:2017fu} and see Sect.~\ref{subsec:eblm_review}), which consist of F/G/K primaries and M-dwarf secondaries. The BEBOP observations reach a precision of a few metres per second. This permits a precise characterisation of the binary orbit, such that by subtracting it from the RV signal we may then search the residuals for an orders of magnitude smaller signal of a circumbinary gas giant planet. 

A fundamental design element of BEBOP is to solely observe SB1s. This means that instead of trying to solve the problem of deconvolving the two spectra in SB2s, such as in the TATOOINE survey, our approach avoids it. 

This paper is structured as follows. In Sect.~\ref{sec:review} we briefly review the present understanding of circumbinary planets, and use this to motivate a radial velocity survey. Second, in Sect.~\ref{sec:bebop_overview} the birth and construction of the BEBOP sample is described, including its roots in the EBLM survey. We then describe the observational strategy in Sect.~\ref{sec:observations}. The subsequent treatment of the data, including the reduction of the spectroscopic data, fitting of radial velocity orbits and model selection is covered in Sect.~\ref{sec:data_treatment}. In Sect.~\ref{sec:physical_parameters} we discuss the calculation of primary and secondary masses, as well as the constraints placed on undetected orbital parameters. In Sect.~\ref{sec:results} we present the results of the survey and the characterisation of tertiary Keplerian signals. In Sect.~\ref{sec:analysis} we analyse the results and compute detection limits for each of our systems. From these detection limits, in Sect.~\ref{sec:abundance} we calculate the completeness of this programme and use it to calculate abundances of circumbinary planets, circumbinary brown dwarfs and tight triple star systems. We compare these values to other surveys for gas giants around single and binary stars in Sect.~\ref{sec:other_survey_comparisons}. Finally, in Sect.~\ref{sec:future} we briefly outline the future of the BEBOP survey, including its recent upgrade to the HARPS spectrograph, before concluding in Sect.~\ref{sec:conclusion}.

\section{Trends seen in transiting circumbinary planets}\label{sec:review}

With not even a dozen confirmed transiting circumbinary planets to date, we only have a cursory knowledge of their population.  To provide context for the BEBOP survey we briefly review some of the trends. For a more in depth discussion of the circumbinary planets discovered to date, the reader is directed to the reviews in \citet{Welsh:2017tr,Martin:2018ya}.

\subsection{Binary orbital periods}\label{subsec:review_binary_orbit_sizes}

\begin{figure}
\begin{center}
\includegraphics[width=0.49\textwidth,trim={0 0 0 0},clip]{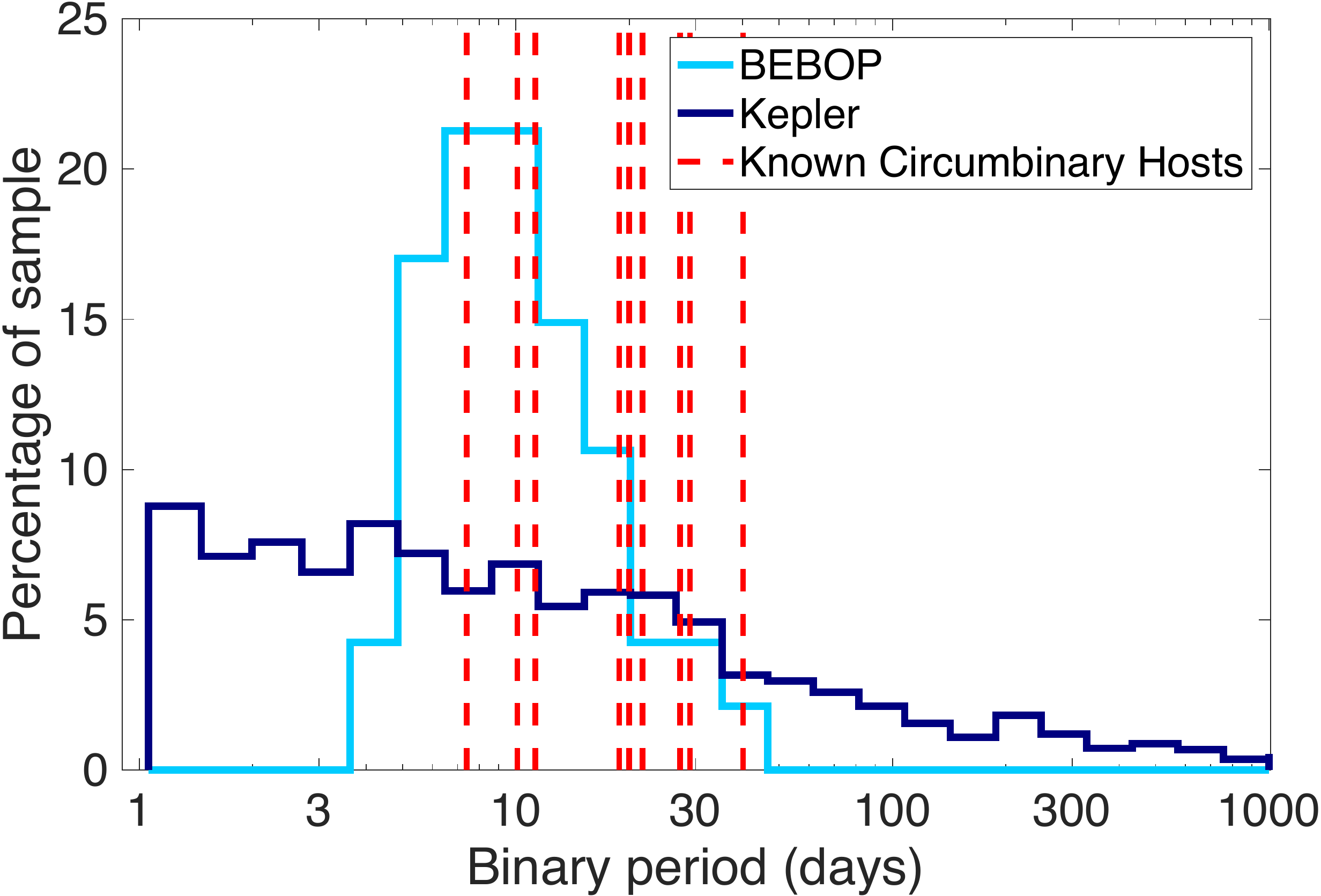}
\caption{In dark blue, a histogram of the {\it Kepler} eclipsing binary catalog for all periods longer than 1 day. Data is taken from \url{http://keplerebs.villanova.edu/} as of May 2017, based on a catalog that is first outlined in \citet{Prsa:2011qw}. Red dashed lines correspond to the nine binaries known to host transiting circumbinary planets. In light blue, a histogram of the 47 BEBOP binaries.}\label{fig:binary_period_histogram}
\end{center}
\end{figure}

\begin{figure*}
\begin{center}
\includegraphics[width=0.99\textwidth,trim={0 0 0 0},clip]{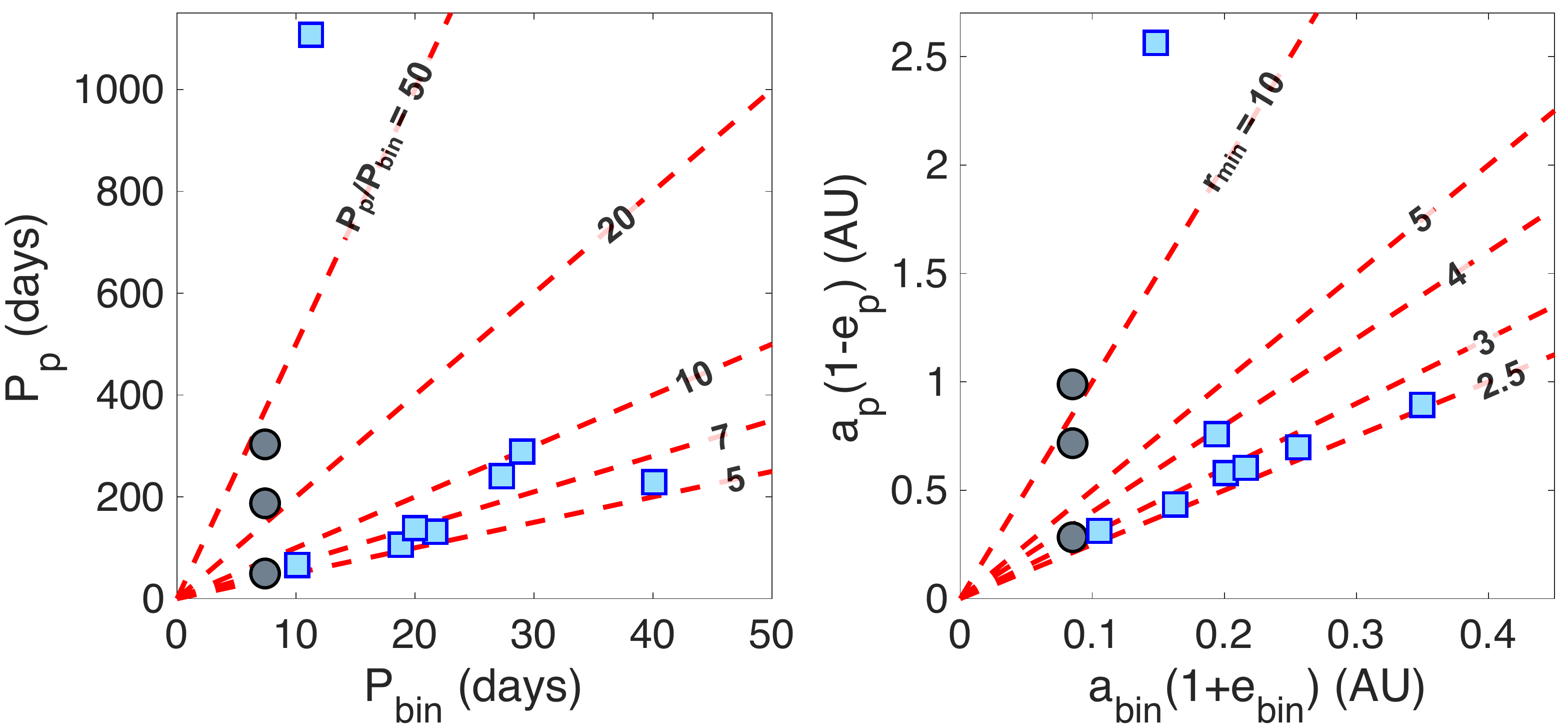}
\caption{ Left: planet and binary orbital periods, with dashed lines of constant period ratio. Right: planet periapse distance and binary apoapse distance, with dashed lines of constant scaled minimum distance, $r_{\rm min} = [a_{\rm p}(1-e_{\rm p})]/[a_{\rm bin}(1+e_{\rm bin})]$. Systems with a single detected planet are shown as blue squares, whereas the three-planet Kepler-47 system is shown as grey circles.}
\label{fig:stability_limit}
\end{center}
\end{figure*}

 All circumbinary planets found as part of the {\it Kepler} mission transit eclipsing binaries. Owing to geometry, the {\it Kepler} eclipsing binary catalog is expectedly biased towards short periods, with a median orbital period of 2.7 days. Planets have only been found around relatively long-period binaries though, with $P_{\rm bin}>7$ days. This is shown in Fig.~\ref{fig:binary_period_histogram}, with a distribution of the {\it Kepler} eclipsing binary catalog and the binary periods known to host circumbinary planets. For comparison, we overlay the distribution of the BEBOP eclipsing binary catalog, the construction of which we discuss in Sect.~\ref{subsec:bebop_birth}.

Tighter binaries are typically accompanied by a third star \citep{Tokovinin:2006la}. This third star is suspected to either disrupt circumbinary planet formation \citep{Munoz:2015uq,Martin:2015iu,Hamers:2016er,Xu:2016qw} or bias the planet sample to long-period, misaligned orbits, both of which would have been missed by the {\it Kepler} transit survey.  Alternate explanations for the dearth of planets around the tightest binaries, which do not invoke the presence of a third star, have also been proposed: tidal expansion of the binary orbit, causing the planet to become unstable \citep{Fleming:2018mf} and UV evaporation of exoplanet atmospheres, shrinking them to an undetectable size \citep{Sanz-Forcada:2014kx}.

\subsection{Planet orbital periods}\label{subsec:review_planet_orbit_sizes}

 Out of the nine known  transiting circumbinary systems, eight contain a planet with a period less than ten times the binary period (plotted in Fig~\ref{fig:stability_limit}, left). The scaled minimum distance between orbits is defined as $r_{\rm min} = [a_{\rm p}(1-e_{\rm p})]/[a_{\rm bin}(1+e_{\rm bin})]$. Seven of the nine systems contain a planet with $r_{\rm min}$ between a narrow range of 2.5 and 3.0, and hence align diagonally in a plot of planet periapse against binary apoapse in Fig.~\ref{fig:stability_limit}, right. A planet cannot orbit too close to the binary lest its orbit becomes unstable through resonant overlap \citep{Mudryk:2006po}, and these seven planets all orbit within a relative distance of 50\% to the stability limit \citep{Dvorak:1986fk,Holman:1999lr,Chavez:2015uq,Quarles:2018ub}\footnote{In reality the stability limit is not a sharp function of orbital width and eccentricity, but it also has subtle dependencies on the binary mass ratio and the mutual inclination, as well as various islands of (in)stability shaped by mean motion resonances.}. The over-density of planets near the stability limit is not believed to be an observational bias \citep{Martin:2014lr,Li:2016ng}, although improved statistics are needed to draw strong conclusions. 

The stability limit coincides closely with where the protoplanetary disc would have been truncated by the inner binary \citep{Artymowicz:1994ty}. Planet formation is believed to be hindered this close to the binary \citep{Meschiari:2012uq,Paardekooper:2012kx,Rafikov:2013xy,Lines:2014ko}. Instead, the favoured explanation for a heightened frequency of planets near the stability limit is formation  in the farther regions of the disc, followed by an inwards migration and then parking near the disc edge \citet{pierens:2013kx,Kley:2014rt,Pierens:2018fe}.

\subsection{Orbital alignment}\label{subsec:review_orbital_alignments}

 The known  transiting circumbinary planets exist on orbits that are coplanar with the binary to within $\sim4^{\circ}$. \citet{Li:2016ng} concludes that this is indicative of the true underlying distribution and not an observational bias. However, the statistics are presently poor, and \citet{Martin:2014lr} demonstrated that highly misaligned systems (more than just a few degrees) have a sparse, hard-to-identify transit signature, and hence could remain hidden in the {\it Kepler} data. 

A  nearly coplanar distribution would be indicative of either a primordially flat environment, or a re-alignment over time of the circumbinary disc \citep{Foucart:2013ys} or the planet itself \citep{Correia:2016tr}. On the other hand, misalignment may be produced by disc-warping \citep{Facchini:2013tr,Lodato:2013lo}, planet-planet scattering \citep{Smullen:2016lo} and tertiary star interactions \citep{Munoz:2015uq,Martin:2015iu,Hamers:2016er}.

\subsection{Planet size and abundance}\label{subsec:review_planet_sizes}

 Only circumbinary planets larger than $3R_{\oplus}$ have been found to date. Some of the larger planets have measured masses  from ETVs, but for most no ETVs are  detectable and hence only an upper limit may be placed. The most massive measured mass is $1.52M_{\rm Jup}$ (Kepler-1647, \citealt{Kostov:2016yq}), but the majority are Saturn mass or smaller. The lack of Earth and super-Earth circumbinary planets is however likely to be an observational bias, owing to the unique challenges of capturing shallow planetary transits  with irregular transit depths, timing and durations \citep{Armstrong:2013rt}.

For circumbinary gas giants the studies by \citet{Martin:2014lr} and \citet{Armstrong:2014yq} provided two important results. First, it was demonstrated that the true abundance was degenerate with the mutual inclination distribution; comparing a coplanar and highly misaligned population, to produce the same number of detections the misaligned population must have a higher planetary abundance as its transit detection rate would be smaller. Second, the minimum abundance of  transiting circumbinary gas giants, corresponding to a near-coplanar distribution, was found to be surprisingly similar to that around single stars \citep{Howard:2010zr,Mayor:2011fj,Santerne:2016lr}.  On a similar note, the imaging survey of \citet{Bonavita:2016gf} determined that the abundance of sub-stellar companions on wide orbits did not differ significantly between single and binary stars. Overall though, these results all require verification due to the presently poor statistics, one of the primary objectives of our survey.

\subsection{Binary mass ratios}\label{subsec:review_mass_ratios}

\begin{figure}
\begin{center}
\includegraphics[width=0.49\textwidth,trim={0 0 0 0},clip]{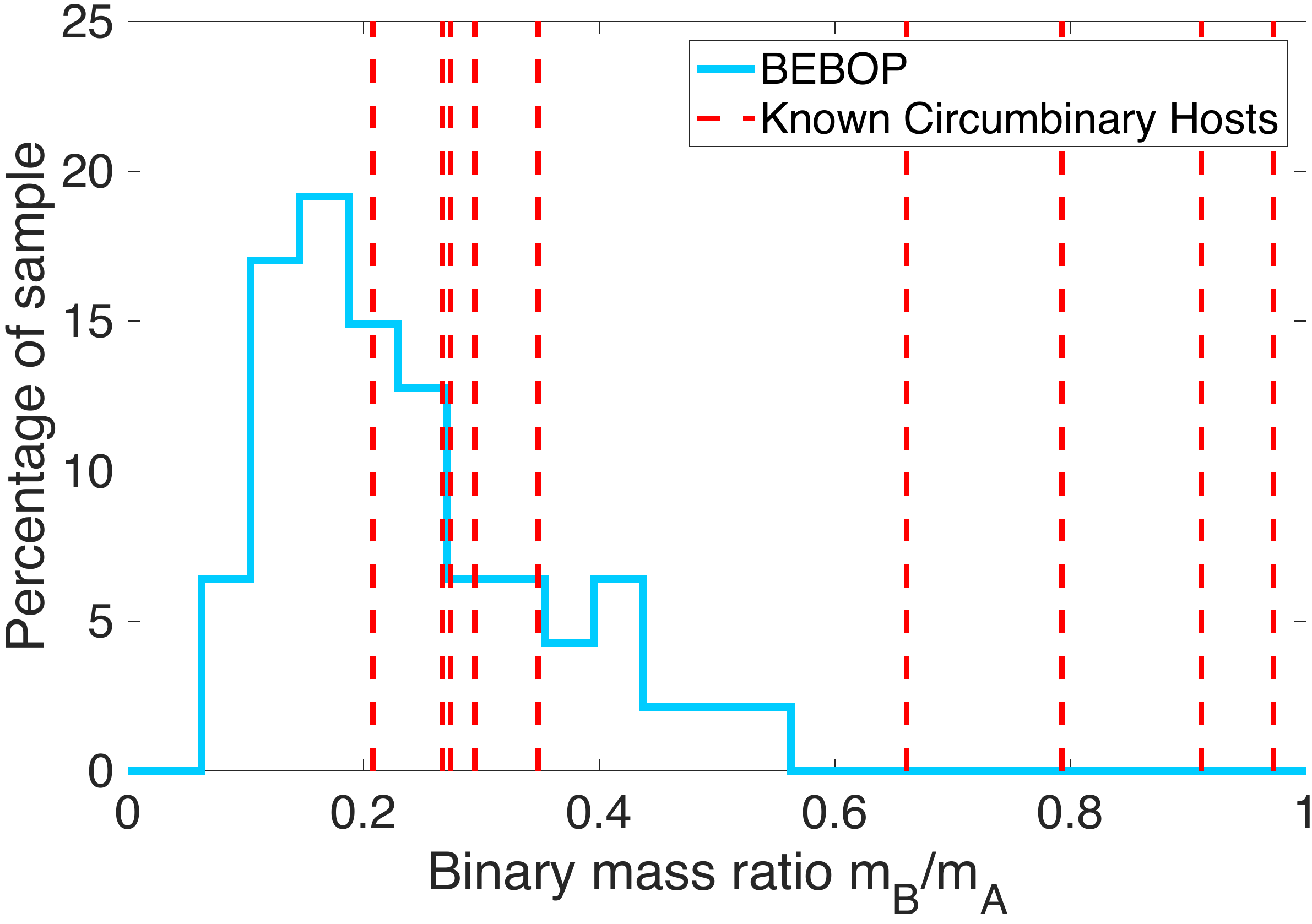}
\caption{In light blue, a histogram of the mass ratio of the 47 BEBOP binaries. The vertical red dashed lines correspond to the mass ratios of the {\it Kepler} binaries known to host transiting circumbinary planets .}
\label{fig:binary_massratio_histogram}
\end{center}
\end{figure}

 In Fig.~\ref{fig:binary_massratio_histogram} we show the mass ratios of the known transiting circumbinary planets (red dashed lines). For comparison, the blue histogram shows the host mass ratios in the BEBOP sample. Planets have been found around binaries of essentially all mass ratios. Whether the abundance of circumbinary planets depends on binary mass ratio remains an open question due primarily to small number statistics\footnote{ An upcoming paper (Martin, under review) investigates {\it Kepler} circumbinary planet population as a function of the mass ratio, although given the small number statistics any inference is only preliminary at this point.}, but for now it is simply re-assuring that the planets known to date are just as commonly found around small mass ratio binaries, like those probed by BEBOP. Finally, we also remind that  the circumbinary brown dwarf HD 202206c orbits a binary with $q=0.08$ \citep{Correia:2005lr,Benedict:2017eg}.

\section{Overview of BEBOP}\label{sec:bebop_overview}

The targets for the BEBOP survey were first discovered and characterised as part of the older EBLM survey for low mass eclipsing binaries. It is for this reason that they are all designated by their EBLM name. This survey has been detailed in a series of papers \citep{Triaud:2013lr,Gomez-Maqueo-Chew:2014jk,vonBoetticher:2017ji,Triaud:2017fu}, so in Sect.~\ref{subsec:eblm_review} it is just briefly reviewed. We then in Sect.~\ref{subsec:bebop_birth} discuss how the BEBOP survey  and its target list was constructed.

\subsection{A brief description of the EBLM survey for low-mass eclipsing binaries}\label{subsec:eblm_review}

Since 2004 the Wide Angle Search for Planets (WASP, \citealt{Pollacco:2006fj,Collier-Cameron:2007fj,Collier-Cameron:2007pb}) has been conducting a ground-based photometric survey of several million stars. Observations span both hemispheres, with sites in La Palma and the South African Astronomical Observatory. The photometric precision and observing baseline are amenable to the detection of Jupiter-sized bodies on orbits of typically less than 10 days,  although some are found with periods up to 40 days.  Detecting hot- and warm-Jupiter planets is the primary objective of WASP.

 However, the WASP survey has also netted a large quantity of astrophysical false-positives. An ever-present challenge is the ambiguity between the transit of a giant exoplanet and the eclipse of a low-mass star. Theoretical and observational studies have demonstrated that the vast range in mass between giant planets ($\sim 0.1M_{\rm Jup}$) and small stars ($\sim 100 M_{\rm Jup}$), including the brown dwarfs in between, only corresponds to a narrow range in radius of $\sim 0.7-2R_{\rm Jup}$, in spite of the very different physical processes taking place \citep{Baraffe:1998ly,Chabrier:2009tr,Baraffe:2003gf,Baraffe:2015lr,Chen:2016lr}. When using the humble precision of WASP (compared with {\it Hubble}, {\it Kepler} etc.), it is therefore almost impossible to distinguish between a giant exoplanet and a small star using photometry alone; spectroscopic reconnaissance is required.

 It is typical for most exoplanet surveys to discard candidates that display RV amplitudes in excess of a few km s$^{-1}$,  and to consider those as false positives. However, this is not the case for the southern (${\rm dec}<+10^{\circ}$) WASP candidates, which are monitored spectroscopically by the Swiss Euler Telescope in La Silla, Chile, using the CORALIE spectrograph \citep{Queloz:2001ty,Wilson:2008lr,Triaud:2017fp}. Systems with semi-amplitudes less than 50 km s$^{-1}$ enter the EBLM  (Eclipsing Binary Low Mass) project. The project started in 2010 as an observational probe of eclipsing binaries with low-mass, M-dwarf secondary stars. The cut of 50 km s$^{-1}$ is designed to concentrate our observational efforts on fully convective secondaries ($< 0.35$ M$_\odot$), for which empirical mass and radius measurements are in short supply. In other words, the EBLM project is a survey of eclipsing SB1s.

An outline of the EBLM project, and some initial results were published in \citet{Triaud:2013lr} and \citet{Gomez-Maqueo-Chew:2014jk}. The spectroscopic orbits of an ensemble of 118 binaries appeared in \citet{Triaud:2017fu}, with this sample due to double in the coming year. %Some EBLM targets have also received follow-up transit observations, both photometric and spectroscopic (i.e. the Rossiter-McLaughlin effect), which will be published shortly. 
The most recent result of the survey is the binary EBLM J0555-57, whose secondary star comes close to the hydrogen-burning limit with a mass of $85\pm4$ M$_{\rm Jup}$ while having a radius of only $0.84^{+0.14}_{-0.04}$ R$_{\rm Jup}$, comparable to Saturn \citep{vonBoetticher:2017ji}.

%A primary goal of EBLM has been to better empirically measure the stellar mass-radius relation in the M-dwarf regime. This is important in both constraining stellar structure and evolution models (REF), but also assisting surveys for exoplanets around M-dwarfs, such as TRAPPIST, SPECULOOS, MEarth and TESS. The EBLM survey also provides observational insights into the distribution and physics of close binaries. In particular, measurements of the eccentricity, spin-orbit (mis)alignment and stellar rotation provide insights into tidal physics. Finally, the EBLM programme was constructed and followed similar to the WASP planet programme, the two may be compared to investigate the similarities and differences between closely-orbiting gas giants and closely-orbiting M-dwarfs.

\subsection{The birth of the BEBOP survey for circumbinary planets}\label{subsec:bebop_birth}

%In 2013, we conceived an additional scientific purpose for the EBLM targets: {\bf take the best targets and conduct more RV observations at a higher precision, in the hunt for circumbinary planets. Analysis of the {\it Kepler} discoveries had uncovered limitations and biases in the transit method, and hence spectroscopy was deemed complementary. 

 A spectroscopic exoplanet survey is complementary to the work already done using transits. This can be seen in the equation for the RV semi-amplitude, defined by

\begin{equation}
\label{eq:RV_K_planet}
K_{\rm c} =  \frac{\left(2\pi G\right)^{1/3}}{\sqrt{1-e_{\rm c}^2}}     \frac{m_{\rm c}\sin I_{\rm c}}{\left(m_{\rm A} + m_{\rm B} + m_{\rm c}\right)^{2/3}}\frac{1}{P_{\rm c}^{1/3}},
\end{equation}
where G is Newton's gravitational constant, $P$ denotes the orbital period, $e$ is the eccentricity, $I$ the inclination compared to the plane of the sky, and $m$ the mass. The subscripts stand for the primary (A), the secondary (B), and the planet (c)\footnote{The language throughout this paper typically refers to tertiary orbiting objects as circumbinary planets, as they are the main goal of the survey. However, we are also sensitive to more massive circumbinary objects such as circumbinary brown dwarfs and  tertiary stars. We therefore use a `c' subscript rather than a `p' subscript to refer to the outer orbit.}. Note that the mass of the secondary star appears in the above equation because the gravitational force of the planet perturbs the barycentre of the inner binary, rather than the primary star alone.  Compared to transit surveys, RVs are sensitive to a wider range of planetary orbits, with a shallower dependence on the planet's period and inclination. 

Furthermore, RVs are sensitive to mass rather than radius. The majority of the {\it Kepler} circumbinary planets do not have masses measured  from ETVs, and the faintness of most {\it Kepler} stars makes them unamenable to spectroscopic follow-up. There also remains some general tension in the community between masses derived photodynamically and spectroscopically \citep{Steffen:2016xc,Rajpaul:2017hj}.

Past radial-velocity surveys have not yielded any confirmed circumbinary planets.  The TATOOINE survey of non-eclipsing SB2s \citep{Konacki:2009lr,Konacki:2010re,Heminiak:2012fk} was the most expansive effort. One of its major successes was an improvement in the precision  of radial velocity measurements of double-lined binaries by at least an order of magnitude. Nevertheless, \citet{Konacki:2009lr} reveal a mean rms across their sample of nearly 20 m s$^{-1}$, which exceeds the formal uncertainties by a factor of a few, and hides most gas giants from identification. We suspect the excess noise originates from an imperfect radial-velocity extraction, as the procedure gets affected by the presence of two sets of lines. Similar effects are seen in ELODIE data \citep{Eggenberger:2004lr}.

 \citet{Konacki:2010re} write that maximal precision can be achieved by monitoring {\it ``single stars, or at best single-lined spectroscopic binaries where the influence of the secondary spectrum can be neglected''}. However, a suitable sample of bright, short-period SB1s was not available when TATOOINE was first constructed.

The BEBOP survey started when we realised that such a binary sample did now exist: the EBLM survey. By construction, the EBLM sample is solely composed of SB1s. Indeed, thanks to their eclipsing configuration we  calculate the true  (not minimum) mass of the secondary and its radius. Together we can robustly estimate the level of contamination produced by the secondary \citep{Triaud:2017fu}. Instead of attempting to build upon the pioneering work of TATOOINE to solve the double-line binary problem, we decided to circumvent it by focusing on single-line binaries. 

By avoiding the  contaminating effect of a secondary set of lines, the identification of a circumbinary planet becomes equivalent to identifying a multi-planet system whose innermost object happens to have a few 100 M$_{\rm Jup}$. We note that hot-Jupiters have dayside temperatures ranging from $\sim800-4600$ K \citep{Triaud:2015fk,Gaudi:2017cx}, and consequently M-dwarfs and hot-Jupiters are similarly located on colour-magnitude diagrams \citep{Triaud:2014kq,Triaud:2014ly}. Surveys for outer companions to  hot-Jupiters are common in the literature (e.g. \citealt{Knutson:2014fg,Bryan:2016qe,Neveu-VanMalle:2016xy}), and the BEBOP survey is conceptually similar.

 In addition to being SB1s, the EBLM targets also have the following beneficial attributes:

%In addition to be eclipsing single-line binaries, there are several additional advantages for choosing the EBLM binaries as the sole targets for the BEBOP programme:

\begin{itemize}
\item Consistency of the sample: the EBLM targets were all discovered and characterised using only WASP photometry and CORALIE spectroscopy, with a consistent set of procedures and sensitivities.
\item  Past EBLM RVs were available to be combined with new measurements taken for BEBOP, which roughly doubles our time baseline, and therefore  improves our sensitivity to long-period outer companions.
\item  Some EBLM targets already had identified stellar activity, and hence could be avoided.
%\item The sample consists of only single-lined spectroscopic binaries. This avoids the potentially difficult task of deconvolving two sets of spectral lines, as was attempted by \citet{konacki05,konacki09,konacki10} in the TATOOINE survey. In particular, since these are eclipsing binaries we know the size of the secondary star, and hence can estimate its magnitude, ensuring they are sufficiently faint to avoid contamination.
\item The radial velocity amplitude of the planet (Eq.~\ref{eq:RV_K_planet}) is a decreasing function of the sum of the primary and secondary masses. Having a low-mass secondary star is therefore beneficial.
\item All of our binaries eclipse, which biases the orbital orientation of any planets to maximise the RV signal.
\item  Another advantage of eclipsing binaries is a positive bias of the transit probability of any discovered circumbinary planet \citep{Borucki:1984fh,Schneider:1994lr,Martin:2014lr,Martin:2015rf,Li:2016ng,Martin:2017qf}. This bias is particularly strong for small mass ratio binaries, which must have inclinations very close to $90^{\circ}$ (further investigation in Martin, under review).
\item  The distribution of EBLM binary periods and mass ratios has significant overlap with the {\it Kepler} binaries known to host circumbinary planets (shown in Fig.~\ref{fig:binary_period_histogram}). 
\item The BEBOP binaries have an average Vmag $=11$, which is roughly 3.3 magnitudes brighter than the  {\it Kepler} circumbinary systems. 
\end{itemize}

% Indeed, }for the TATOOINE survey, \citet{Konacki:2009lr} reveal a mean rms across their sample of nearly 20 m s$^{-1}$, which exceeds the formal uncertainties by a factor of a few, and hides most gas giants from identification. We suspect the excess noise originates from an imperfect radial-velocity extraction, as the procedure gets affected by the presence of two sets of lines. Similar effects are seen in ELODIE data \citep{Eggenberger:2004lr}.

%  Binaries displaying the The best binaries, by merit of obtainable precision and a lack of stellar activity, would receive up to 20 additional observations in the hunt for circumbinary planets. This was the spawn of the BEBOP survey. Since we only see the light from the primary star in the EBLM binaries, this is the only star in which we may detect an RV wobble induced by a circumbinary planet. This wobble would have a semi-amplitude given by 

%In 2013, we conceived an additional scientific purpose for the EBLM targets. The best binaries, by merit of obtainable precision and a lack of stellar activity, would receive up to 20 additional observations in the hunt for circumbinary planets. This was the spawn of the BEBOP survey. Since we only see the light from the primary star in the EBLM binaries, this is the only star in which we may detect an RV wobble induced by a circumbinary planet. This wobble would have a semi-amplitude given by 

\subsection{Sample construction}\label{subsec:bebop_sample_construction}

The BEBOP binaries are selected from the EBLM sample, with the following protocol: 

\begin{itemize}
\item The BEBOP binaries comply with a difference of four visual magnitudes between the primary and secondary stars, such that we avoid secondary contamination of the primary's spectrum\footnote{There was one slight exception to this cut: EBLM J0425-46, for which the difference in visual magnitudes is only 3.85. However, even with the a slightly heightened threat of spectral contamination, the 30 CORALIE observations yielded an eccentric k1 fit, with a $\chi^2_{\rm red}=0.73$ statistic, indicating a perfect fit. Evidently there was not wide-spread spectral contamination in this target. This target is discussed further in Sect.~\ref{subsec:spectral_contamination}}. Almost all of the EBLM binaries naturally fulfill this criterion.

\item We only keep binaries on whose primary we reach a precision of 70 m s$^{-1}$ or better during a 30 minute observation, which is the typical exposure time used for the WASP planet survey \citep{Triaud:2011qy}. %This criterion was partially the product of available observing time; it led to a manageable sample size of almost 50 targets. This criterion also assured that around all of the chosen binaries we could probe circumbinary objects within the planetary domain. 
This is sufficient to reach the planetary domain. For instance, a  hypothetical $3M_{\rm Jup}$  planet at $P_{\rm c}=50$ days around a $m_{\rm A}+m_{\rm B}=1.2M_{\odot}$ binary produces a detectable radial velocity signal with semi-amplitude $K_{\rm c}=146$ m s$^{-1}$.

\item We exclude systems that display signs of stellar activity, as seen in an abnormal variation of the span of the bisector \citep{Queloz:2001lr,Figueira:2013lr}. While stellar activity does not prevent the large-amplitude binary orbit to be characterised, it becomes a hindrance for detecting small-amplitude planets, sometimes mimicking their signal. The identification of stellar activity in the EBLM binaries is outlined in \citet{Triaud:2017fu}.

\end{itemize}

Some of the EBLM binaries were already known to exist inside a triple star system. Outer stellar companions are thought to truncate and shorten the lifetime of the protoplanetary disc \citep{Kraus:2012ad,Daemgen:2013af,Cheetham:2015jr}, and generally be detrimental for the formation and survival of circumbinary planets \citep{Munoz:2015uq,Martin:2015iu,Hamers:2016er,Xu:2016qw}. However, such triple systems are kept in our sample for two reasons. First, the searches for circumbinary planets around the {\it Kepler} eclipsing binaries were done so without any a priori knowledge of a tertiary companion. Indeed, one example is known of Kepler-64 \citep{Schwamb:2013kx,Kostov:2013lr}\footnote{Also known as PH-1 \citep{Schwamb:2013kx}.} which has an outer stellar companion, which is itself a binary, albeit at a large separation of $\sim 1000$ AU. The second reason to keep triple star systems is to avoid introducing a confirmation bias into our survey.

The BEBOP binaries are typically longer period than the EBLM sample from which they were chosen. This was {\it not} chosen to match the trend seen in the {\it Kepler} results that the tightest binaries do not host planets (Fig.~\ref{fig:binary_period_histogram}), as this would also introduce a confirmation bias. Instead, this long-period selection is a function of the obtainable RV precision. Our binaries are all (or close to it) tidally synchronised (or pseudo-synchronised if eccentric), and hence the rotation period equals the orbital period. Consequently, the tightest binaries are also the fastest rotators, which have the worst RV precision due to broadened spectral lines. An example of this can be seen by comparing EBLM J1146-42 and EBLM J1525+03. The two targets have a similar visual magnitude ($Vmag=10.29$ and 10.74) and primary mass ($M_{\rm A}=1.35M_{\odot}$ and $1.23M_{\odot}$). However, the RV precision is significantly different ($\sigma_{1800s}=9$ m s$^{-1}$  and 48 m s$^{-1}$), which we attribute to different orbital periods ($P_{\rm bin}=10.5$ days and 3.82 days). In fact, this 3.82-day period for EBLM J1525+03 is the shortest in the sample, and also corresponds to one of our worst precisions.

The BEBOP sample tallies 47 binaries, which we present in this paper. In Table~\ref{tab:observables} we list some of the fundamental observational and physical parameters of these binaries. The calculation of the primary and secondary masses is discussed in Sect.~\ref{subsec:orbit_fitting}. The primary visual magnitudes are all taken for the NOMAD survey, except for EBLM J1934-42 which did not have available data. For this exceptional case the \citet{Baraffe:2015lr} models were used at an age of 1 Gyr. For the secondary visual magnitudes \citet{Baraffe:2015lr} models were used in all cases. The mid-times of primary and secondary eclipse ($T_{\rm pri}$ and $T_{\rm sec}$), are calculated based on the CORALIE spectroscopy alone and not the WASP photometry. The different $\sigma$ values are the observational precisions, and are discussed in Sect.~\ref{subsec:errorbars}.

\section{Observational strategy}\label{sec:observations}

All spectroscopic observations were taken at the Swiss Euler Telescope in La Silla, Chile, using the CORALIE spectrograph. CORALIE \citep{Queloz:2001ty,Marmier:2013lr} is a thermally stabilised, fibre-fed echelle spectrograph with a resolving power of $R=55,000$. 

The goal was to collect 20 observations of 30 minutes length on each binary. The flexible observing schedule of the Swiss Telescope allows for observations to be spread out over the year. This is important for probing long-period planets, like the ones we expect to find. The {\it Kepler}-discovered circumbinary planets have periods between 49.5 and 1107 days, with a median of 184 days. 

Observations were instructed to be separated by at least half the binary's orbital period. This means that  the 20 observations would span at least $10P_{\rm bin}$,  which would be long enough to cover at least one orbital revolution of a planet in eight of the nine {\it Kepler} circumbinary systems. Typically though, the observations were spaced over a longer timespan.

Other constraints on the observations were as follows:

\begin{itemize}
\item Separation between the target star and the Moon by at least $70^{\circ}$. This conservative criterion avoids contamination of the  spectrum by the gentle Sun's light reflected off the delicate Lunar surface.
\item Avoidance of primary eclipses of the binary. When the secondary M dwarf passes in front of the primary star the radial velocity signal is slightly distorted by the Rossiter-McLaughlin effect \citep{Holt:1893fk,Rossiter:1924il,McLaughlin:1924uq,Queloz:2000rt,Triaud:2013lr}. We did not instruct observers to avoid secondary eclipses of the binary, as the faintness of the secondary stars makes these phenomena negligible.
\item Generally good, clear observing conditions were required. This meant an airmass of the target better than 1.5  and a seeing better than 2.0 arcseconds.
\end{itemize}

\begin{figure*}
\begin{center}
\includegraphics[width=0.99\textwidth,trim={0 0 0 0},clip]{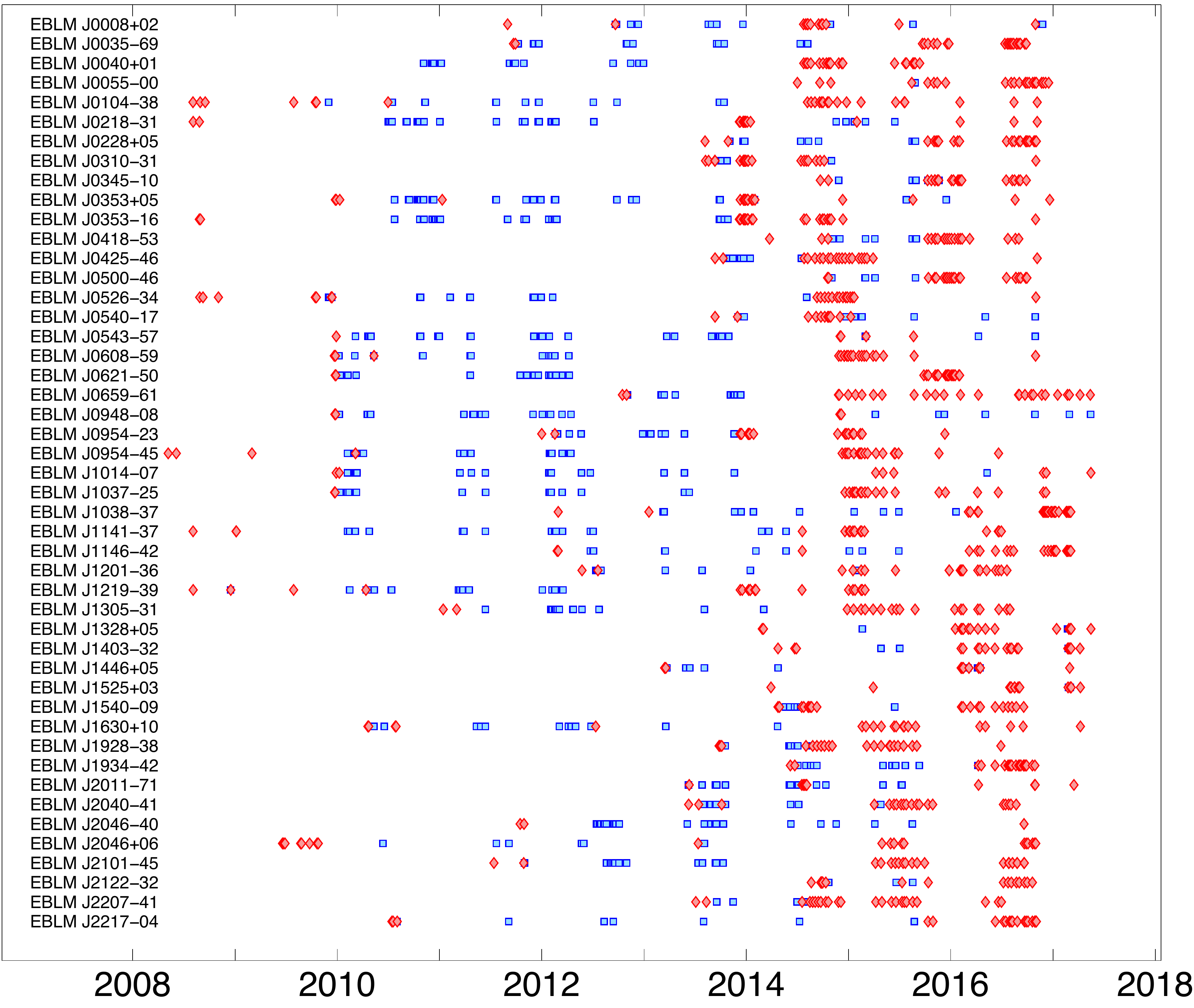}
\caption{Time-series of 1519 radial velocity observations of the 47 eclipsing binaries in the BEBOP programme. Red diamonds are observations for 1700 seconds or longer. Blue squares are for shorter observations. It is seen that all binaries typically receive two long observations initially before a series of short observations as part of the EBLM programme. A series of long observations typically commenced near the end of 2013 as part of the BEBOP survey.}
\label{fig:observation_calendar}
\end{center}
\end{figure*}

Since the BEBOP sample was constructed from the existing EBLM programme, we included all available radial velocity data except those likely affected by the Rossiter-McLaughlin effect. Measurements from the EBLM programme were also removed if deemed outliers, which is explained in Sect.~\ref{subsec:outliers}.

In Fig.~\ref{fig:observation_calendar} we show the calendar of observations on the 47 eclipsing binaries. The red diamonds correspond to long observations (1700+ seconds). These typically correspond to the BEBOP programme since late 2013, and a couple of initial observations dating back as far as 2008. The blue squares are for shorter observations, which in most cases were taken under the guise of the EBLM programme.  A small number of BEBOP targets did not receive a full quota of 20 long observations, owing to limitations in available observing time, but most exceeded this.

\section{Radial velocity data treatment}\label{sec:data_treatment}

The radial velocity data was treated in the same way as in \citet{Triaud:2017fu}. We therefore only provide a summary of the methods used here, and refer the reader to that paper for a more thorough discussion.

\subsection{Reduction of spectroscopic data}\label{subsec:reduction}

The CORALIE Data Reduction Software (DRS) is similar to that used with the HARPS, HARPS-North and SOPHIE instruments. A cross correlation function (CCF) is created between the observed spectrum and a numerical mask \citep{Baranne:1996qa}.  The CCF is a weighted average spectral line, which contains characteristics of individual absorption lines such as their width and asymmetries, but with a heightened signal to noise. The CCF is binned in 0.5 km s$^{-1}$ increments, owing to the $R=55,000$ resolving power of the spectrograph. Two different spectral type masks were used: G2 and K5. These were chosen based on the spectral type of the primary star, which in our sample ranges between K2 and F0. \citet{Dumusque:2012lr} demonstrated that the closeness of the spectral type mask largely affects only the absolute radial velocity and not the radial velocity variations. Only having two spectral type masks therefore does not hinder our analysis.

The CCF of each measurement was compared with a Thorium-Argon spectrum, which was used as a wavelength-calibration reference \citep{Lovis:2007ec}. This accounted for variations in the instrument which would otherwise impose a drift on the measured radial velocity of the star. The main source of variation was the pressure of the spectrograph, which followed the ambient atmospheric pressure because CORALIE, unlike HARPS and HARPS-North, is not pressurised. In 2014 a Fabry-P\'erot unit was added to provide even more precise calibrations.

\subsection{Outlier removal}\label{subsec:outliers}

For each CCF, which is approximately Gaussian, we measure the span of the bisector slope.  The bisector is calculated by tracing vertically the midpoint of the CCF at each value of flux intensity. The span of the bisector slope is the difference between the bisector at the top and bottom of the CCF \citep{Queloz:2001lr}. The bisector therefore reflects any asymmetries in the absorption lines. We remove any observations with bisector positions  more than three interquartile ranges below the first quartile or above the third quartile. Such outliers may be from the wrong star being observed accidentally, or an abnormally low signal to noise observation. A visual inspection was also done of the data to outliers in the  CCF's Full Width Half Maximum (FWHM). 

\subsection{Calculation of radial velocity uncertainties}\label{subsec:errorbars}

In Table~\ref{tab:observables} three different $\sigma$ radial velocity uncertainties are listed. The first value, $\sigma_{\rm 1800s}$, is mean the photon noise uncertainty for all observations of 1800 seconds, which was the typical observation length during the BEBOP programme. The value $\sigma_{\rm median}$ is the median photon noise precision for all observations, i.e. both the 1800 second observations taken for BEBOP and earlier, shorter observations taken in the EBLM programme. 

After removing observations with significant outlier bisector values there may still remain some variation in the bisector. We consider such asymmetries to be a source of error, which is shown in Table~\ref{tab:observables} as $\sigma_{\rm add}$. This value is calculated by

\begin{equation}
\sigma_{\rm add} = \sqrt{\frac{\delta_{\rm bis}^2}{4} - \left<\sigma_{\gamma}^2 \right>},
\end{equation}
where $\delta_{\rm bis}$ is the root mean square of the variation of the bisector measurements around their mean and $\left<\sigma_{\gamma}\right>$ is the mean photon noise error. If $\left<\sigma_{\gamma}\right> > \delta_{\rm bis}^2/4$ then we take $\sigma_{\rm add}=0$, which was the case for 30 of the 47 binaries. Otherwise, we add $\sigma_{\rm add}$ in quadrature to the radial velocity measurements.

Finally, the CORALIE spectrograph was historically stable to a precision of 6 m s$^{-1}$ \citep{Marmier:2014rz}. A recent change of the optical fibres from circular to hexagonal improved the stability to 3 m s$^{-1}$ \citep{Triaud:2017fp}. To each data point we add 6 m s$^{-1}$ of Gaussian noise in quadrature. Choosing 6 m s$^{-1}$ and not 3 m s$^{-1}$ was considered conservative, and also reasonable since most of the observations occurred before the fibre upgrade.

\subsection{Orbit fitting}\label{subsec:orbit_fitting}

To fit orbits to the spectroscopic data we use the {\sc Yorbit} genetic algorithm, which has been developed over the years at the University of Geneva and implemented in numerous radial velocities using CORALIE and HARPS (e.g. \citealt{Bonfils:2011lr,Mayor:2011fj,Marmier:2013lr}). Only static Keplerian orbits are fitted, i.e. orbital variations induced by gravitational interactions between orbits are ignored. This is a reasonable assumption except for very tight triple star systems, and in Sect.~\ref{subsec:nbody} we briefly discuss one such example. More details on {\sc Yorbit} may be found in \citet{Bouchy:2016fu}.

When {\sc Yorbit} is run to search for a single Keplerian orbit it will inevitably first fit that of the inner binary, as its signal is orders of magnitude higher than any potential circumbinary orbit. This binary orbit is characterised by six parameters: the period, $P$,  semi-amplitude, $K$, eccentricity, $e$, time of periapsis passage, $T_0$, mass function $f(m)$ and the argument of periapsis, $\omega$. Error bars are calculated for each of these parameters by running 5000 Monte Carlo simulations.

\subsection{Model selection}\label{subsec:model_selection}

For each target we fitted five different types of model to the spectroscopic data. These are listed below, along with the number of parameters shown in parenthesis. 

\begin{enumerate}
\item k1: a single Keplerian (6)
\item k1d1: a single Keplerian plus a linear drift (7)
\item k1d2: a single Keplerian plus a quadratic drift (8)
\item k1d3: a single Keplerian plus a cubic drift (9)
\item k2: a pair of Keplerians (12)
\end{enumerate}

 Models more complex than a single Keplerian are likely indicative of a tertiary companion. This tertiary companion will have its own Keplerian orbit, but if the observational timespan only covers a small fraction of this outer period then the orbit will be sufficiently modeled by a drift. A drift could alternatively be explained by an instrumental variation, but for CORALIE the temperature stabilization and nightly pressure calibrations have historically avoided this. A third explanation would be long-term stellar activity, although the binaries were all vetted for heightened activity, as described in \citet{Triaud:2017fu} based on the proceedures of \citet{Queloz:2001lr,Figueira:2013lr}.

When attempting to fit a two-Keplerian model the genetic algorithm was restricted to searching for periods greater than  four times the inner binary period.  Numerous stability studies (e.g. \citealt{Dvorak:1986fk,Holman:1999lr,Chavez:2015uq,Quarles:2018ub}) show that circumbinary planets would be unstable with shorter orbits. Aside from this minimum period, no further restrictions are applied to the fitting. In particular, we search for and are sensitive to binaries and planets of all eccentricities. Upon the discovery of any candidate triple systems the orbital stability is then tested more carefully.

%This minimised the return of unphysical solutions. 

All targets in the BEBOP programme have been observed at least 16 times, with a median count of 32, which is more than there are free parameters in any of the models.

{\sc Yorbit} will always retain small eccentricities like most fitting procedures \citep{Lucy:1971uq}. Therefore, for each of the above types of model we also test a fit where  eccentricity is forced to zero. This allows us to test if the eccentricity fitted by {\sc Yorbit} is significant. We are therefore left with a total of ten tested models, where the number of parameters for the forced circular model is always two less than the corresponding eccentric model, as both $e$ and $\omega$ are removed. Throughout this paper we use ``(circ)'' and ``(ecc)'' to distinguish between forced circular and freely eccentric models.  Note that when testing the k2 (circ) model only the binary eccentricity is forced to zero, not that of the planet.

For all ten models {\sc Yorbit} outputs a  $\chi^2$ statistic, which is a weighted sum of the square of the residuals. The reduced $\chi^2$ statistic is calculated by normalising over the number of free parameters:

\begin{equation}
\label{eq:red_chi2}
\chi^2_{\rm red}=\frac{\chi^2}{n_{\rm obs}-k},
\end{equation}
where $n_{\rm obs}$ is the number of spectroscopic observations and $k$ is the number of model parameters. A value of $\chi^2_{\rm red}=1$ is indicative of an optimal fit.

To choose the most appropriate model between the ten possibilities we follow the same procedure as in \citet{Triaud:2017fu}. For this we calculate the Bayesian Information Criterion \citep{Schwarz:1978zz,Kass:1995rt}, henceforth referred to as the BIC, according to

\begin{equation}
{\rm BIC} = \chi^2 + k\ln \left(n_{\rm obs}\right).
\end{equation}
The BIC is defined such that it naturally punishes complex models (large $k$), and hence has an inbuilt Ockham's razor, in selecting the most parsimonious explanation possible. 

In our process of model selection we start with the simplest model, which is k1 (circ) with only four parameters, and calculate the BIC. We then calculate the BIC for other models in order of complexity. To choose the next  most complex model we demand that the BIC improves (decreases) by at least six. This is believed to be strong evidence for the more complex model \citep{Kass:1995rt}.

We allow the model selection to  ``jump'' levels of complexity if the BIC improves by $n\times6$, where $n$ is the number of ranks of complexity moved through. For example, if BIC = 40 for k1 (circ) with four parameters, BIC = 38 for k1d1 (circ) with five parameters and BIC = 25 for k1d2 (circ) with six parameters, then the chosen model would be k1d2 (circ), even though there was only a marginal BIC improvement between k1 (circ) and k1d1 (circ). 

This process is done for k1, k1d1 and k1d2 models, in both circular and eccentric flavours. These were deemed the ``base'' models. The k1d3 and k2 models were deemed ``complex'' models. Complex models are only tested if $\chi^2_{\rm red}>2$ for the best base model. This criterion was an additional means of penalising overly complex models. Since the aim of the survey is to find circumbinary planets, i.e. k2 models, this cautious approach minimises false discoveries. Note that sometimes complex models were tested but ultimately a base model was chosen.

\section{Calculating physical parameters}\label{sec:physical_parameters}

\subsection{Primary and secondary masses}\label{subsec:masses}

Because the BEBOP sample only contains single-lined binaries the primary and secondary masses are not directly measured, but rather only the mass function is directly measured. Primary masses are calculated the same way as in \citet{Triaud:2017fu}, based on photometric colour fitting methodology outlined in \citet{Maxted:2014qw}.

Knowing both the primary mass and the mass function, the secondary mass is calculated by solving numerically

\begin{equation}
\label{eq:mB}
f(m) = \frac{\left(m_{\rm B}\sin I_{\rm bin}\right)^3}{\left(m_{\rm A} + m_{\rm B}\right)^2} = \frac{P_{\rm bin}K^3_{\rm pri}}{2\pi G}.
\end{equation}

The error in $m_{\rm B}$ is calculated as

\begin{equation}
\label{eq:mB_error}
\frac{\delta m_{\rm B}}{m_{\rm B}} = \frac{1}{3}\left(\frac{\delta f(m)}{f(m)} + 2\frac{\delta m_{\rm A}}{m_{\rm A}} + 3\frac{\delta \sin I_{\rm bin}}{\sin I_{\rm bin}} \right),
\end{equation}
where $\delta$ indicates a $1\sigma$ uncertainty and $\delta f(m)$ is a direct output from {\sc Yorbit} and $\delta m_{\rm A}$ is calculated based on \citet{Torres:2010uq,Maxted:2014qw}. The uncertainty in the binary inclination is calculated as $\delta \sin I_{\rm bin} = R_{\rm A}/a_{\rm pri}$, where $a_{\rm pri}$ is the semi-major axis of the binary and $R_{\rm A}$ is the radius of the primary star, as calculated based on \citet{Gray:2008fj}. When $m_{\rm B}$ is calculated in Eq.~\ref{eq:mB} we take $I_{\rm bin}=90^{\circ}$ because  of the existence of the binary eclipses. By adding this small inclination uncertainty we reflect our ignorance of the impact parameter of the eclipse. This typically adds $20\%$ or less to the error in $m_{\rm B}$. 
Based on the primary and secondary masses the semi-major axis is calculated using Kepler's third law.

\subsection{Calculating upper limits on undetected orbital parameters}\label{subsec:upper_limits}

We follow the same methodology as for the EBLM survey \citep{Triaud:2017fu} to constrain upper limits on orbital parameters which we do not have the precision to directly measure. For all binaries where a forced circular solution was chosen by the BIC model selection we constrain the eccentricity to within 0 and an upper limit. This upper limit is calculated by adding the fitted eccentricity for that model to the $1\sigma$ uncertainty on that eccentricity. The same procedure is done for drifts in the radial velocity, e.g. if k1 (ecc) was the chosen model then the upper limit on the coefficient of linear drift was taken as the fitted value in k1d1 (ecc) plus its $1\sigma$ uncertainty.

\section{Spectroscopic results}\label{sec:results}

\subsection{Chosen models}\label{subsec:chosen_models}

Table~\ref{tab:BIC} shows all of the information pertaining to the model selection. The BIC for the selected model is highlighted in bold font.
In Table~\ref{tab:model_count} we count how many binaries were fitted by each of the ten possible models. The same table appears in \citet{Triaud:2017fu} (table 2 in that paper) for the EBLM survey. The most noticeable difference is that here we report 6 binaries fitted with k1 (circ) and 25 with k1 (ecc). Contrastingly, for the EBLM paper 58 binaries were fitted with k1 (circ) and 39 with k1 (ecc). The heightened percentage of binaries fitted with k1 (ecc) in the BEBOP survey reflects the additional long-exposure radial velocity measurements, which heighten our sensitivity to eccentricities as small as 0.001.

\begin{table}
\caption{Number of binaries fitted with each model}
\label{tab:model_count}
\centering % centering table
\tiny
\begin{tabular}{|cc|}
\hline
\rowcolor{gray!50}
name & count \\
\hline\hline %inserting double-line
\multicolumn{2}{|c|}{Base models} \\
\hline % inserts single-line
k1 (circ) & 6 \\
k1 (ecc) & 25 \\
k1d1 (circ) & 0 \\
k1d1 (ecc) & 4 \\
k1d2 (circ) & 2 \\
k1d2 (ecc) & 2 \\
\hline % inserts single-line
\multicolumn{2}{|c|}{Complex models} \\
\hline % inserts single-line
k1d3 (circ) & 1 \\
k1d3 (ecc) & 2 \\
k2 (circ) & 2 \\
k2 (ecc) & 3\\
\hline % inserts single-line
\end{tabular}
\end{table}

\subsection{Inner binary parameters}\label{subsec:inner_binary_parameters}

The orbital parameters and masses for the inner binary are all listed in Table~\ref{tab:params}, all taken from the chosen model indicated in Table~\ref{tab:BIC}. The $1\sigma$ uncertainty for each parameter is given in parenthesis, corresponding to the last two digits of the measured value. For example, for EBLM J0008+02 the period is $P_{\rm bin}=4.7222824(48)$ days, which can be otherwise written as $P_{\rm bin}=4.7222824\pm0.0000048$ days. For some systems upper limits are provided for the eccentricity and coefficients of drift according to Sect.~\ref{subsec:upper_limits}.  The quantity $\omega_{\rm bin}$ is undefined when a forced circular model is chosen.

\subsection{Discovered or potential tertiary bodies}\label{subsec:tertiaries}

For five of our binaries the selected model is a pair of Keplerian orbits. Unfortunately from a planetary perspective, all of the characterised tertiary orbits are within the stellar regime. The smallest characterised tertiary mass is $m_{\rm c}\sin I_{\rm c}=0.1207M_{\odot}$ for EBLM J2011-71.

In Table~\ref{tab:triple} we provide parameters for all five characterised triple star systems. Compared with the EBLM release in \citet{Triaud:2017fu}, there is an additional system: EBLM J1038-37. This system was included in \citet{Triaud:2017fu} but the BIC selection criteria characterised it as a single, eccentric binary plus a cubic drift. That prior characterisation was based on 13 observations taken over 3.89 years, with a median precision of 131 m s$^{-1}$. The double Keplerian characterisation presented in this paper is based on 33 observations taken over 5.01 years, with a median precision of 74 m s$^{-1}$. This is an example of the improved orbital fits provided by the BEBOP survey in comparison with the original EBLM survey.

It is interesting to note that all of the minimum masses of the tertiary stars are all significantly smaller than the primary masses. However, we caution drawing too much from this result as a more massive tertiary would have diluted the already shallow WASP eclipse depth of the secondary star, and hence such systems may not have been detected in the first place. None of the tertiary stars are bright enough for us to directly observe their flux.

\section{Analysis}\label{sec:analysis}

\subsection{Residuals as a function of orbital phase -- a test for spectral contamination}\label{subsec:spectral_contamination}

\begin{figure}
\begin{center}
\includegraphics[width=0.49\textwidth,trim={0 0 0 0},clip]{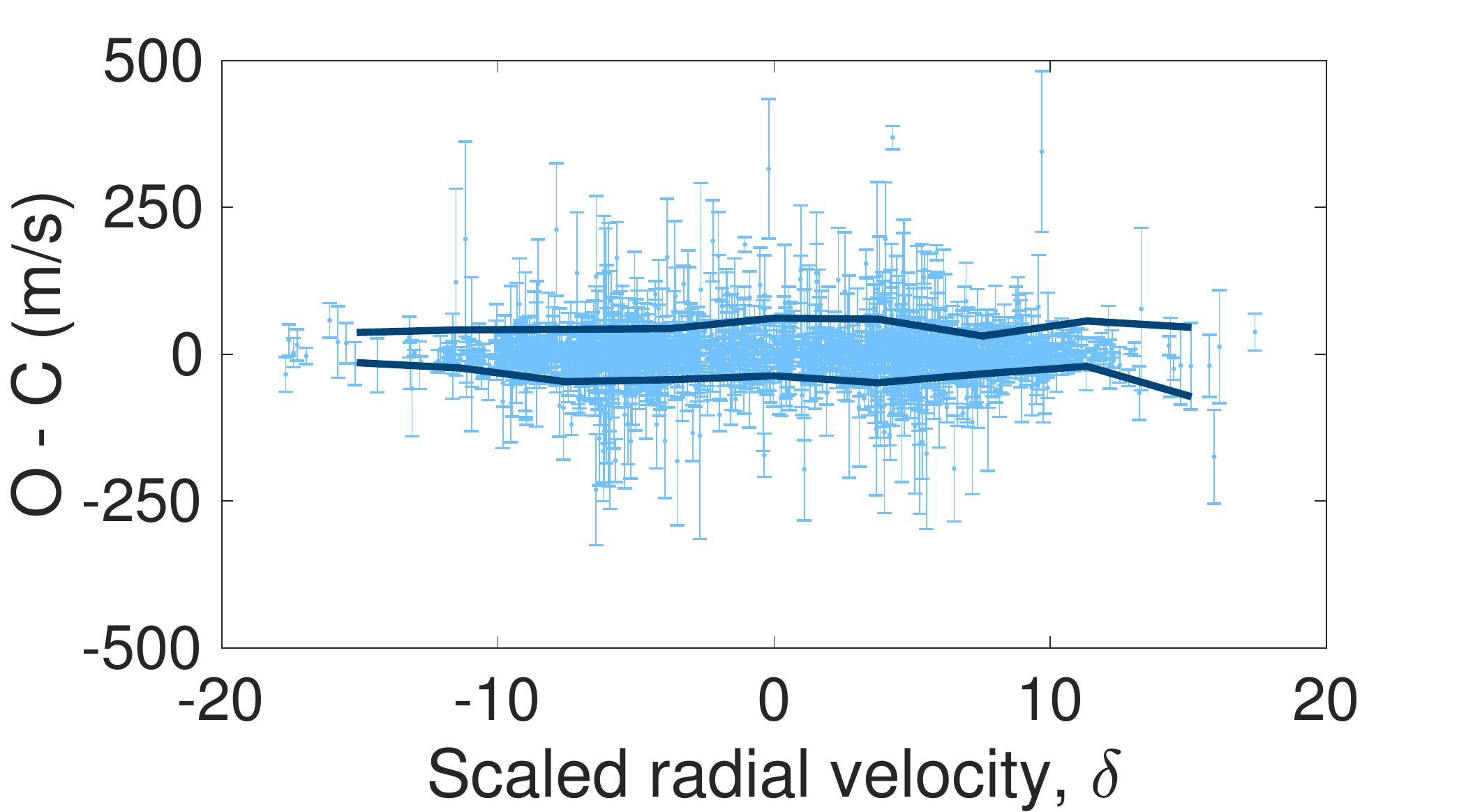}
\caption{Radial velocity residuals (``Observed minus Calculated'' or O - C for short) of the best-fitting model to all BEBOP binaries for all 1519 observations, stacked on top of each other as a function of the scaled radial velocity value $\delta$, defined according to Eq.~\ref{eq:delta}. The $\delta$ value says how far the observed radial velocity observation was from the systemic velocity, scaled by the average FWHM for that target. The two roughly horizontal dark blue lines are the root mean squared (RMS) values of the residuals, split into positive and negative values. If the RMS were maximised near $\delta=0$ then this would be indicative of wide-spread spectral contamination from the secondary star, but this is evidently not the case.}
\label{fig:contamination_test}
\end{center}
\end{figure}

The BEBOP binary sample was constructed to avoid cross contamination between two sets of stellar spectral lines by only choosing binaries with faint secondary stars. However, if there were spectral contamination then it would only be expected to affect the observations at certain binary orbital phases. The vulnerable orbital phases correspond to the primary star's radial velocity equalling the system's systemic velocity, as it will be also equal to the secondary star's radial velocity (which we do not directly measure). At this point the primary and secondary spectral lines overlap, and hence the chance of contamination is maximised. It was said in Sect.~\ref{sec:observations} that observations were taken at orbital phases that avoid the primary eclipse, such that we do not observe a Rossiter McLaughlin effect which would skew the radial velocities. This fortuitously helps us avoid spectral contamination, as the eclipse corresponds to an overlapping of the primary and secondary radial velocity signals.

The width of the spectral lines, quantified by the full width half maximum (FWHM), demarcates the range of radial velocity values that may be contaminated. A larger FWHM means that spectral contamination may occur at a greater difference in the primary and secondary radial velocities.

To test if there is wide-spread spectral contamination in our sample we analyse the residuals of the best-fitting model to each binary as a function of a scaled radial velocity value, for all 47 binaries. This scaled radial velocity value, which we denote by $\delta$, is calculated by

\begin{equation}
\label{eq:delta}
\delta = \frac{RV-RV_0}{FWHM}\frac{m_{\rm A}}{m_{\rm B}},
\end{equation}
where $RV$ is an individual radial velocity measurement on the primary star and $RV_0$ is the systemic radial velocity mid-point for the system. The mass ratio factor $m_{\rm A}/m_{\rm B}$ converts the radial velocities from the measured values on the primary star to the larger but not directly measured values on the secondary star. The $\delta$ value denotes how far the potentially contaminant secondary star radial velocity signal is away from $RV_0$.

The results are plotted in Fig.~\ref{fig:contamination_test}. In blue we  plot the residuals (O - C) of all 1519 data points for the 47 BEBOP targets on top of each other as a function of $\delta$. The two  dark blue lines are the root mean squared (RMS) values at ten different values of $\delta$, calculated separately for positive and negative values of O - C. If our sample were affected by stellar contamination then there would be a significant increase in the residuals near $\delta=0$, and hence the red RMS curves would deviate significantly from zero. This is not the case.

\begin{figure}
\begin{center}
\includegraphics[width=0.49\textwidth,trim={0 0 0 0},clip]{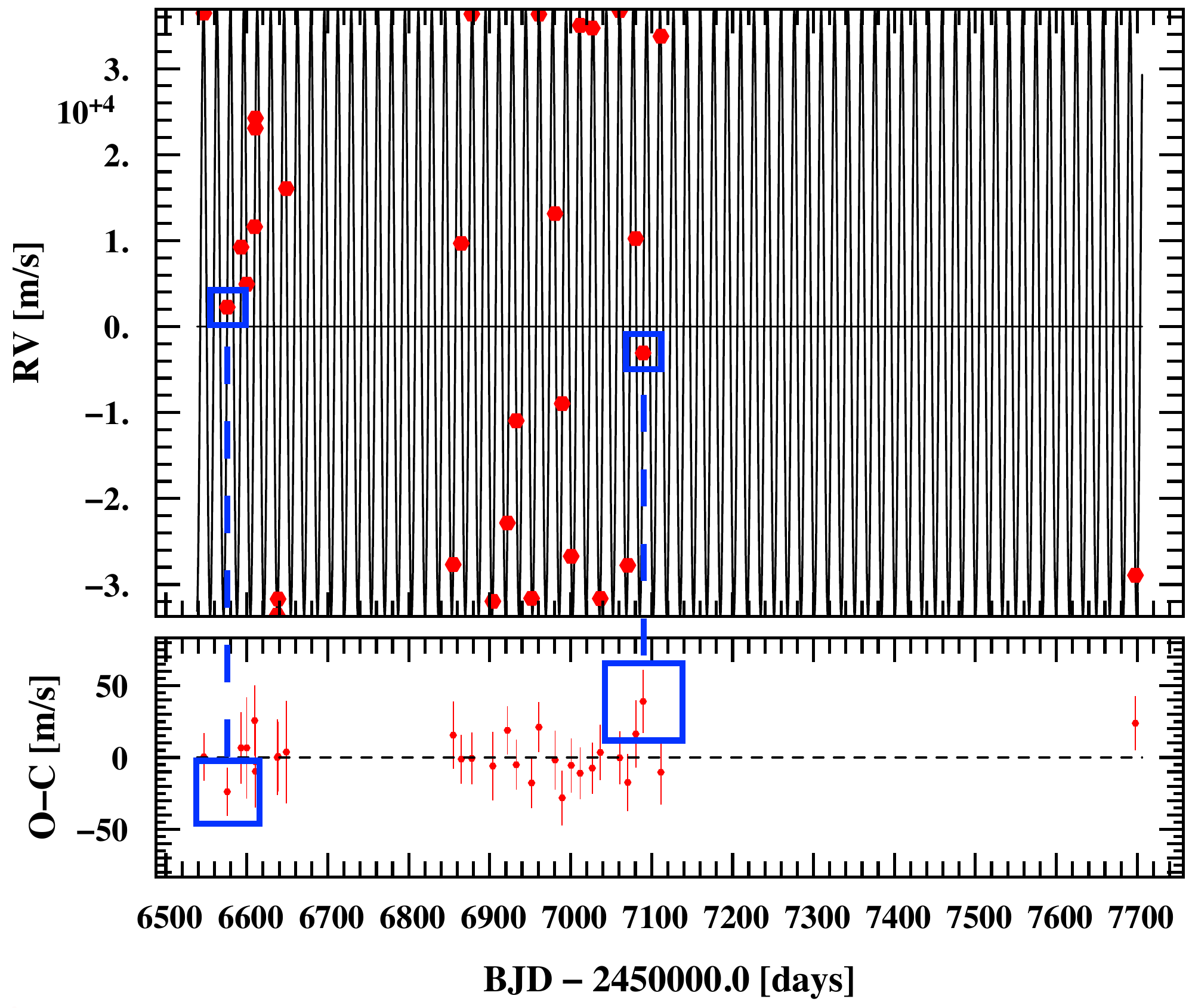}
\caption{Top: radial velocities of the target EBLM J0425-46 over 3.15 years and the selected eccentric single Keplerian model. Error bars are typically 10-20 m s$^{-1}$ and too small to see at this scale. Bottom: residuals to the fitted model, with $1\sigma$ error bars. Blue boxes highlight two marginal outliers, with a dashed line connecting the radial velocity measurement, which is near 0 m s$^{-1}$, and the residual. Both outliers are less than $2\sigma$ from the model.}
\label{fig:0425_residuals}
\end{center}
\end{figure}

It was noted in Sect.~\ref{subsec:bebop_birth} in the construction of the BEBOP sample that one target - EBLM J0425-46 - has a visual magnitude difference of 3.85 between the primary and secondary stars, which is slightly smaller than the threshold of 4. Its mass ratio is the highest of the sample: $m_{\rm B}/m_{\rm A}=0.527$. This may put it at risk of spectral contamination from the relatively ``bright'' secondary star. 

In Fig.~\ref{fig:0425_residuals} we plot the 30 radial velocity measurements over the and their residuals to the fitted eccentric k1 model. This is the model that was chosen as the most appropriate by the BIC selection method, with $\chi^2_{\rm red}=0.73$. We highlight with blue boxes in this plot two very marginal outliers of the fit, showing that they correspond to the two radial velocity measurements closest to the systemic velocity (0 m s$^{-1}$ on this plot). Having outliers here is consistent with spectral contamination. However, each value is less than two standard deviations away from the fit, and hence not statistically signifiant, and the fit is overall very good to the data ($\chi^2_{\rm red}<1$).

%In Sect.~\ref{sec:future} we discuss future plans for expanding this approach of precision radial velocities on single lined spectroscopic binaries.

\subsection{Calculating detection limits}\label{subsec:detection_limits}

\begin{figure}
\begin{center}
\includegraphics[width=0.49\textwidth,trim={0 0 0 0},clip]{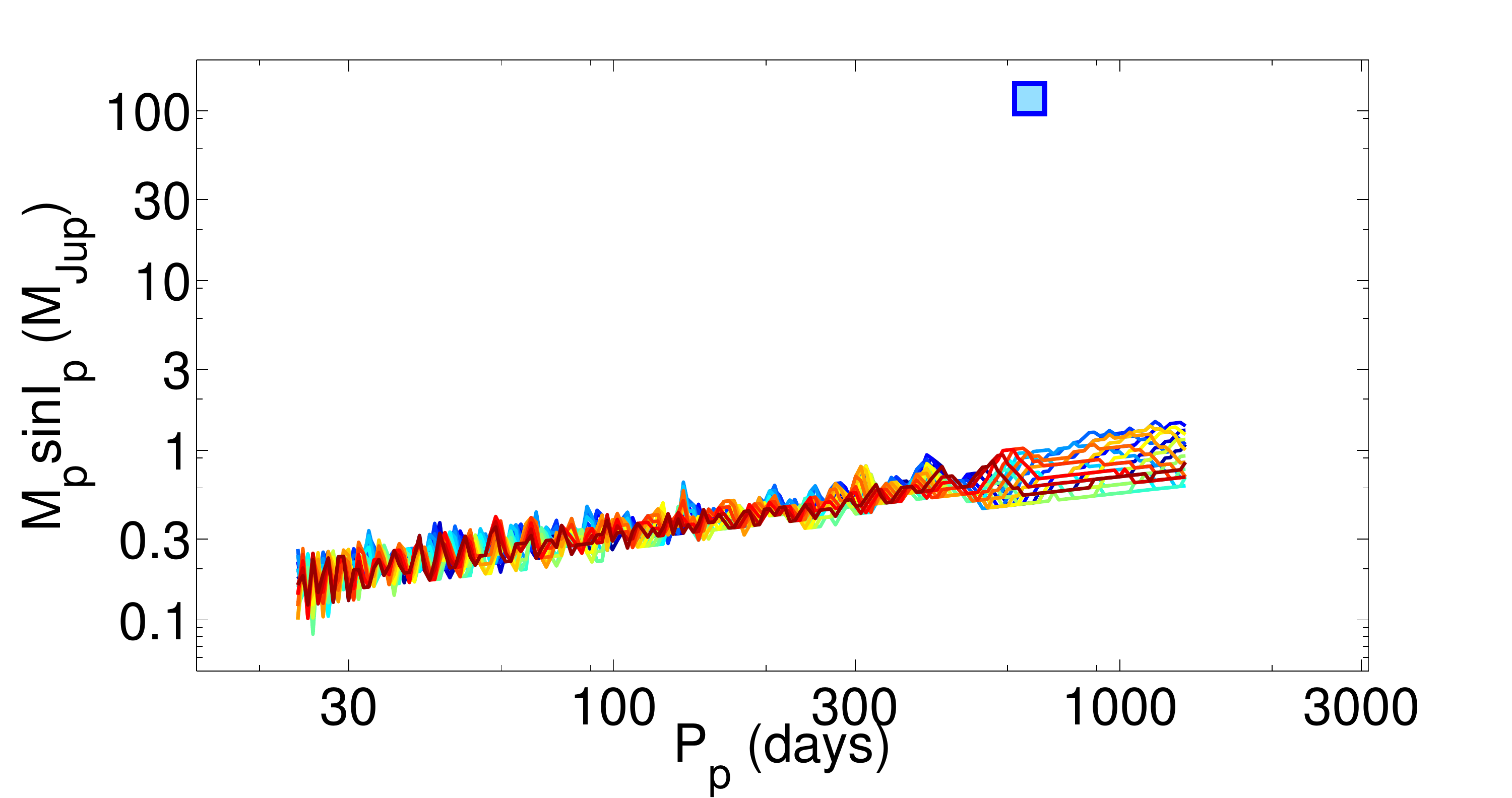}
\caption{Example of the detection limits calculated for EBLM J2011-71, which is one of the BEBOP targets with the highest precision. Each line represents the smallest detectable minimum circumbinary planet mass ($m_{\rm c} \sin I_{\rm c}$) as a function of the planet's period ($P_{\rm c}$) for a different orbital phase of the planet. There are 20 orbital phases tested, all equally spaced. The blue square near the top of the plot is the detected circumbinary object in the EBLM J2011-71, which is in fact not a circumbinary planet but rather a low-mass tertiary star.}
\label{fig:detection_limit_example}
\end{center}
\end{figure}

\begin{figure*}  
\begin{center}  
	\begin{subfigure}[b]{0.49\textwidth}
		\includegraphics[width=\textwidth]{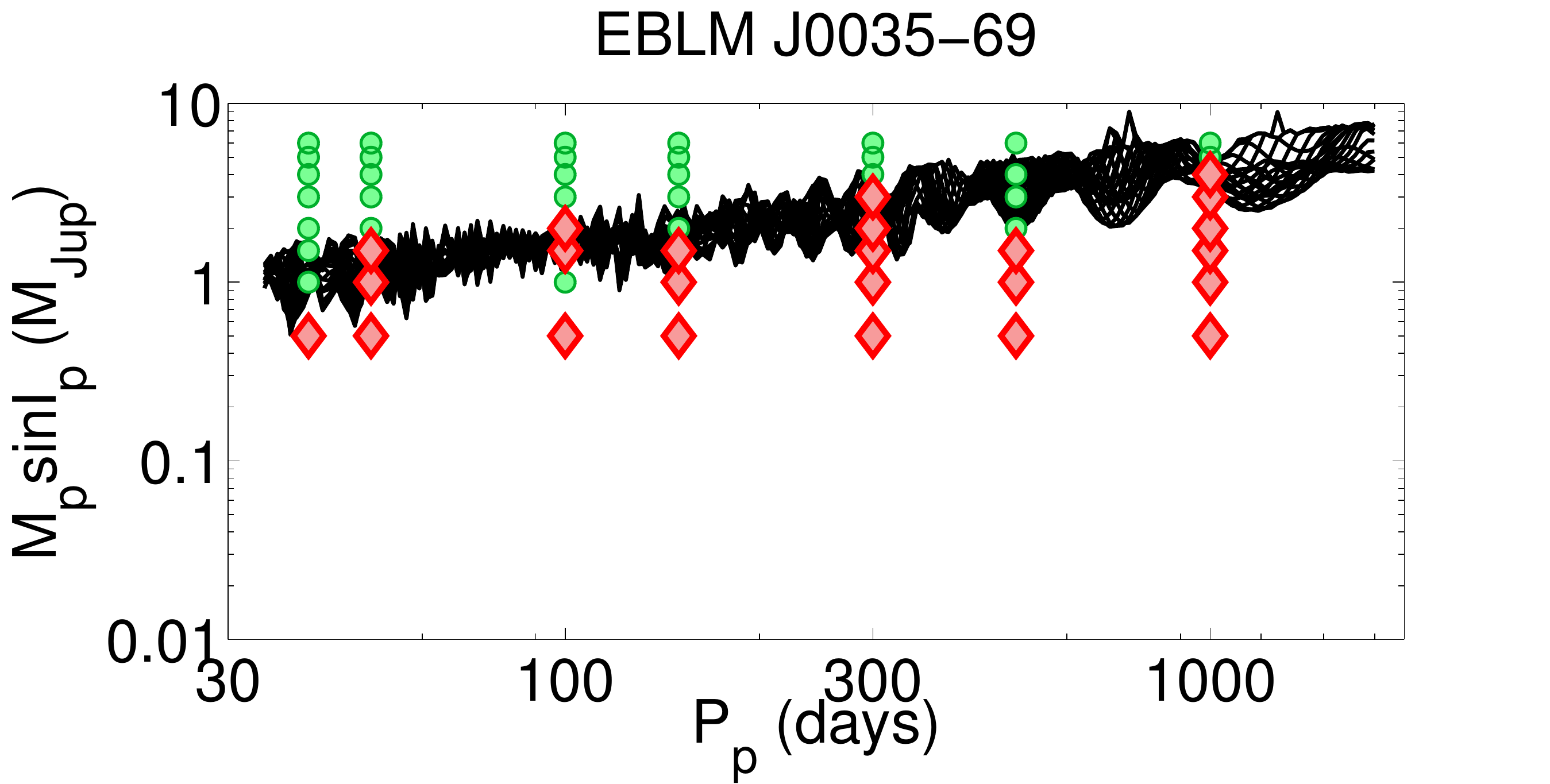}  
	\end{subfigure}
	\begin{subfigure}[b]{0.49\textwidth}
		\includegraphics[width=\textwidth]{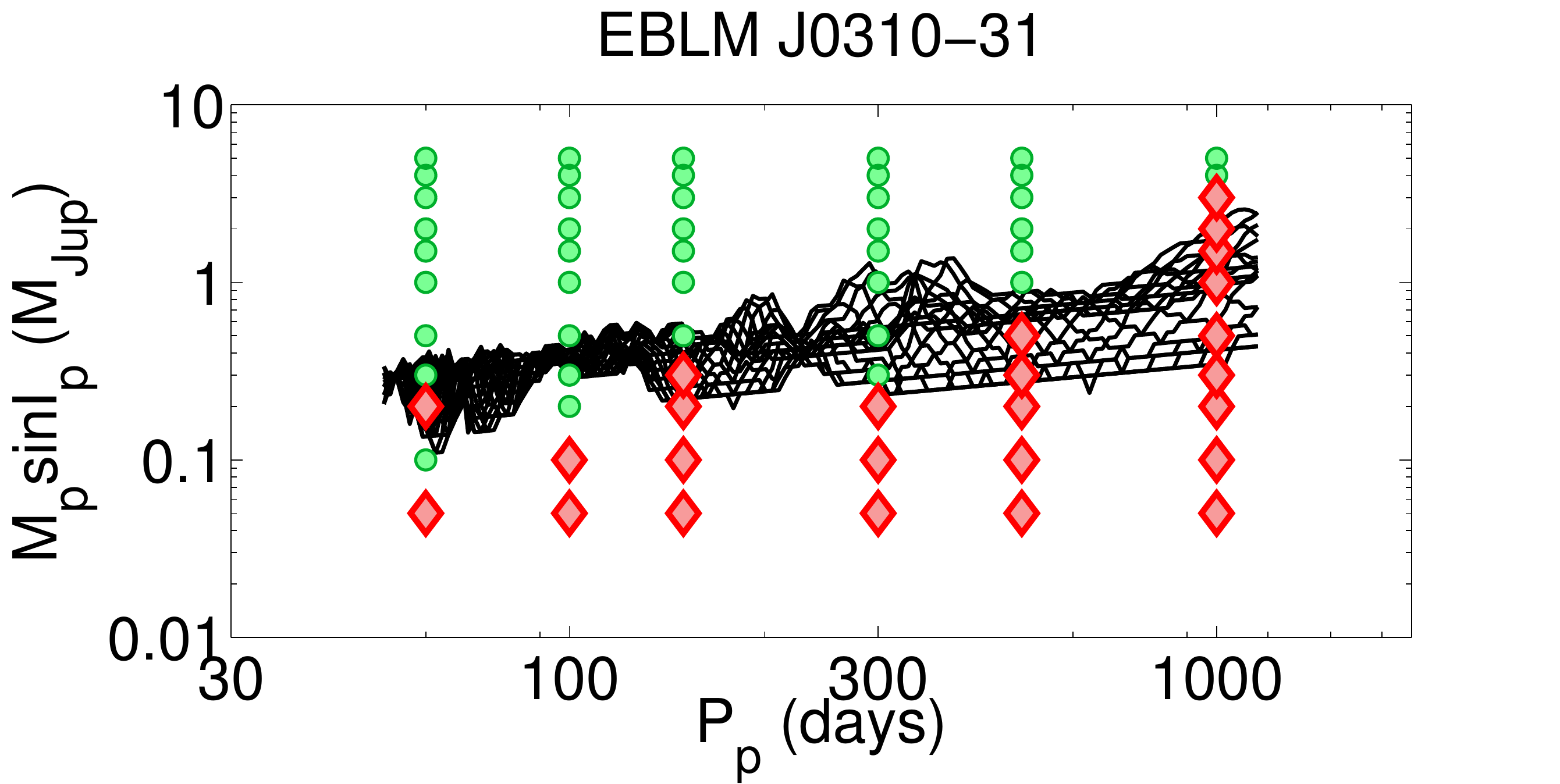}  
	\end{subfigure}
	\caption{Detection limit curves in black for two targets in the BEBOP programme: J0035-69 (left) and J0310-31 (right). Each curve represents the smallest $m_{\rm c}\sin I_{\rm c}$ for a putative circumbinary planet at a given period, according to the periodogram measure of detectability described in Sect.~\ref{subsec:detection_limits}. For each target there are 20 black curves, one for each of the tested orbital phases of the injected planet. At seven discrete periods a series of tests were run to recover n-body simulated circumbinary planets with the {\sc Yorbit} genetic algorithm. At these periods a green circle indicates a successful recovery, whereas a red diamond indicates a failure.}
\label{fig:nbody_test}  
\end{center}  
\end{figure*} 

%Studies such as \citet{Mayor:2011fj} were done using single stars, so there was no a priori signal. That paper nevertheless undertook a cleaning of each star by systematically fitting and removing sinusoids until the GLS periodogram was devoid of peaks above a 1\% FAP. Such periodicities could be produced by stellar activity, the observing cadence or, occasionally, unidentified planets.
%Since BEBOP is a survey around binary stars, there is guaranteed to be a radial velocity signal from the inner binary 

There are two main factors that determine the detectability of a putative planet: its minimum mass, $m_{\rm c}\sin I_{\rm c}$, and its period, $P_{\rm c}$. Not only are these the main contributors to the amplitude of the radial velocity signal, but the observational timespan needs to cover a significant portion of the planetary period. The eccentricity of a planet also increases $K_{\rm c}$, but \citet{Endl:2002az} demonstrate that it has a minimum effect on detectability for $e_{\rm c}\lesssim 0.5$.  Therefore, whilst our search for circumbinary planets is sensitive to any eccentricity, when quantifying the detectability we only consider circular planetary orbits.

To calculate detection limits for each binary we introduce and attempt to retrieve artificial Doppler signals. We follow a similar procedure to that described in \citet{Konacki:2009lr}, which is based on methods that are regularly employed to calculate the occurrence rates of planets by long-term Doppler surveys on single stars (e.g. \citealt{Cumming:2008ys,Mayor:2011fj,Bonfils:2013fk}).

We first start by defining  what  makes a hypothetical planet detectable. For this, we use the Generalised Lomb Scargle (GLS) periodogram, which identifies periodic signals of varying strengths within data. We define a putative planet as ``detectable" when a GLS periodogram displays a signal with a strength rising above a False Alarm Probability (FAP) of 1\%.

Importantly, the injection of the synthetic Keplerian sinusoid must be done to data that has already been cleansed of any existing periodic signals. To do so, the main signal which we remove is that of the binary, which is multiple orders of magnitude larger than that from any planet. Additionally, in some systems we have evidence for an outer stellar companion. These additional signals can add power to the GLS periodogram and need to be removed as well. Therefore, for calculating detection limits we use the residuals to the best-fitting models, as determined in Sect.~\ref{sec:results}. On these residuals no periodic signal with a period longer than $4P_{\rm bin}$ (the rough stability limit, \citealt{Holman:1999lr}) was discovered above a 1\% FAP.

On the cleansed data for each of the 47 binaries we insert and retrieve artificial circumbinary signals in the following way. We create a grid of planet periods that is uniform in $\log n_{\rm c}$, where $n_{\rm c}=2\pi /P_{\rm c}$ is the orbital frequency of the planet. The minimum period tested is $4P_{\rm bin}$, as shorter period planets would be unstable. The maximum period tested is equal to $4\Delta T$, where $\Delta T$ is the observing timespan of the observations for a given binary. We note that for EBLM J2046-40 the outer triple star has a  period of 5557 days, which was discovered using a timespan of 1801 days, which is roughly a third in length.

For each period we insert Keplerian sinusoids with increasing radial velocity semi-amplitudes, $K_{\rm c}$. Following that, we attempt to retrieve the artificial signal using the periodogram. The value of $K_{\rm c}$ is directly correlated to $m_{\rm c}\sin I_{\rm c}$, and this takes into account our calculated values of $m_{\rm A}$ and $m_{\rm B}$. As soon as the injected sinusoid produces a signal above a 1\% FAP we define this as the minimum detectable mass for the binary at that period. For each period this process is repeated for 20 different planetary orbital phases, equally spaced between $0^{\circ}$ and $360^{\circ}$, since some phases may be better illuminated by the observations than others. 

%An extreme example of this effect is illustrated in Fig. BLAH. There are two sinusoids of these same amplitude and period, and in both cases nine observations have been made at the same times. The blue data points have been taken near the radial velocity maxima and minima, and hence would reveal the planet. The red points, on the other hand, have been unfortunately poorly timed to coincide with the radial velocity zero point, and hence are indistinguishable from a flat line.  

In Fig.~\ref{fig:detection_limit_example} we show example detection limit curves for one of the most precise BEBOP targets: EBLM J2011-71. This system is known to contain a tertiary M dwarf star, which is demarcated on the plot well above the detection limit curve. In Appendix~\ref{app:RV_plots} of this paper detection limit curves are provided for all BEBOP targets.

\subsection{Genetic algorithm detection of n-body simulated radial velocity signals}\label{subsec:nbody_test}

Tests are run to verify that this periodogram-based definition of detectability matches our ability to detect planets with the {\sc Yorbit} genetic algorithm. For two targets, J0035-69 and J0310-31, we construct a coarse grid of circumbinary planet periods and minimum masses. The grids are chosen to straddle either side of the detection limit curves which were calculated in Sect.~\ref{subsec:detection_limits}. All other planetary parameters are set to zero, except the inclination which is taken at $90^{\circ}$. The binary parameters are the measured values. The reason for choosing these two targets in particular is that they have different binary parameters (including the mass ratio, period and eccentricity) and also precisions at either extreme of our programme (5-6 m s$^{-1}$ for J0310-31 and 50-60 m s$^{-1}$ for J0035-69).

At each grid point an n-body code\footnote{A fourth order Runge-Kutta code with a fixed 0.05 hour time step, which meant that any non-conservation of energy was negligible.} is run to simulate the radial velocity signal of the hypothetical circumbinary system, including both the large-amplitude binary signal and the much smaller planetary signal. The radial velocity measurements are simulated at the same epochs as the actual observations were for each target, and are given the same uncertainty. Importantly, the n-body simulation does not assume static Keplerians, and hence any dynamical perturbations by the binary on the planet's orbit are naturally included. Contrastingly, the periodogram analysis assumes static orbits.

The {\sc Yorbit} code is then run on the simulated radial velocities to search for a two-Keplerian solution. Owing to its large amplitude, the binary signal is always found easily.  A second signal will always be fitted but we only consider the detection of the simulated planet to be successful if the {\sc Yorbit}-found orbit has a period within 10\% of the n-body simulated planet period. The threshold of 10$\%$ is admittedly somewhat arbitrary, but felt to be sufficient for this simple demonstration. A more thorough study of the effects of n-body interactions on RV detectability is beyond the scope of this paper.

In Fig.~\ref{fig:nbody_test} we show the results of these tests. There is a close connection between the {\sc Yorbit} detectability and the periodogram detectability. Recall that the range in the periodogram detectability is a result of testing 20 different orbital phases, whereas for {\sc Yorbit} only a single phase is tested. There are a few exceptional cases where the {\sc Yorbit} detectability is not a monotonic function of $m_{\rm c}\sin I_{\rm c}$. There are two explanations for this. First, there is an element of randomness in any genetic algorithm. Second, the observational errors are redrawn from a normal distribution for each n-body test, and hence will randomly impact some simulations more than others.

Based on these tests, we conclude that the periodogram means of determining detectability sufficiently replicates how we actually detect circumbinary planets using {\sc Yorbit}. Non-Keplerian effects, even near the stability limit for one of our most precise targets, are seemingly a negligible hindrance on detectability.

\subsection{Evidence for n-body interactions}\label{subsec:nbody}

\begin{figure}  
\begin{center}  
	\begin{subfigure}[b]{0.49\textwidth}
		\caption{}
		\label{fig:1146_periodogram}
		\includegraphics[width=\textwidth]{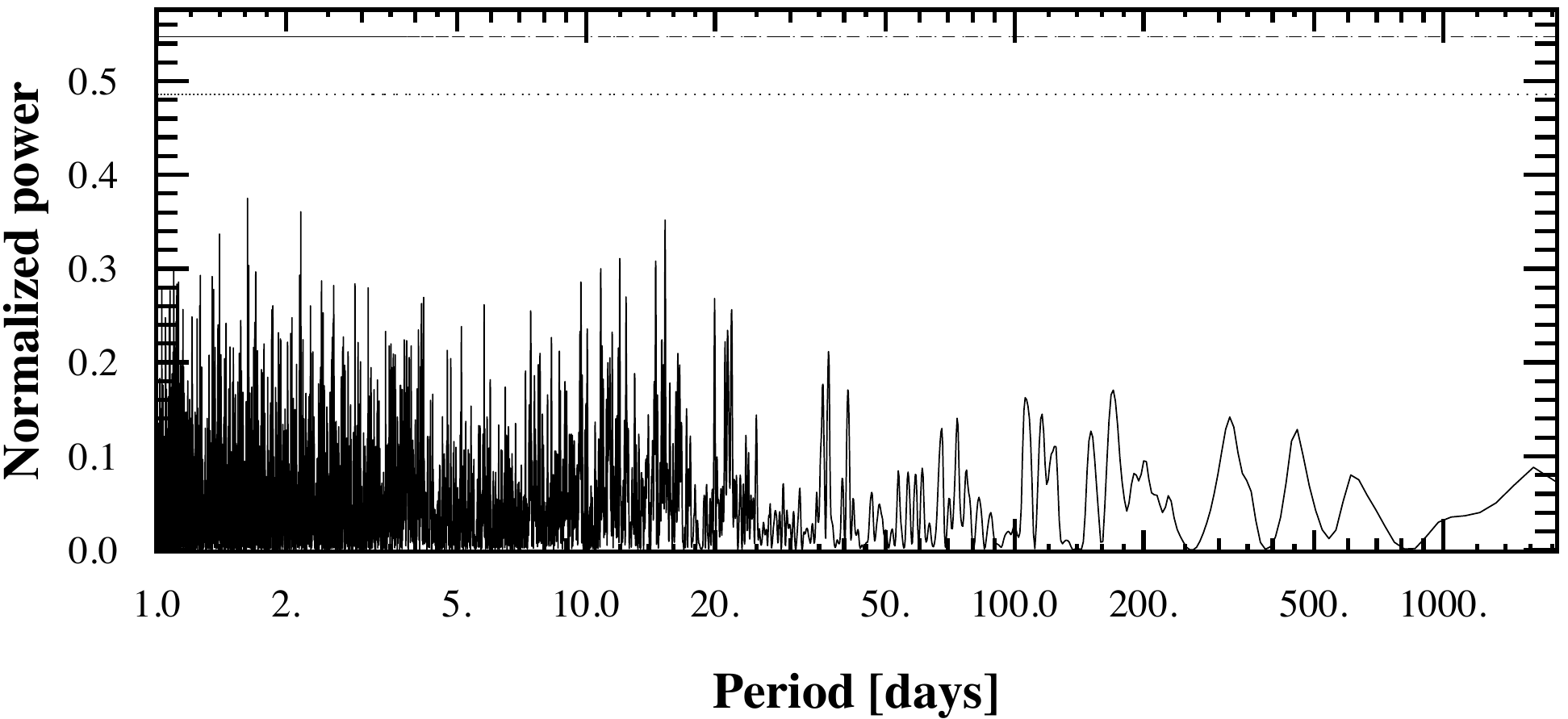}  
	\end{subfigure}
	\begin{subfigure}[b]{0.49\textwidth}
		\caption{}
		\label{fig:1146_bisector}
		\includegraphics[width=\textwidth]{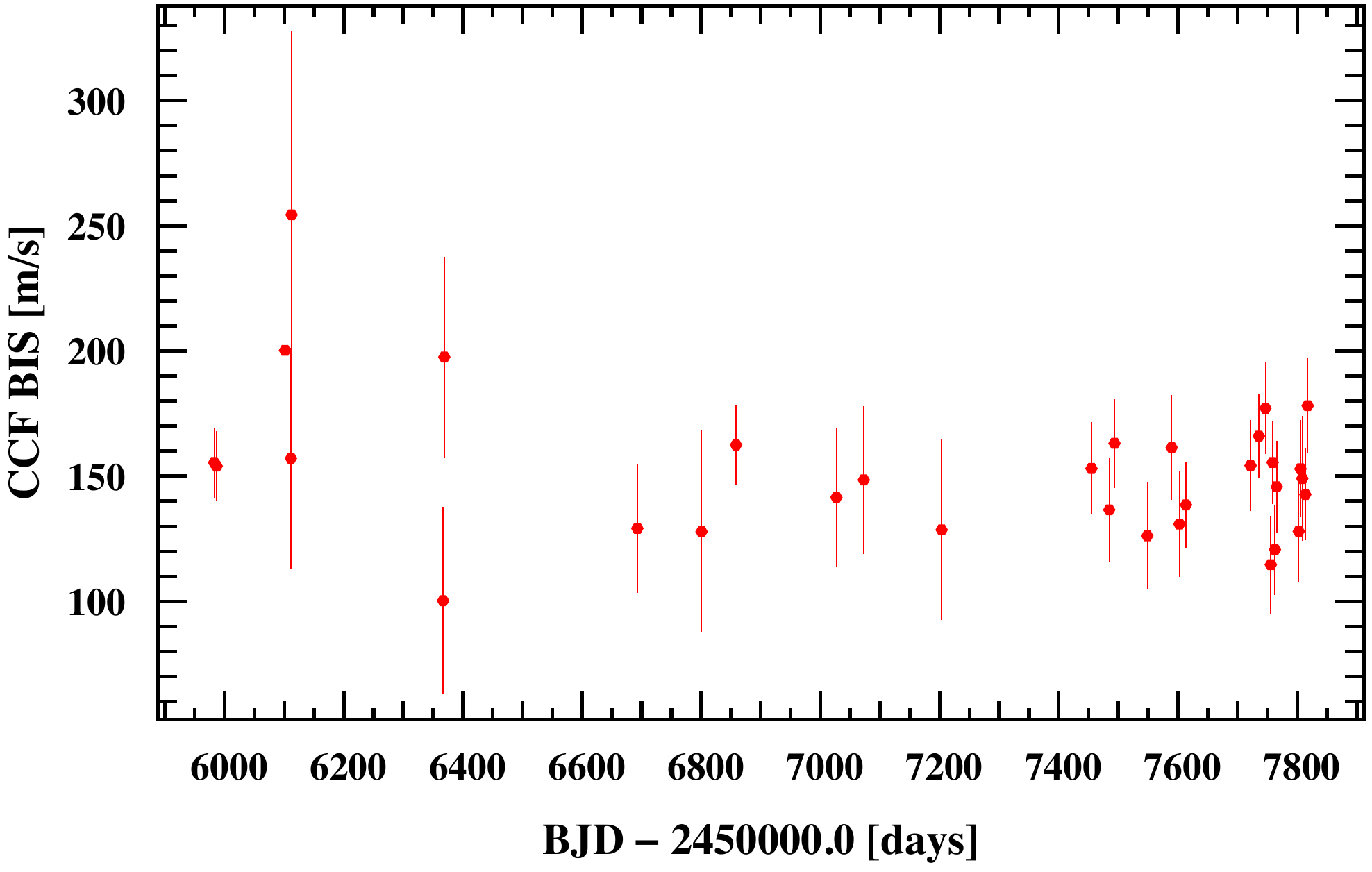}  
	\end{subfigure}
	\begin{subfigure}[b]{0.49\textwidth}
		\caption{}
		\label{fig:1146_activity}
		\includegraphics[width=\textwidth]{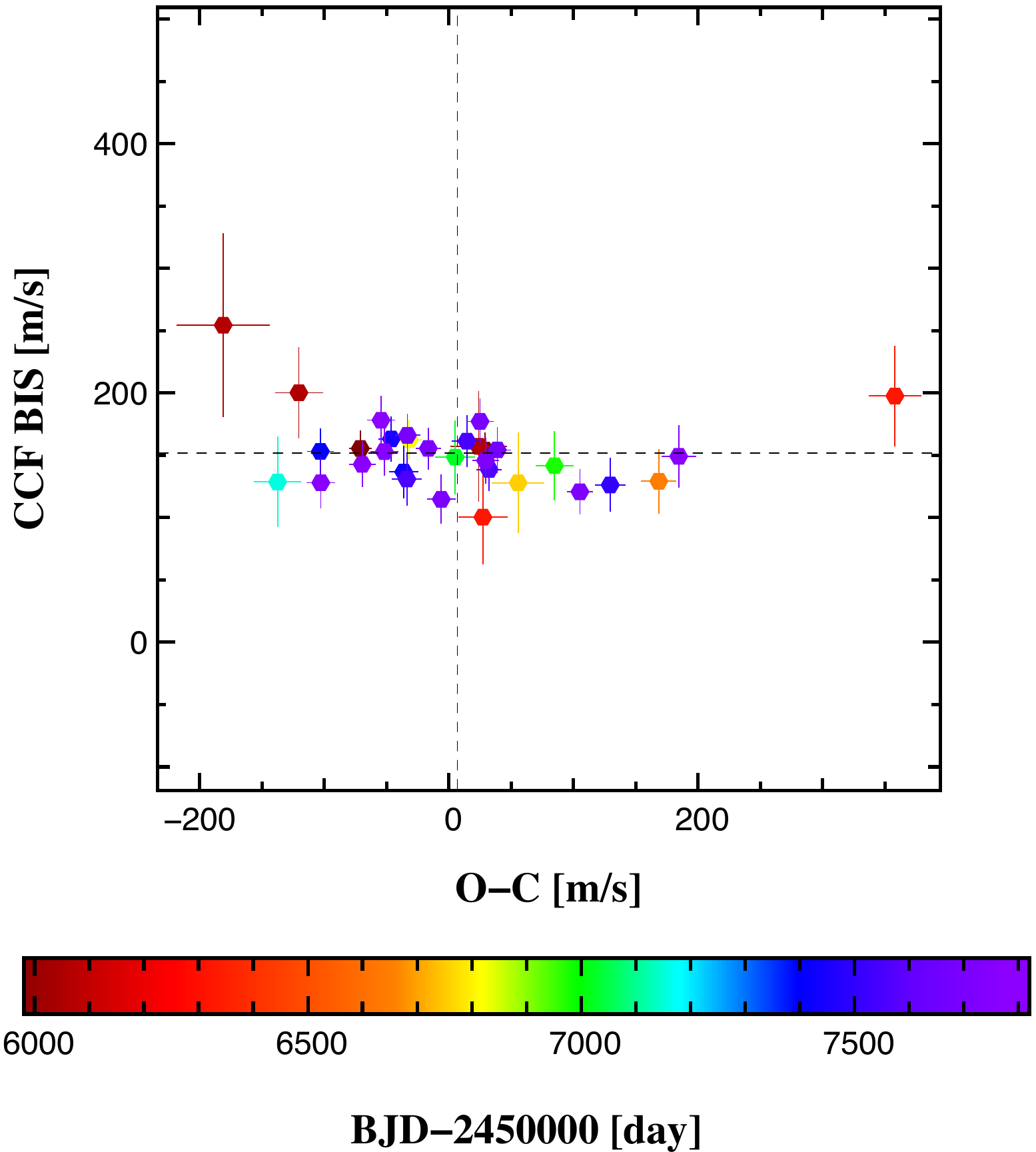}  
	\end{subfigure}
	\caption{Data for the EBLM J1146-42 triple star system. a) Periodogram of the residuals to a double Keplerian fit. The dash-dotted horizontal line at the top of the plot corresponds to a false alarm probability of 1 \%. The dotted line below is for 10 \%. There is therefore no statistically significant periodicity within the residuals. See Appendix~\ref{app:RV_plots} for more plots for this system. b) Bisector for each radial velocity measurement. c) Radial velocity residuals to a double Keplerian fit as a function of the bisector. A negative correlation would be a marker of stellar activity, but is not apparent here.}
\label{fig:1146}  
\end{center}  
\end{figure} 

 Only in one of our targets do we see likely evidence of n-body interactions: EBLM J1146-42. As seen in Table~\ref{tab:BIC} and in the orbit plots in Appendix~\ref{app:RV_plots}, there are significant residuals to the double Keplerian orbital fit: a scatter of $\sim \pm200$ m s$^{-1}$ and $\chi_{\rm red}^2=77.96$. In Fig.~\ref{fig:1146_periodogram} we show the periodogram of the residuals to the k2 fit, which demonstrates a lack of any significant periodicities in the data. Indeed, attempts were made to fit additional orbits and drift parameters but none resulted in a significant improvement to the fit. 

Stellar activity cannot produce residuals of that magnitude\footnote{Stellar activity: it's a trap that can easily be mistaken for a planetary orbit, however seemingly not in this case.}, and this target shows no signs of such activity \citep{Triaud:2017fu}. In Fig.~\ref{fig:1146_bisector} we show a constant bisector of the radial velocity measurements. Figure~\ref{fig:1146_activity} plots the residuals to the k2 fit as a function of the bisector. The classic indicator of activity is a negative correlation on this plot \citep{Queloz:2001lr,Figueira:2013lr}, which is not apparent here.
Furthermore, our spectra show no sign of contamination. This is expected since there is a 5.14 magnitude difference between the primary and secondary star flux. The tertiary body is  even less massive than the secondary unless its orbital plane is misaligned by more than $55^{\circ}$. %Furthermore, there is no evidence of the tertiary in our spectra.
%For the tertiary body to be 60\% as massive as the primary, at which point we may expect spectral contamination to be significant (see Sect.~\ref{subsec:spectral_contamination}), it would have to be misaligned by at least $63^{\circ}$. 

Alternatively, the  large residuals may be products of n-body interactions between the inner and outer orbits. Such interactions are not accounted for in the {\sc Yorbit}-determined orbits, which are assumed to be static Keplerians. It is a future study to analyse such interactions in a means similar to \citet{Correia:2010vn}. This will hopefully yield a direct measurement of additional parameters such as the mutual inclination between the inner and outer orbits.

\section{The abundance of circumbinary objects}\label{sec:abundance}

\subsection{Calculating the completeness of the programme}\label{subsec:abundance_completeness}

\begin{figure*}
\begin{center}
\includegraphics[width=0.99\textwidth,trim={0 0 0 0},clip]{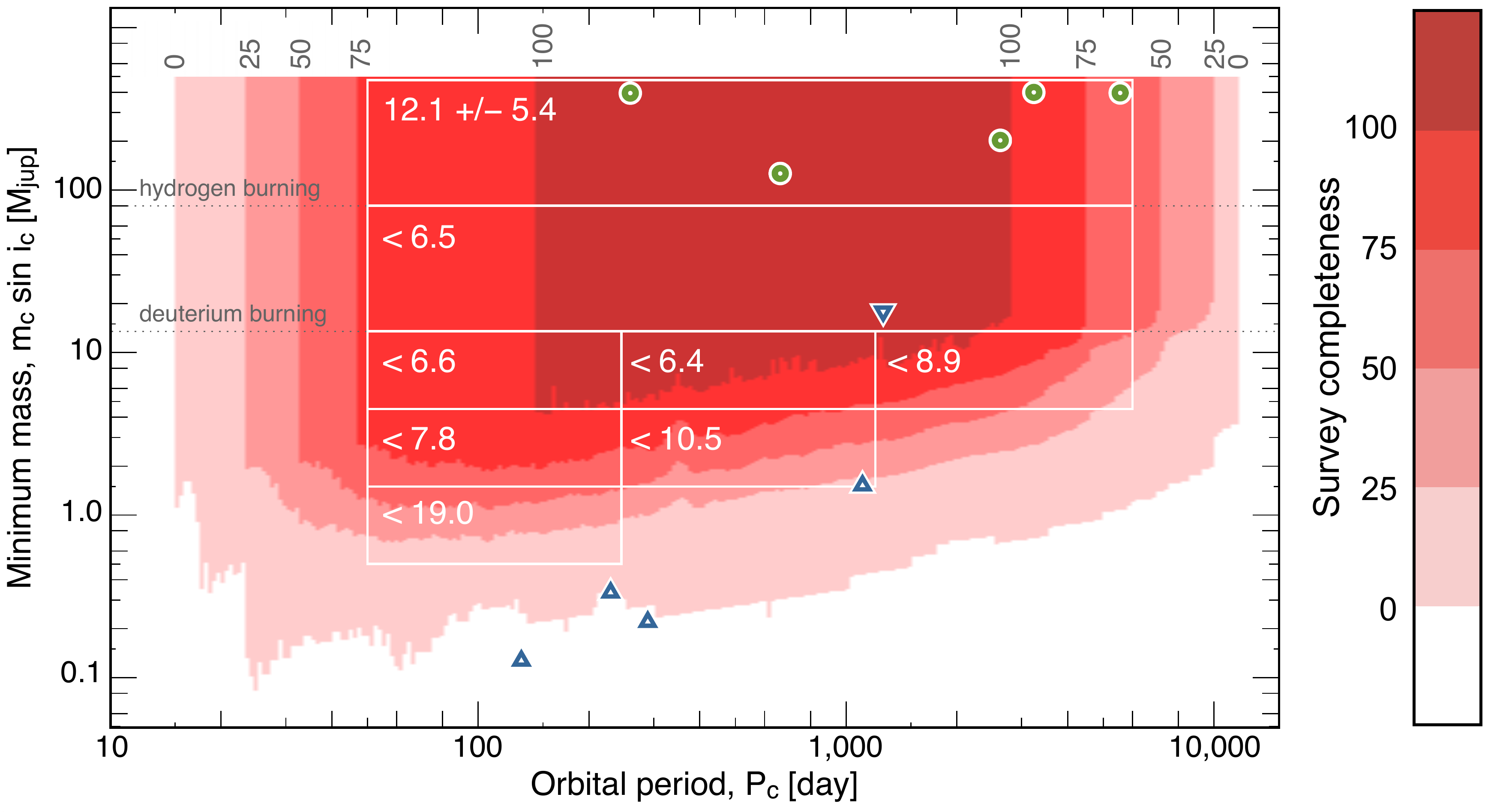}
\caption{Completeness of the BEBOP radial velocity survey of 47 low-mass eclipsing binaries, as a function of the circumbinary minimum mass and period. Six different colour contours indicate the programme completeness between 0\% (white) and 100\% (dark red). The green circles near the top of the plot correspond to the five  BEBOP triples, i.e. binaries with well-characterised tertiary stellar companions. The upright blue triangles in the bottom half of the plot represent the four {\it Kepler} transiting circumbinary planets with published masses: Kepler-16, -34, -35 and -1647. The inverted blue triangle represents the circumbinary brown dwarf HD 202206 ($m_{\rm c} = 17.9M_{\rm Jup}$, $P_{\rm c} = 1261$ days) discovered using a combination of RVs and astrometry \citep{Correia:2005lr,Benedict:2017eg}. There are eight white boxes covering different parameter spaces within which we constrain the abundance of tertiary objects. The number in each box is the $2\sigma$ upper limit to the circumbinary abundance,  given as a per cent. An exception is the top box which covers triple star systems. Since we have detections in this box, we derive an actual value for the abundance and its $1\sigma$ uncertainty.}
\label{fig:completeness}
\end{center}
\end{figure*}

We define the completeness of our programme as a function of planet period and minimum mass, $C(P_{\rm c},m_{\rm c}\sin I_{\rm c})$, as the fraction of target binaries for which a planet with such parameters is detectable. This is calculated using all of the detection limits curves calculated based on Sect.~\ref{subsec:detection_limits}. For every binary 20 detection limit curves were calculated, each corresponding to different planetary phases. In calculating $C(P_{\rm c},m_{\rm c}\sin I_{\rm c})$ these 20 curves are treated as if they were 20 individual targets. This is the same approach as was used in earlier studies such as \citet{Cumming:2008ys,Mayor:2011fj}. 

The completeness of our programme is shown in Fig.~\ref{fig:completeness}. White corresponds to 0\% completeness, meaning that none of our targets are sensitive to planets of such minimum mass and period. We then use red gradient contours to denote increasing completeness, with dark red corresponding to 100\%, meaning that all of our targets at all 20 planetary phases are sensitive to planets at such a period and minimum mass.

Our program lacks completeness at short periods less than 50 days. This is a consequence of the stability limit restriction $P_{\rm c}\gtrsim 4P_{\rm bin}$, although we can alternatively say that we are complete down to the stability limit. At longer periods  there is a drop in completeness due to the observational timespan, which varies between targets. For periods between roughly 50 and 3000 days the completeness contours follow a rough power law $m_{\rm c}\sin I_{\rm c} \propto P_{\rm c}^{1/3}$. This is expected based on the radial velocity semi-amplitude equation (Eq.~\ref{eq:RV_K_planet}). 

\begin{figure}
\begin{center}
\includegraphics[width=0.49\textwidth,trim={0 0 0 0},clip]{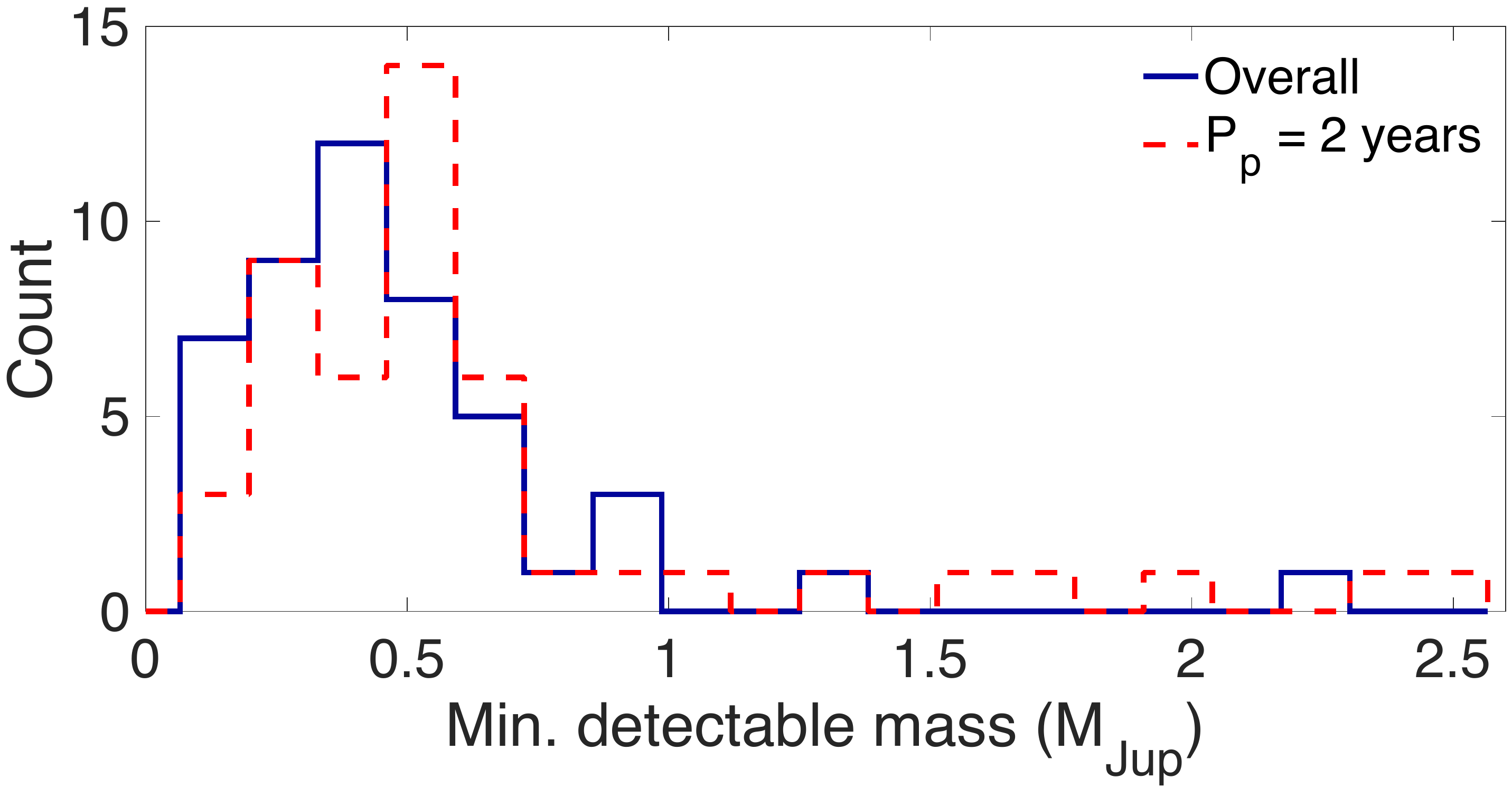}
\caption{ Histogram of the smallest detectable planet mass for the 47 eclipsing binaries in the BEBOP survey. The navy blue solid line calculates the minimum across all possible circumbinary periods and phases. The red dashed line calculates across all possible phases but has a fixed orbital period of 2 years. Note that one target, EBLM J0500-46, is excluded in the histogram for $P_{\rm p}= 2$ years because its observing timespan is too short to be sensitive to planets at this period.}
\label{fig:smallest_detectable_mass}
\end{center}
\end{figure}

 In Fig.~\ref{fig:smallest_detectable_mass} we calculate for all 47 eclipsing binaries the smallest detectable planet mass. The solid navy line is calculated across all planet phases and periods, whereas the red dashed line is calculated across all phases but assumes a period of 2 years. It is seen that for all 47 targets we have the ability to detect a circumbinary object less massive than $2.5 M_{\rm Jup}$. At smaller masses, for 30/47 of the targets we are sensitive down to $0.5 M_{\rm Jup}$. If we consider planets with periods up to 2 years, then 25/47 of our targets are sensitive to $0.5 M_{\rm Jup}$ mass planets. 

The smallest detectable planet across the entire survey is $0.082 M_{\rm Jup}$, corresponding to EBLM J2011-07. However, we note that this system contains a tertiary stellar companion at a close period of 663 days. This third star does not hinder the detectability of interior planets, but may have inhibited any from forming in the first place \citep{Munoz:2015uq,Martin:2015iu,Hamers:2016er,Xu:2016qw}. The next smallest detectable mass is $0.110 M_{\rm Jup}$ for EBLM J0310-31, for which no tertiary star has been found. For each target's detectability curve in Appendix~\ref{app:RV_plots} we identify the smallest detectable planet mass and the corresponding orbital period, denoting it with a yellow star.

 For more massive objects, i.e. circumbinary brown dwarfs and triple star systems, Fig.~\ref{fig:completeness} shows that we have essentially 100\% completeness, aside for very short and very long orbital periods. We impose an upper limit of $500M_{\rm Jup}$ on the completeness of tertiary star masses. This is because more massive stars would likely be bright enough to produce detectable spectral lines, whereas our entire sample consists of solely single-lined binaries. A smaller effect would be that brighter stars would dilute the already small eclipse depths of the M-dwarf secondary stars, potentially hindering the initial discovery of the inner binary in the EBLM programme.  We elaborate upon circumbinary brown dwarfs and triple stars in Sect.~\ref{subsec:triples}.

%We have near 100\% completeness to circumbinary brown dwarfs and triple star systems within this period range, something which we elaborate upon in Sect.~\ref{subsec:triples}.

%An additional feature is a small but noticeable bump in the completeness map at a period of 1 year. This is a well-known characteristic detection hinderance in radial velocity surveys. Stars are generally only observable at night for part of the year from one observatory. For planets with an orbital period close to one year, there will be gaps in the phase coverage that persist even over long observational baselines, making them harder to detect.

\subsection{Constraining the abundance of circumbinary planets}\label{subsec:abundance_CBPs}

%Blue squares denote all confirmed and characterised tertiary bodies. They are all have minimum masses greater than $100M_{\rm Jup}$ and hence are triple star systems, not circumbinary planets or even circumbinary brown dwarfs. Based on how many targets were found to have a radial velocity drift, there are inevitably many more tertiary components which will fill out this parameter space with longer baseline radial velocities, or alternatively complementary observations such as astrometry and direct imaging.

 %It is the result of stars generally not being observable continuously throughout the year. Consider an example star that is only observable for six months of the year. A planet with a 30 day period can have its entire period covered within one observing season. If we now consider a much longer planet of say 500 days, then after one observing season only a small portion of its RV signal will be covered. However, the next six months of observability will correspond to a different portion of the planetary orbit. After a few years all planetary phases can be covered.

%If we now have a planet with a one year period, every year the six months of observability will correspond to the same part 50\% of the planetary phase, whilst the other 50\% will never be covered. By having this resonance between the observability and orbital period, the planet's signal is never completely covered, and hence it is harder to detect.

%\subsection{Constraining the abundance of circumbinary planets}\label{subsec:abundance}

Since we do not have any confirmed discoveries of tertiary objects in the planetary domain, we can only place upper limits on their abundance. For this we use the same process as \citet{He:2017lq}. They conducted a survey of planets transiting brown dwarfs, and similarly had no confirmed detections with a comparable number of targets. The upper limit is calculated as

\begin{equation}
\eta_{\rm CBP} = \frac{1-(1-\kappa)^{1/n_{\rm stars}}}{C(P_{\rm c},m_{\rm c}\sin I_{\rm c})},
\end{equation}
where $n_{\rm stars}=47$ is the number of stars in the BEBOP survey and $\kappa$ is the desired confidence interval, e.g. $\kappa=0.95$ for a $2\sigma$ confidence interval.

This upper limit is calculated within various parameter spaces, demarcated by the lower six white boxes in Fig.~\ref{fig:completeness}. The period bounds are roughly evenly separated in log space: 50, 245, 1200 and 6000 days. The first periods are chosen to illicit an easy comparison with the work of \citet{Santerne:2016lr} for single stars, which we do in Sect.~\ref{subsec:santerne_comparison}. The planet minimum mass bounds are chosen to span sub-Jupiter masses up to the deuterium burning limit, which marks the lower bound of the brown dwarf regime. The values are 0.5, 1.5, 4.5 and 13.5 $M_{\rm Jup}$, which have roughly even log spacings.

Within each parameter space the completeness $C(P_{\rm c},m_{\rm c}\sin I_{\rm c})$ is taken as the mean value within the box. In Table ~\ref{tab:CBP_abundance} we show the abundance constraints at 50, $1\sigma$ and $2\sigma$ confidence for each of these planet parameter spaces. In Fig.~\ref{fig:completeness} to be conservative we only show the $2\sigma$ constraint.

A promising result is that the abundance constraints are only a weak function of orbital period. For example, we place a $<6.6$\% $2\sigma$ constraint on super-Jupiter circumbinary planets between 50 and 245 days, and only a marginally inferior constraint of $<8.9\%$ $2\sigma$ for planets of the same mass but periods between 1200 and 6000 days. This is in contrast to the transit method, which more strictly favours short orbital periods, and consequently no abundance constraints have been made for periods greater than 300 days \citep{Armstrong:2014yq}. It was predicted by \citet{pierens:2013kx} that the most massive circumbinary planets would be far from the stability limit, not close like the sub-Jupiter planets (see Fig.~\ref{fig:stability_limit}). Kepler-1647 (the top right upwards triangle in Fig.~\ref{fig:completeness}) follows this trend, and indeed would have been the easiest planet to detect in our programme (see Sect.~\ref{subsec:kepler_CBP_detectability}). Whilst our results at present are unable to confirm the predictions of \citet{pierens:2013kx}, it is apparent that a radial velocity survey is well-suited for such a task.

\begin{table*}
\caption{Constraints on the abundances of circumbinary planets within different minimum mass and period parameter spaces}
\label{tab:CBP_abundance}
\centering % centering table
%\tiny
\begin{tabular}{|cccc|c|ccc|}
\hline
%\rowcolor{gray!50}
$P_{\rm c,min}$ & $P_{\rm c,max}$ & $m_{\rm c,min}\sin I_{\rm c}$ & $m_{\rm c,max}\sin I_{\rm c}$  & average \% & \multicolumn{3}{c|}{$\eta$ \%} \\
 (days)& (days)&($M_{\rm Jup}$)  &($M_{\rm Jup}$) & completeness & 50\% conf. & $1\sigma$ conf. & $2\sigma$ conf. \\
\hline %inserting double-line
50 & 245 & 0.5 & 1.5 & 19.0 & $<$4.5 & $<$6.4 & $<$19.0\\ 
50 & 245 & 1.5 & 4.5 & 81.1 & $<$1.8 & $<$2.6 & $<$7.8\\
50 & 245 & 4.5 & 13.5 & 94.2 & $<$1.6 & $<$2.2 & $<$6.6\\
245 & 1200 & 1.5 & 4.5 & 59.0 & $<$2.5 & $<$3.6 & $<$10.5\\
245 & 1200 & 4.5 & 13.5 & 96.5 & $<$1.5 & $<$2.2 & $<$6.4\\
1200 & 6000 & 4.5 & 13.5 & 69.4 & $<$2.1 & $<$3.0 & $<$8.9 \\
\hline
50 & 6000 & 13.5 & 80 & 95.0 & $<$1.5 & $<$2.2 & $<$6.5\\
50 & 6000 & 80 & 475 & 95.0 & \multicolumn{3}{|c|}{12.1 $\pm$ 5.4  ($1\sigma$ conf.)}\\
\hline % inserts single-line
\end{tabular}
\end{table*}

\subsection{Circumbinary brown dwarfs and tight triple star systems}\label{subsec:triples}

%\begin{table*}
%\caption{Constraints on the abundances of circumbinary brown dwarfs and moderately tight, low-mass tertiary stars}
%\label{tab:tertiary_abundance}
%\centering % centering table
%%\tiny
%\begin{tabular}{|c|cccc|c|c|}
%\hline
%%\rowcolor{gray!50}
%Type & $P_{\rm c,min}$ & $P_{\rm c,max}$ & $m_{\rm C,min}\sin I_{\rm C}$ & $m_{\rm C,max}\sin I_{\rm C}$  & average \% & abundance ($\eta$) \% \\
%& (days)& (days)&($M_{\rm Jup}$)  &($M_{\rm Jup}$) & completeness &  \\
%\hline %inserting double-line
%Circumbinary brown dwarfs & 150 & 2500 & 13 & 80 & 100.0 & $< 1.5$ (6.2)\\ 
%Moderately tight, low mass tertiary stars & 250 & 6000 & 80 & 400 & 95.3 & $2.1\pm(0.6)$\\
%
%\hline % inserts single-line
%\end{tabular}
%\end{table*}

Figure~\ref{fig:completeness} shows that whilst the BEBOP survey only has partial completeness within the circumbinary planetary mass domain, it has practically 100\% completeness for circumbinary brown dwarfs and tertiary stars with moderately long periods. We therefore use this information to calculate the abundances of such objects. Since we have not detected any circumbinary brown dwarfs, we can only place an upper limit. For triple stars though we have five well-characterised systems, and hence can calculate an actual abundance.

The top two boxes in Fig.~\ref{fig:completeness} correspond to the abundance calculations for closely-orbiting low-mass triple star systems and circumbinary brown dwarfs. These results are also included in Table~\ref{tab:CBP_abundance}.

We define brown dwarfs as bodies with masses within the deuterium-burning regime: 13.5 to 80$M_{\rm Jup}$. Our BEBOP survey has almost 100\% completeness for such objects on orbits between 50 and 6000 days, but with no confirmed discoveries. Using the same method as for the cirucmbinary planets, we constrain the abundance within this period range to be $\eta_{\rm BD}< 6.5\%$ to $2\sigma$ confidence. The known circumbinary brown dwarf HD 202206 \citep{Correia:2005lr,Benedict:2017eg} interestingly falls within this parameter space, with the planet just above the deuterium burning limit.

Our result here should be considered preliminary on account of the size of the BEBOP sample. Brown dwarfs companions to single Sun-like stars are inherently rare, at a rate of $<1\%$. This has led to the coining of the phrase ``brown dwarf desert'' to represent the paucity of companion masses in between the planetary and stellar domains \citep{Marcy:2000yf,Grether:2006kx,Sahlmann:2011qy,Kraus:2011ig,Cheetham:2015jr}. If this rarity extends to brown dwarfs around binary stars, then to have not discovered one in a sample of 47 binaries is not surprising. 

Beyond the brown dwarf regime we use our five characterised triple star systems to calculate the tertiary abundance between 50 and 6000 days for minimum masses between 80 and 475 $M_{\rm Jup}$. This period range corresponds to the rough limits of detectability of our program, as tighter triples would be unstable and wider triples would not be well-characterised by our observational timespan. The mass range is chosen to be equal in log space to that for the brown dwarfs, whilst remaining less than the $500M_{\rm Jup}$ upper limit which we impose on the programme.

We follow the work of \citet{Mayor:2011fj} to calculate the abundance as

\begin{equation}
\label{eq:tertiary_abundance}
\eta_{\rm triple} = \frac{1}{n_{\rm stars}} \sum_{i=1}^{n_{\rm det}}\frac{1}{C_i(P_{\rm c},m_{\rm c} \sin I_{\rm c})},
\end{equation}
where $n_{\rm det}=5$ is the number of detected triple stars, $C_i(P_{\rm c},m_{\rm c} \sin I_{\rm c})$ is the completeness level in the parameter space for each of the five triples. Using Poisson statistics the $1\sigma$ uncertainty is calculated as

\begin{equation}
\label{eq:tertiary_abundance_uncertainty}
\sigma = 2\frac{\eta_{\rm triple}}{\sqrt{n_{\rm det}}}.
\end{equation}
The fraction of 50 - 6000 day  tertiary stellar companions (between 80 and 475 $M_{\rm Jup}$) is calculated to be $12.1 \pm 5.4$, within a $1\sigma$ confidence interval. 

It is a future task to compare our work with complementary surveys of triple star systems. These include the work by \citet{Tokovinin:2006la} to directly image distant stellar companions to tight spectroscopic binaries, the studies of eclipse timing variations observed by {\it Kepler} \citep{Borkovits:2015op} and the imaging survey for sub-stellar companions around binaries by \citet{Bonavita:2016gf}. A goal is to construct a mass distribution of tertiary objects, that can ultimately be compared with that of secondary objects (e.g. \citealt{Grether:2006kx,Triaud:2017fu}).

Our results for circumbinary brown dwarfs and triple stars will be improved in the future by analysing the entire EBLM programme, for which 118 binaries have been published in \citet{Triaud:2017fu} and the entire sample numbers over 220. Whilst the measurements on the binaries not selected for BEBOP are too imprecise to aid the abundance calculations for circumbinary planets, almost all of these binaries permit the detection of circumbinary brown dwarfs and triple star systems. 

For many binaries the observational baseline of a few years only permits us to see a drift in the radial velocity residuals to a single Keplerian fit. For these targets we will take new observations in the coming years to extend this baseline and better constrain the period and mass of the outer body.

\section{Comparisons with other surveys}\label{sec:other_survey_comparisons}

\subsection{Detectability of known {\it Kepler} planets}\label{subsec:kepler_CBP_detectability}

Masses can be derived for transiting circumbinary planets if they induce detectable eclipse timing variations in their host binary. This has been done for four of the discovered systems: Kepler-16, -34, -35 and -1647. These planets are shown in Fig.~\ref{fig:completeness} as upright blue triangles. Three of the planets are just below our limits of detectability. Furthermore, the other circumbinary planets without measured masses are most likely smaller, and hence would have been even tougher to find. 

Only a planet with Kepler-1647's properties ($1.27M_{\rm Jup}$, 1108 day period) falls within the completeness of the BEBOP CORALIE programme (14.1\% for these parameters). We estimate the probability that we would have found such a planet in our survey using the equation

\begin{equation}
D=1-0.9^{n_{\rm stars}C(P_{\rm c},m_{\rm c}\sin I_{\rm c})},
\end{equation}
where we assume that around each binary there is a 10\% chance of a gas giant planet existing, according to the abundance studies of \citet{Martin:2014lr,Armstrong:2014yq}. For Kepler-1647 $D=50\%$.

 Overall, we have demonstrated the ability of our survey to detect planetary-mass circumbinary objects, typically smaller than $1M_{\rm Jup}$. However, the masses of the known {\it Kepler} transiting planets are almost all sub-Saturn, unfortunately placing them slightly below the detection threshold for most of our targets.

%Overall, our comparisons with the known {\it Kepler} transiting planets seems to explain why our non-detection of circumbinary planets is not a surprise. 

%\begin{table*}
%\caption{Detectability of the ten published circumbinary planets with respect to the BEBOP RV survey}
%\label{tab:kepler_planets}
%\centering % centering table
%%\tiny
%\begin{tabular}{|cccccc|}
%\hline
%%\rowcolor{gray!50}
%{\it Kepler} number & $R_{\rm c} (R_{\rm Jup})$& $m_{\rm c}^*(M_{\rm Jup})$& $P_{\rm c}$ (days) & completeness \% & $D$ \% \\
%\hline %inserting double-line
%16&  0.75 & 0.43 & 229 & 3.0 & 13.7\\
%34 &  0.76 & 0.45 & 289 & 1.9 & 9.0\\
%35 &  0.73 & 0.39 & 131 & 2.7 & 12.6\\
%38 & 0.40 & 0.06 & 106 & 0 & 0\\
%47b&  0.27 & 0.02 & 50 & 0 & 0\\
%47c &  0.42 & 0.07 & 303 & 0 & 0\\
%64 &  0.56 & 0.18 & 139 & 0 & 0\\
%413 &  0.40 & 0.06 & 66 & 0 & 0\\
%453&  0.56& 0.18 & 241& 0 & 0\\
%1647 &  1.08 &1.27 & 1108 & 14.2 & 50.4\\
%\hline % inserts single-line
%\multicolumn{6}{l}{\footnotesize{{\bf *} $m_{\rm c}$ calculated from $R_{\rm c}$ assuming a Jupiter density of 1.33 g/cm$^3$.}}\\
%\end{tabular}
%\end{table*}

\subsection{Comparison with Armstrong et al. (2014) circumbinary abundance calculations}\label{subsec:armstrong_comparison}

\begin{figure*}  
\begin{center}  
	\begin{subfigure}[b]{0.49\textwidth}
		%\caption{50 days $ < P_{\rm c} < $300 days, $0.76M_{\rm Jup} < m_{\rm c} < 13.5M_{\rm Jup}$}
		\includegraphics[width=\textwidth,trim={3cm 2cm 0 2cm}]{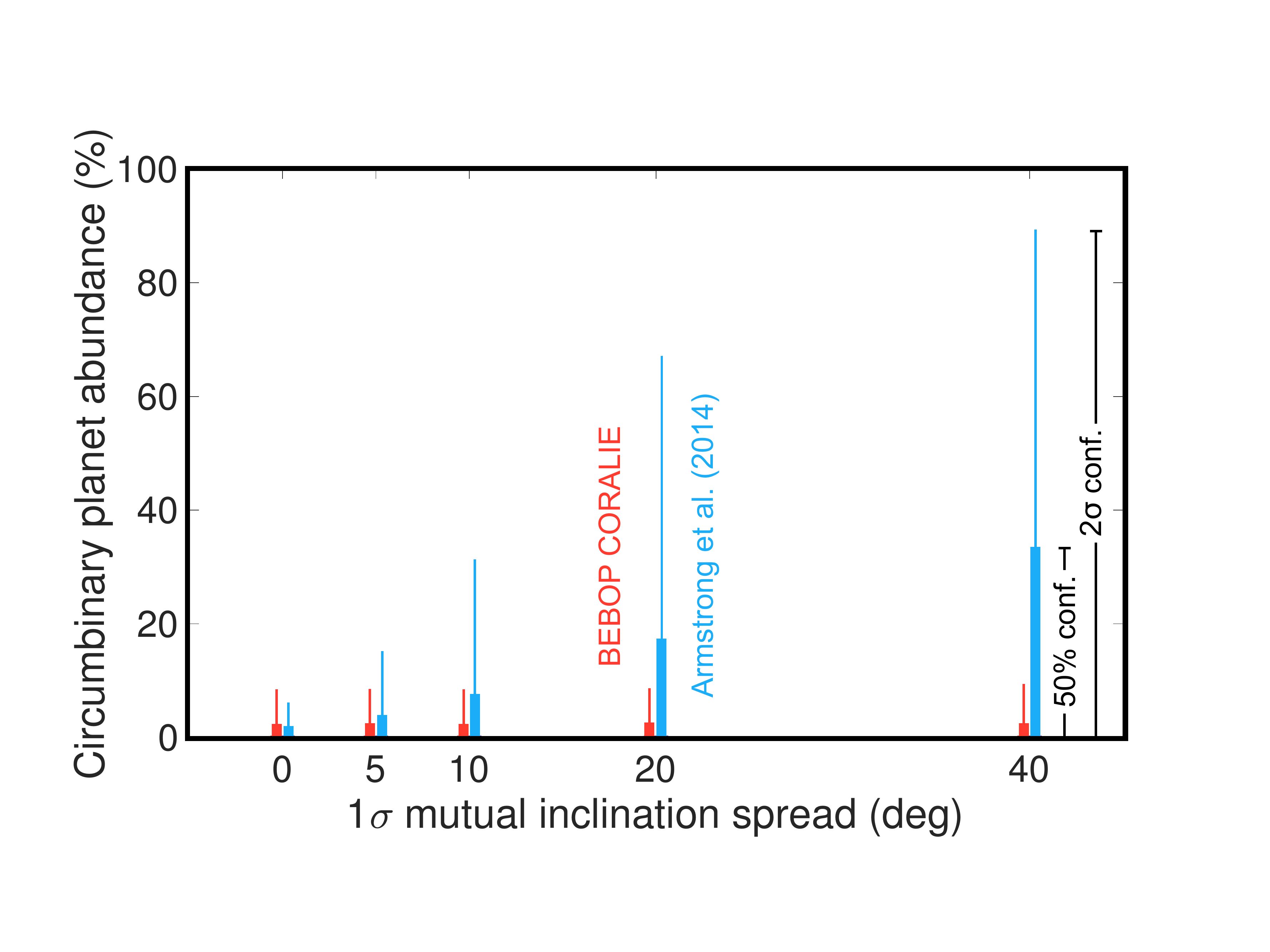}  
	\end{subfigure}
	\begin{subfigure}[b]{0.49\textwidth}
		%\caption{50 days $ < P_{\rm c} < $300 days, $0.76M_{\rm Jup} < m_{\rm c} < 13.5M_{\rm Jup}$ (zoomed)}
		\includegraphics[width=\textwidth,,trim={3cm 2cm 0 0}]{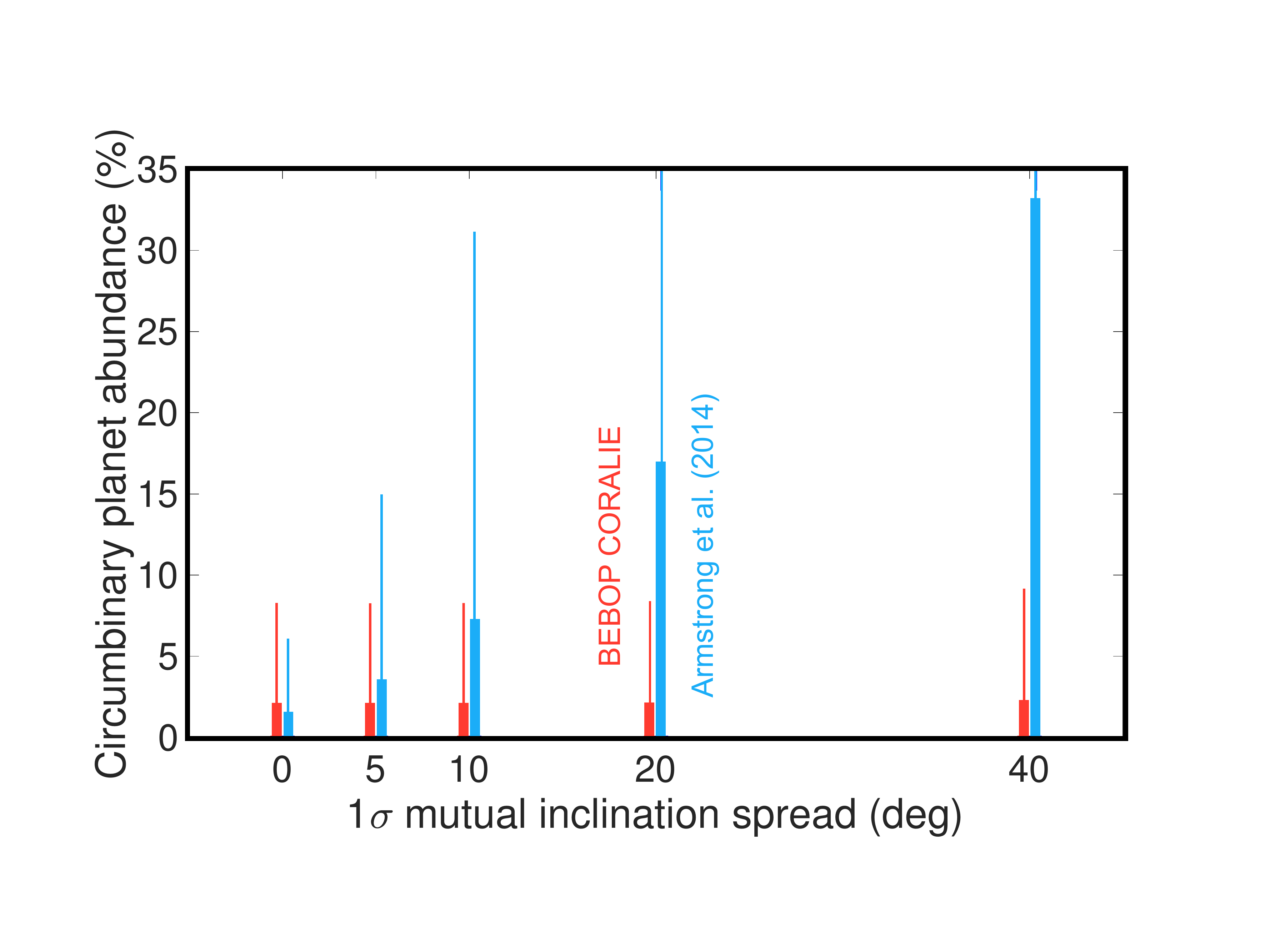}  
	\end{subfigure}
%	\begin{subfigure}[b]{0.49\textwidth}
%		\caption{50 days $ < P_{\rm c} < $300 days, $1M_{\rm Jup} < m_{\rm c} < 3M_{\rm Jup}$}
%		\includegraphics[width=\textwidth]{paper_figures/abundance_50_300_1_3-eps-converted-to.pdf}  
%	\end{subfigure}
%	\begin{subfigure}[b]{0.49\textwidth}
%		\caption{300 days $ < P_{\rm c} < $1200 days, $1M_{\rm Jup} < m_{\rm c} < 3M_{\rm Jup}$}
%		\includegraphics[width=\textwidth]{paper_figures/abundance_300_1200_1_3-eps-converted-to.pdf}  
%	\end{subfigure}
	\caption{Percentage abundance of massive circumbinary planets with periods between 50 and 300 days, as a function of the underlying spread of planetary inclinations according to a Gaussian distribution convolved with an isotropic distribution of $\cos \Delta I$. In all cases the thick, shorter lines are limits at a 50\% confidence interval and longer, thinner lines are for a $2\sigma$ confidence interval. At each value of $\Delta I$ there are two lines: BEBOP in red on the left and  \citet{Armstrong:2014yq} in blue on the right. The lower mass limit is $0.76M_{\rm Jup}$, which corresponds to $10R_{\oplus}$, assuming a Jupiter density (so we can roughly compare between our RV survey and the transit survey of \citet{Armstrong:2014yq}). The upper limit is the deuterium-burning limit of $13.5M_{\rm Jup}$. The plot on the right is the same as the left but zoomed in the vertical axis, better showing how the BEBOP abundance constraints are nearly independent on the inclination.}
\label{fig:abundance}  
\end{center}  
\end{figure*} 

Using different approaches, \citet{Armstrong:2014yq} and \cite{Martin:2014lr} estimated that the abundance of circumbinary gas giants is roughly 10\%. This is compatible with the upper limits we derive at the end of our BEBOP survey with CORALIE. \citet{Armstrong:2014yq} remarked that no circumbinary planets $> 1R_{\rm Jup}$ were detected\footnote{The largest transiting circumbinary, Kepler-1647, had not been discovered at that time.}, with the authors implying a low abundance for masses greater than Jupiter's. However, the mass-radius relation is roughly flat within a range of approximately 1 to 100$M_{\rm Jup}$ \citep{Baraffe:2015lr}. This means that \citet{Armstrong:2014yq}, based on radius, were not particularly sensitive to the frequency of circumbinary gas giants as a function of mass, something that can instead be tested thanks to radial-velocities. Our constrained abundances from Sect.~\ref{subsec:abundance_CBPs} are consistent with a $\sim 10\%$ gas giant abundance, although given our limited sensitivity to {\it Kepler}-like circumbinary planets, our comparisons can only be preliminary at this time.

The limits we place on planet occurence rates have implications on the distribution of planetary orbital inclination with respect to their binary host. {\it Kepler} is mostly sensitive to coplanar configuration. Should planetary orbital planes follow a distribution with mutual inclinations with the binary of several degrees, most planets would go undetected. Each {\it Kepler} detection would therefore imply a much more abundant population than if all circumbinary systems were coplanar.  

\citet{Armstrong:2014yq} derived circumbinary abundances within various radius bins. Since we are typically sensitive to planets more massive than Jupiter, we choose to compare with their results for $>10R_{\oplus}$. We convert the lower bound to mass using a Jupiter density: $0.76 M_{\rm Jup}$. We take the upper bound to be $13.5M_{\rm Jup}$, so we are considering all massive planetary objects. We use a period range of 50 to 300 days to match \citet{Armstrong:2014yq}.

Within this parameter space we calculate upper limits on the abundance of circumbinary planets as a function of $\sigma_{\Delta I}$, the standard deviation of the mutual inclination distribution. We calculate those following \citet{Armstrong:2014yq}, where the Gaussian distribution of $\Delta I$ is convolved with a uniform distribution in $\cos {\Delta I}$. For each value of $\sigma _{\Delta I}$ the abundance is calculated by increasing the mass limits by a factor $\sin \sigma _{\Delta I}$. The results are shown in Fig.~\ref{fig:abundance}. 

%We can immediately notice that the Doppler method is much less sensitive to orbital inclination than the transit method. Indeed our results appear to rule-out $\sigma_{\Delta I} > 10^\circ$. While highly inclined circumbinary planets may exist, the population is dominated with orbital configuration following modest $\Delta I$.
%The spread is characterised by a Gaussian distribution with a standard deviation of $\sigma _{\Delta I}=$ 0, 5, 10, 20 and 40$^{\circ}$. Following \citet{Armstrong:2014yq}, the results of which we compare to, the Gaussian distribution is convolved with a uniform distribution in $\cos {\Delta I}$. For each value of $\sigma _{\Delta I}$ the abundance is calculated by increasing the mass limits by a factor $\sin \sigma _{\Delta I}$.

%In Fig.~\ref{fig:abundance} the mass bounds of $0.76M_{\rm Jup}$ and $2.55M_{\rm Jup}$. These is a bit different to the mass bounds used in Table~\ref{tab:CBP_abundance} and Fig.~\ref{fig:abundance}. The reason for the change is to have a rough comparison with the results of \citet{Armstrong:2014yq}, who give abundance constraints for planets larger than $10R_{\oplus}$. Assuming a Jupiter density, $0.76M_{\rm Jup}$ and $2.55M_{\rm Jup}$ correspond to planetary radii of $10R_{\oplus}$ and $15R_{\oplus}$, respectively. The period range of 50 to 300 days is also chosen to match \citet{Armstrong:2014yq}.

The radial velocity-derived abundances are significantly less affected by mutual inclinations than the transit results from \citet{Armstrong:2014yq}. Whilst the detectability of transits is a sensitive function of the misalignment \citep{Martin:2015rf,Martin:2017qf}, the radial velocity detectability is a more shallow $K_{\rm c}\propto \sin {\Delta I}$ (Eq.~\ref{eq:RV_K_planet} and knowing $I_{\rm bin}\approx 90^{\circ}$).  The comparison in Fig.~\ref{fig:abundance} shows that within this parameter space   the upper limits placed by \citet{Armstrong:2014yq} are only more constraining for a strictly coplanar system. For $5^{\circ}$ and above the BEBOP results place tighter constraints on the presence of planets more massive than Jupiter. 

 Based on a preliminary comparison with the work of \citet{Armstrong:2014yq} in Fig.~\ref{fig:abundance}, it appears that there does not exist a numerous population of misaligned giant circumbinary planets, which would have evaded transit detection but have been spotted by BEBOP. We estimate that the spread of the mutual inclination is less than $10^{\circ}$, as otherwise the high planetary abundance would have practically guaranteed a BEBOP discovery.  This result is compatible with the {\it Kepler} analysis by \citet{Li:2016ng}.

For smaller planets, between 0.5 and 1$M_{\rm Jup}$ and periods between 50 and 300 days, we can only place a rudimentary $2\sigma$ abundance constraint of $<29\%$. In Sect.~\ref{sec:future} we briefly discuss the recent upgrade of the BEBOP survey to the HARPS instrument on the 3.6 metre ESO telescope, and the SOPHIE instrument on the 1.93 metre OHP telescope, and how this will advance our constraints on circumbinary abundances.

\subsection{Comparison with Santerne et al. (2016) single star abundance calculations}\label{subsec:santerne_comparison}

The \citet{Santerne:2016lr} SOPHIE radial velocity survey of {\it Kepler} transit candidates is one of the most comprehensive works in the literature on gas giant abundances around single stars. A comparison between the populations of planets orbiting one and two stars would shed light on the fundamental process of planet formation and evolution. 

\citet{Santerne:2016lr} calculates the gas giant abundance between 50 and 245 days to be $3.69\pm0.84\%$. In this work they also re-analyse the data from \citet{Mayor:2011fj} within these period ranges, calculating a remarkably similar gas giant abundance of $3.85\pm0.85\%$. The sample in \citet{Santerne:2016lr} has planets with masses between $\sim0.3-9.3M_{\rm Jup}$. If we compare our work over the mass range $0.5-13.5M_{\rm Jup}$, i.e. including all planetary mass bodies down to our rough sensitivity limit, we calculated a $2\sigma$ constrained circumbinary abundance of $<9\%$ for $50-245$ day planets. This value is compatible with the \citet{Mayor:2011fj,Santerne:2016lr} results, but we currently lack the precision to know whether the circumbinary abundance is truly greater or smaller.

\section{Future prospects}\label{sec:future}

This initial CORALIE survey had the capacity to detect  Jupiter-mass planets on two-year orbital periods for roughly half of our sample of binary stars. Our preliminary understanding of typical circumbinary masses, based on the {\it Kepler} results, was insufficient to know whether high-mass circumbinary planets were particularly abundant. Our results imply that such high-mass circumbinaries planets are indeed not frequent. Furthermore, constrained circumbinary abundances are compatible with high-mass planets orbiting single stars, but our detection capabilities  inhibit a statistically strong comparison. The results of this initial survey are primarily limited by the stability of the CORALIE instrument, which is pressurised but not thermalised, and the photon noise typical to a one metre class telescope. 

We sought to extend the BEBOP survey to the HARPS spectrograph on the ESO 3.6 metre telescope. We first conducted a short pilot programme that demonstrated that HARPS can reach radial-velocities with $\sim 1-2$ m s$^{-1}$ rms on single-line eclipsing binaries, implying a sensitivity to planets with masses as low as Neptune's. Those results are due to be published shortly. Building on this, we proposed and were awarded a large programme on HARPS, with the first data being collected in April 2018. An extension has also been granted to the northern hemisphere, with a  three-year large programme using the OHP 1.93 metre telescope with the SOPHIE spectrograph. These new observations will enhance the survey described in these pages, reaching an order of magnitude deeper in mass, extending the period range, and covering a greater number of systems. Our future observations will enable proper comparisons between the properties of planets orbiting single stars, to planets orbiting binary stars. We will report our results in a series of BEBOP papers, of which this is the first instalment.

Despite concentrating our current efforts on single-line binaries, we are also motivated in observing double-line binaries. This is important as it would provide a larger and brighter sample of binaries, but also a greater range in mass ratio. Measuring planet abundances as function of inner binary mass ratio would be insightful (Martin, under review). In Sect.~\ref{subsec:spectral_contamination} we showed no strong evidence that our secondary stars contaminate the measurement of radial-velocities. Our target most at risk of spectral contamination, EBLM J0425-46 ($q = 0.527$), which has a visual magnitude difference of 3.85, still produced a statistically perfect fit ($\chi_{\rm red}^2=0.73$). In the future, we will investigate at which flux ratios contamination becomes an issue. We will then learn how to deal with it.

\section{Conclusion}\label{sec:conclusion}

Our BEBOP radial velocity survey for circumbinary planets has started by using the  CORALIE spectrograph.  Such planets possess significant intrigue, yet the statistics of a mere dozen or so confirmed cases mean that our insights to date are only preliminary. This first paper reports the results from eight years monitoring 47 single-lined eclipsing binaries. We had two primary intentions. The first was to verify whether high-mass ($\gtrsim 1M_{\rm Jup}$) circumbinary planets were abundant, potentially owing to an unknown misaligned population, and to make an opportunistic discovery. Whilst we made no planetary detection, we  have succeeded to place constraints on the presence of  giant circumbinary planets out to orbital periods of a few years. The precision of our upper limits indicates that most massive circumbinary planets likely occupy orbits close to coplanar, with a standard deviation on the mutual inclination likely less than $\sim10^\circ$. In the process of planet-hunting, we also characterise five triple star systems. Two of these have orbits tighter than 2 AU, including one with evidence for non-Keplerian dynamical interactions.

Our second main goal was to test whether single-line binaries were amenable to circumbinary planet detection, a change in direction from the classically-observed double-line systems. We successfully demonstrated that the light from the faint secondary star  can be ignored without  inducing spectral contamination. This enabled statistically perfect binary orbital fits with a precision as small as 5 m s$^{-1}$, and a sensitivity to circumbinary planets down to 0.1 $M_{\rm Jup}$ in the best cases. Our initial programme is opening opportunities to seek smaller planets with more sensitive equipment.

%Importantly, our observations successfully demonstrate that single line binaries present a viable means of probing sub-Jupiter mass circumbinary planets, opening opportunities to seek light planets with more sensitive equipment. 
%A recently approved extension of BEBOP to the HARPS spectrograph will push our sensitivity towards Neptune mass circumbinary planets, furthering our understanding of this new genre of extra-solar planet.

\paragraph{\textbf{Nota Bene}}
We used the Barycentric Julian Dates in our analysis. Our results are based on the equatorial solar and jovian radii and masses taken from Allen's Astrophysical Quantities \citep{Cox:2000vn}.

\begin{acknowledgements} 

The authors would like to attract attention to the help and kind attention of the ESO staff at La Silla  and to the dedication of the many technicians and observers from the University of Geneva and EPFL, to upkeep the telescope, acquire the data that we present here, and bring the fondue cheese all the way from Switzerland. We would also like to acknowledge that the {\it Euler} Swiss Telescope at La Silla is a project funded by the Swiss National Science Foundation (SNSF).

Over the time required to collect and analyse the data,  DVM was supported by the SNSF, the University of Chicago and the Universit{\'e} de G{\`e}neve. AHMJT has received funding from the SNSF, the University of Toronto and the University of Cambridge. PM gratefully acknowledges support provided by the UK Science and Technology Facilities Council through grant number ST/M001040/1. This project has received funding from the European Research Council (ERC) under the European Union's Horizon 2020 research and innovation programme (grant agreement n° 803193/BEBOP).

We also thank Fran\c{c}ois Bouchy, Dan Fabrycky and Rosemary Mardling for reviewing an earlier version of this work that appeared in DVM's PhD thesis.

Finally, we kindly thank our editor and anonymous referees for reading our paper and making comments that led to significant improvements.

This publication makes use of data products from two projects, which were obtained through the  \href{http://simbad.u-strasbg.fr/simbad/}{Simbad} and \href{http://vizier.u-strasbg.fr/viz-bin/VizieR}{VizieR} services hosted at the \href{http://cds.u-strasbg.fr}{CDS-Strasbourg}:
\begin{itemize}
\item The Two Micron All Sky Survey (2MASS), which is a joint project of the University of Massachusetts and the Infrared Processing and Analysis Center/California Institute of Technology, funded by the National Aeronautics and Space Administration and the National Science Foundation \citep{Skrutskie:2006kx}.
\item The Naval Observatory Merged Astrometric Dataset (NOMAD), which is project of the US Naval Observatory \citep{Monet:2003fk}.
\end{itemize}
\end{acknowledgements}

\bibliographystyle{aa}
\bibliography{library_arXiv_V1.bib}

\rowcolors{2}{gray!25}{white}

\appendix

\section{Data tables of the BEBOP target list, model selection and derived orbital parameters}\label{app:tables}

\rowcolors{2}{gray!25}{white}

\onecolumn

\begin{landscape}
\tiny
\begin{longtable}{ccccccccccccccc}
\caption{Observational and calculated parameters of the inner binaries}\label{tab:observables}\\
\hline
name & RA & dec & $P_{\rm bin}$ & $m_{\rm A}$ & Vmag & $m_{\rm B}$ & Vmag & $T_{\rm pri}$ & $T_{\rm sec}$ & total & $\sigma_{\rm 1800s}$ & $\sigma_{\rm median}$ & $\sigma_{\rm add}$ & timespan  \\
& [deg] & [deg] & [days] & $M_{\odot}$ & A &$M_{\odot}$&B & [BJD - 2,455,000] & [BJD - 2,455,000] & observations & [m s$^{-1}$] & [m s$^{-1}$] & [m s$^{-1}$] & [years] \\
\hline
\endfirsthead
\hline 
name & RA & dec & $P_{\rm bin}$ & $m_{\rm A}$ & Vmag & $m_{\rm B}$ & Vmag & $T_{\rm pri}$ & $T_{\rm sec}$ & total & $\sigma_{\rm 1800s}$ & $\sigma_{\rm median}$ & $\sigma_{\rm add}$ & timespan  \\
& [deg] & [deg] & [days] & $M_{\odot}$ &A& $M_{\odot}$& B& [BJD - 2,455,000] & [BJD - 2,455,000] & observations & [m s$^{-1}$] & [m s$^{-1}$] & [m s$^{-1}$] & [years] \\
\hline
\endhead
\hline
\multicolumn{15}{l}{Table continues next page...}\\
\hline
\endfoot
\hline
\endlastfoot 
EBLM J0008+02 & 00 08 57.97 & +02 56 42.0 & 4.72 & 1.6 & 10.06 & 0.183 & 20.39 & 1808.2212 & 1806.3285 & 29 & 15 & 19 & 10 & 5.23\\
EBLM J0035-69 & 00 35 40.39 & -69 48 52.2 & 8.41 & 1.17 & 12.38 & 0.198 & 20.19 & 2082.113 & 2079.1783 & 38 & 52 & 68 & 0 & 5.02\\
EBLM J0040+01 & 00 40 01.50 & +01 05 40.3 & 7.23 & 0.81 & 11.4 & 0.102 & 21.35 & 1733.7657 & 1730.4681 & 39 & 11 & 18 & 8 & 4.85\\
EBLM J0055-00 & 00 55 13.72 & -00 07 54.0 & 11.4 & 1.12 & 10.96 & 0.279 & 18.34 & 2555.8339 & 2550.5422 & 24 & 12 & 10 & 0 & 2.46\\
EBLM J0104-38 & 01 04 19.13 & -38 18 30.7 & 8.26 & 1.38 & 11.26 & 0.274 & 19.48 & 1319.6732 & 1323.7962 & 36 & 25 & 35 & 0 & 8.26\\
EBLM J0218-31 & 02 18 13.24 & -31 05 17.3 & 8.88 & 1.17 & 9.96 & 0.359 & 16.35 & 1217.5183 & 1213.0763 & 47 & 10 & 17 & 16 & 8.25\\
EBLM J0228+05 & 02 28 08.87 & +05 35 47.7 & 6.63 & 1.53 & 10.24 & 0.18 & 20.52 & 2338.3823 & 2335.0649 & 33 & 25 & 30 & 0 & 3.24\\
EBLM J0310-31 & 03 10 22.62 & -31 07 35.7 & 12.6 & 1.26 & 9.34 & 0.408 & 15.94 & 1744.1012 & 1747.9902 & 27 & 5 & 5 & 4 & 3.23\\
EBLM J0345-10 & 03 45 13.09 & -10 58 24.2 & 6.06 & 1.21 & 11.2 & 0.525 & 16.61 & 2403.2691 & 2406.2885 & 28 & 47 & 56 & 0 & 2.02\\
EBLM J0353+05 & 03 53 08.94 & +05 36 33.3 & 6.86 & 1.19 & 11.18 & 0.179 & 19.42 & 1190.8562 & 1194.2856 & 53 & 10 & 16 & 0 & 6.99\\
EBLM J0353-16 & 03 53 54.52 & -16 57 15.3 & 11.8 & 1.17 & 10.52 & 0.222 & 18.7 & 1482.3867 & 1476.5373 & 48 & 8 & 11 & 0 & 8.18\\
EBLM J0418-53 & 04 18 05.11 & -53 48 05.2 & 13.7 & 1.11 & 11.45 & 0.135 & 21.98 & 2259.1678 & 2252.3181 & 28 & 19 & 21 & 0 & 2.44\\
EBLM J0425-46 & 04 25 31.70 & -46 13 07.7 & 16.6 & 1.19 & 10.98 & 0.627 & 14.83 & 1915.273 & 1907.4629 & 30 & 13 & 14 & 13 & 3.15\\
EBLM J0500-46 & 05 00 32.88 & -46 11 21.3 & 8.28 & 1.19 & 12.03 & 0.181 & 20.6 & 2393.9691 & 2391.0235 & 27 & 41 & 44 & 0 & 1.95\\
EBLM J0526-34 & 05 26 39.07 & -34 36 59.4 & 10.2 & 1.35 & 11.18 & 0.338 & 18.56 & 901.2815 & 905.5906 & 37 & 18 & 20 & 0 & 8.18\\
EBLM J0540-17 & 05 40 43.58 & -17 32 44.8 & 6 & 1.2 & 11.31 & 0.171 & 21.34 & 1891.2198 & 1888.2173 & 20 & 14 & 20 & 0 & 3.13\\
EBLM J0543-57 & 05 43 51.45 & -57 09 48.5 & 4.46 & 1.23 & 11.68 & 0.16 & 21.64 & 944.6231 & 942.3912 & 37 & 20 & 27 & 4 & 6.84\\
EBLM J0608-59 & 06 08 31.95 & -59 32 28.1 & 14.6 & 1.2 & 11.73 & 0.325 & 19.3 & 1203.9735 & 1210.6009 & 37 & 25 & 26 & 0 & 6.86\\
EBLM J0621-50 & 06 21 56.64 & -50 55 32.4 & 4.96 & 1.23 & 11.95 & 0.42 & 18.37 & 1394.9468 & 1392.4649 & 41 & 60 & 69 & 40 & 6.11\\
EBLM J0659-61 & 06 59 07.78 & -61 50 24.1 & 4.24 & 1.16 & 11.36 & 0.456 & 17.6 & 2197.4146 & 2195.2968 & 42 & 70 & 78 & 10 & 4.58\\
EBLM J0948-08 & 09 48 49.45 & -08 29 36.4 & 5.38 & 1.41 & 9.32 & 0.675 & 13.32 & 1040.2326 & 1037.7108 & 30 & 12 & 22 & 16 & 7.39\\
EBLM J0954-23 & 09 54 52.89 & -23 19 55.7 & 7.57 & 1.44 & 10.71 & 0.107 & 23.01 & 1584.8243 & 1588.5539 & 38 & 15 & 19 & 29 & 3.94\\
EBLM J0954-45 & 09 54 58.68 & -45 17 26.2 & 8.07 & 1.69 & 9.83 & 0.412 & 17.26 & 1258.3051 & 1254.9709 & 40 & 24 & 27 & 46 & 8.12\\
EBLM J1014-07 & 10 14 45.10 & -07 13 33.5 & 4.56 & 1.3 & 9.71 & 0.241 & 18.11 & 1115.3922 & 1113.5413 & 28 & 31 & 46 & 46 & 7.38\\
EBLM J1037-25 & 10 37 06.93 & -25 34 17.6 & 4.94 & 1.26 & 10.17 & 0.26 & 17.83 & 1460.3457 & 1457.9879 & 36 & 27 & 35 & 32 & 6.96\\
EBLM J1038-37 & 10 38 24.51 & -37 50 18.1 & 5.02 & 1.17 & 13.26 & 0.173 & 22.81 & 2268.9118 & 2266.401 & 33 & 69 & 74 & 0 & 5.01\\
EBLM J1141-37 & 11 41 12.18 & -37 47 29.6 & 5.15 & 1.22 & 9.58 & 0.354 & 16.17 & 1269.1192 & 1266.5454 & 32 & 18 & 28 & 62 & 7.9\\
EBLM J1146-42 & 11 46 50.49 & -42 36 59.4 & 10.5 & 1.35 & 10.29 & 0.536 & 15.43 & 2322.4973 & 2327.646 & 32 & 9 & 10 & 4 & 5.02\\
EBLM J1201-36 & 12 01 46.86 & -36 26 49.0 & 9.11 & 1.19 & 10.82 & 0.101 & 22.95 & 2103.826 & 2107.5601 & 32 & 16 & 18 & 0 & 4.16\\
EBLM J1219-39 & 12 19 21.03 & -39 51 25.6 & 6.76 & 0.95 & 10.32 & 0.0996 & 21.23 & 1201.7261 & 1198.5692 & 41 & 8 & 11 & 0 & 6.57\\
EBLM J1305-31 & 13 05 05.91 & -31 26 13.3 & 10.6 & 1.1 & 11.94 & 0.288 & 18.81 & 1890.1521 & 1895.2356 & 34 & 29 & 50 & 0 & 5.54\\
EBLM J1328+05 & 13 28 15.46 & +05 35 39.4 & 7.25 & 1.01 & 11.67 & 0.341 & 17.73 & 2417.5834 & 2413.9743 & 19 & 25 & 24 & 0 & 3.21\\
EBLM J1403-32 & 14 03 40.20 & -32 33 27.3 & 11.9 & 1.06 & 12.08 & 0.27 & 18.9 & 2465.6567 & 2460.0625 & 23 & 25 & 24 & 0 & 2.95\\
EBLM J1446+05 & 14 46 23.16 & +05 19 09.3 & 7.76 & 1.04 & 11.3 & 0.196 & 19.94 & 2223.2012 & 2219.3277 & 18 & 13 & 19 & 0 & 3.96\\
EBLM J1525+03 & 15 25 43.02 & +03 07 51.6 & 3.82 & 1.23 & 10.74 & 0.144 & 21.82 & 2577.0884 & 2575.1775 & 16 & 48 & 46 & 53 & 3.03\\
EBLM J1540-09 & 15 40 08.99 & -09 29 02.2 & 26.3 & 1.18 & 11 & 0.444 & 16.06 & 2208.8855 & 2196.6415 & 27 & 10 & 10 & 4 & 2.4\\
EBLM J1630+10 & 16 30 25.64 & +10 09 29.8 & 11 & 1.07 & 12.01 & 0.238 & 18.54 & 1251.4374 & 1256.4529 & 37 & 28 & 39 & 0 & 6.97\\
EBLM J1928-38 & 19 28 58.85 & -38 08 27.2 & 23.3 & 0.98 & 11.21 & 0.268 & 17.67 & 1926.5214 & 1937.3808 & 34 & 13 & 14 & 0 & 2.75\\
EBLM J1934-42 & 19 34 25.69 & -42 23 11.6 & 6.35 & 0.97 & 12.42 & 0.178 & 20.66 & 2440.0983 & 2436.9221 & 33 & 36 & 43 & 0 & 2.39\\
EBLM J2011-71 & 20 11 19.73 & -71 40 02.4 & 5.87 & 1.41 & 9.3 & 0.285 & 17.02 & 2002.2966 & 2005.2002 & 26 & 4 & 6 & 0 & 3.77\\
EBLM J2040-41 & 20 40 41.58 & -41 31 59.8 & 14.5 & 1.13 & 11.5 & 0.165 & 21.02 & 2019.7014 & 2014.1477 & 32 & 32 & 27 & 0 & 3.2\\
EBLM J2046-40 & 20 46 38.09 & -40 32 19.2 & 37 & 1.07 & 11.49 & 0.193 & 19.99 & 1421.6718 & 1430.159 & 30 & 10 & 22 & 0 & 4.93\\
EBLM J2046+06 & 20 46 43.88 & +06 18 09.7 & 10.1 & 1.28 & 9.87 & 0.192 & 19.01 & 1361.5185 & 1365.816 & 28 & 15 & 15 & 0 & 7.37\\
EBLM J2101-45 & 21 01 02.24 & -45 06 57.4 & 25.6 & 1.29 & 10.5 & 0.523 & 15.7 & 2107.4564 & 2096.0603 & 36 & 18 & 21 & 0 & 5.19\\
EBLM J2122-32 & 21 22 57.86 & -32 29 17.1 & 18.4 & 1.19 & 10.63 & 0.592 & 14.92 & 2211.8669 & 2217.6097 & 24 & 12 & 10 & 0 & 2.16\\
EBLM J2207-41 & 22 07 28.13 & -41 48 55.7 & 14.8 & 1.21 & 10.39 & 0.121 & 22.13 & 2006.3888 & 2013.4725 & 28 & 9 & 9 & 0 & 2.99\\
EBLM J2217-04 & 22 17 58.13 & -04 51 52.6 & 8.16 & 0.95 & 12.18 & 0.208 & 19.22 & 1960.8946 & 1956.9827 & 31 & 39 & 50 & 0 & 6.3\\

\end{longtable}
\end{landscape}

\begin{landscape}
\tiny
\begin{longtable}{lccc|cc|cc|cc|c|cc|cc|c}
\caption{Bayesian Information Criterion (BIC) with selected model in bold}\label{tab:BIC}\\
\hline 
\multicolumn{4}{c|}{System} & \multicolumn{7}{c|}{Base Models} &  \multicolumn{5}{c}{Complex Models} \\
\hline
\multicolumn{4}{c|}{} & \multicolumn{2}{c|}{k1} & \multicolumn{2}{c|}{k1d1} &  \multicolumn{2}{c|}{k1d2} & &\multicolumn{2}{c|}{k1d3} & \multicolumn{2}{c}{k2} &\\
name & num. & chosen & flag & circ & ecc & circ & ecc & circ & ecc & $\chi_{\rm red}^2$ & circ & ecc  & circ & ecc & $\chi_{\rm red}^2$  \\
 & obs. & model && 4 params. & 6 params. & 5 params. & 7 params. & 6 params. & 8 params. & & 7 params. & 9 params.  & 10 params. & 12 params. &  \\
\hline
\endfirsthead
\hline
\multicolumn{4}{c|}{System} & \multicolumn{7}{c|}{Base Models} &  \multicolumn{5}{c}{Complex Models} \\
\hline
\multicolumn{4}{c|}{} & \multicolumn{2}{c|}{k1} & \multicolumn{2}{c|}{k1d1} &  \multicolumn{2}{c|}{k1d2} & &\multicolumn{2}{c|}{k1d3} & \multicolumn{2}{c|}{k2} &\\
name & num. & chosen & flag & circ & ecc & circ & ecc & circ & ecc & $\chi_{\rm red}^2$ & circ & ecc  & circ & ecc & $\chi_{\rm red}^2$  \\
 & obs. & model && 4 params. & 6 params. & 5 params. & 7 params. & 6 params. & 8 params. & & 7 params. & 9 params.  & 10 params. & 12 params. &  \\
\hline
\endhead
\hline
\multicolumn{14}{l}{Table continues next page...}\\
\hline
\endfoot
\hline
\endlastfoot 
EBLM J0008+02 & 29 & k1d3 (ecc) & drift &577368 &14236 &550131 &1278 &532922 &70 & 2.06 &549712 & {\bf61} &532223 &55 & 1.54\\
EBLM J0035-69 & 38 & k1 (ecc) &  &95240 & {\bf54} &90375 &58 &131052 &61 & 1.02& -- & -- & -- & -- & -- \\
EBLM J0040+01 & 39 & k1 (ecc) &  &40986 & {\bf82} &40970 &87 &40894 &93 & 1.81& -- & -- & -- & -- & -- \\
EBLM J0055-00 & 24 & k1 (ecc) &  &104183 & {\bf51} &102390 &46 &115297 &48 & 1.79& -- & -- & -- & -- & -- \\
EBLM J0104-38 & 36 & k1d1 (ecc) & drift &334 &313 &112 & {\bf55} &738 &56 & 1.04& -- & -- & -- & -- & -- \\
EBLM J0218-31 & 47 & k1d2 (circ) & drift &2020 &1939 &425 &394 & {\bf84} &89 & 1.48& -- & -- & -- & -- & -- \\
EBLM J0228+05 & 33 & k1 (circ) &  & {\bf55} &47 &58 &50 &63 &54 & 1.4& -- & -- & -- & -- & -- \\
EBLM J0310-31 & 27 & k1 (ecc) &  &15578899 & {\bf45} &15574230 &49 &15586085 &43 & 1.2& -- & -- & -- & -- & -- \\
EBLM J0345-10 & 28 & k1 (ecc) &  &153 & {\bf41} &157 &45 &175 &49 & 0.96& -- & -- & -- & -- & -- \\
EBLM J0353+05 & 53 & k1d3 (ecc) & drift &57424 &57030 &1896 &2508 &263 &280 & 5.09 &150 & {\bf70} &111 &73 & 0.77\\
EBLM J0353-16 & 48 & k1 (ecc) &  &2808 & {\bf137} &2811 &149 &2940 &161 & 2.71 &2988 &167 &2467 &119 & 2.02\\
EBLM J0418-53 & 28 & k1 (circ) &  & {\bf40} &45 &107 &121 &267 &143 & 1.13& -- & -- & -- & -- & -- \\
EBLM J0425-46 & 30 & k1 (ecc) &  &108304 & {\bf38} &108065 &40 &108070 &42 & 0.73& -- & -- & -- & -- & -- \\
EBLM J0500-46 & 27 & k1 (ecc) &  &66139 & {\bf44} &74349 &49 &69532 &92 & 1.17& -- & -- & -- & -- & -- \\
EBLM J0526-34 & 37 & k1 (ecc) &  &279887 & {\bf59} &267350 &63 &296897 &76 & 1.22& -- & -- & -- & -- & -- \\
EBLM J0540-17 & 20 & k1d3 (circ) & drift &11534 &11093 &1039 &816 &244 &187 & 13.57 & {\bf35} &39 &40 &45 & 1.08\\
EBLM J0543-57 & 37 & k2 (circ) & triple &243088 &235202 &36182 &35330 &28508 &27115 & 934.02 &17669 &17043 & {\bf69} &67 & 1.21\\
EBLM J0608-59 & 37 & k1 (ecc) &  &285417 & {\bf52} &280796 &56 &282062 &60 & 0.99& -- & -- & -- & -- & -- \\
EBLM J0621-50 & 41 & k1 (circ) &  & {\bf64} &60 &67 &64 &168 &69 & 1.32& -- & -- & -- & -- & -- \\
EBLM J0659-61 & 42 & k1d2 (circ) & drift &216 &221 &143 &146 & {\bf72} &73 & 1.39& -- & -- & -- & -- & -- \\
EBLM J0948-08 & 30 & k1d2 (ecc) & drift &154658 &3170 &103082 &240 &116841 & {\bf54} & 1.22& -- & -- & -- & -- & -- \\
EBLM J0954-23 & 38 & k1 (ecc) &  &1708 & {\bf40} &1710 &44 &1745 &47 & 0.57& -- & -- & -- & -- & -- \\
EBLM J0954-45 & 40 & k1 (ecc) &  &394834 & {\bf58} &394836 &61 &394868 &65 & 1.05& -- & -- & -- & -- & -- \\
EBLM J1014-07 & 28 & k1d1 (ecc) & drift &67846 &409 &67771 & {\bf44} &67890 &40 & 0.97& -- & -- & -- & -- & -- \\
EBLM J1037-25 & 36 & k1 (ecc) &  &86037 & {\bf65} &85731 &63 &86405 &71 & 1.46& -- & -- & -- & -- & -- \\
EBLM J1038-37 & 33 & k2 (circ) & triple &14963 &12333 &986 &1246 &196 &388 & 6.47 &437 &77 & {\bf52} &54 & 0.75\\
EBLM J1141-37 & 32 & k1 (circ) &  & {\bf46} &51 &48 &53 &50 &56 & 1.14& -- & -- & -- & -- & -- \\
EBLM J1146-42 & 32 & k2 (ecc) & triple &83846377 &4837974 &8253823 &6837451 &10039944 &7622437 & 16902.88 &9618945 &6886228 &371898 & {\bf1601} & 77.96\\
EBLM J1201-36 & 32 & k1 (ecc) &  &72874 & {\bf47} &72861 &50 &73024 &58 & 1& -- & -- & -- & -- & -- \\
EBLM J1219-39 & 41 & k1 (ecc) &  &117367 & {\bf151} &116621 &164 &116840 &199 & 3.68 &122718 &195 &88208 &134 & 3.07\\
EBLM J1305-31 & 34 & k1 (ecc) &  &10294 & {\bf67} &9306 &94 &11116 &103 & 1.63& -- & -- & -- & -- & -- \\
EBLM J1328+05 & 19 & k1 (ecc) &  &143 & {\bf31} &144 &34 &150 &39 & 1.01& -- & -- & -- & -- & -- \\
EBLM J1403-32 & 23 & k1 (ecc) &  &75007 & {\bf38} &75944 &42 &85119 &44 & 1.15& -- & -- & -- & -- & -- \\
EBLM J1446+05 & 18 & k1d1 (ecc) & drift &431 &358 &73 & {\bf32} &214 &37 & 1.1& -- & -- & -- & -- & -- \\
EBLM J1525+03 & 16 & k1 (circ) &  & {\bf24} &29 &25 &25 &26 &26 & 1.08& -- & -- & -- & -- & -- \\
EBLM J1540-09 & 27 & k1 (ecc) &  &587240 & {\bf41} &588126 &44 &588684 &48 & 0.99& -- & -- & -- & -- & -- \\
EBLM J1630+10 & 37 & k1d2 (ecc) & drift &216655 &161 &207098 &59 &209752 & {\bf51} & 0.77& -- & -- & -- & -- & -- \\
EBLM J1928-38 & 34 & k1 (ecc) &  &145570 & {\bf63} &145481 &98 &149531 &154 & 1.51& -- & -- & -- & -- & -- \\
EBLM J1934-42 & 33 & k1 (circ) &  & {\bf44} &51 &96 &103 &1357 &198 & 1.05& -- & -- & -- & -- & -- \\
EBLM J2011-71 & 26 & k2 (ecc) & triple &1953861 &1672329 &2021269 &1678558 &1823813 &2099610 & 5017.9 &1528686 &1394468 &80319 & {\bf61} & 1.56\\
EBLM J2040-41 & 32 & k1 (ecc) &  &212000 & {\bf33} &208614 &40 &245024 &42 & 0.46& -- & -- & -- & -- & -- \\
EBLM J2046-40 & 30 & k2 (ecc) & triple &806786 &54394 &391292 &76554 &343239 &24868 & 1129.11 &394428 &33466 &273865 & {\bf49} & 0.47\\
EBLM J2046+06 & 28 & k1 (ecc) &  &1859295 & {\bf57} &1828192 &65 &1850631 &67 & 1.7& -- & -- & -- & -- & -- \\
EBLM J2101-45 & 36 & k1 (ecc) &  &112860 & {\bf47} &112855 &55 &112655 &64 & 0.86& -- & -- & -- & -- & -- \\
EBLM J2122-32 & 24 & k1 (ecc) &  &14047751 & {\bf52} &15257023 &48 &15453715 &62 & 1.83& -- & -- & -- & -- & -- \\
EBLM J2207-41 & 28 & k1 (ecc) &  &47555 & {\bf49} &46398 &54 &61847 &51 & 1.32& -- & -- & -- & -- & -- \\
EBLM J2217-04 & 31 & k1d1 (ecc) & drift &5713 &76 &5716 & {\bf41} &5722 &45 & 0.72& -- & -- & -- & -- & -- \\

\end{longtable}
\end{landscape}

\begin{landscape}
\tiny
%\begin{longtable}{ c | c | c | c | c | c | c | c | c | c | c | c | c | c }
\begin{longtable}{cccccccccccccc}
\caption{Orbital parameters for the inner binary from selected models}\label{tab:params}\\
\hline
name & $P_{\rm bin}$ & $a_{\rm bin}$ & $K_{\rm pri}$ & $e_{\rm bin}$ & $\omega_{\rm bin}$ & $T_{\rm peri}$ & $f(m)$ & $m_{\rm A}$ & $m_{\rm B}$ & lin & quad  & cubic  \\
 & [day] & [AU] & [km s$^{-1}$] &  & [deg] & [BJD-2,455,000] & [$10^{-3} M_{\odot}$] & [$M_{\odot}$] & [$M_{\odot}$] & [m s$^{-1}$/yr] & [m s$^{-1}$$^2$/yr] & [m s$^{-1}$$^3$/yr]  \\
\hline
\endfirsthead
\hline 
name & $P_{\rm bin}$ & $a_{\rm bin}$ & $K_{\rm pri}$ & $e_{\rm bin}$ & $\omega_{\rm bin}$ & $T_{\rm peri}$ & $f(m)$ & $m_{\rm A}$ & $m_{\rm B}$ & lin & quad  & cubic  \\
 & [day] & [AU] & [km s$^{-1}$] &  & [deg] & [BJD-2,455,000] & [$10^{-3} M_{\odot}$] & [$M_{\odot}$] & [$M_{\odot}$] & [m s$^{-1}$/yr] & [m s$^{-1}$$^2$/yr] & [m s$^{-1}$$^3$/yr]  \\
\hline
\endhead
\hline
\multicolumn{14}{c}{Table continues next page...}\\
\hline
\endfoot
\hline
\endlastfoot 
EBLM J0008+02&4.7222824(48)&0.0668(17)&16.2623(75)&0.24428(41)&-51.03(12)&1806.6388(15)&1.9187(58)&1.60(11)&0.183(28)&-321.975000(19)&70.6(1.2)&-4.6395(80)\\
EBLM J0035-69&8.414600(12)&0.0899(20)&17.327(13)&0.24561(76)&-12.76(18)&2080.3731(39)&4.131(22)&1.170(70)&0.198(20)&$<$7.6&$--$&$--$\\
EBLM J0040+01&7.2348346(71)&0.0710(18)&11.8613(51)&0.06989(50)&6.13(34)&1732.2392(68)&1.2418(36)&0.810(60)&0.102(10)&$<$3.5&$--$&$--$\\
EBLM J0055-00&11.391756(44)&0.1108(25)&21.1723(38)&0.05718(24)&-12.83(23)&2552.7839(74)&11.147(15)&1.120(70)&0.279(25)&$<$16.5&$--$&$--$\\
EBLM J0104-38&8.2560776(35)&0.0945(27)&20.6334(56)&0.00226(29)&-114.8(7.2)&1323.23(17)&7.514(13)&1.38(11)&0.274(33)&20.7(1.6)&$--$&$--$\\
EBLM J0218-31&8.8841013(44)&0.0967(25)&27.7863(55)&$<$0.00041&$--$&1215.29728(33)&19.747(12)&1.170(80)&0.359(36)&-64.037500(27)&12.96(87)&$--$\\
EBLM J0228+05&6.634731(14)&0.0826(20)&14.2380(72)&$<$0.0023&$--$&2336.72359(71)&1.9841(30)&1.53(10)&0.180(23)&$<$7.2&$--$&$--$\\
EBLM J0310-31&12.6427888(77)&0.1260(36)&27.8655(25)&0.308544(81)&-174.182(18)&1746.22173(62)&24.394(15)&1.26(10)&0.408(41)&$<$5.3&$--$&$--$\\
EBLM J0345-10&6.061353(10)&0.0782(23)&42.473(11)&0.00432(39)&-132.4(4.1)&2405.580(68)&48.118(93)&1.210(90)&0.525(66)&$<$20&$--$&$--$\\
EBLM J0353+05&6.8620280(20)&0.0785(30)&16.3051(26)&0.00119(16)&-108.4(7.3)&1193.93(14)&3.0819(30)&1.19(13)&0.179(26)&221(41)&-28.30(27)&3.2463(18)\\
EBLM J0353-16&11.7611990(82)&0.1130(27)&16.6461(30)&0.00575(18)&43.5(2.1)&1480.882(67)&5.6204(61)&1.170(80)&0.222(21)&$<$1.7&$--$&$--$\\
EBLM J0418-53&13.699304(60)&0.1205(26)&10.3769(49)&$<$0.0016&$--$&2255.7430(10)&1.5860(23)&1.110(70)&0.135(11)&$<$30&$--$&$--$\\
EBLM J0425-46&16.587983(32)&0.1553(37)&35.1967(42)&0.04775(13)&16.24(18)&1912.1139(85)&74.682(57)&1.190(80)&0.627(51)&$<$8.4&$--$&$--$\\
EBLM J0500-46&8.284445(33)&0.0890(22)&15.890(12)&0.2314(11)&-8.65(17)&2392.3129(34)&3.171(21)&1.190(80)&0.181(20)&$<$39&$--$&$--$\\
EBLM J0526-34&10.1909062(41)&0.1095(32)&23.5979(54)&0.12648(26)&-163.78(11)&903.8849(28)&13.543(21)&1.35(11)&0.338(38)&$<$5.3&$--$&$--$\\
EBLM J0540-17&6.0048865(88)&0.0718(20)&16.1980(70)&$<$0.001&$--$&1889.71855(32)&2.6442(34)&1.200(90)&0.171(22)&750.85(71)&261.5333(45)&-86.64360(10)\\
EBLM J0543-57&4.4638354(32)&0.0592(18)&16.6484(65)&$<$0.0021&$--$&943.50713(44)&2.1342(25)&1.23(10)&0.160(25)&$--$&$--$&$--$\\
EBLM J0608-59&14.6085430(10)&0.1346(32)&21.6190(67)&0.15581(29)&117.58(12)&1204.7895(44)&14.740(29)&1.200(80)&0.325(29)&$<$2.4&$--$&$--$\\
EBLM J0621-50&4.9638422(21)&0.0673(23)&37.552(20)&$<$0.0019&$--$&1393.70583(40)&27.235(43)&1.23(11)&0.420(62)&$<$7.1&$--$&$--$\\
EBLM J0659-61&4.2356391(27)&0.0601(19)&43.595(18)&$<$0.0014&$--$&2196.35571(31)&36.360(44)&1.160(90)&0.456(64)&-20.894(28)&71.809(30)&$--$\\
EBLM J0948-08&5.37980047(96)&0.0768(25)&50.3033(82)&0.04926(19)&4.38(18)&1039.0369(27)&70.693(77)&1.41(11)&0.675(93)&-184.716(22)&17.228(10)&$--$\\
EBLM J0954-23&7.574636(17)&0.0873(23)&8.6874(67)&0.04208(77)&-106.5(1.0)&1588.235(21)&0.5132(24)&1.44(11)&0.107(14)&$<$7.3&$--$&$--$\\
EBLM J0954-45&8.0726547(62)&0.1009(40)&27.887(14)&0.29456(47)&63.329(92)&1257.9885(18)&15.831(56)&1.69(19)&0.412(62)&$<$5&$--$&$--$\\
EBLM J1014-07&4.5574647(34)&0.0621(21)&23.699(15)&0.20565(70)&-44.59(28)&1113.9254(30)&5.891(27)&1.30(12)&0.241(39)&98.7(5.3)&$--$&$--$\\
EBLM J1022-40&14.06601(34)&0.1220(35)&12.2072(55)&0.25759(41)&84.02(15)&2773.5328(48)&2.3916(73)&1.070(90)&0.153(15)&$--$&$--$&$--$\\
EBLM J1037-25&4.9365622(21)&0.0652(20)&24.812(13)&0.12136(50)&-73.28(26)&1458.1667(34)&7.641(25)&1.26(10)&0.260(38)&$<$14.3&$--$&$--$\\
EBLM J1038-37&5.021630(12)&0.0633(16)&17.633(17)&$<$0.0035&$--$&2267.6564(10)&2.8523(81)&1.170(80)&0.173(23)&$--$&$--$&$--$\\
EBLM J1141-37&5.1476901(27)&0.0679(22)&32.296(17)&$<$0.0013&$--$&1267.83232(47)&17.967(29)&1.22(10)&0.354(50)&$<$16&$--$&$--$\\
EBLM J1146-42&10.466689(19)&0.1157(42)&34.220(29)&0.05250(74)&104.0(1.0)&2322.863(29)&43.28(21)&1.35(14)&0.536(67)&$--$&$--$&$--$\\
EBLM J1201-36&9.113151(14)&0.0930(24)&8.7401(48)&0.15239(63)&-158.76(20)&2106.2138(49)&0.6086(24)&1.190(90)&0.101(11)&$<$2.2&$--$&$--$\\
EBLM J1219-39&6.7600023(28)&0.0711(18)&10.8293(36)&0.05563(31)&21.21(32)&1200.5443(61)&0.8854(17)&0.950(70)&0.100(11)&$<$4.5&$--$&$--$\\
EBLM J1305-31&10.619131(15)&0.1055(27)&22.402(11)&0.03736(46)&-153.52(79)&1893.473(23)&12.344(35)&1.100(80)&0.288(28)&$<$15&$--$&$--$\\
EBLM J1328+05&7.251561(15)&0.0811(24)&30.6808(71)&0.00396(40)&-24.5(5.3)&2415.28(11)&21.698(41)&1.010(70)&0.341(48)&$<$11&$--$&$--$\\
EBLM J1403-32&11.908774(68)&0.1122(29)&20.9367(65)&0.10868(42)&-64.20(27)&2460.7492(86)&11.124(26)&1.060(80)&0.270(25)&$<$12&$--$&$--$\\
EBLM J1446+05&7.763054(10)&0.0824(25)&18.3173(66)&0.00252(39)&-49(12)&2220.20(26)&4.943(11)&1.040(90)&0.196(22)&-67.7(3.9)&$--$&$--$\\
EBLM J1525+03&3.821911(42)&0.0532(17)&15.846(24)&$<$0.0037&$--$&2576.1329(21)&1.5757(73)&1.23(11)&0.144(23)&$<$84&$--$&$--$\\
EBLM J1540-09&26.338293(42)&0.2037(57)&23.1814(41)&0.12033(16)&62.861(74)&2207.3253(53)&33.259(36)&1.18(10)&0.444(37)&$<$3.5&$--$&$--$\\
EBLM J1630+10&10.9637957(69)&0.1056(28)&19.4121(66)&0.18079(37)&111.38(12)&1251.8849(31)&7.905(19)&1.070(80)&0.238(22)&23.881(14)&3.1079(73)&$--$\\
EBLM J1928-38&23.322855(71)&0.1720(47)&17.2688(45)&0.07351(23)&137.24(19)&1929.196(12)&12.344(19)&0.980(80)&0.268(21)&$<$21.8&$--$&$--$\\
EBLM J1934-42&6.352513(16)&0.0703(18)&18.6212(89)&$<$0.00099&$--$&2438.51022(65)&4.2498(61)&0.970(70)&0.178(19)&$<$25.47&$--$&$--$\\
EBLM J2011-71&5.8727035(34)&0.0760(26)&23.6643(18)&0.03099(13)&-106.41(21)&2004.9486(34)&8.0518(51)&1.41(13)&0.285(42)&$--$&$--$&$--$\\
EBLM J2040-41&14.456256(31)&0.1266(31)&12.4620(40)&0.22645(32)&-36.818(95)&2015.5312(37)&2.6788(59)&1.130(80)&0.165(15)&$<$8.1&$--$&$--$\\
EBLM J2046-40&37.01426(32)&0.2350(58)&11.986(12)&0.47316(54)&155.771(59)&1424.1437(45)&4.515(27)&1.070(80)&0.193(14)&$--$&$--$&$--$\\
EBLM J2046+06&10.1077862(61)&0.1041(31)&15.5479(60)&0.34361(34)&108.922(81)&1361.7645(22)&3.2601(89)&1.28(11)&0.192(22)&$<$4.6&$--$&$--$\\
EBLM J2101-45&25.576836(40)&0.2072(54)&25.5107(65)&0.09100(25)&19.89(14)&2103.155(10)&43.451(69)&1.29(10)&0.523(43)&$<$7.9&$--$&$--$\\
EBLM J2122-32&18.421323(26)&0.1655(52)&35.5386(99)&0.40520(13)&-135.260(27)&2216.6569(15)&65.462(98)&1.19(11)&0.592(57)&$<$13.1&$--$&$--$\\
EBLM J2207-41&14.774903(59)&0.1296(33)&8.6936(26)&0.06841(37)&118.11(29)&2007.398(12)&0.9988(21)&1.210(90)&0.121(12)&$<$7.4&$--$&$--$\\
EBLM J2217-04&8.1552180(55)&0.0833(25)&19.9394(95)&0.04693(56)&47.16(67)&1960.005(15)&6.676(21)&0.950(80)&0.208(22)&-20.7(2.7)&$--$&$--$\\

\end{longtable}
\end{landscape}

\begin{landscape}
\begin{table}
\caption{Orbital parameters from the selected models for k2 fits}\label{tab:triple}
\centering % centering table
\tiny
\begin{tabular}{cccccccccccccc}
\rowcolor{gray!50}
\hline
name & $P$ & $a$ & $K$ & $e$ & $\omega$ & $T{\rm peri}$ & $f(m)$ & $m_{\rm A}$ & $m_{\rm B}$ & $m_{\rm c}\sin I_{\rm c}$ \\
 & [day] & [AU] & [km s$^{-1}$] &  & [deg] & [BJD-2,455,000] & [$10^{-3} M_{\odot}$] & [$M_{\odot}$] & [$M_{\odot}$] & [$M_{\odot}$] \\
\hline
EBLM J0543-57 inner binary &4.4638354(32)&0.0592(18)&16.6484(65)&$<$0.0021&$--$&943.50713(44)&2.1342(25)&1.23(10)&0.160(25)&--\\
EBLM J0543-57 tertiary orbit &3237(27)&5.18(18)&4.101(21)&0.4056(33)&18.84(99)&2409.7(44)&17.67(42)&-- &-- &0.381(26)\\
\hline
EBLM J1038-37 inner binary &5.021630(12)&0.0633(16)&17.633(17)&$<$0.0035&$--$&2267.6564(10)&2.8523(81)&1.170(80)&0.173(23)&--\\
EBLM J1038-37 tertiary orbit &2624.954110(11)&4.3(1.4)&2.67(11)&0.55(13)&136(13)&3549.150158(10)&3.0(1.7)&-- &-- &0.193(45)\\
\hline
EBLM J1146-42 inner binary &10.466689(19)&0.1157(42)&34.220(29)&0.05250(74)&104.0(1.0)&2322.863(29)&43.28(21)&1.35(14)&0.536(67)&--\\
EBLM J1146-42 tertiary orbit &259.225(58)&1.045(37)&7.462(37)&0.2006(35)&51.5(1.4)&2355.38(29)&10.49(16)&-- &-- &0.377(30)\\
\hline
EBLM J2011-71 inner binary &5.8727035(34)&0.0760(26)&23.6643(18)&0.03099(13)&-106.41(21)&2004.9486(34)&8.0518(51)&1.41(13)&0.285(42)&--\\
EBLM J2011-71 tertiary orbit &662.99(58)&1.815(61)&1.9865(27)&0.1012(30)&33.89(77)&2536.2(34)&0.5302(26)&-- &-- &0.1207(84)\\
\hline
EBLM J2046-40 inner binary &37.01426(32)&0.2350(58)&11.986(12)&0.47316(54)&155.771(59)&1424.1437(45)&4.515(27)&1.070(80)&0.193(14)&--\\
EBLM J2046-40 tertiary orbit &5557.046803(46)&7.24(59)&3.772(41)&0.500(21)&134.1(3.0)&2262(45)&20.1(2.3)&-- &-- &0.378(33)\\

%[0.5ex]
\hline % inserts single-line

\hline % inserts single-line
\end{tabular}
\end{table}
\end{landscape}

\twocolumn

\section{Radial velocities and detection limits for the 47 BEBOP systems}\label{app:RV_plots}

\begin{figure}
\begin{center}
\subcaption*{EBLM J0008+02: chosen model = k1d3 (ecc) \newline \newline $m_{\rm A} = 1.6M_{\odot}$, $m_{\rm B} = 0.183M_{\odot}$, $P = 4.722$ d, $e = 0.244$}
\begin{subfigure}[b]{0.49\textwidth}
\includegraphics[width=\textwidth,trim={0 10cm 0 1.2cm},clip]{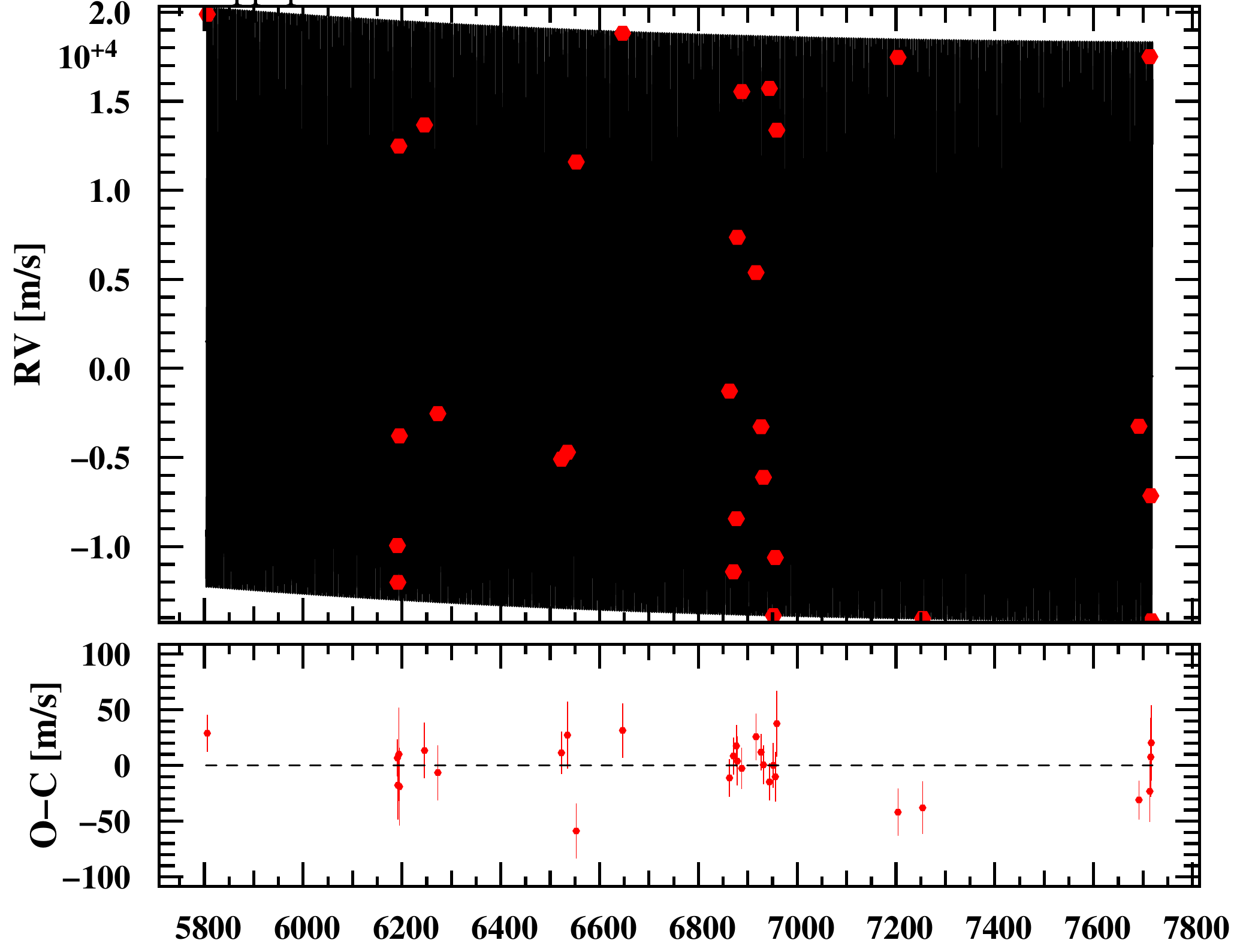}
\end{subfigure}
\begin{subfigure}[b]{0.49\textwidth}
\includegraphics[width=\textwidth,trim={0 0 2cm 0},clip]{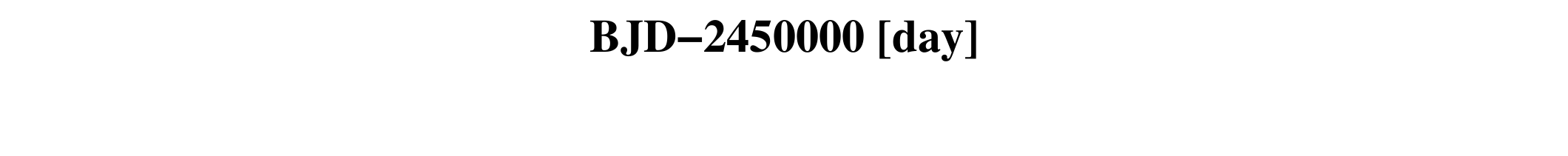}
\end{subfigure}
Radial velocities folded on binary phase
\begin{subfigure}[b]{0.49\textwidth}
\includegraphics[width=\textwidth,trim={0 0.5cm 0 0},clip]{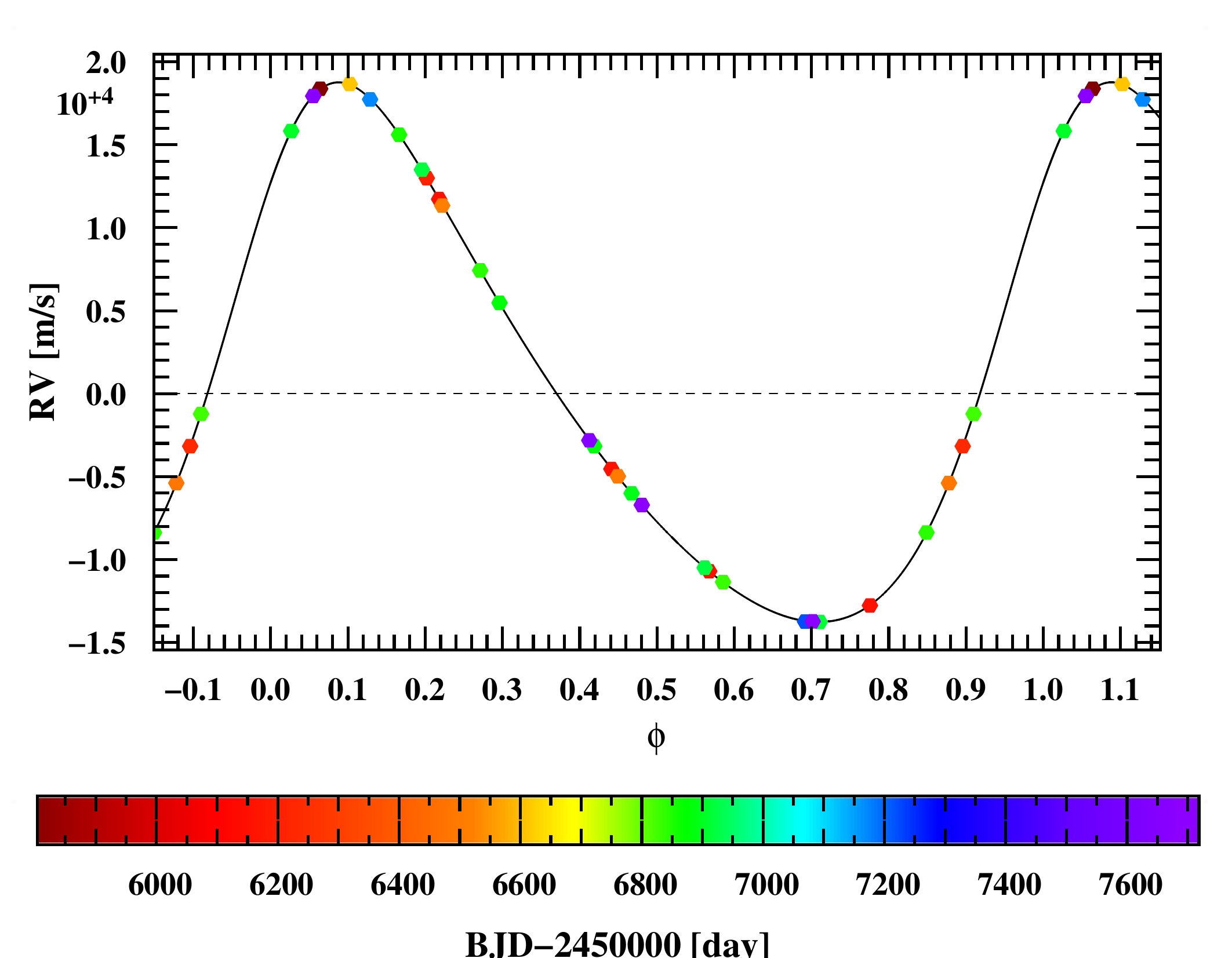}
\end{subfigure}
\begin{subfigure}[b]{0.49\textwidth}
\includegraphics[width=\textwidth,trim={0 0 2cm 0},clip]{orbit_figures/BJD_bar.pdf}
\end{subfigure}
Detection limits
\begin{subfigure}[b]{0.49\textwidth}
\vspace{0.5cm}
\includegraphics[width=\textwidth,trim={0 0 0 0},clip]{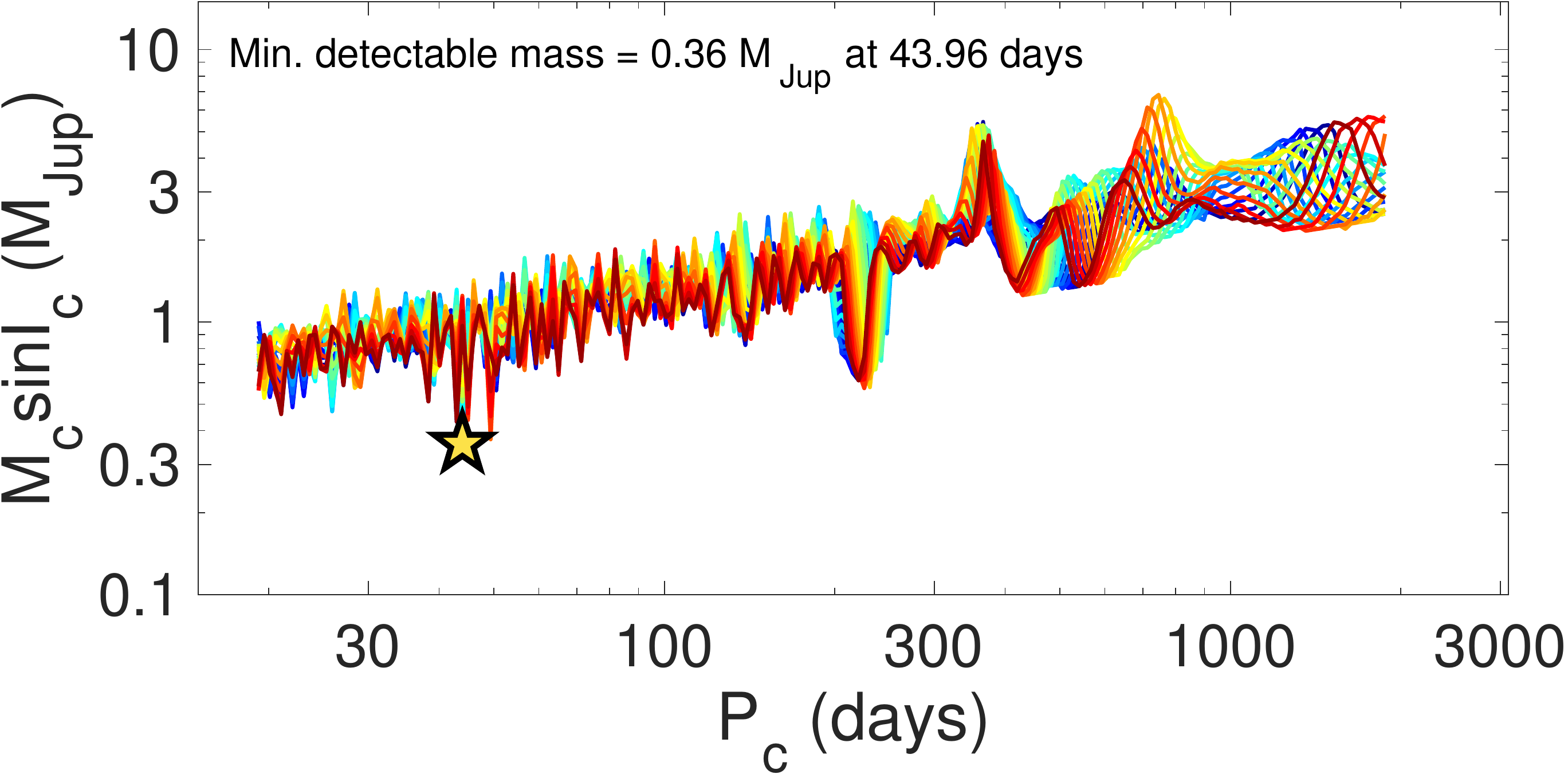}
\end{subfigure}
\end{center}
\end{figure}
\begin{figure}
\begin{center}
\subcaption*{EBLM J0035-69: chosen model = k1 (ecc) \newline \newline $m_{\rm A} = 1.17M_{\odot}$, $m_{\rm B} = 0.198M_{\odot}$, $P = 8.415$ d, $e = 0.246$}
\begin{subfigure}[b]{0.49\textwidth}
\includegraphics[width=\textwidth,trim={0 10cm 0 1.2cm},clip]{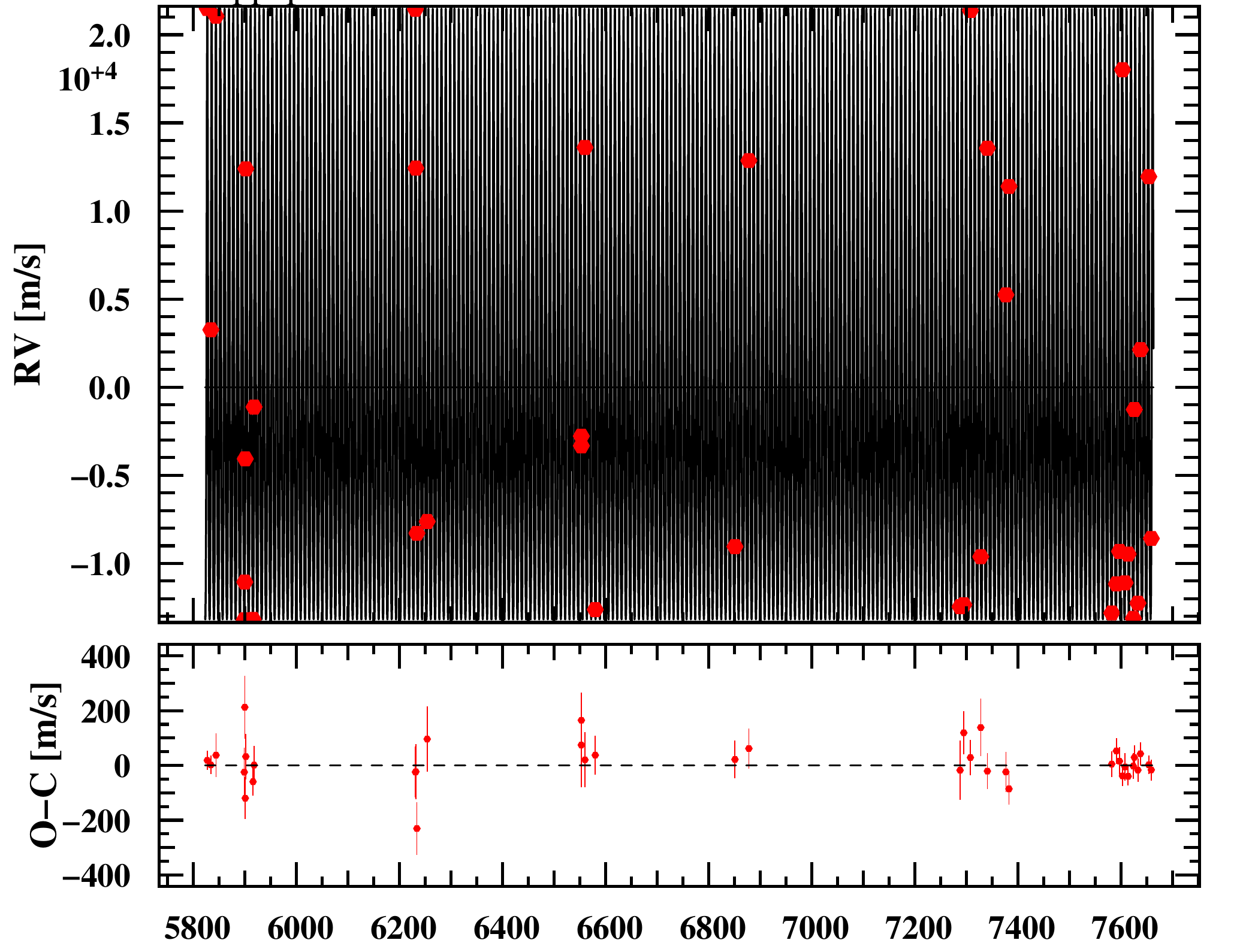}
\end{subfigure}
\begin{subfigure}[b]{0.49\textwidth}
\includegraphics[width=\textwidth,trim={0 0 2cm 0},clip]{orbit_figures/BJD_bar.pdf}
\end{subfigure}
Radial velocities folded on binary phase
\begin{subfigure}[b]{0.49\textwidth}
\includegraphics[width=\textwidth,trim={0 0.5cm 0 0},clip]{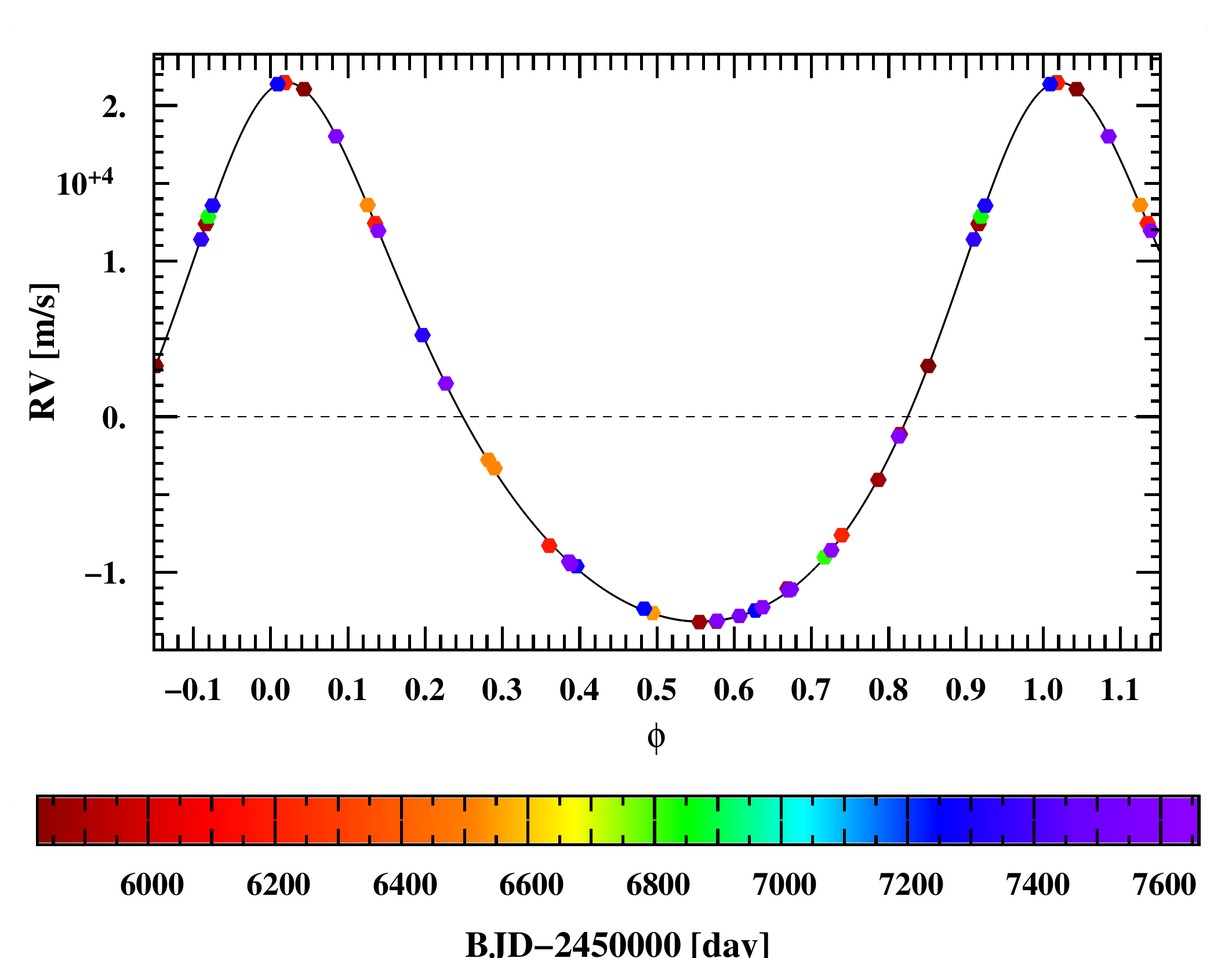}
\end{subfigure}
\begin{subfigure}[b]{0.49\textwidth}
\includegraphics[width=\textwidth,trim={0 0 2cm 0},clip]{orbit_figures/BJD_bar.pdf}
\end{subfigure}
Detection limits
\begin{subfigure}[b]{0.49\textwidth}
\vspace{0.5cm}
\includegraphics[width=\textwidth,trim={0 0 0 0},clip]{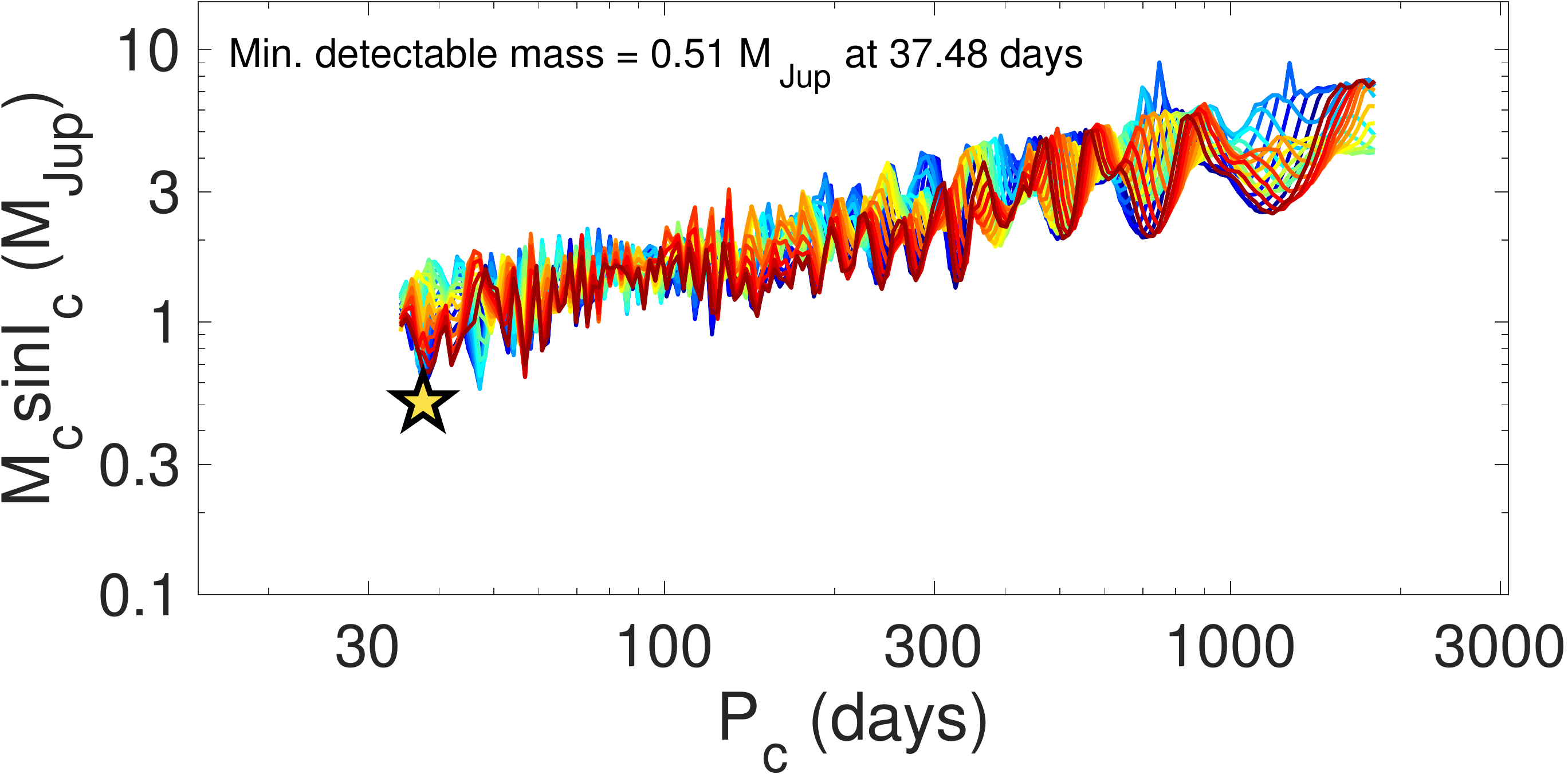}
\end{subfigure}
\end{center}
\end{figure}
\begin{figure}
\begin{center}
\subcaption*{EBLM J0040+01: chosen model = k1 (ecc) \newline \newline $m_{\rm A} = 0.81M_{\odot}$, $m_{\rm B} = 0.102M_{\odot}$, $P = 7.235$ d, $e = 0.07$}
\begin{subfigure}[b]{0.49\textwidth}
\includegraphics[width=\textwidth,trim={0 10cm 0 1.2cm},clip]{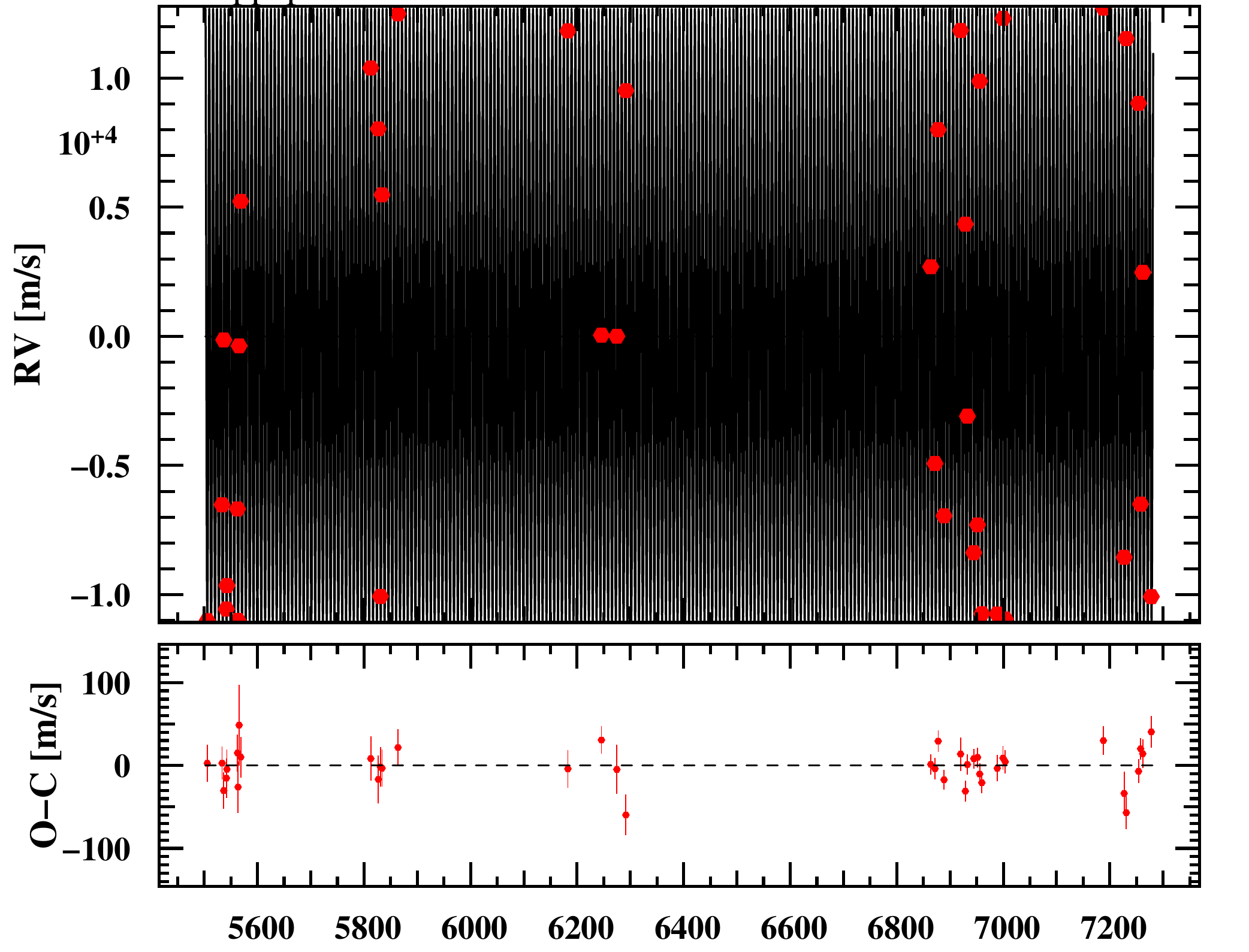}
\end{subfigure}
\begin{subfigure}[b]{0.49\textwidth}
\includegraphics[width=\textwidth,trim={0 0 2cm 0},clip]{orbit_figures/BJD_bar.pdf}
\end{subfigure}
Radial velocities folded on binary phase
\begin{subfigure}[b]{0.49\textwidth}
\includegraphics[width=\textwidth,trim={0 0.5cm 0 0},clip]{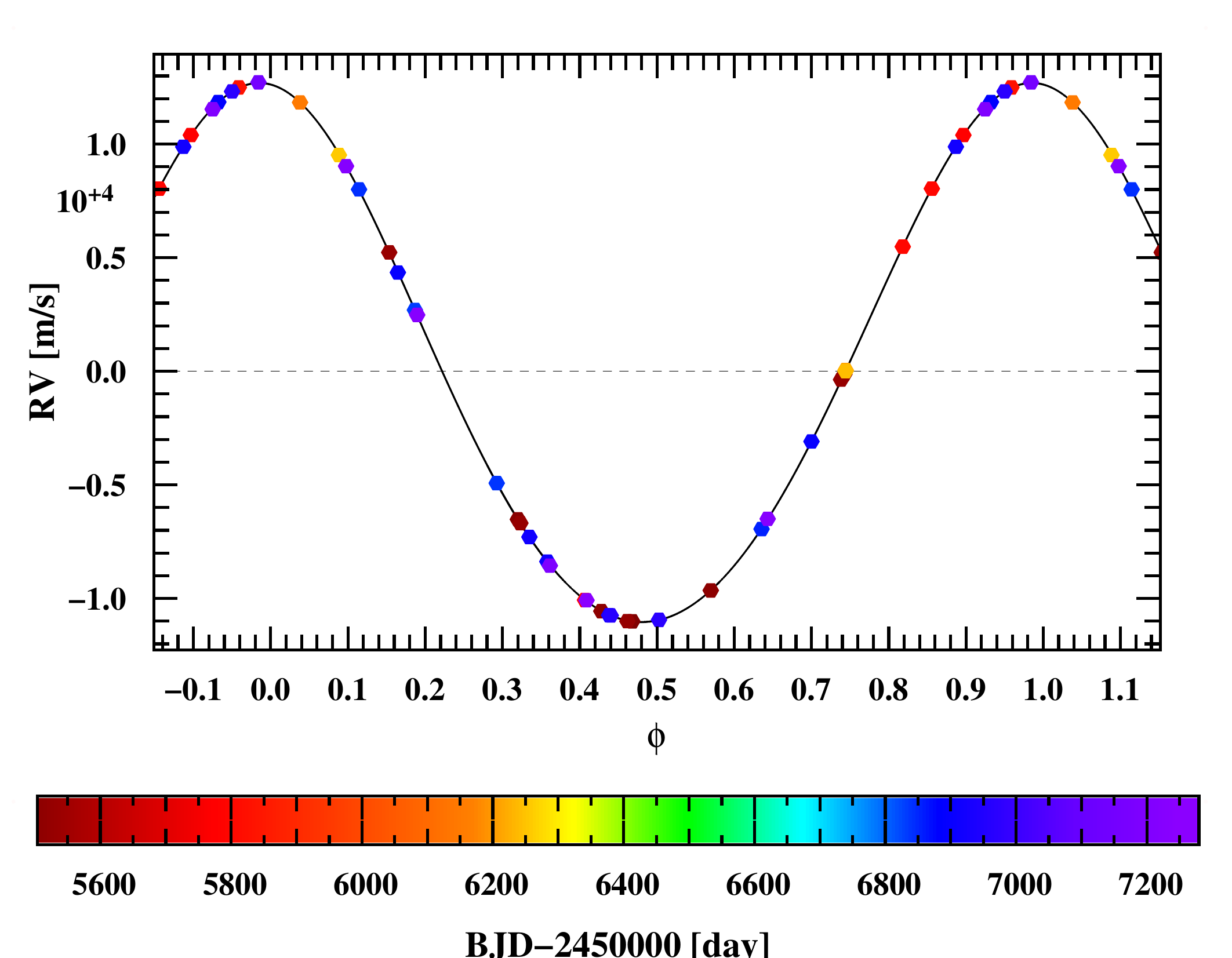}
\end{subfigure}
\begin{subfigure}[b]{0.49\textwidth}
\includegraphics[width=\textwidth,trim={0 0 2cm 0},clip]{orbit_figures/BJD_bar.pdf}
\end{subfigure}
Detection limits
\begin{subfigure}[b]{0.49\textwidth}
\vspace{0.5cm}
\includegraphics[width=\textwidth,trim={0 0 0 0},clip]{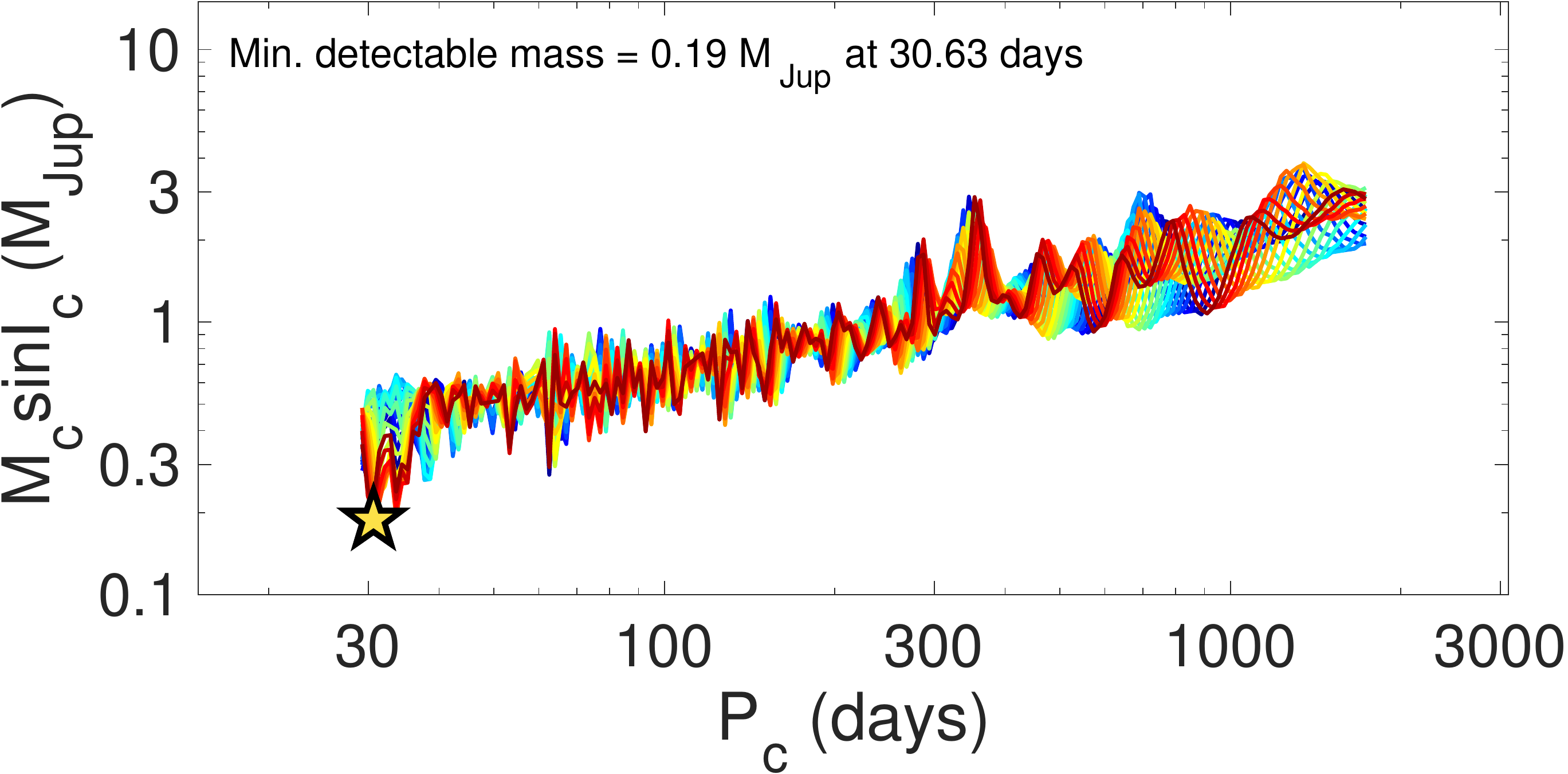}
\end{subfigure}
\end{center}
\end{figure}
\begin{figure}
\begin{center}
\subcaption*{EBLM J0055-00: chosen model = k1 (ecc) \newline \newline $m_{\rm A} = 1.12M_{\odot}$, $m_{\rm B} = 0.279M_{\odot}$, $P = 11.392$ d, $e = 0.057$}
\begin{subfigure}[b]{0.49\textwidth}
\includegraphics[width=\textwidth,trim={0 10cm 0 1.2cm},clip]{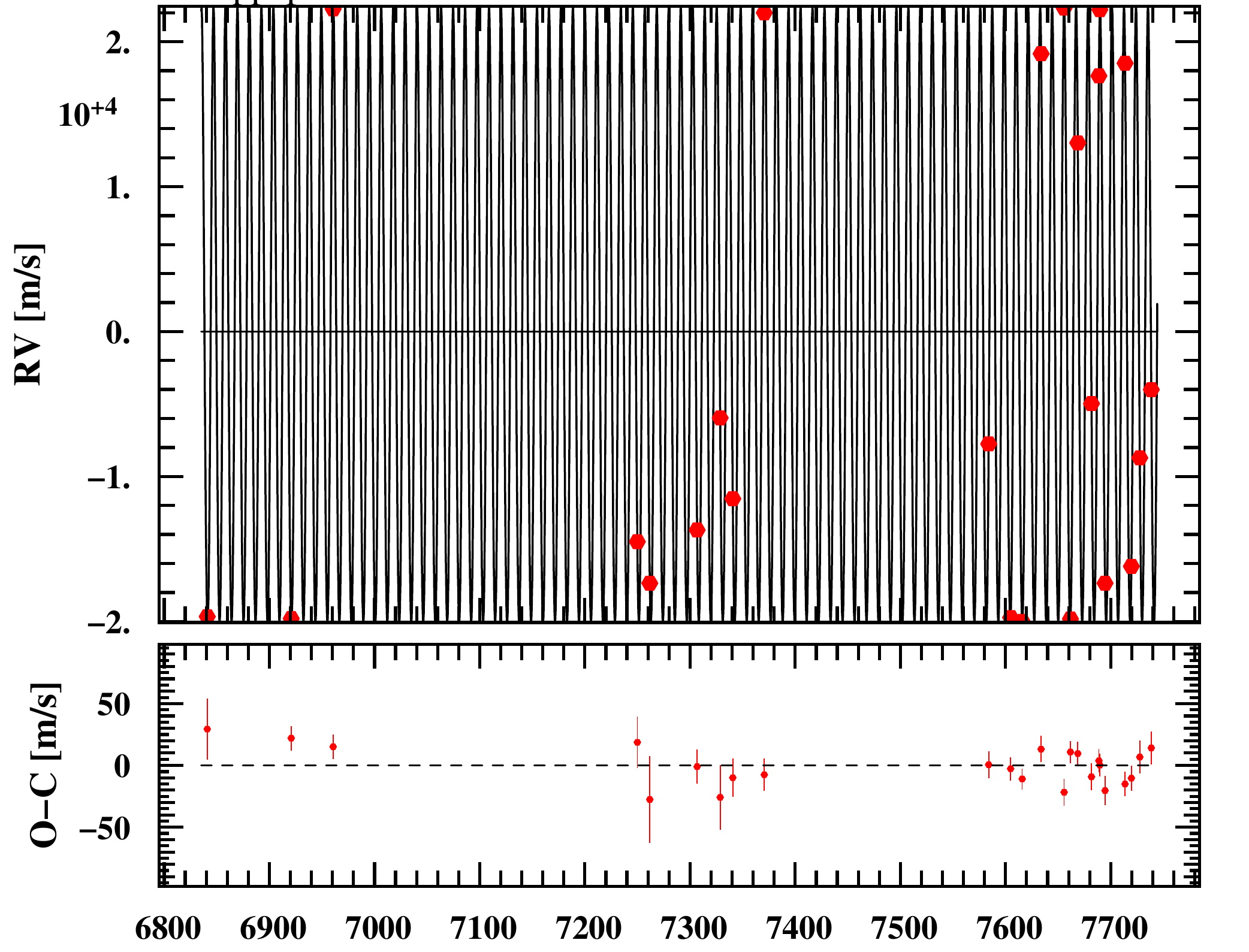}
\end{subfigure}
\begin{subfigure}[b]{0.49\textwidth}
\includegraphics[width=\textwidth,trim={0 0 2cm 0},clip]{orbit_figures/BJD_bar.pdf}
\end{subfigure}
Radial velocities folded on binary phase
\begin{subfigure}[b]{0.49\textwidth}
\includegraphics[width=\textwidth,trim={0 0.5cm 0 0},clip]{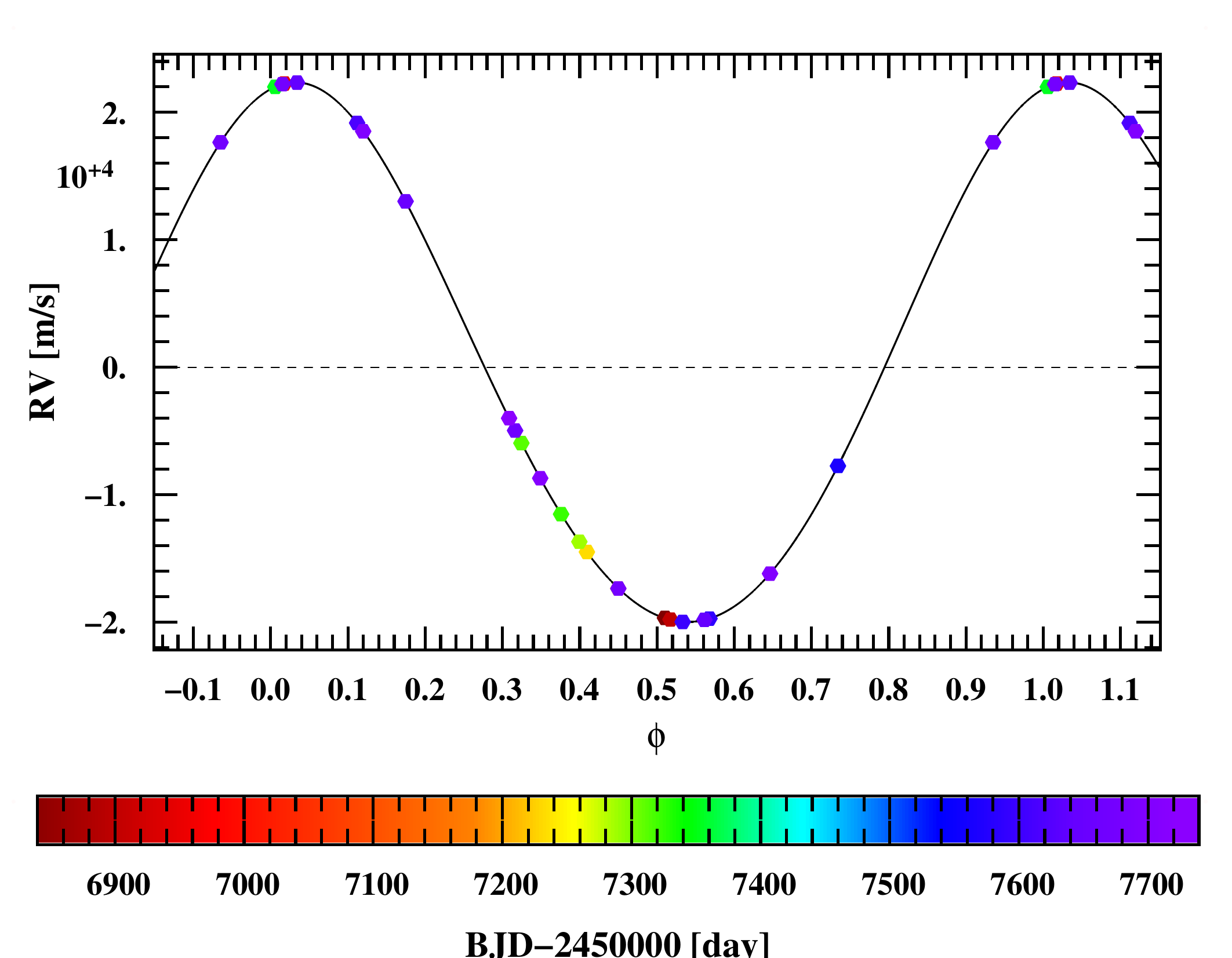}
\end{subfigure}
\begin{subfigure}[b]{0.49\textwidth}
\includegraphics[width=\textwidth,trim={0 0 2cm 0},clip]{orbit_figures/BJD_bar.pdf}
\end{subfigure}
Detection limits
\begin{subfigure}[b]{0.49\textwidth}
\vspace{0.5cm}
\includegraphics[width=\textwidth,trim={0 0 0 0},clip]{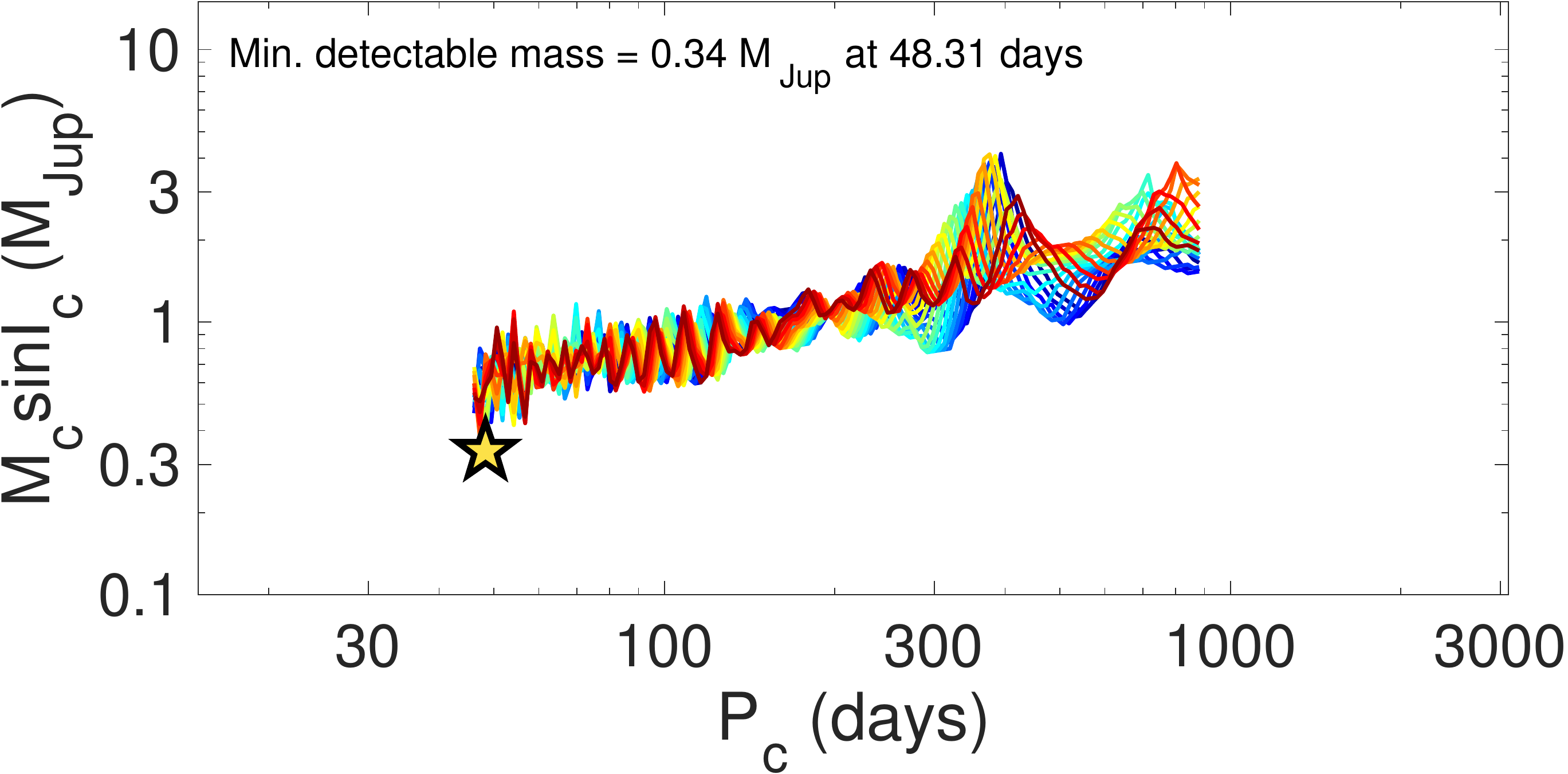}
\end{subfigure}
\end{center}
\end{figure}
\begin{figure}
\begin{center}
\subcaption*{EBLM J0104-38: chosen model = k1d1 (ecc) \newline \newline $m_{\rm A} = 1.38M_{\odot}$, $m_{\rm B} = 0.274M_{\odot}$, $P = 8.256$ d, $e = 0.002$}
\begin{subfigure}[b]{0.49\textwidth}
\includegraphics[width=\textwidth,trim={0 10cm 0 1.2cm},clip]{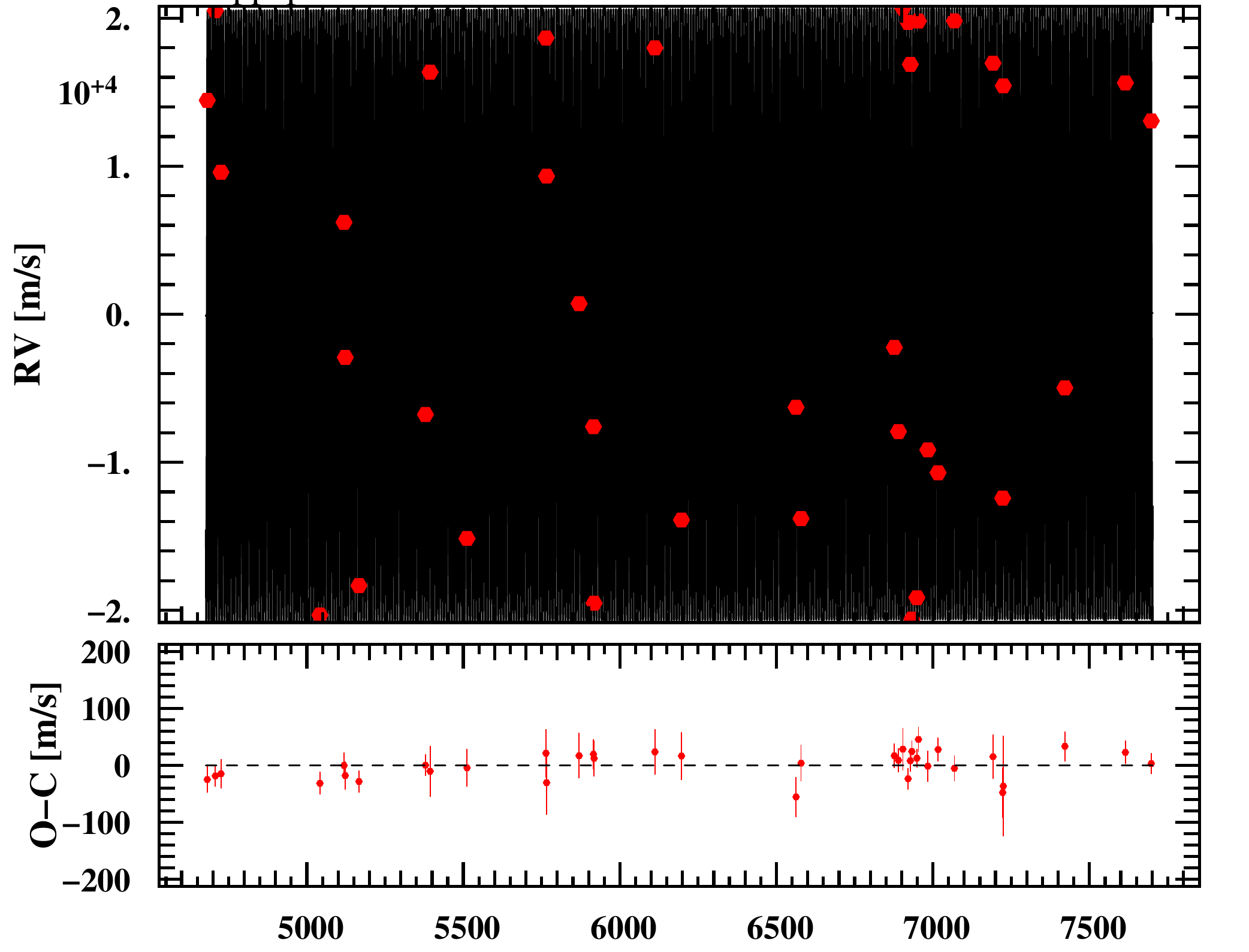}
\end{subfigure}
\begin{subfigure}[b]{0.49\textwidth}
\includegraphics[width=\textwidth,trim={0 0 2cm 0},clip]{orbit_figures/BJD_bar.pdf}
\end{subfigure}
Radial velocities folded on binary phase
\begin{subfigure}[b]{0.49\textwidth}
\includegraphics[width=\textwidth,trim={0 0.5cm 0 0},clip]{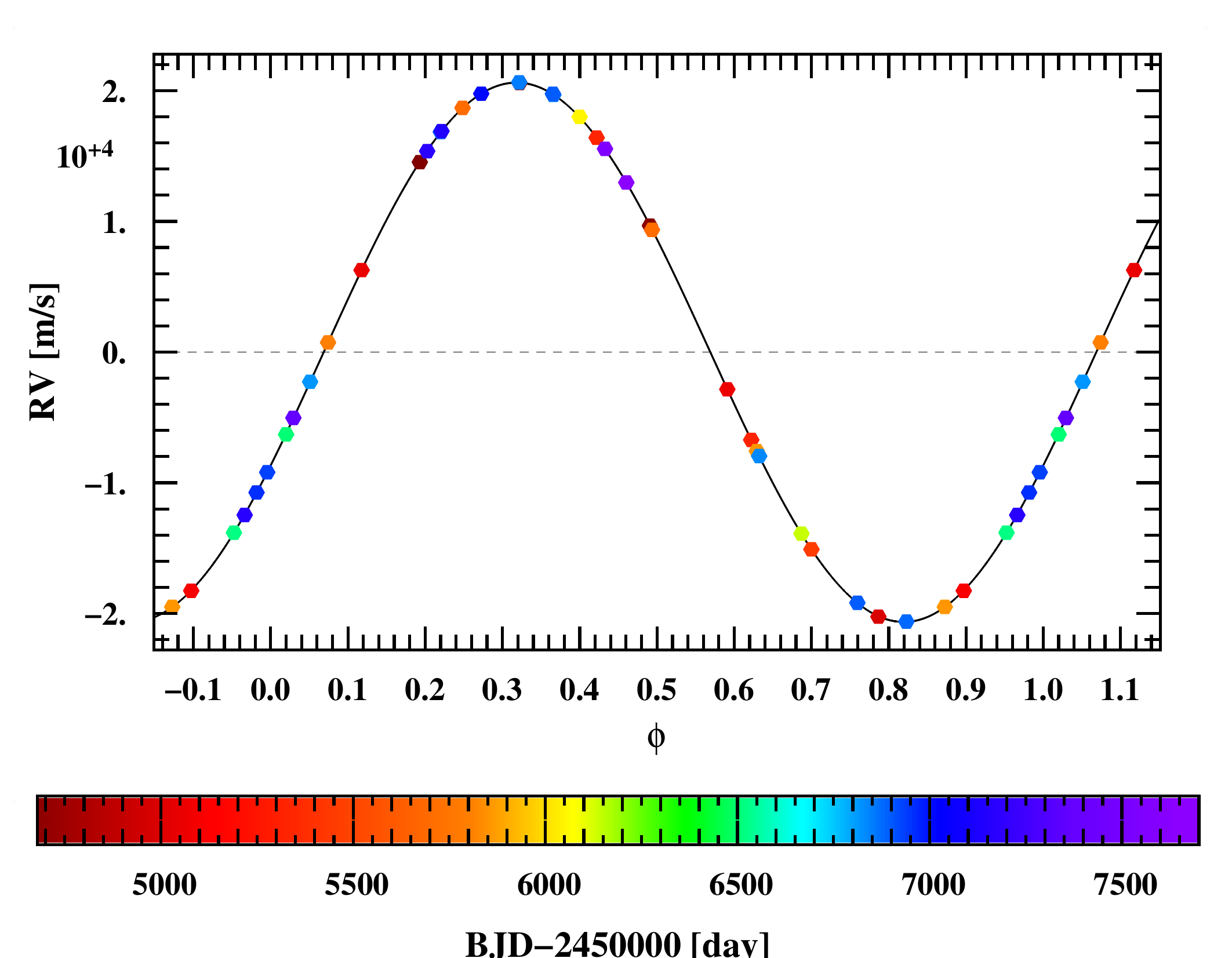}
\end{subfigure}
\begin{subfigure}[b]{0.49\textwidth}
\includegraphics[width=\textwidth,trim={0 0 2cm 0},clip]{orbit_figures/BJD_bar.pdf}
\end{subfigure}
Detection limits
\begin{subfigure}[b]{0.49\textwidth}
\vspace{0.5cm}
\includegraphics[width=\textwidth,trim={0 0 0 0},clip]{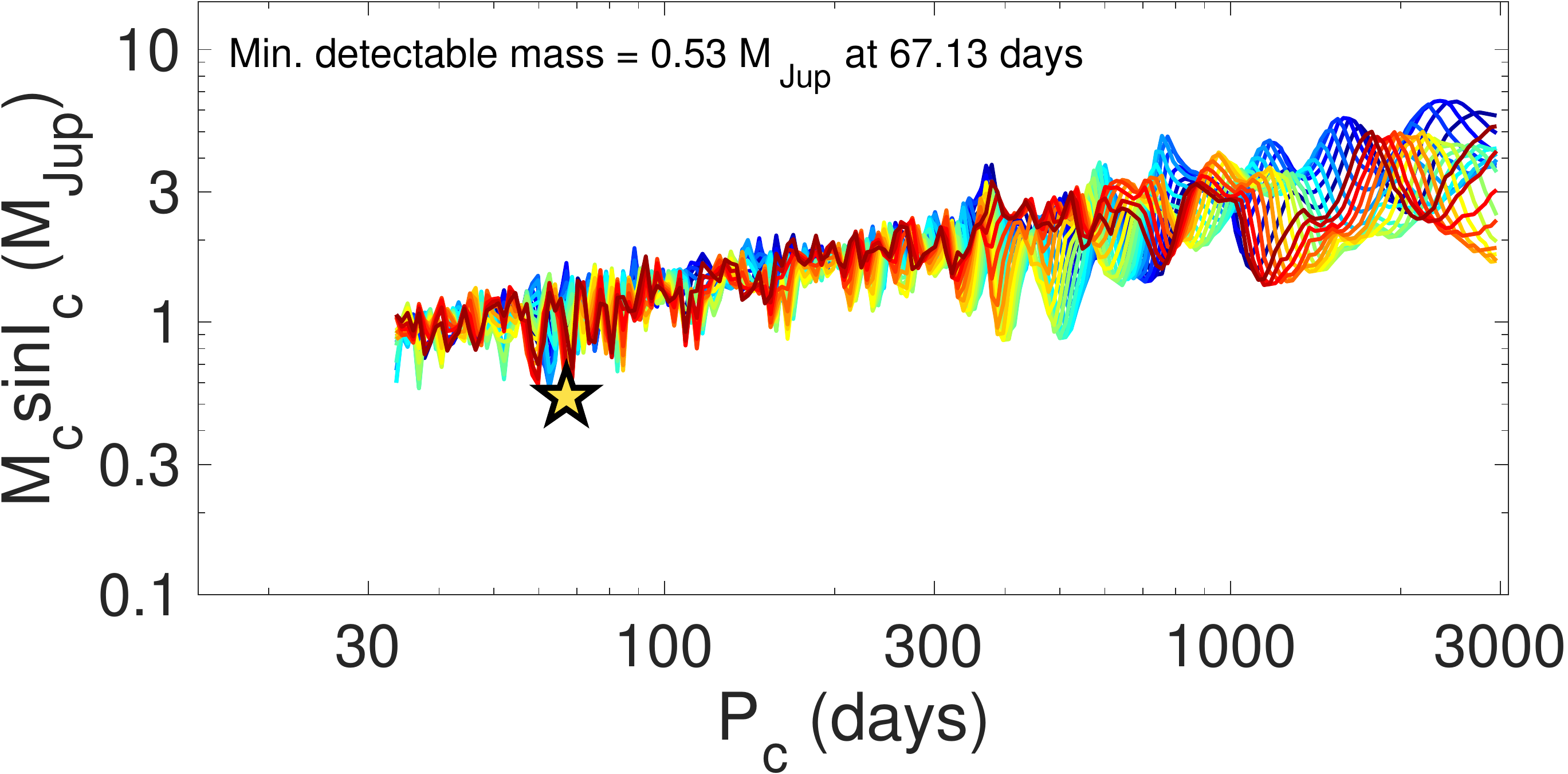}
\end{subfigure}
\end{center}
\end{figure}
\begin{figure}
\begin{center}
\subcaption*{EBLM J0218-31: chosen model = k1d2 (circ) \newline \newline $m_{\rm A} = 1.17M_{\odot}$, $m_{\rm B} = 0.359M_{\odot}$, $P = 8.884$ d, $e = 0$}
\begin{subfigure}[b]{0.49\textwidth}
\includegraphics[width=\textwidth,trim={0 10cm 0 1.2cm},clip]{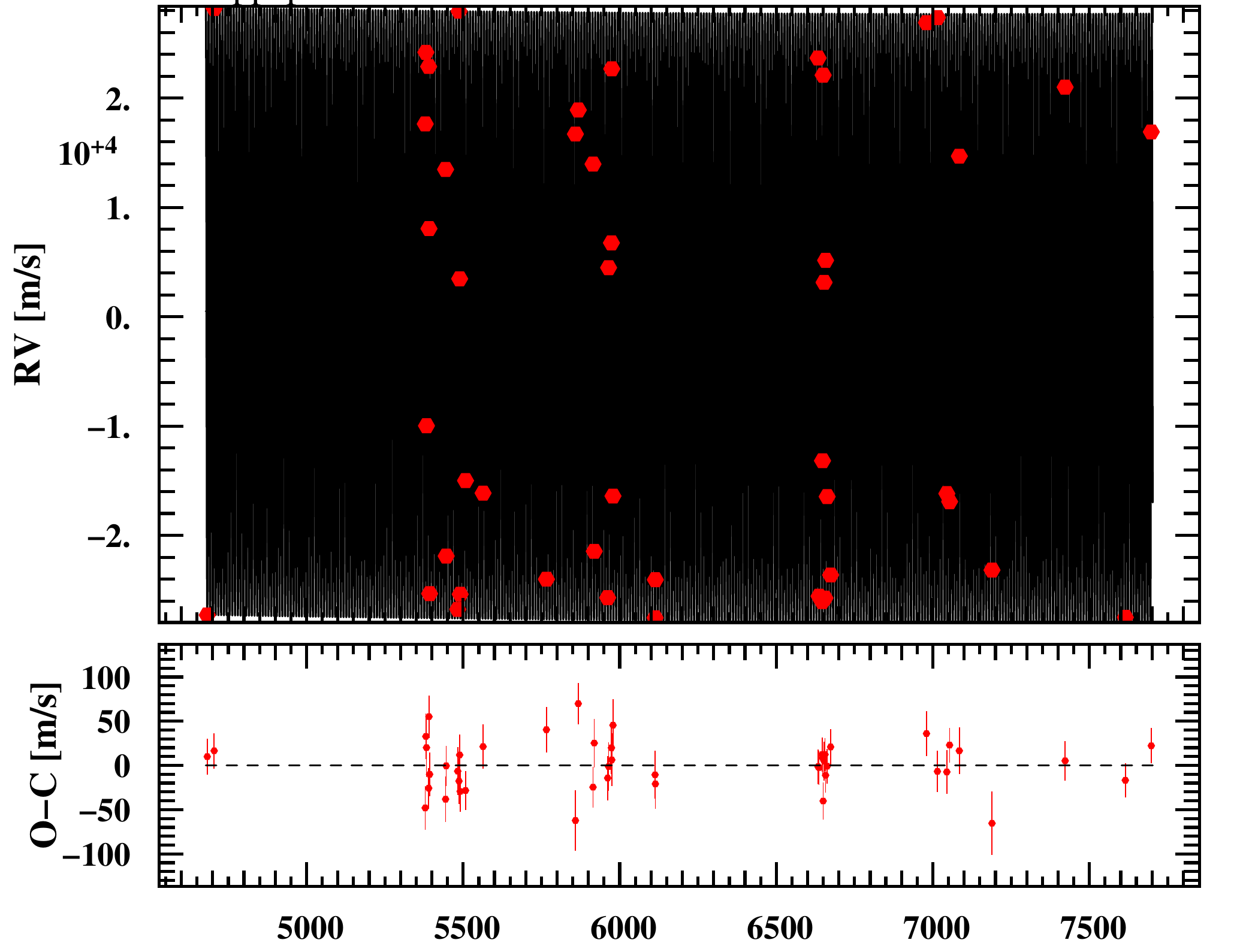}
\end{subfigure}
\begin{subfigure}[b]{0.49\textwidth}
\includegraphics[width=\textwidth,trim={0 0 2cm 0},clip]{orbit_figures/BJD_bar.pdf}
\end{subfigure}
Radial velocities folded on binary phase
\begin{subfigure}[b]{0.49\textwidth}
\includegraphics[width=\textwidth,trim={0 0.5cm 0 0},clip]{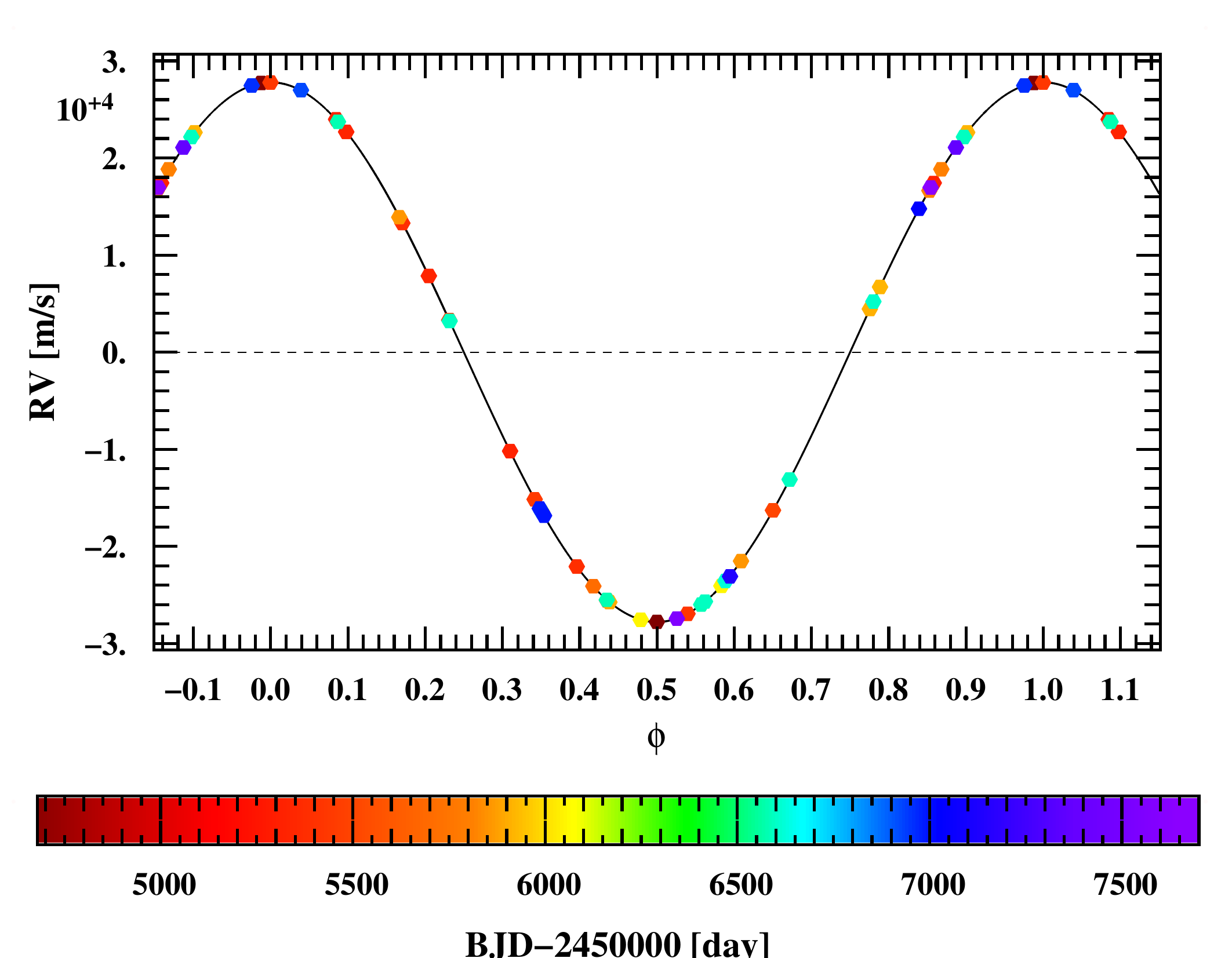}
\end{subfigure}
\begin{subfigure}[b]{0.49\textwidth}
\includegraphics[width=\textwidth,trim={0 0 2cm 0},clip]{orbit_figures/BJD_bar.pdf}
\end{subfigure}
Detection limits
\begin{subfigure}[b]{0.49\textwidth}
\vspace{0.5cm}
\includegraphics[width=\textwidth,trim={0 0 0 0},clip]{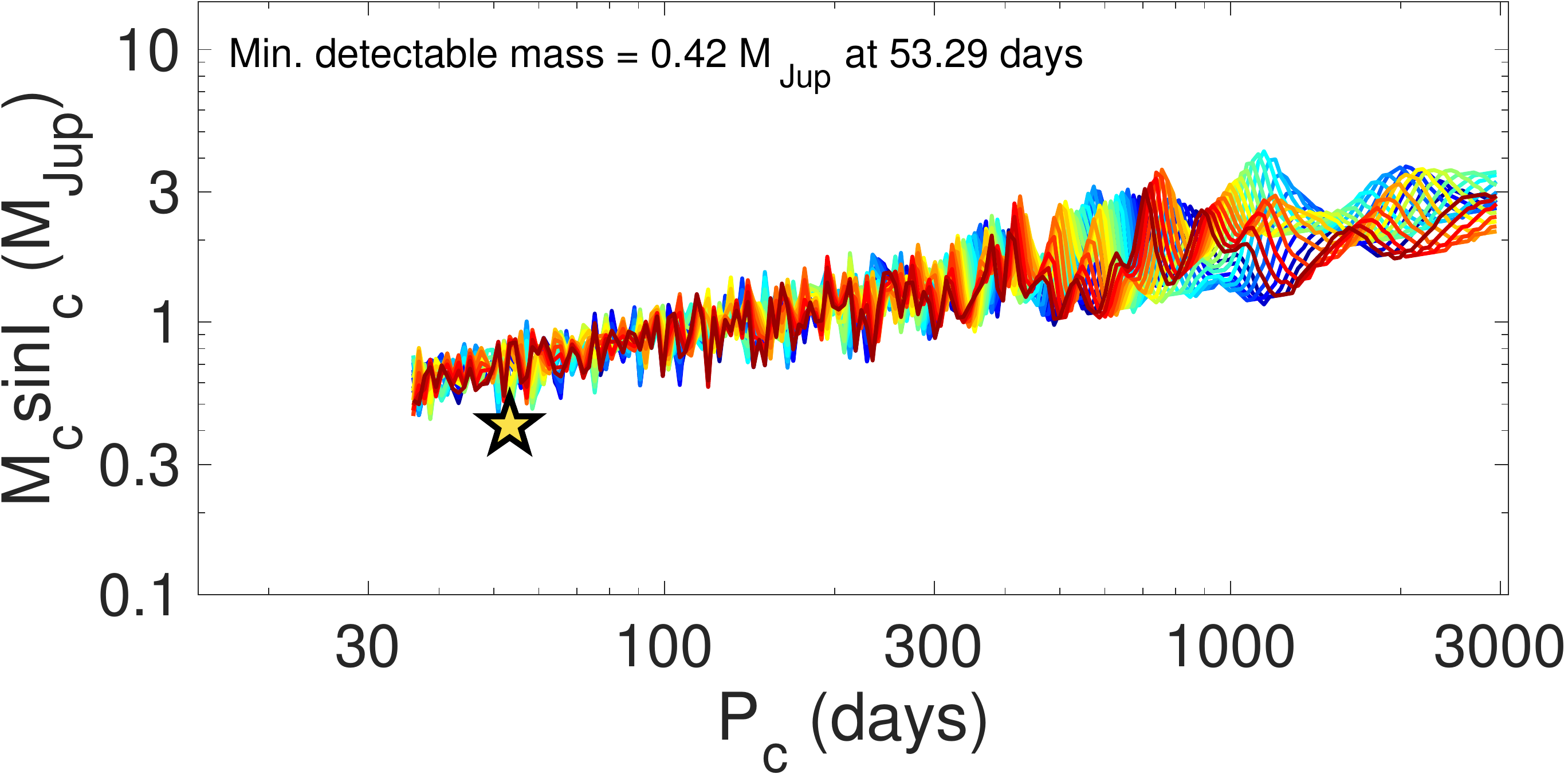}
\end{subfigure}
\end{center}
\end{figure}
\begin{figure}
\begin{center}
\subcaption*{EBLM J0228+05: chosen model = k1 (circ) \newline \newline $m_{\rm A} = 1.53M_{\odot}$, $m_{\rm B} = 0.18M_{\odot}$, $P = 6.635$ d, $e = 0$}
\begin{subfigure}[b]{0.49\textwidth}
\includegraphics[width=\textwidth,trim={0 10cm 0 1.2cm},clip]{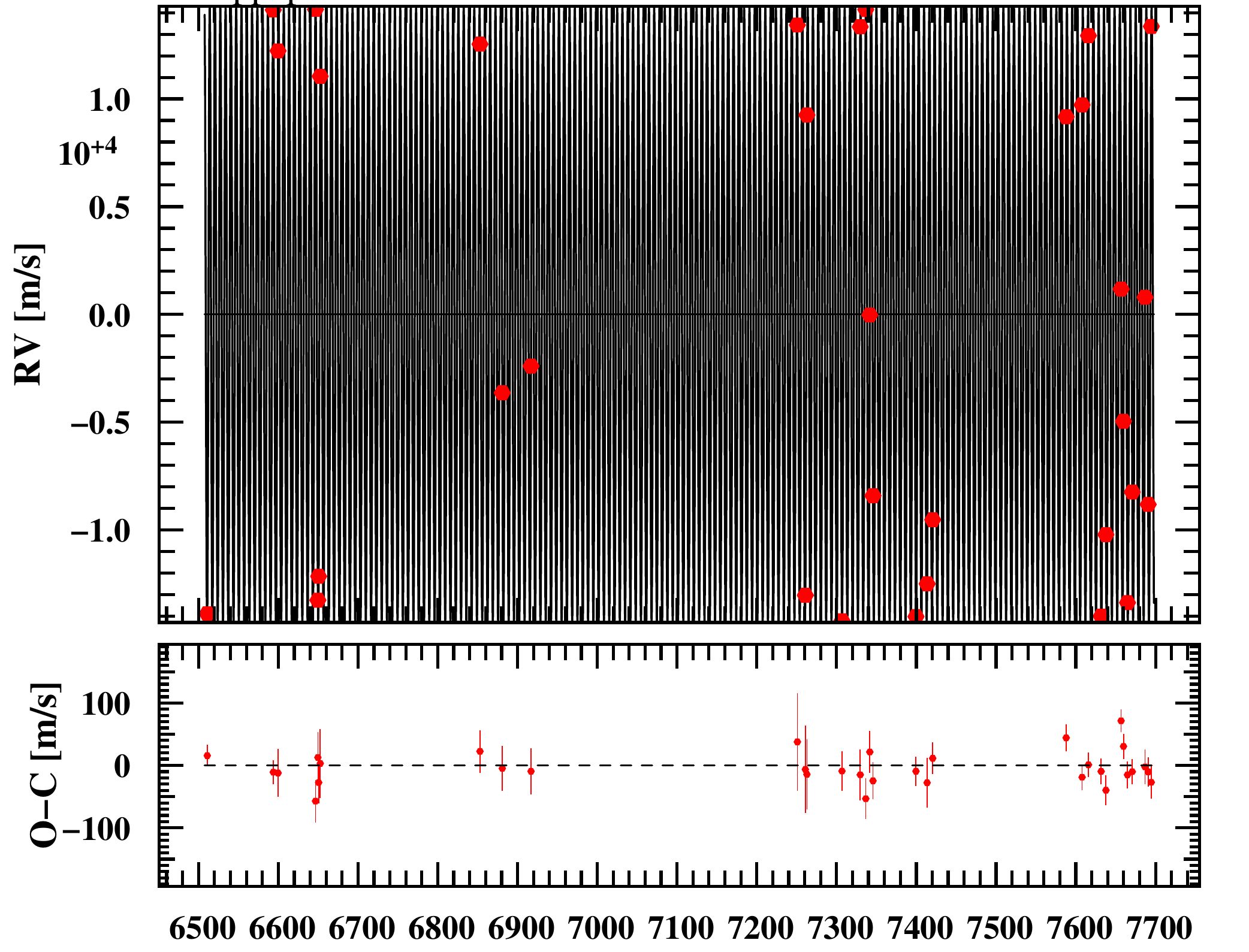}
\end{subfigure}
\begin{subfigure}[b]{0.49\textwidth}
\includegraphics[width=\textwidth,trim={0 0 2cm 0},clip]{orbit_figures/BJD_bar.pdf}
\end{subfigure}
Radial velocities folded on binary phase
\begin{subfigure}[b]{0.49\textwidth}
\includegraphics[width=\textwidth,trim={0 0.5cm 0 0},clip]{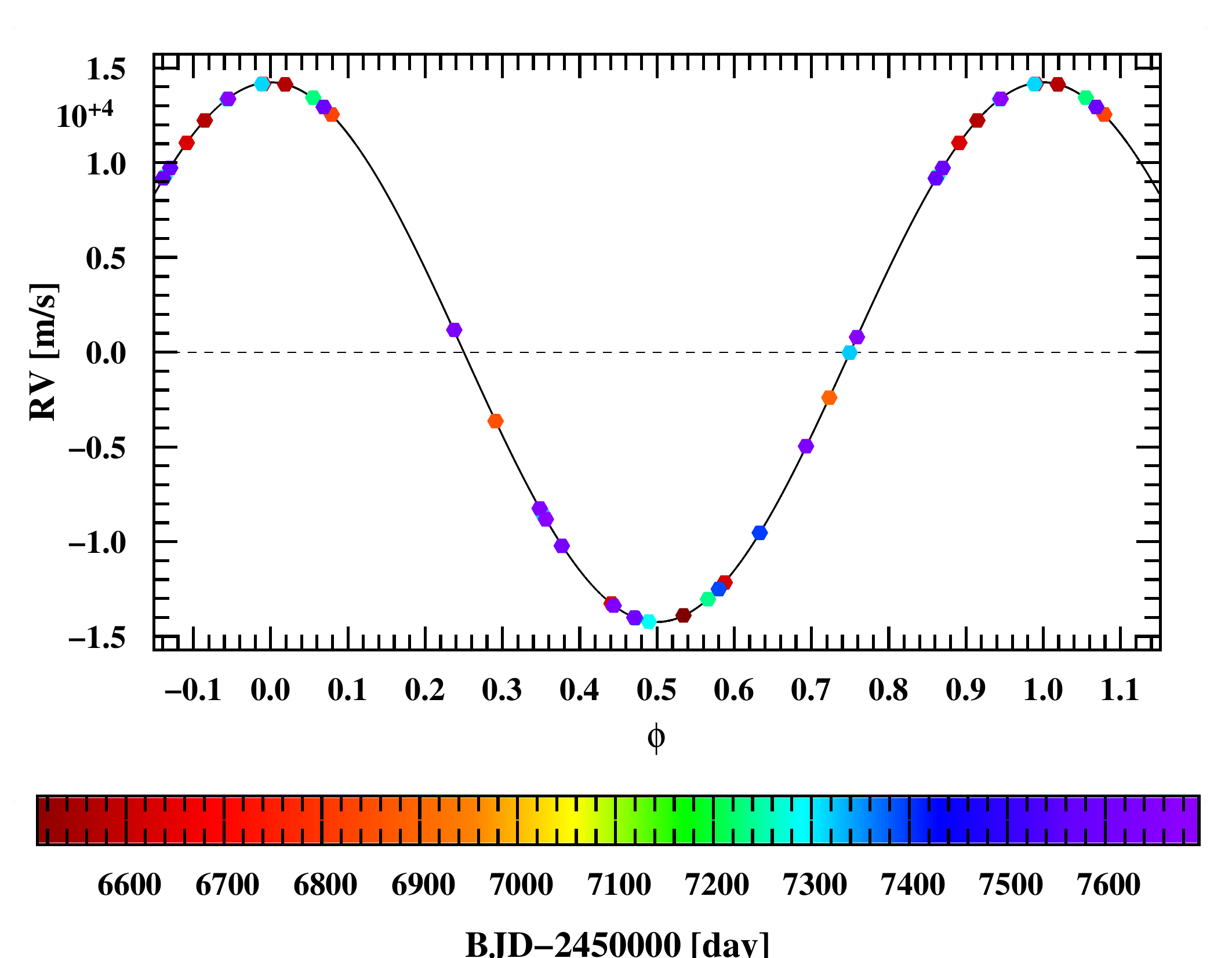}
\end{subfigure}
\begin{subfigure}[b]{0.49\textwidth}
\includegraphics[width=\textwidth,trim={0 0 2cm 0},clip]{orbit_figures/BJD_bar.pdf}
\end{subfigure}
Detection limits
\begin{subfigure}[b]{0.49\textwidth}
\vspace{0.5cm}
\includegraphics[width=\textwidth,trim={0 0 0 0},clip]{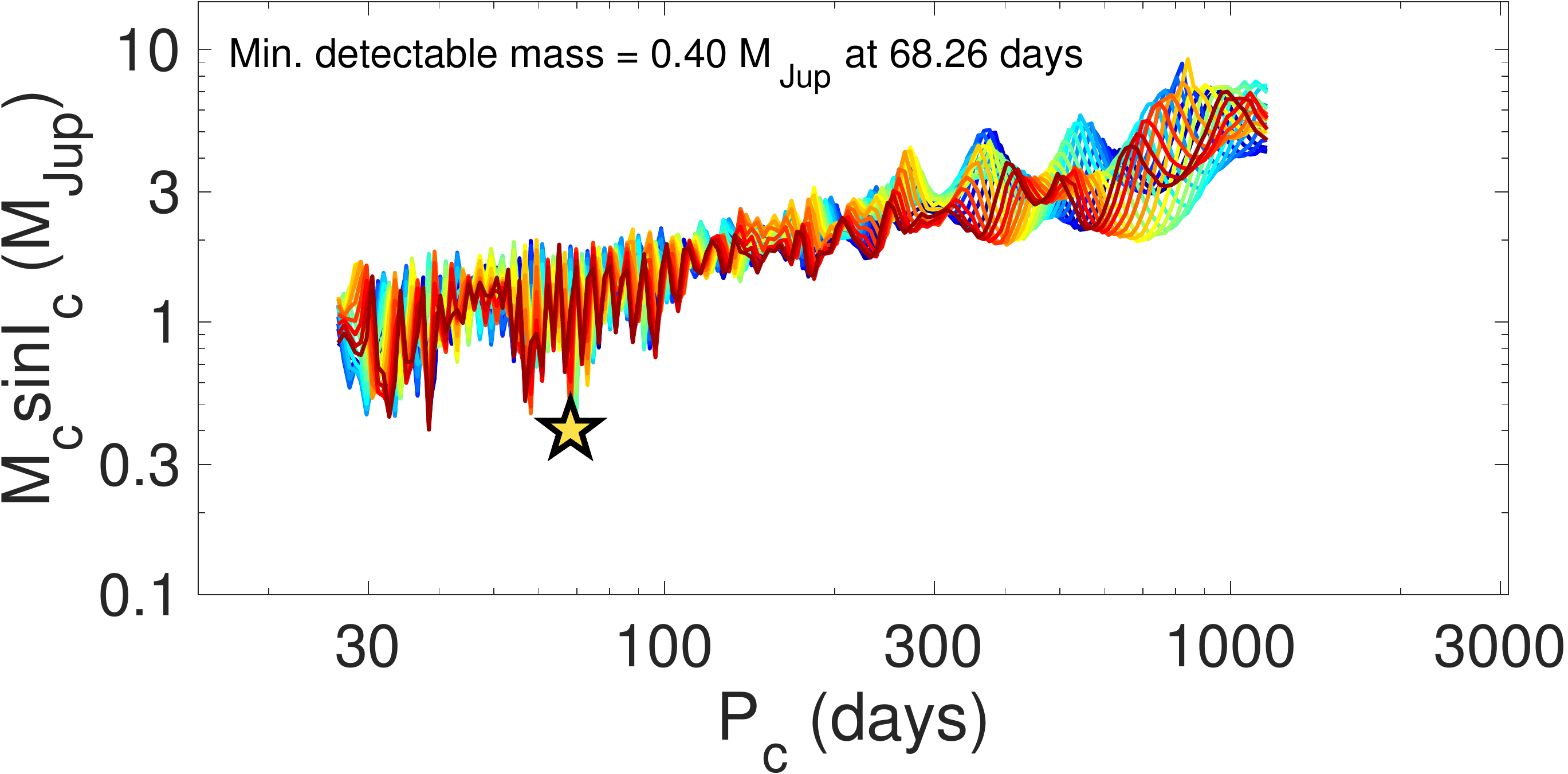}
\end{subfigure}
\end{center}
\end{figure}
\begin{figure}
\begin{center}
\subcaption*{EBLM J0310-31: chosen model = k1 (ecc) \newline \newline $m_{\rm A} = 1.26M_{\odot}$, $m_{\rm B} = 0.408M_{\odot}$, $P = 12.643$ d, $e = 0.309$}
\begin{subfigure}[b]{0.49\textwidth}
\includegraphics[width=\textwidth,trim={0 10cm 0 1.2cm},clip]{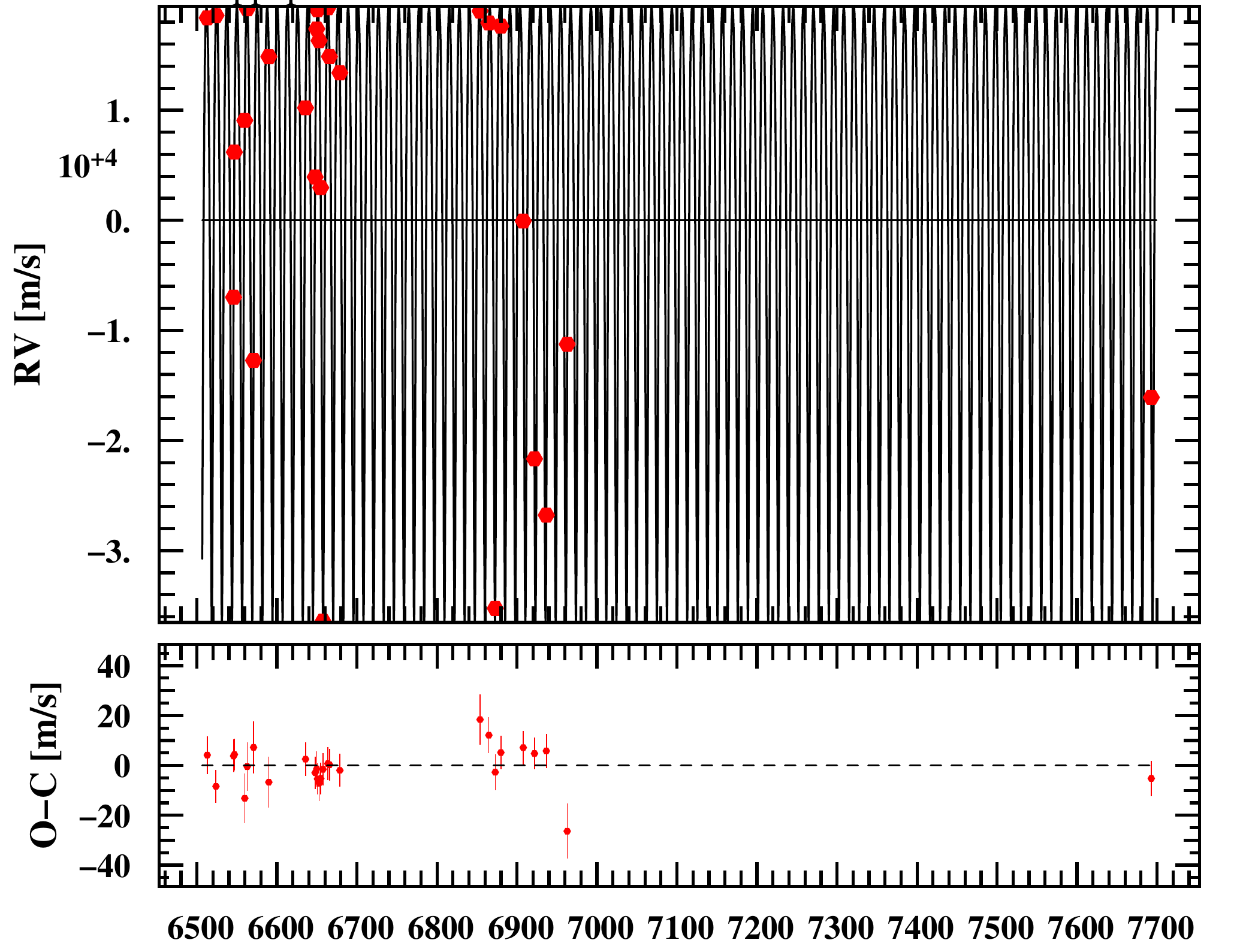}
\end{subfigure}
\begin{subfigure}[b]{0.49\textwidth}
\includegraphics[width=\textwidth,trim={0 0 2cm 0},clip]{orbit_figures/BJD_bar.pdf}
\end{subfigure}
Radial velocities folded on binary phase
\begin{subfigure}[b]{0.49\textwidth}
\includegraphics[width=\textwidth,trim={0 0.5cm 0 0},clip]{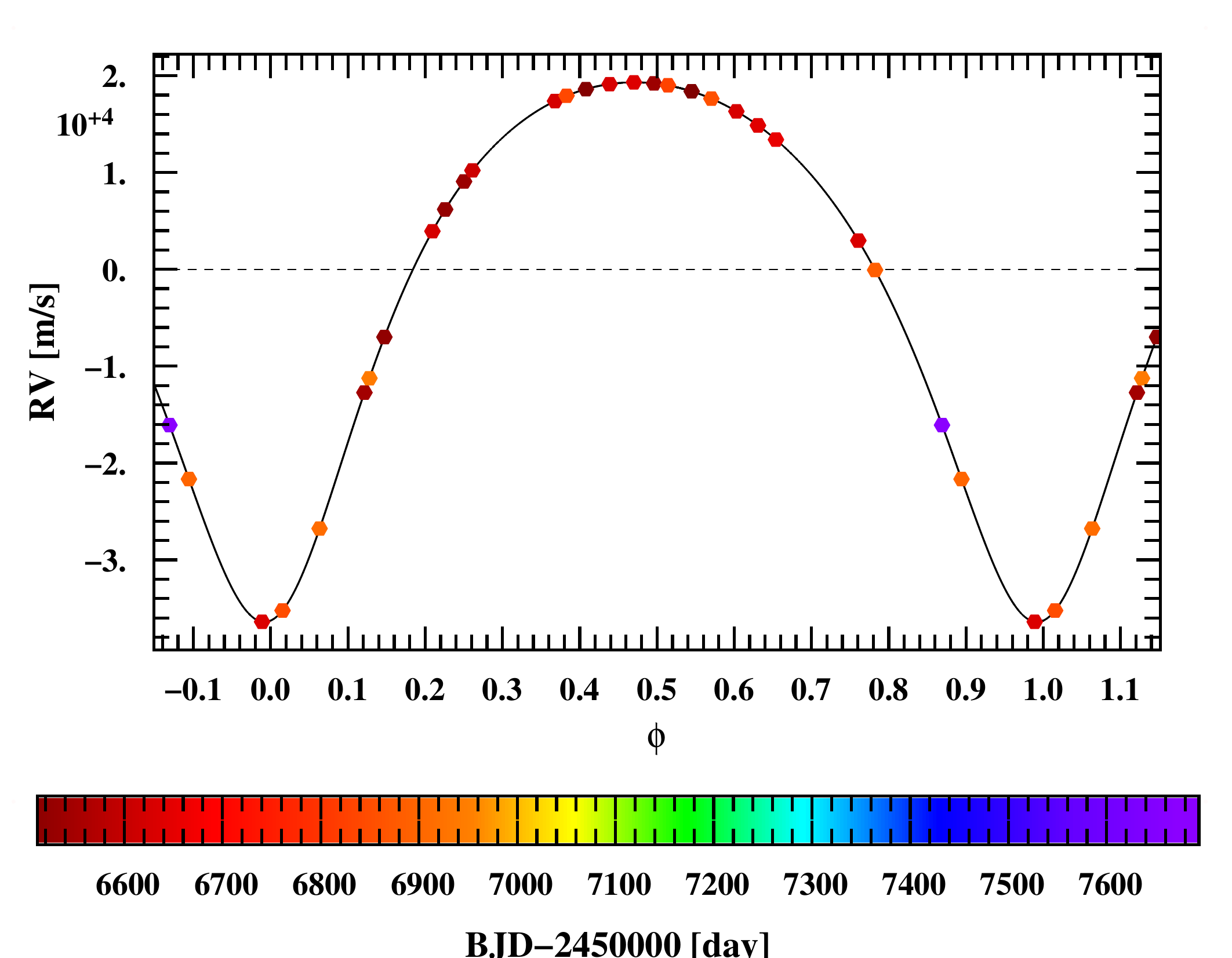}
\end{subfigure}
\begin{subfigure}[b]{0.49\textwidth}
\includegraphics[width=\textwidth,trim={0 0 2cm 0},clip]{orbit_figures/BJD_bar.pdf}
\end{subfigure}
Detection limits
\begin{subfigure}[b]{0.49\textwidth}
\vspace{0.5cm}
\includegraphics[width=\textwidth,trim={0 0 0 0},clip]{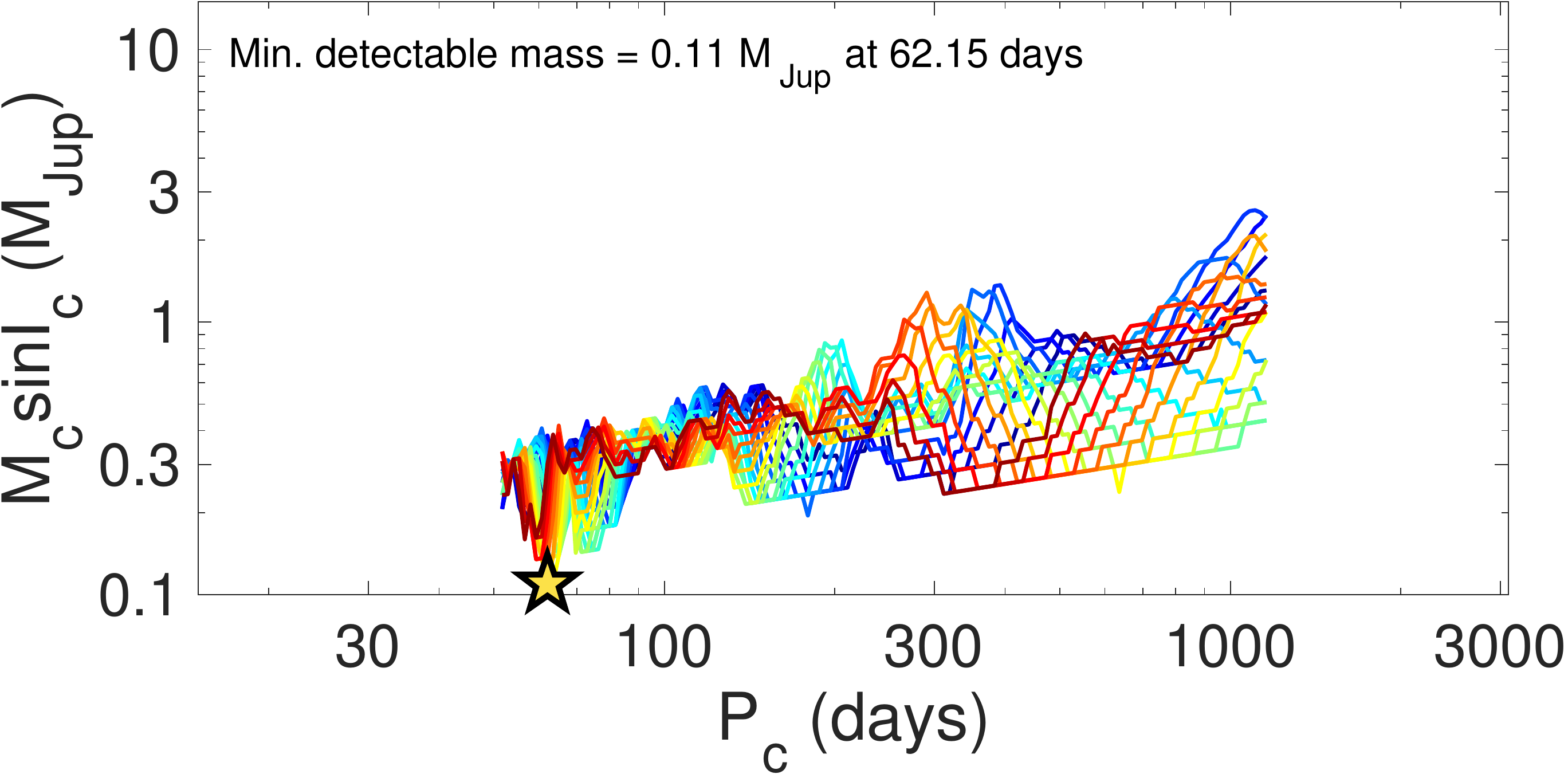}
\end{subfigure}
\end{center}
\end{figure}
\begin{figure}
\begin{center}
\subcaption*{EBLM J0345-10: chosen model = k1 (ecc) \newline \newline $m_{\rm A} = 1.21M_{\odot}$, $m_{\rm B} = 0.525M_{\odot}$, $P = 6.061$ d, $e = 0.004$}
\begin{subfigure}[b]{0.49\textwidth}
\includegraphics[width=\textwidth,trim={0 10cm 0 1.2cm},clip]{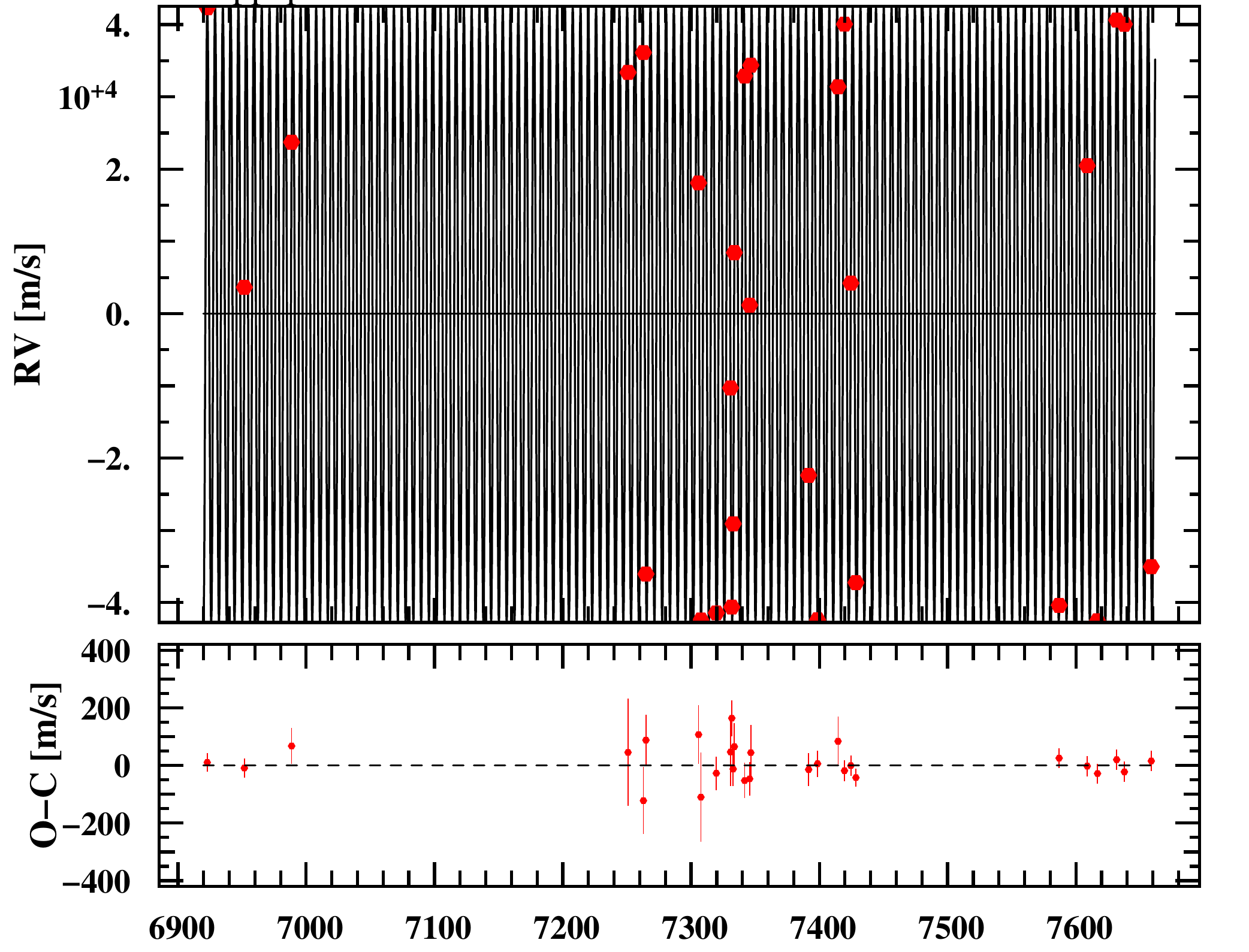}
\end{subfigure}
\begin{subfigure}[b]{0.49\textwidth}
\includegraphics[width=\textwidth,trim={0 0 2cm 0},clip]{orbit_figures/BJD_bar.pdf}
\end{subfigure}
Radial velocities folded on binary phase
\begin{subfigure}[b]{0.49\textwidth}
\includegraphics[width=\textwidth,trim={0 0.5cm 0 0},clip]{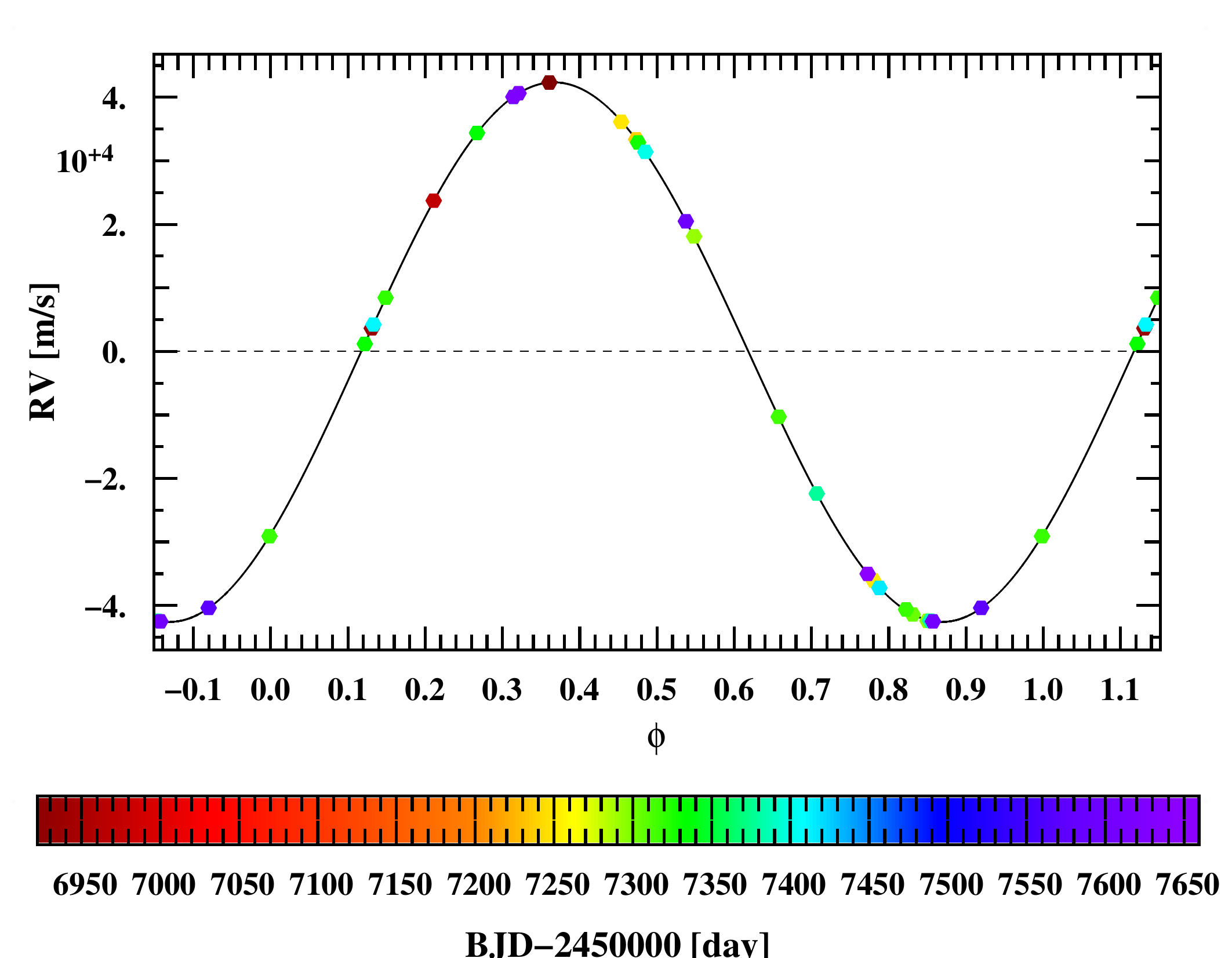}
\end{subfigure}
\begin{subfigure}[b]{0.49\textwidth}
\includegraphics[width=\textwidth,trim={0 0 2cm 0},clip]{orbit_figures/BJD_bar.pdf}
\end{subfigure}
Detection limits
\begin{subfigure}[b]{0.49\textwidth}
\vspace{0.5cm}
\includegraphics[width=\textwidth,trim={0 0 0 0},clip]{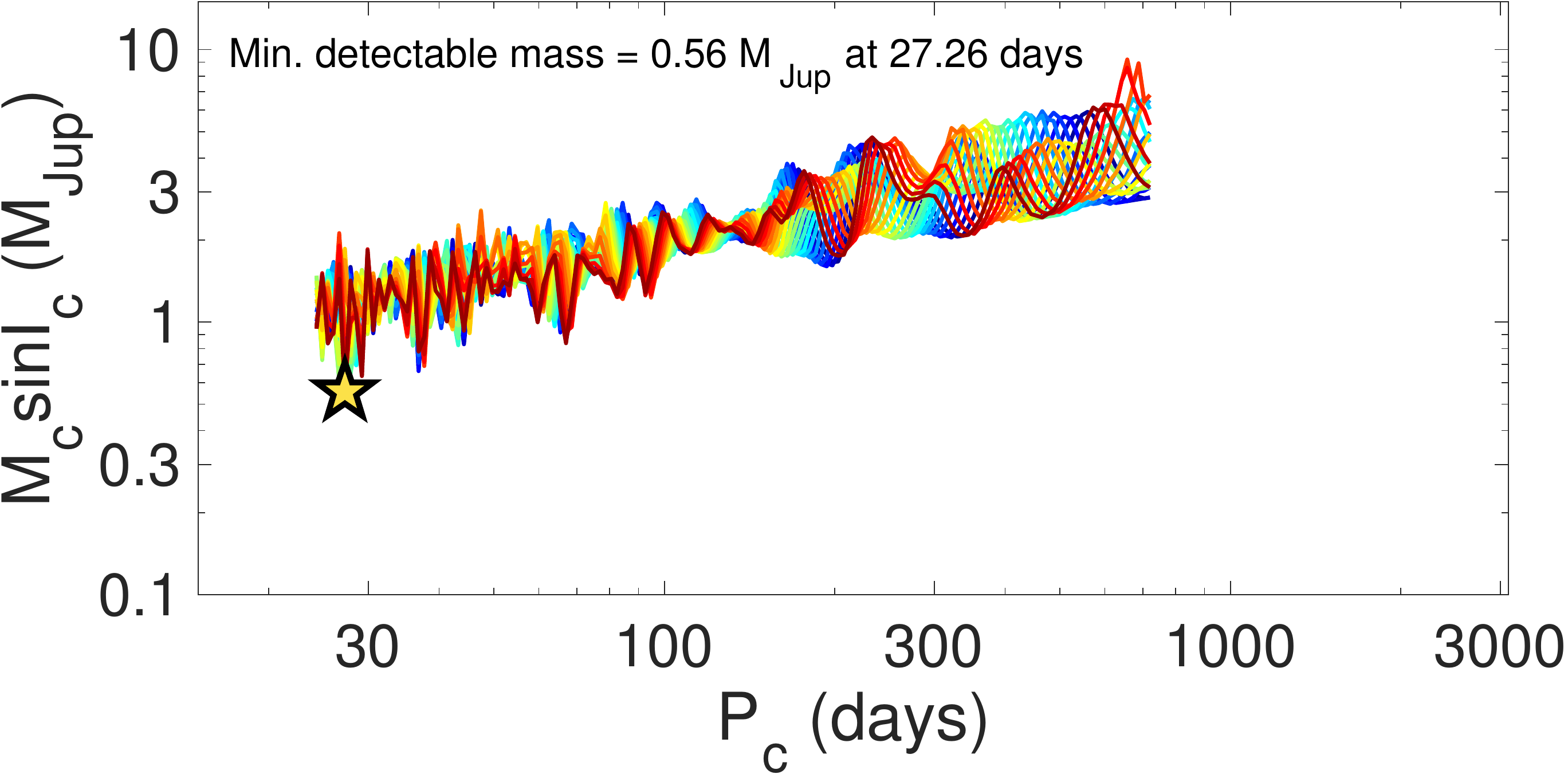}
\end{subfigure}
\end{center}
\end{figure}
\begin{figure}
\begin{center}
\subcaption*{EBLM J0353+05: chosen model = k1d3 (ecc) \newline \newline $m_{\rm A} = 1.19M_{\odot}$, $m_{\rm B} = 0.179M_{\odot}$, $P = 6.862$ d, $e = 0.001$}
\begin{subfigure}[b]{0.49\textwidth}
\includegraphics[width=\textwidth,trim={0 10cm 0 1.2cm},clip]{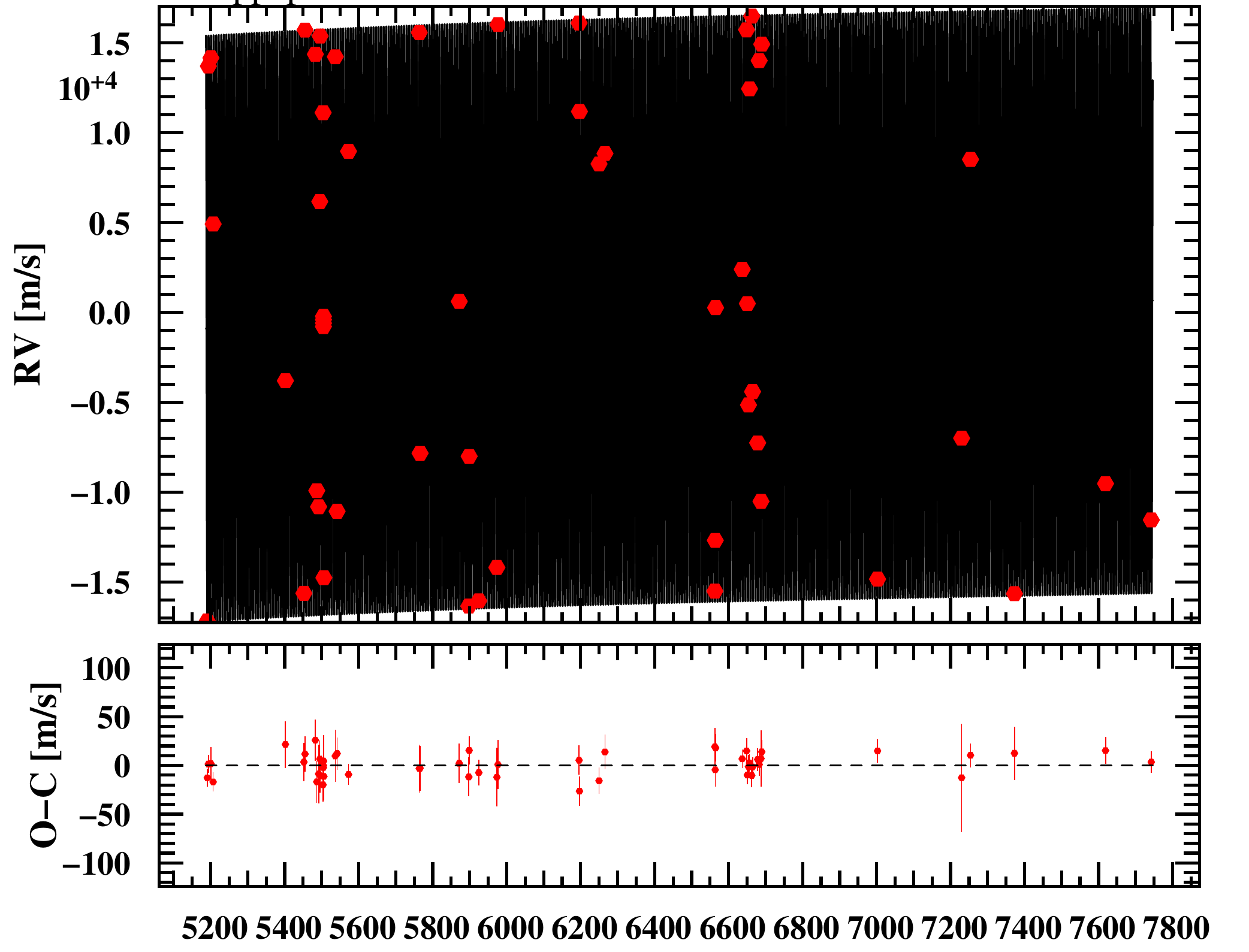}
\end{subfigure}
\begin{subfigure}[b]{0.49\textwidth}
\includegraphics[width=\textwidth,trim={0 0 2cm 0},clip]{orbit_figures/BJD_bar.pdf}
\end{subfigure}
Radial velocities folded on binary phase
\begin{subfigure}[b]{0.49\textwidth}
\includegraphics[width=\textwidth,trim={0 0.5cm 0 0},clip]{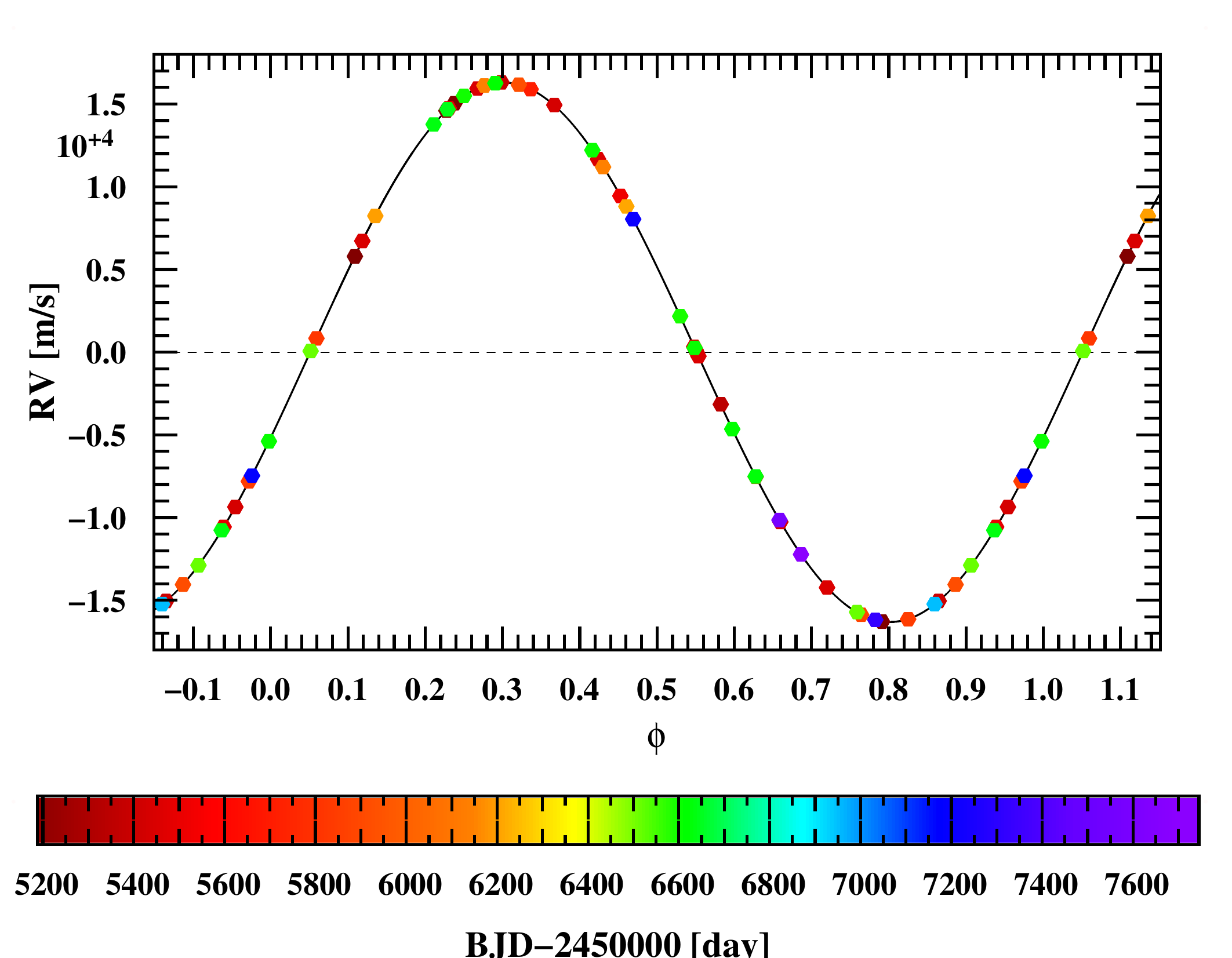}
\end{subfigure}
\begin{subfigure}[b]{0.49\textwidth}
\includegraphics[width=\textwidth,trim={0 0 2cm 0},clip]{orbit_figures/BJD_bar.pdf}
\end{subfigure}
Detection limits
\begin{subfigure}[b]{0.49\textwidth}
\vspace{0.5cm}
\includegraphics[width=\textwidth,trim={0 0 0 0},clip]{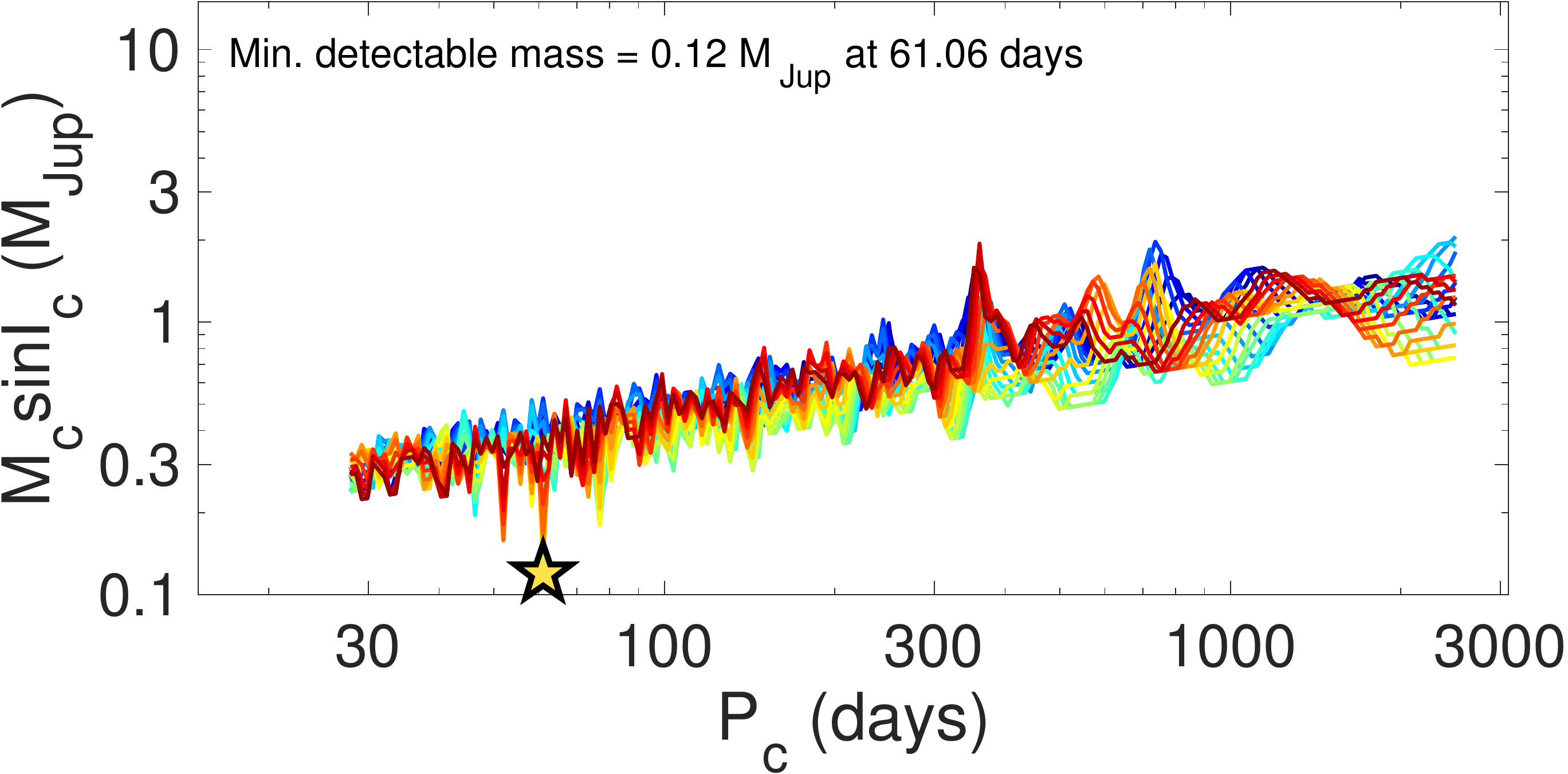}
\end{subfigure}
\end{center}
\end{figure}
\begin{figure}
\begin{center}
\subcaption*{EBLM J0353-16: chosen model = k1 (ecc) \newline \newline $m_{\rm A} = 1.17M_{\odot}$, $m_{\rm B} = 0.222M_{\odot}$, $P = 11.761$ d, $e = 0.006$}
\begin{subfigure}[b]{0.49\textwidth}
\includegraphics[width=\textwidth,trim={0 10cm 0 1.2cm},clip]{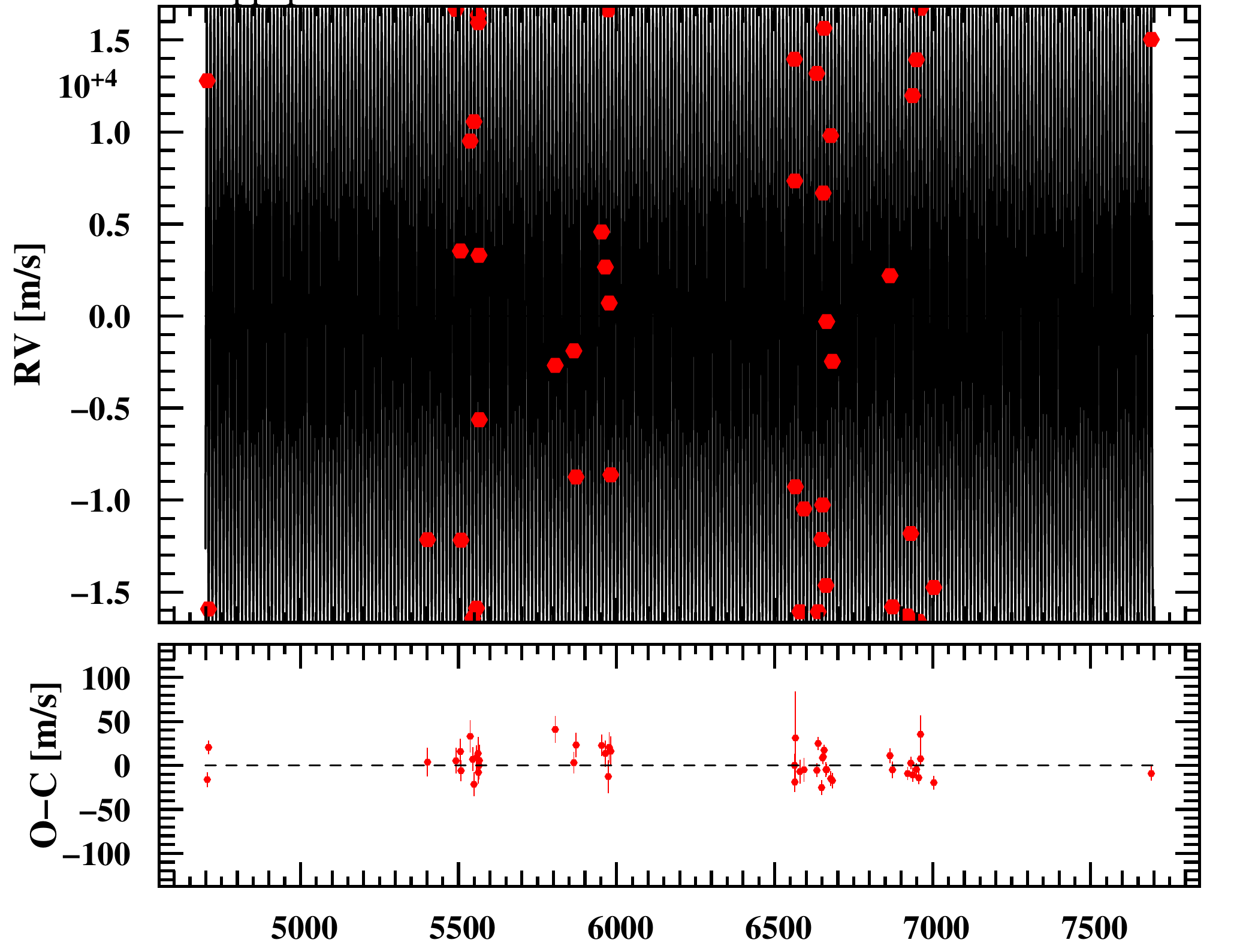}
\end{subfigure}
\begin{subfigure}[b]{0.49\textwidth}
\includegraphics[width=\textwidth,trim={0 0 2cm 0},clip]{orbit_figures/BJD_bar.pdf}
\end{subfigure}
Radial velocities folded on binary phase
\begin{subfigure}[b]{0.49\textwidth}
\includegraphics[width=\textwidth,trim={0 0.5cm 0 0},clip]{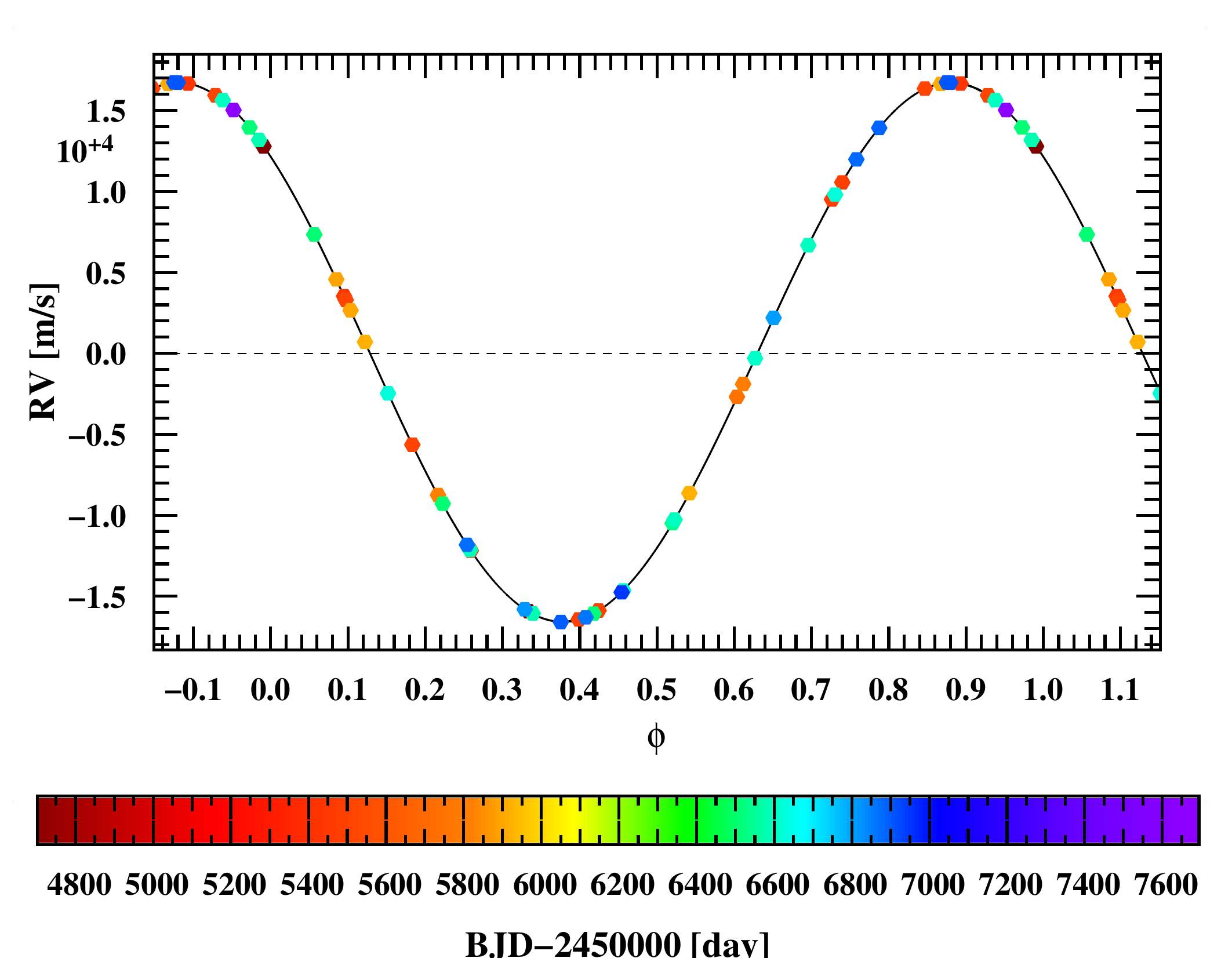}
\end{subfigure}
\begin{subfigure}[b]{0.49\textwidth}
\includegraphics[width=\textwidth,trim={0 0 2cm 0},clip]{orbit_figures/BJD_bar.pdf}
\end{subfigure}
Detection limits
\begin{subfigure}[b]{0.49\textwidth}
\vspace{0.5cm}
\includegraphics[width=\textwidth,trim={0 0 0 0},clip]{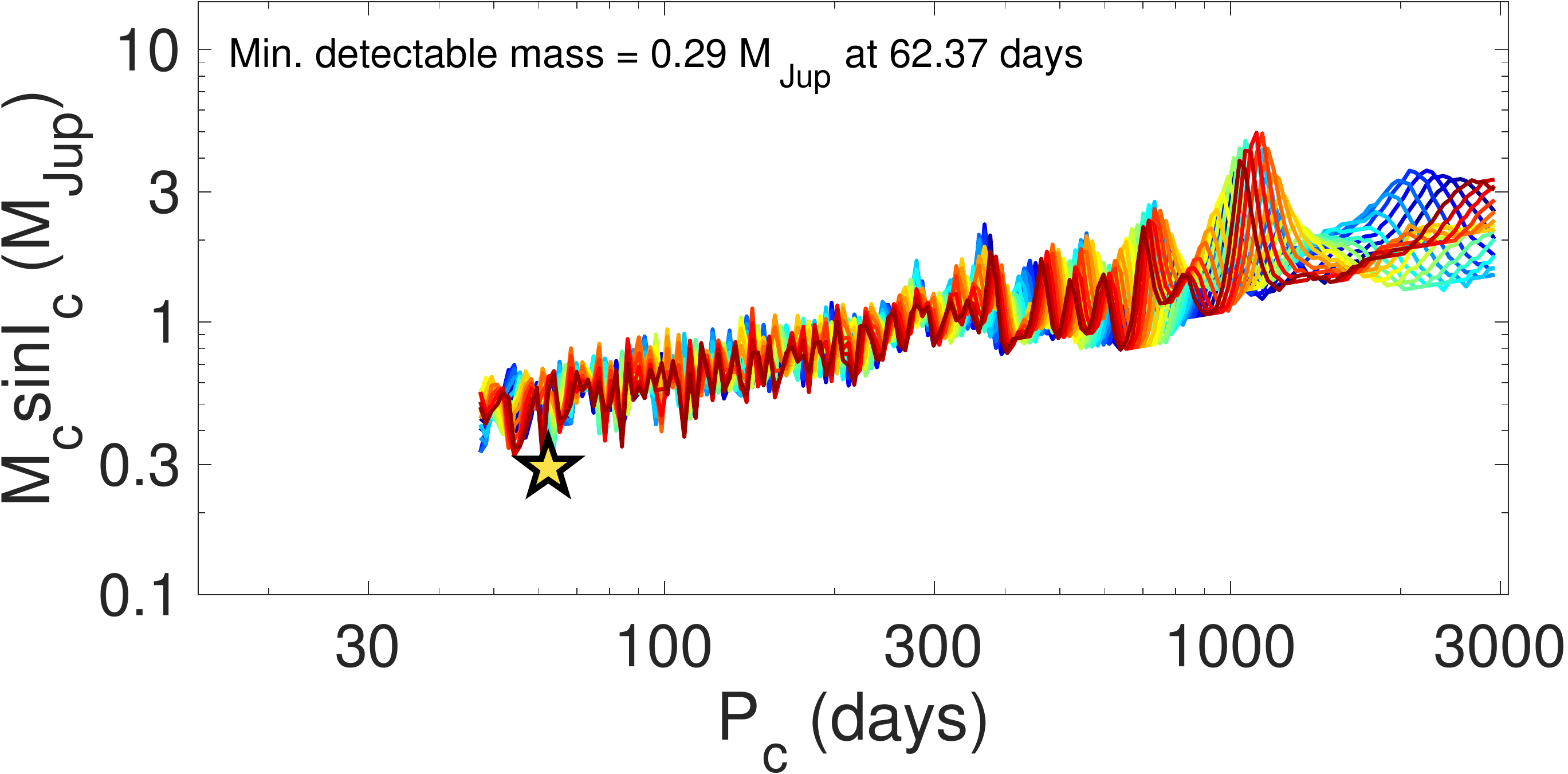}
\end{subfigure}
\end{center}
\end{figure}
\begin{figure}
\begin{center}
\subcaption*{EBLM J0418-53: chosen model = k1 (circ) \newline \newline $m_{\rm A} = 1.11M_{\odot}$, $m_{\rm B} = 0.135M_{\odot}$, $P = 13.699$ d, $e = 0$}
\begin{subfigure}[b]{0.49\textwidth}
\includegraphics[width=\textwidth,trim={0 10cm 0 1.2cm},clip]{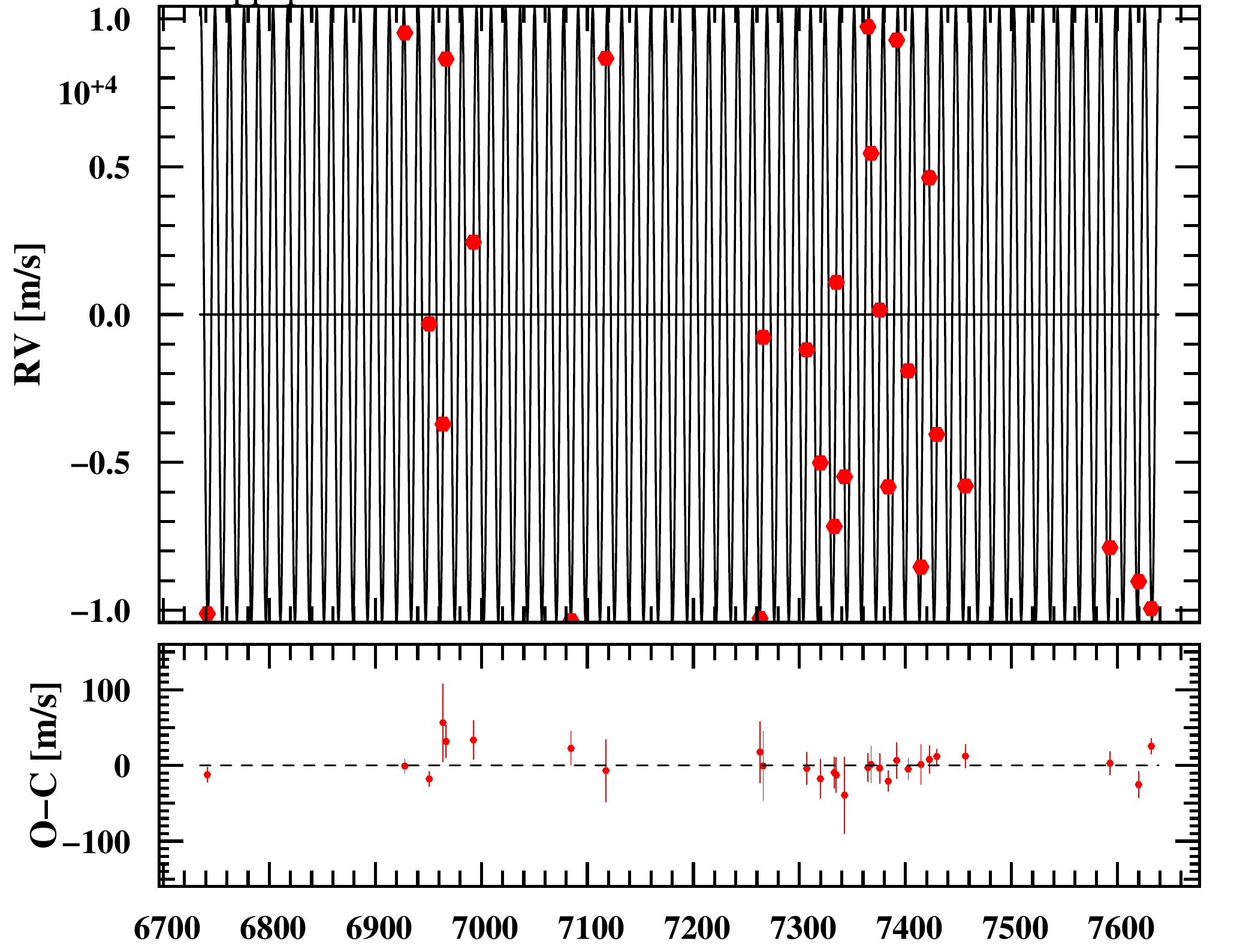}
\end{subfigure}
\begin{subfigure}[b]{0.49\textwidth}
\includegraphics[width=\textwidth,trim={0 0 2cm 0},clip]{orbit_figures/BJD_bar.pdf}
\end{subfigure}
Radial velocities folded on binary phase
\begin{subfigure}[b]{0.49\textwidth}
\includegraphics[width=\textwidth,trim={0 0.5cm 0 0},clip]{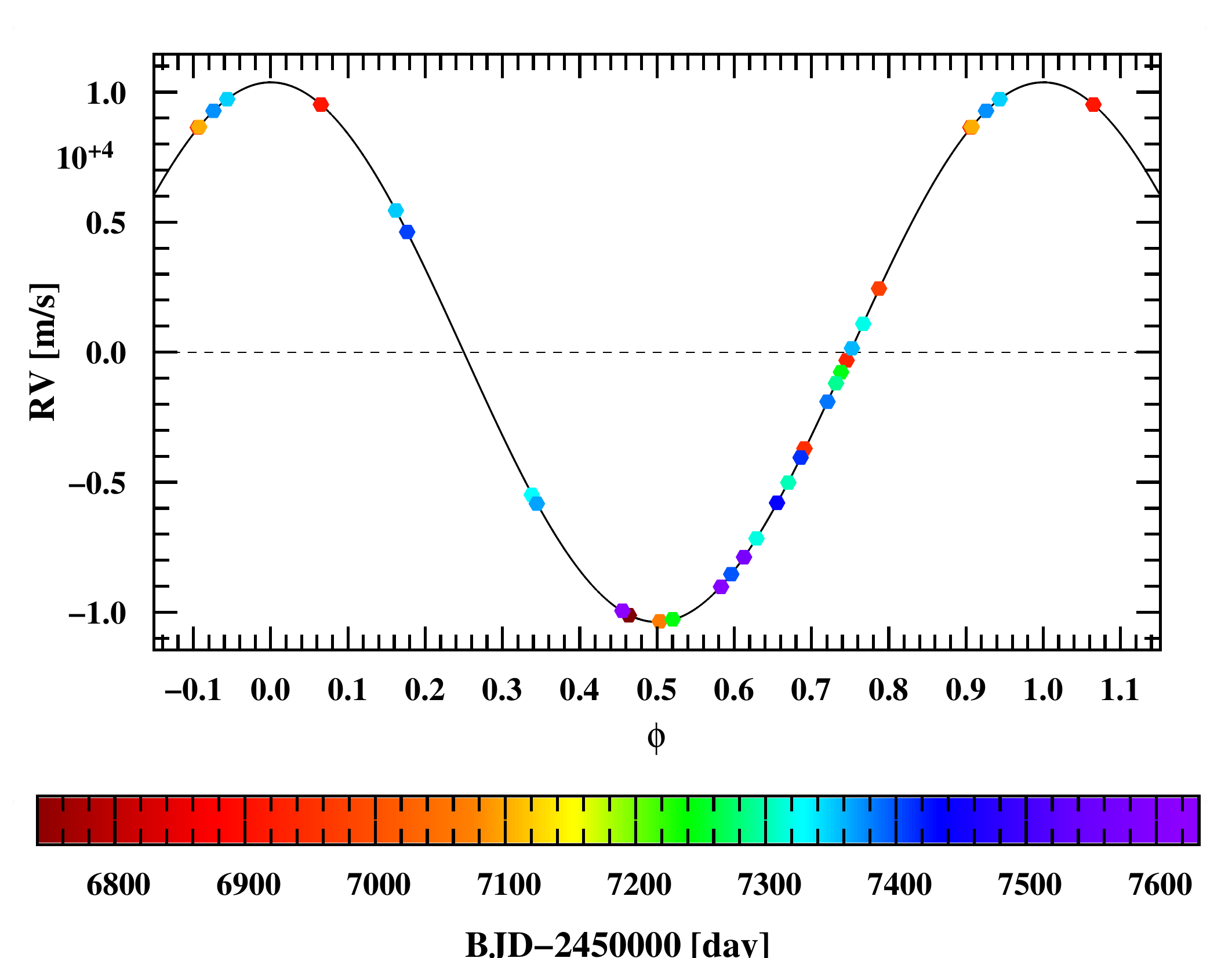}
\end{subfigure}
\begin{subfigure}[b]{0.49\textwidth}
\includegraphics[width=\textwidth,trim={0 0 2cm 0},clip]{orbit_figures/BJD_bar.pdf}
\end{subfigure}
Detection limits
\begin{subfigure}[b]{0.49\textwidth}
\vspace{0.5cm}
\includegraphics[width=\textwidth,trim={0 0 0 0},clip]{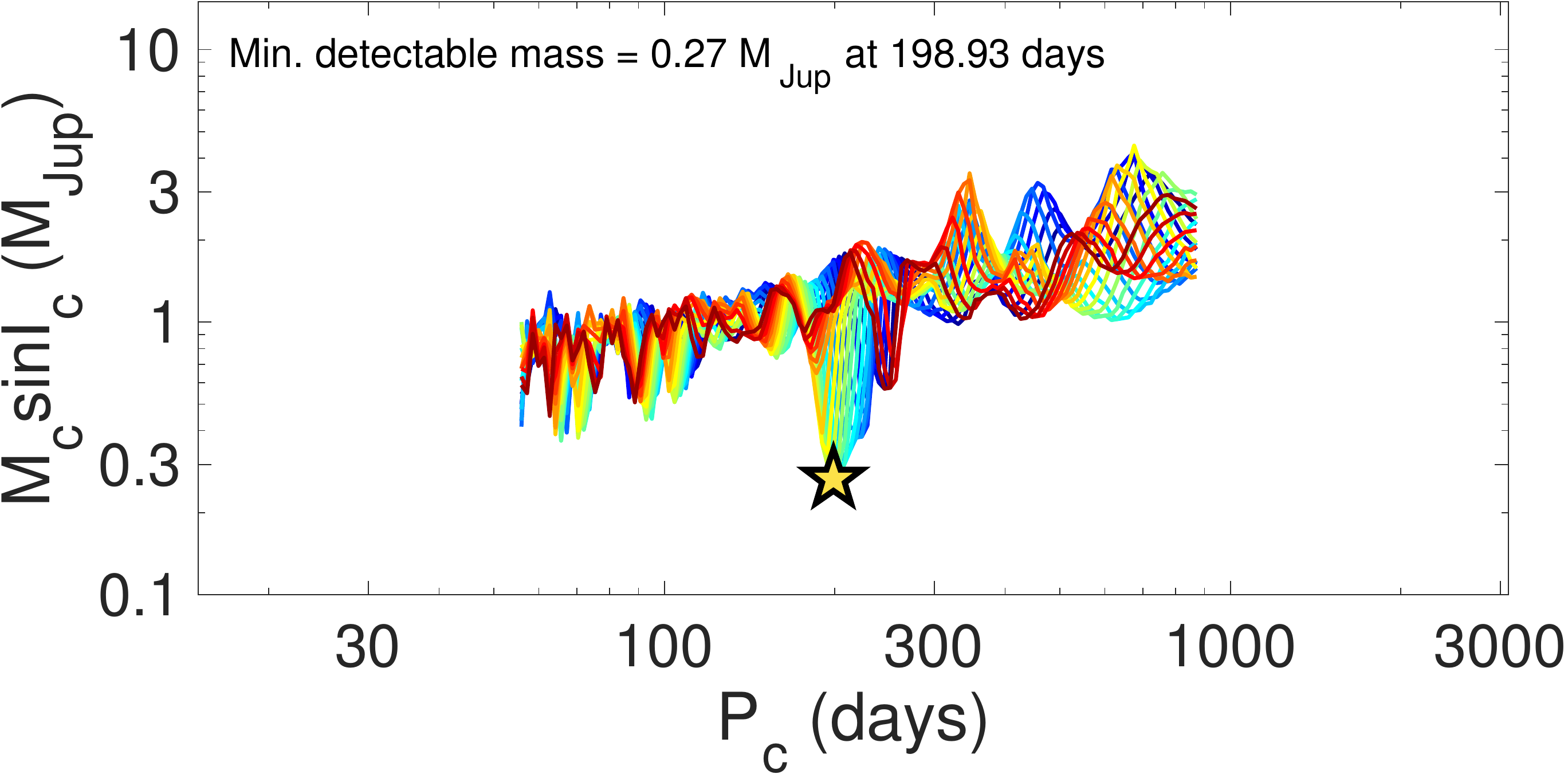}
\end{subfigure}
\end{center}
\end{figure}
\begin{figure}
\begin{center}
\subcaption*{EBLM J0425-46: chosen model = k1 (ecc) \newline \newline $m_{\rm A} = 1.19M_{\odot}$, $m_{\rm B} = 0.627M_{\odot}$, $P = 16.588$ d, $e = 0.048$}
\begin{subfigure}[b]{0.49\textwidth}
\includegraphics[width=\textwidth,trim={0 10cm 0 1.2cm},clip]{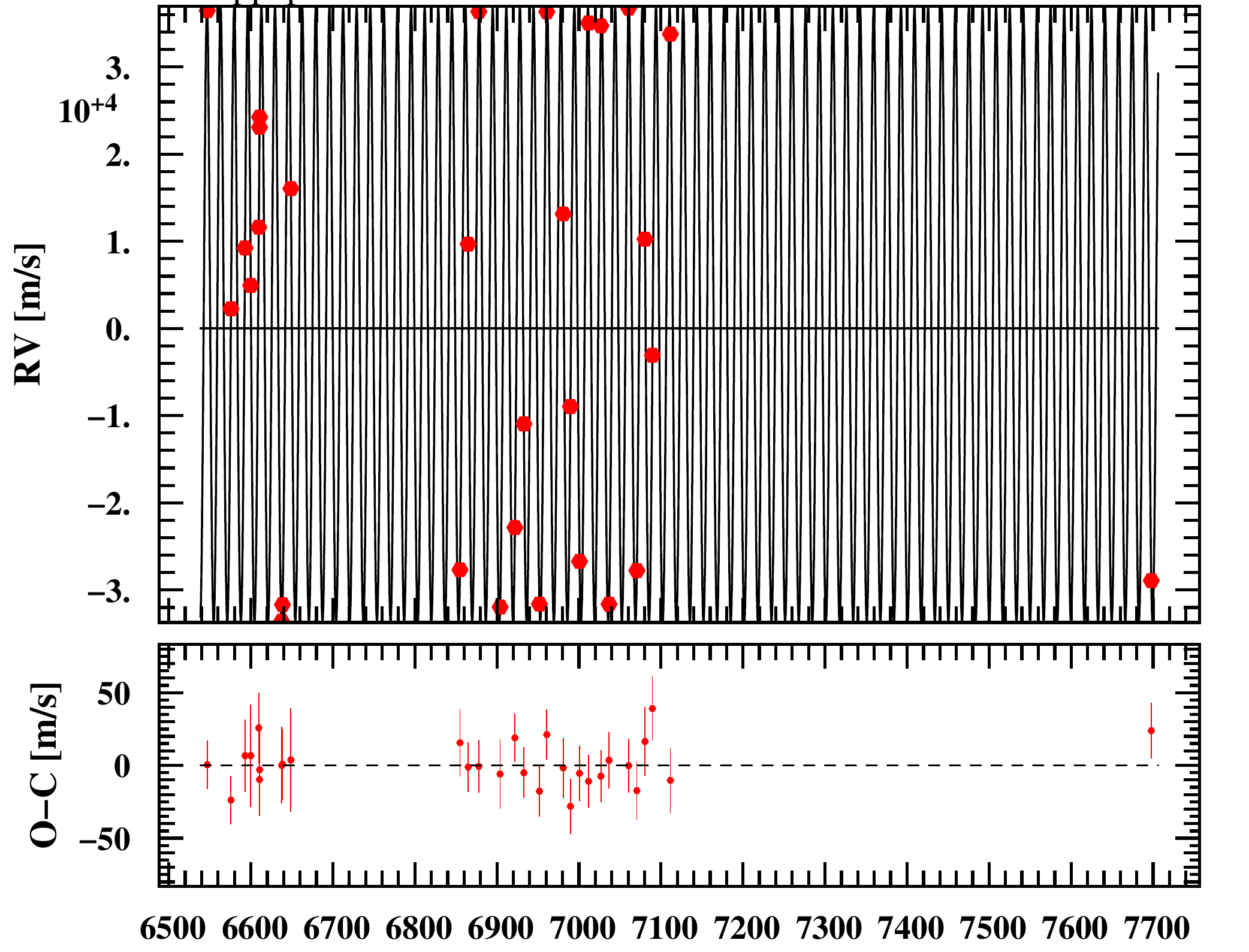}
\end{subfigure}
\begin{subfigure}[b]{0.49\textwidth}
\includegraphics[width=\textwidth,trim={0 0 2cm 0},clip]{orbit_figures/BJD_bar.pdf}
\end{subfigure}
Radial velocities folded on binary phase
\begin{subfigure}[b]{0.49\textwidth}
\includegraphics[width=\textwidth,trim={0 0.5cm 0 0},clip]{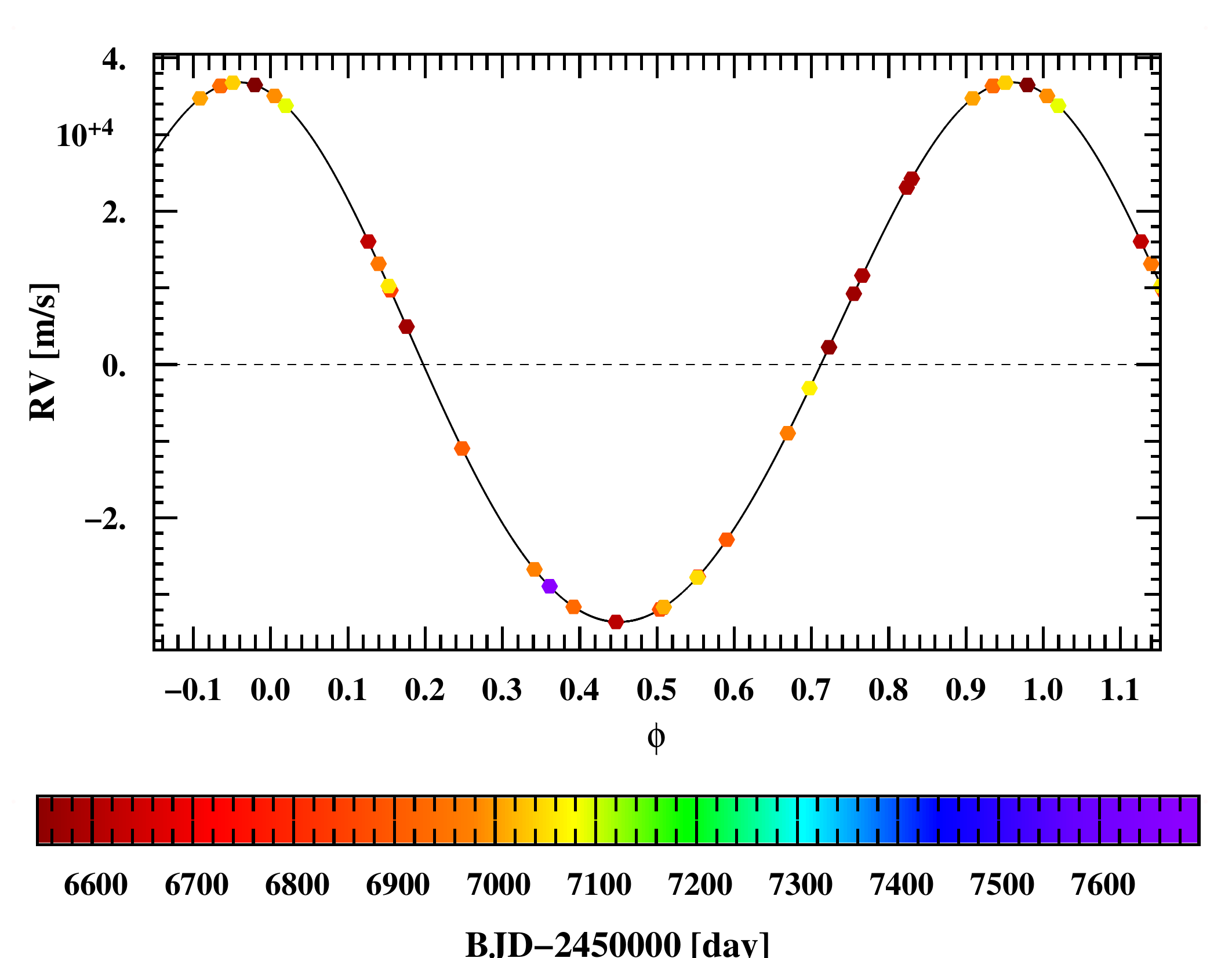}
\end{subfigure}
\begin{subfigure}[b]{0.49\textwidth}
\includegraphics[width=\textwidth,trim={0 0 2cm 0},clip]{orbit_figures/BJD_bar.pdf}
\end{subfigure}
Detection limits
\begin{subfigure}[b]{0.49\textwidth}
\vspace{0.5cm}
\includegraphics[width=\textwidth,trim={0 0 0 0},clip]{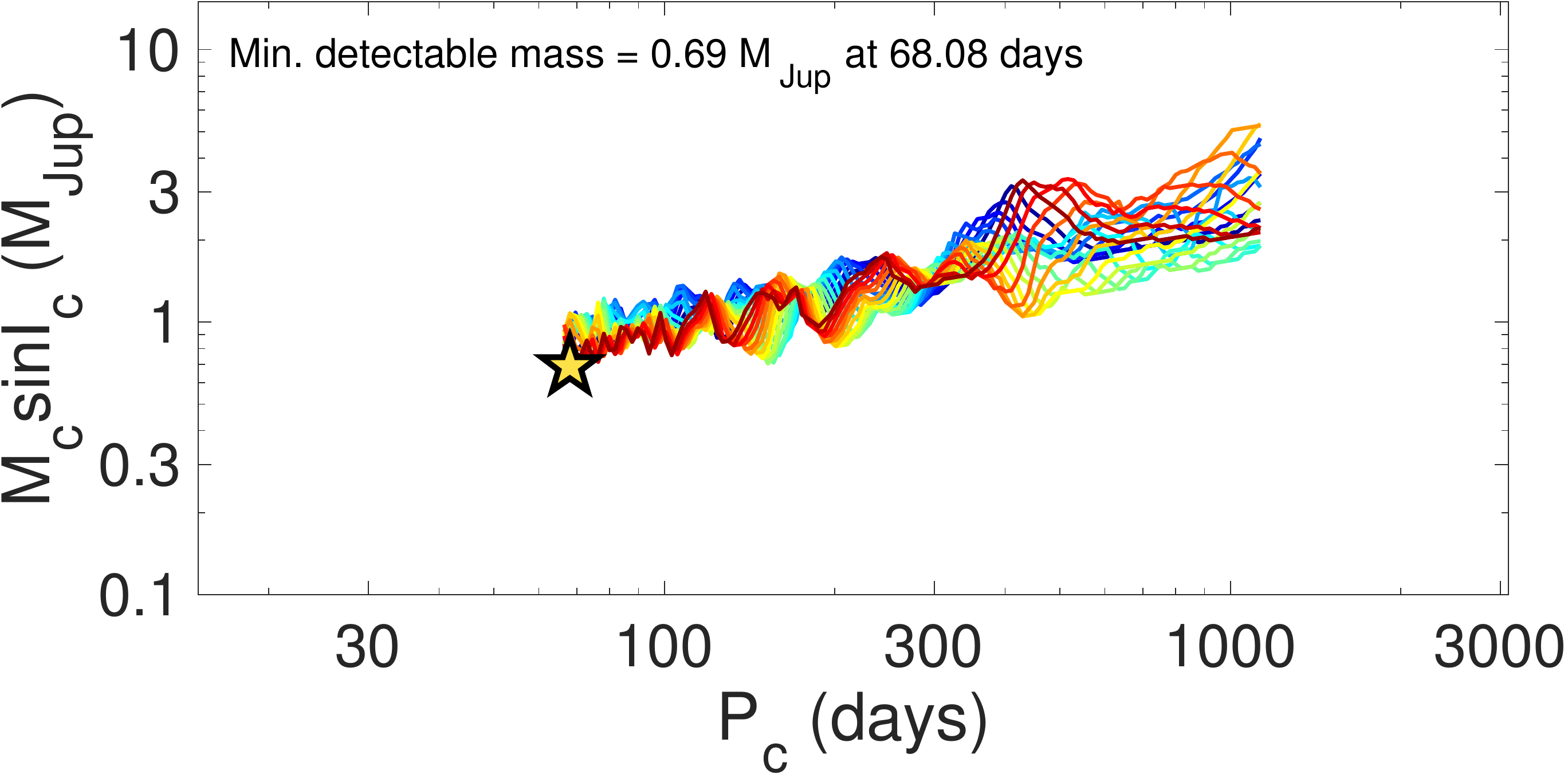}
\end{subfigure}
\end{center}
\end{figure}
\begin{figure}
\begin{center}
\subcaption*{EBLM J0500-46: chosen model = k1 (ecc) \newline \newline $m_{\rm A} = 1.19M_{\odot}$, $m_{\rm B} = 0.181M_{\odot}$, $P = 8.284$ d, $e = 0.231$}
\begin{subfigure}[b]{0.49\textwidth}
\includegraphics[width=\textwidth,trim={0 10cm 0 1.2cm},clip]{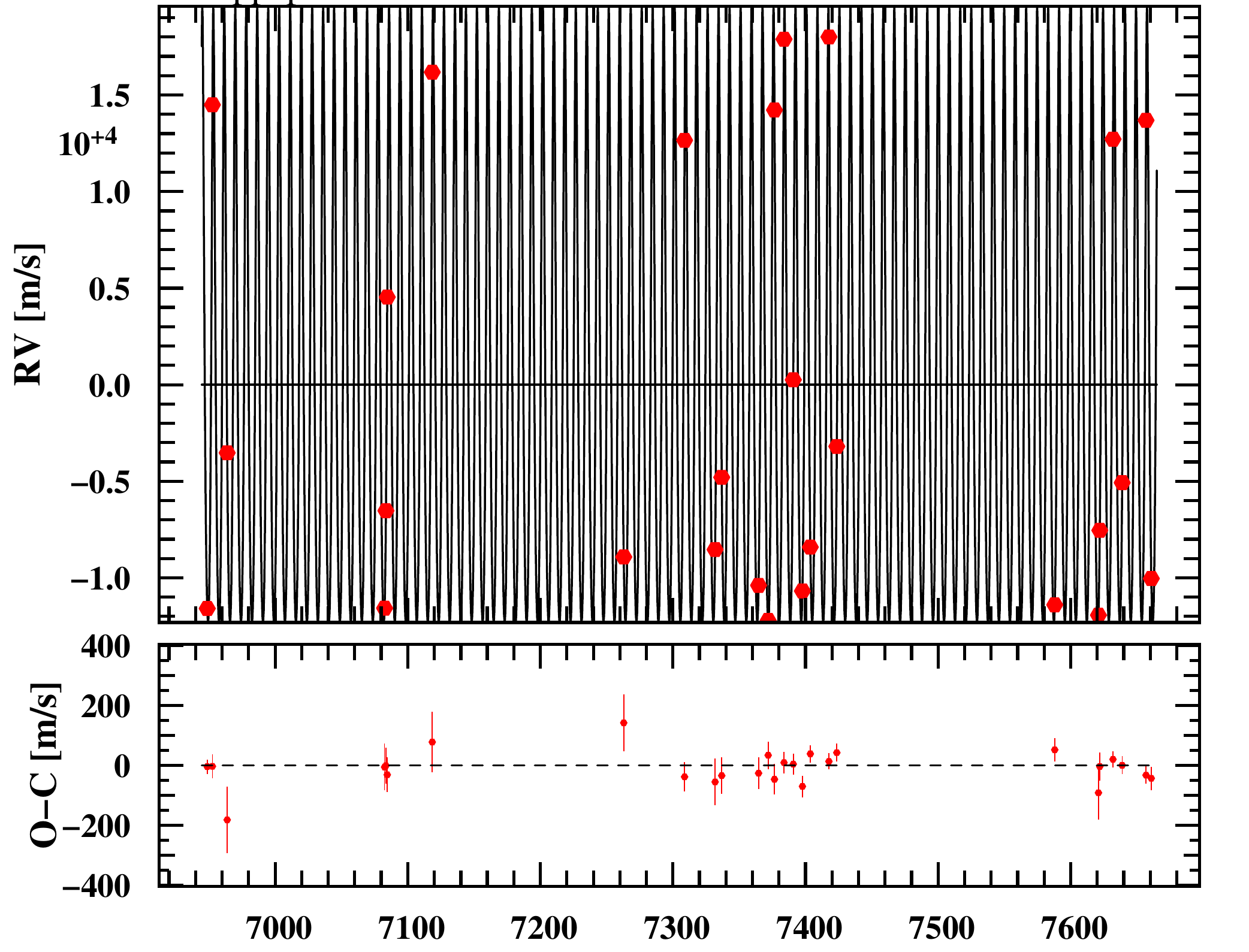}
\end{subfigure}
\begin{subfigure}[b]{0.49\textwidth}
\includegraphics[width=\textwidth,trim={0 0 2cm 0},clip]{orbit_figures/BJD_bar.pdf}
\end{subfigure}
Radial velocities folded on binary phase
\begin{subfigure}[b]{0.49\textwidth}
\includegraphics[width=\textwidth,trim={0 0.5cm 0 0},clip]{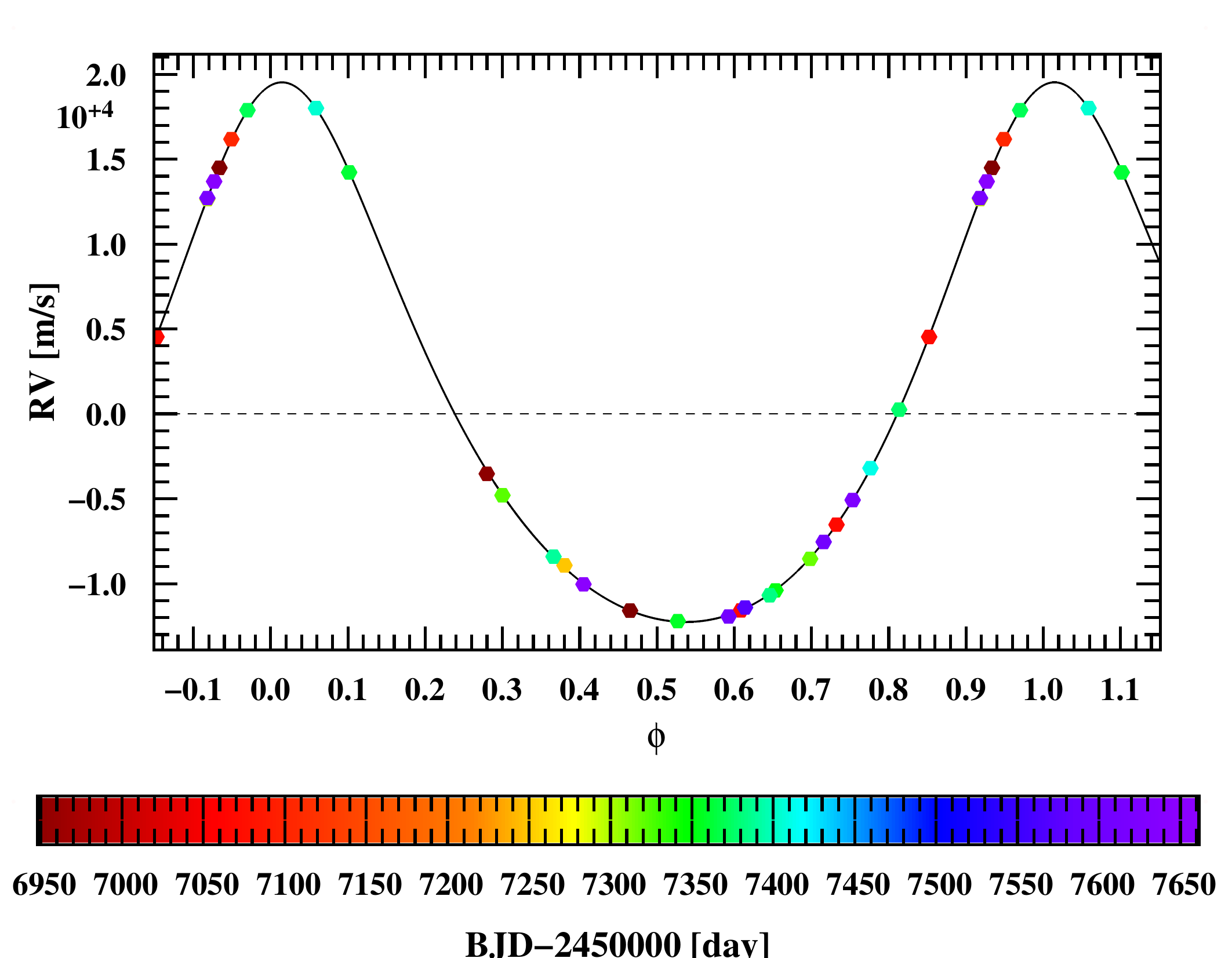}
\end{subfigure}
\begin{subfigure}[b]{0.49\textwidth}
\includegraphics[width=\textwidth,trim={0 0 2cm 0},clip]{orbit_figures/BJD_bar.pdf}
\end{subfigure}
Detection limits
\begin{subfigure}[b]{0.49\textwidth}
\vspace{0.5cm}
\includegraphics[width=\textwidth,trim={0 0 0 0},clip]{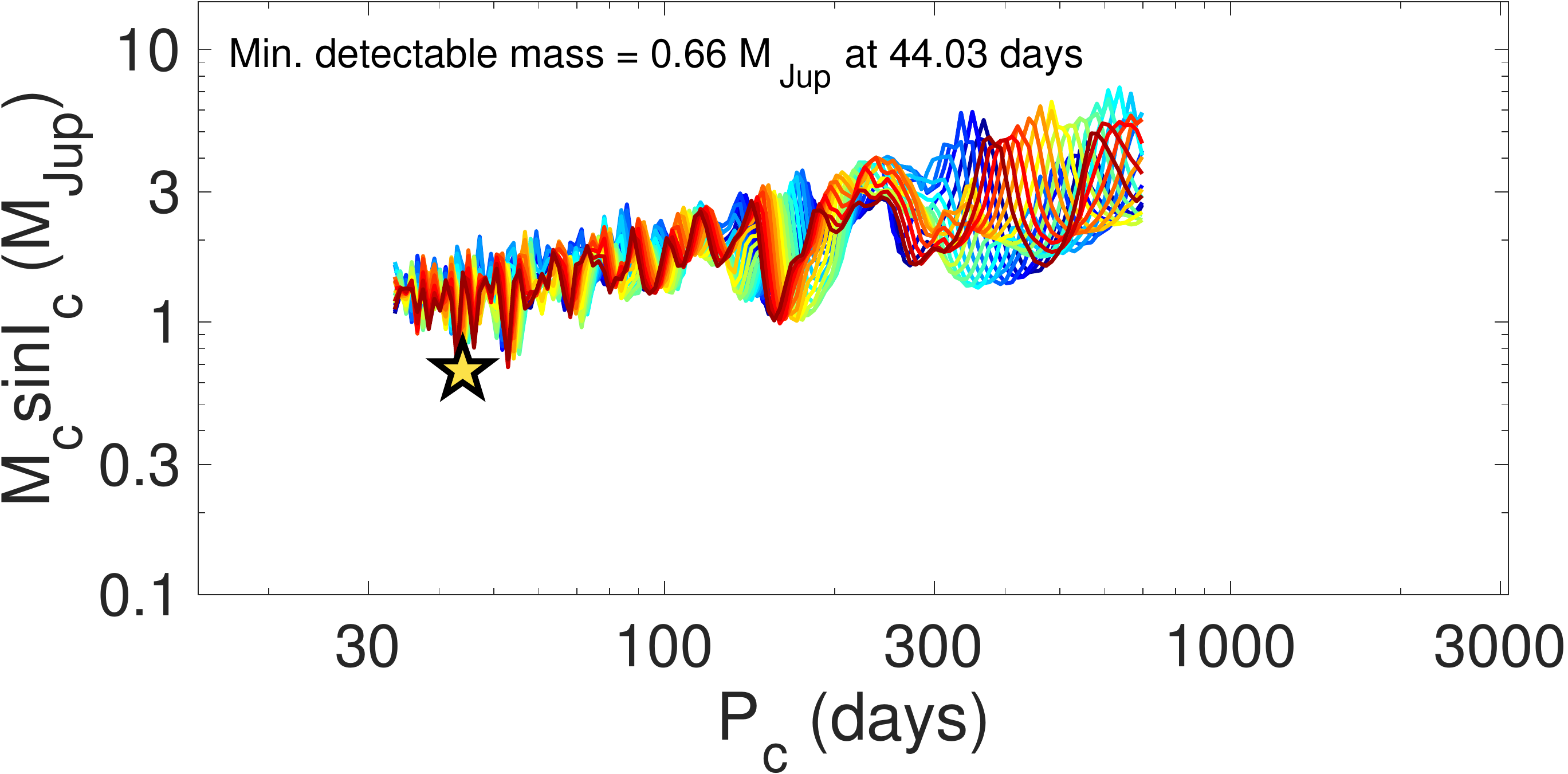}
\end{subfigure}
\end{center}
\end{figure}
\begin{figure}
\begin{center}
\subcaption*{EBLM J0526-34: chosen model = k1 (ecc) \newline \newline $m_{\rm A} = 1.35M_{\odot}$, $m_{\rm B} = 0.338M_{\odot}$, $P = 10.191$ d, $e = 0.126$}
\begin{subfigure}[b]{0.49\textwidth}
\includegraphics[width=\textwidth,trim={0 10cm 0 1.2cm},clip]{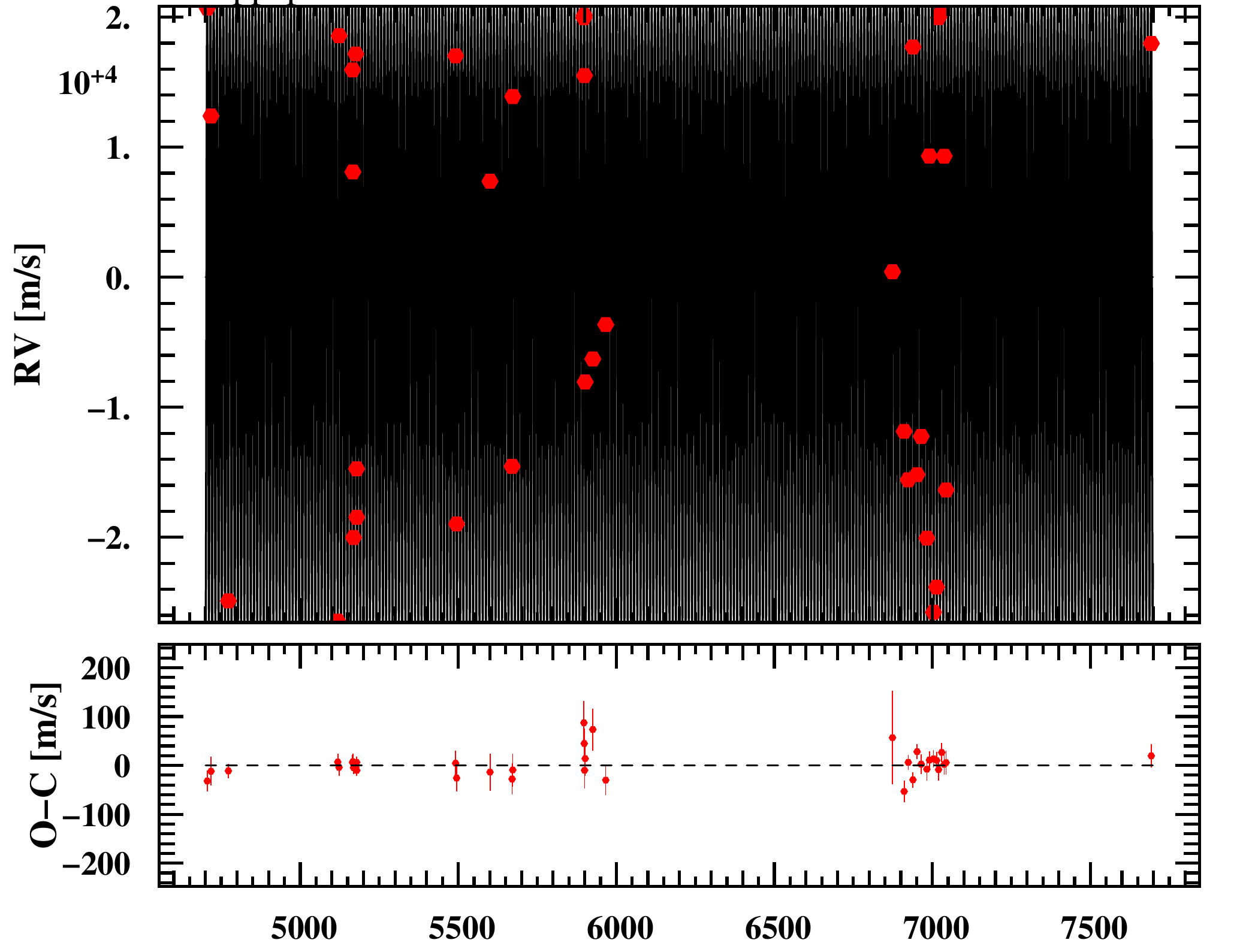}
\end{subfigure}
\begin{subfigure}[b]{0.49\textwidth}
\includegraphics[width=\textwidth,trim={0 0 2cm 0},clip]{orbit_figures/BJD_bar.pdf}
\end{subfigure}
Radial velocities folded on binary phase
\begin{subfigure}[b]{0.49\textwidth}
\includegraphics[width=\textwidth,trim={0 0.5cm 0 0},clip]{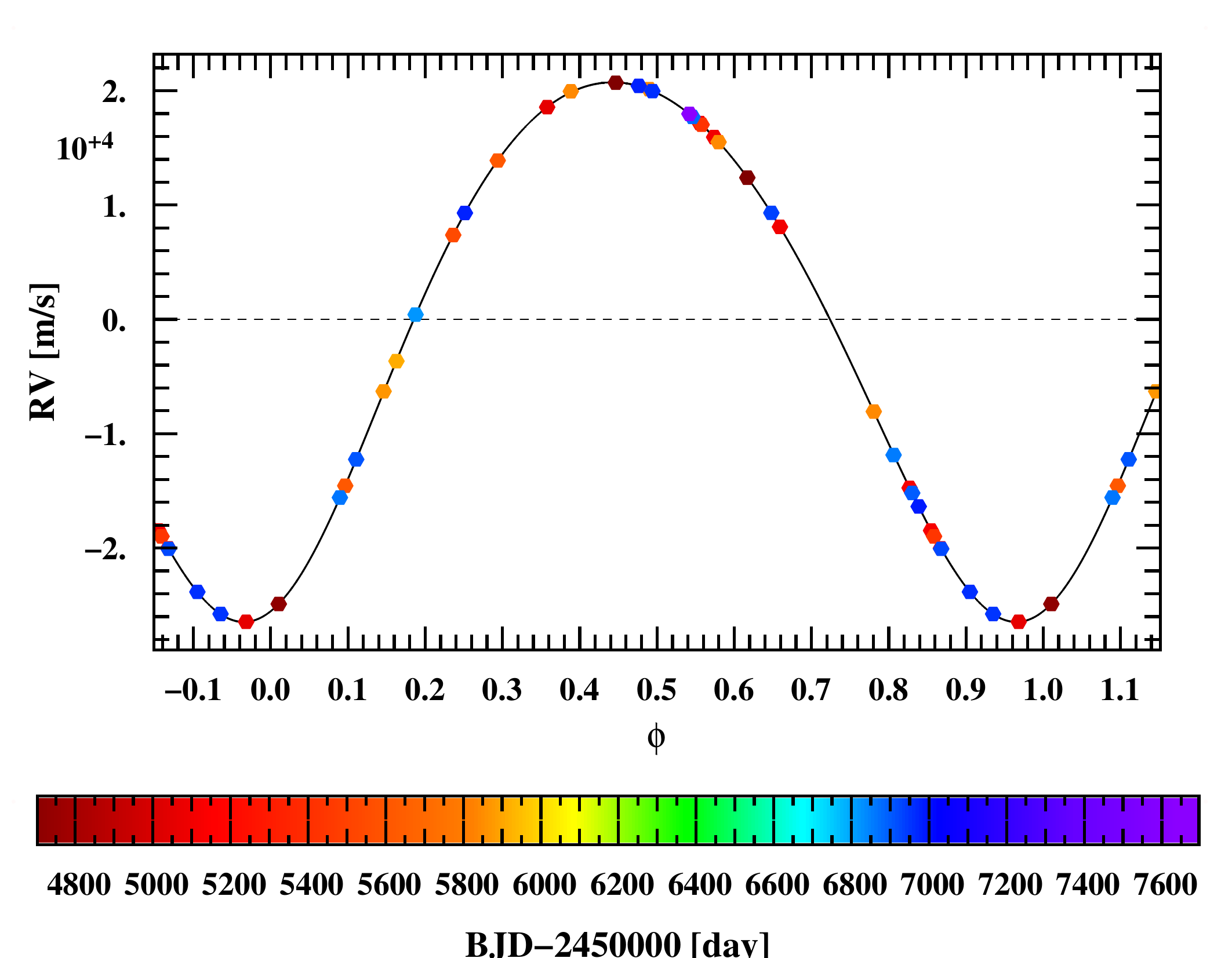}
\end{subfigure}
\begin{subfigure}[b]{0.49\textwidth}
\includegraphics[width=\textwidth,trim={0 0 2cm 0},clip]{orbit_figures/BJD_bar.pdf}
\end{subfigure}
Detection limits
\begin{subfigure}[b]{0.49\textwidth}
\vspace{0.5cm}
\includegraphics[width=\textwidth,trim={0 0 0 0},clip]{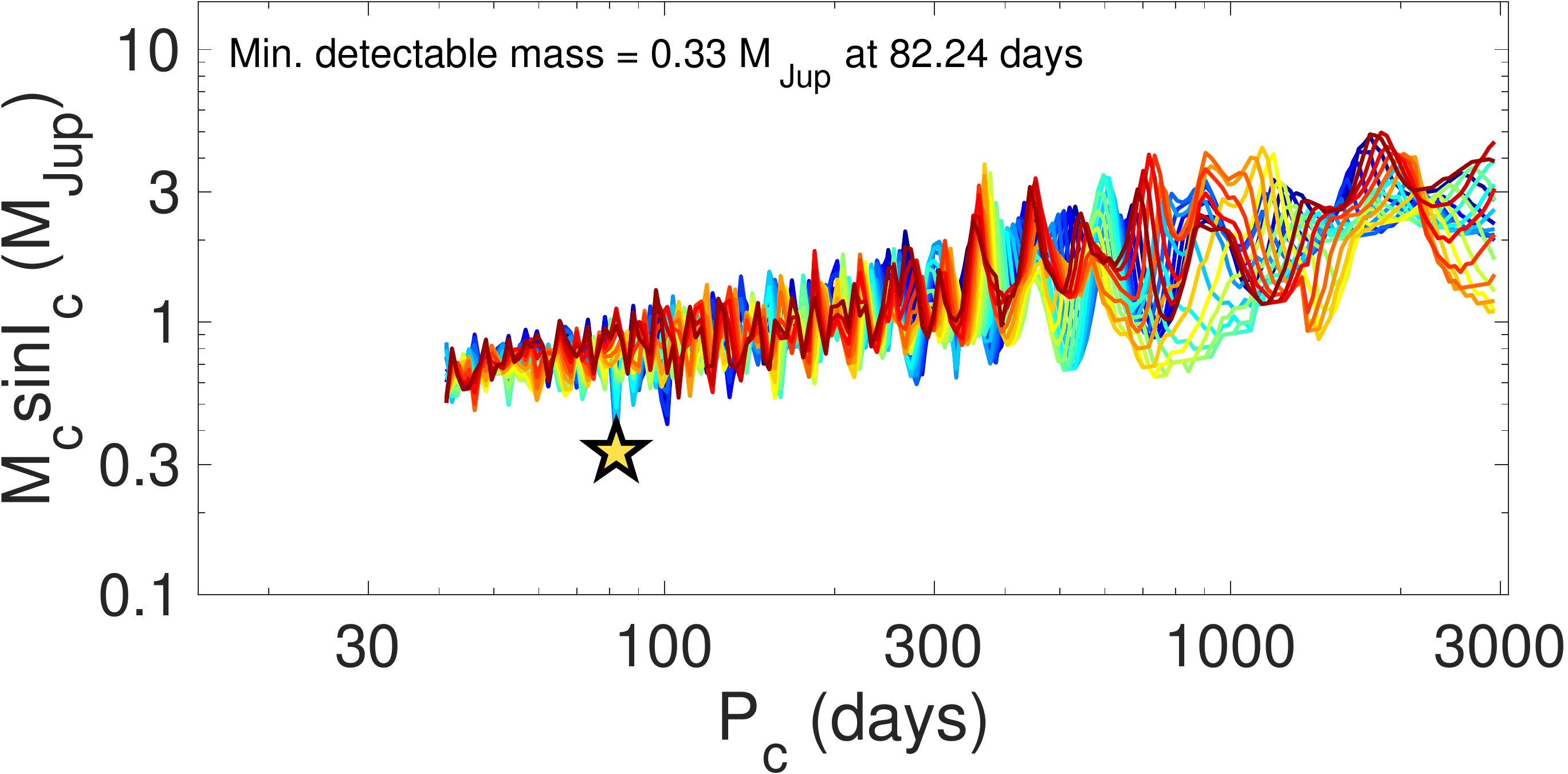}
\end{subfigure}
\end{center}
\end{figure}
\begin{figure}
\begin{center}
\subcaption*{EBLM J0540-17: chosen model = k1d3 (circ) \newline \newline $m_{\rm A} = 1.2M_{\odot}$, $m_{\rm B} = 0.171M_{\odot}$, $P = 6.005$ d, $e = 0$}
\begin{subfigure}[b]{0.49\textwidth}
\includegraphics[width=\textwidth,trim={0 10cm 0 1.2cm},clip]{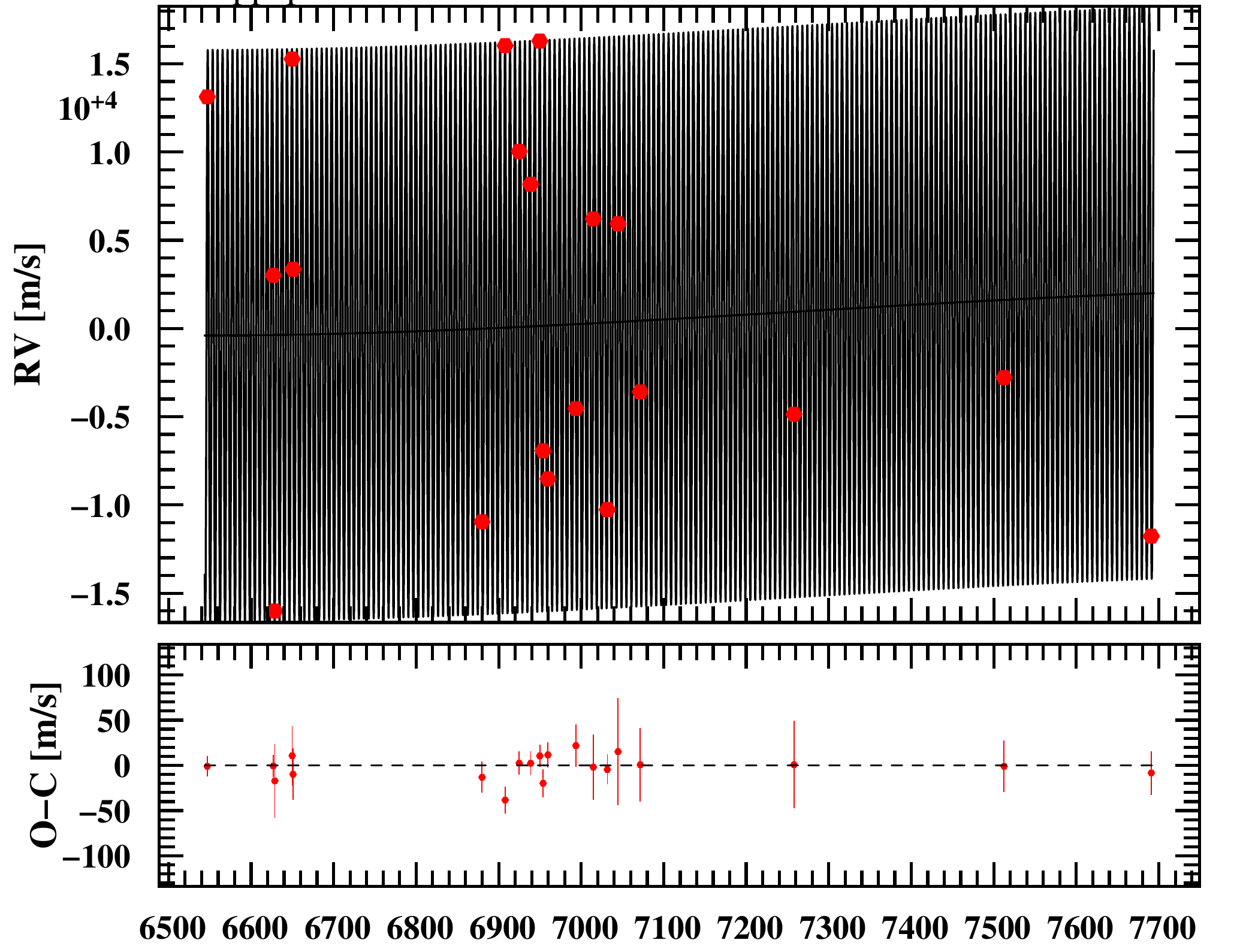}
\end{subfigure}
\begin{subfigure}[b]{0.49\textwidth}
\includegraphics[width=\textwidth,trim={0 0 2cm 0},clip]{orbit_figures/BJD_bar.pdf}
\end{subfigure}
Radial velocities folded on binary phase
\begin{subfigure}[b]{0.49\textwidth}
\includegraphics[width=\textwidth,trim={0 0.5cm 0 0},clip]{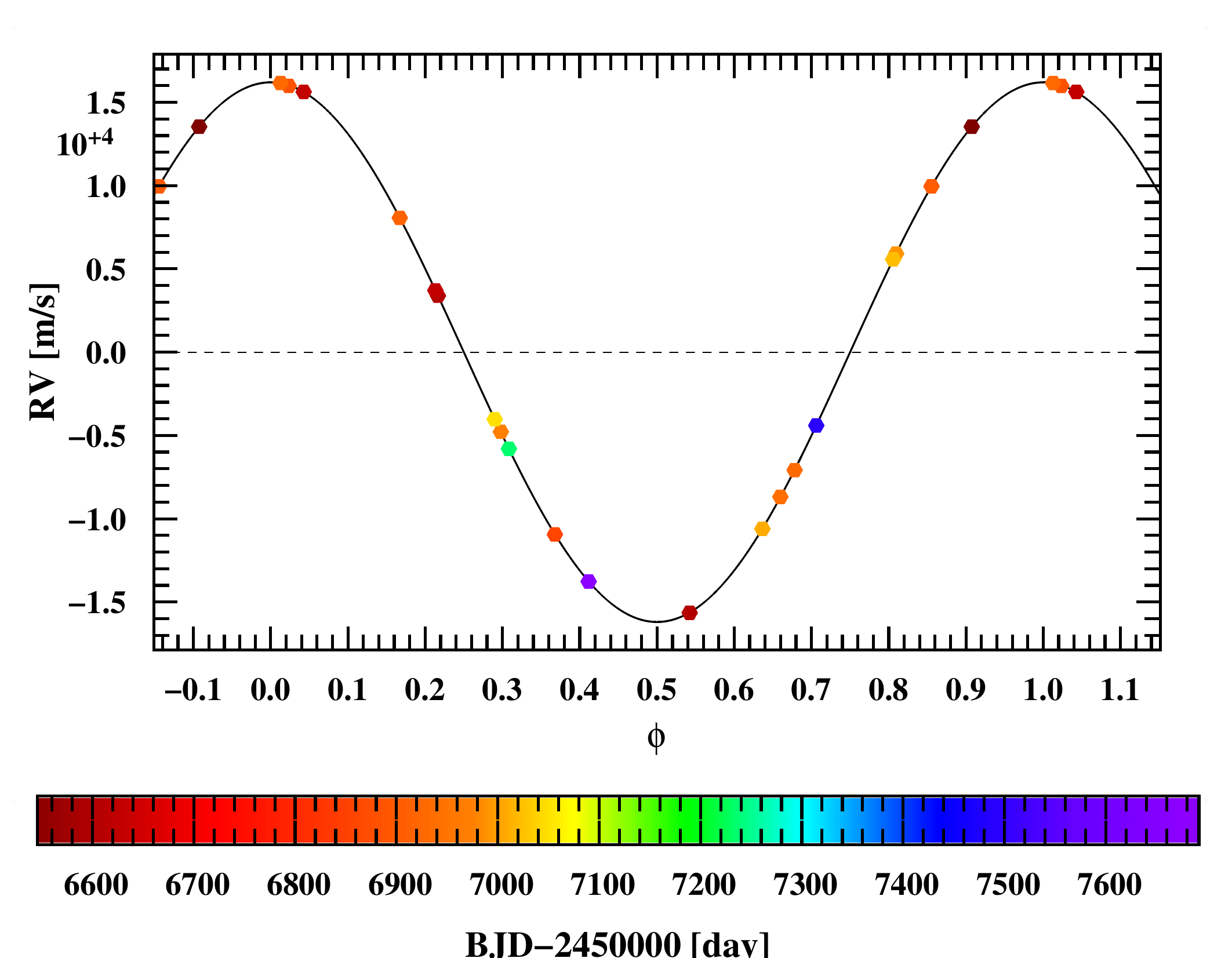}
\end{subfigure}
\begin{subfigure}[b]{0.49\textwidth}
\includegraphics[width=\textwidth,trim={0 0 2cm 0},clip]{orbit_figures/BJD_bar.pdf}
\end{subfigure}
Detection limits
\begin{subfigure}[b]{0.49\textwidth}
\vspace{0.5cm}
\includegraphics[width=\textwidth,trim={0 0 0 0},clip]{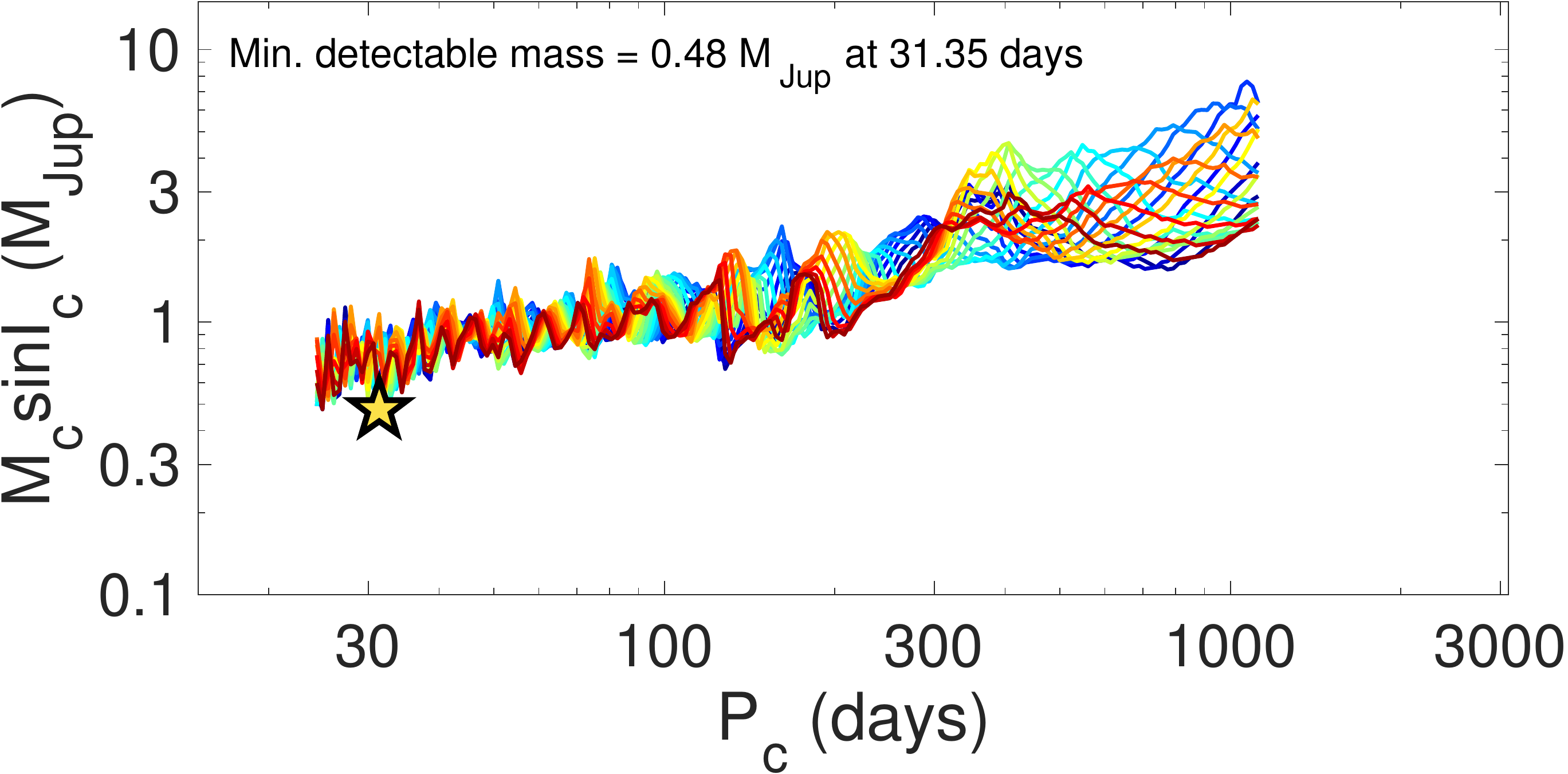}
\end{subfigure}
\end{center}
\end{figure}
\begin{figure}
\begin{center}
\subcaption*{EBLM J0543-57: chosen model = k2 (circ) \newline \newline $m_{\rm A} = 1.23M_{\odot}$, $m_{\rm B} = 0.16M_{\odot}$, $P = 4.464$ d, $e = 0$}
\begin{subfigure}[b]{0.49\textwidth}
\includegraphics[width=\textwidth,trim={0 10cm 0 1.2cm},clip]{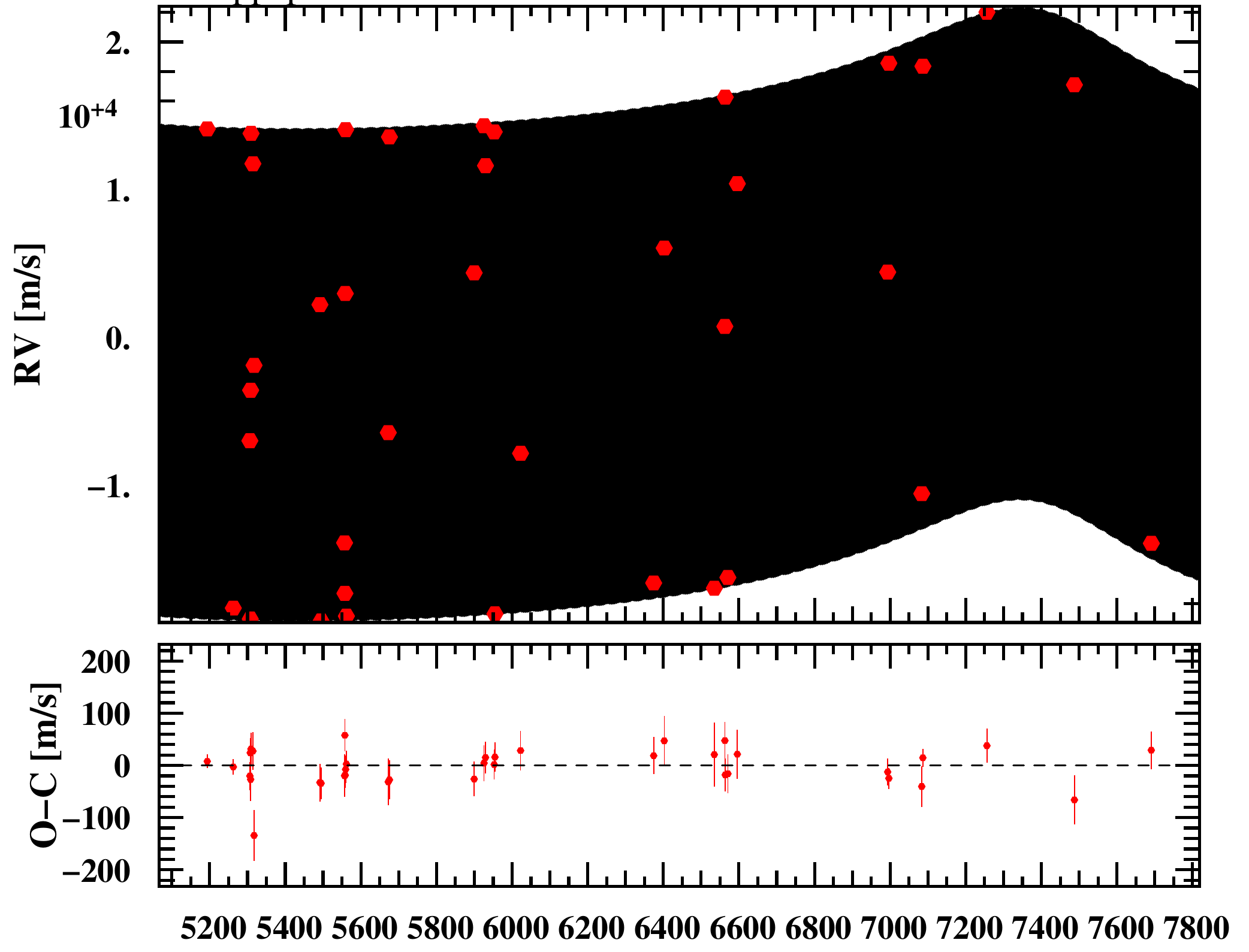}
\end{subfigure}
\begin{subfigure}[b]{0.49\textwidth}
\includegraphics[width=\textwidth,trim={0 0 2cm 0},clip]{orbit_figures/BJD_bar.pdf}
\end{subfigure}
Radial velocities folded on binary phase
\begin{subfigure}[b]{0.49\textwidth}
\includegraphics[width=\textwidth,trim={0 0.5cm 0 0},clip]{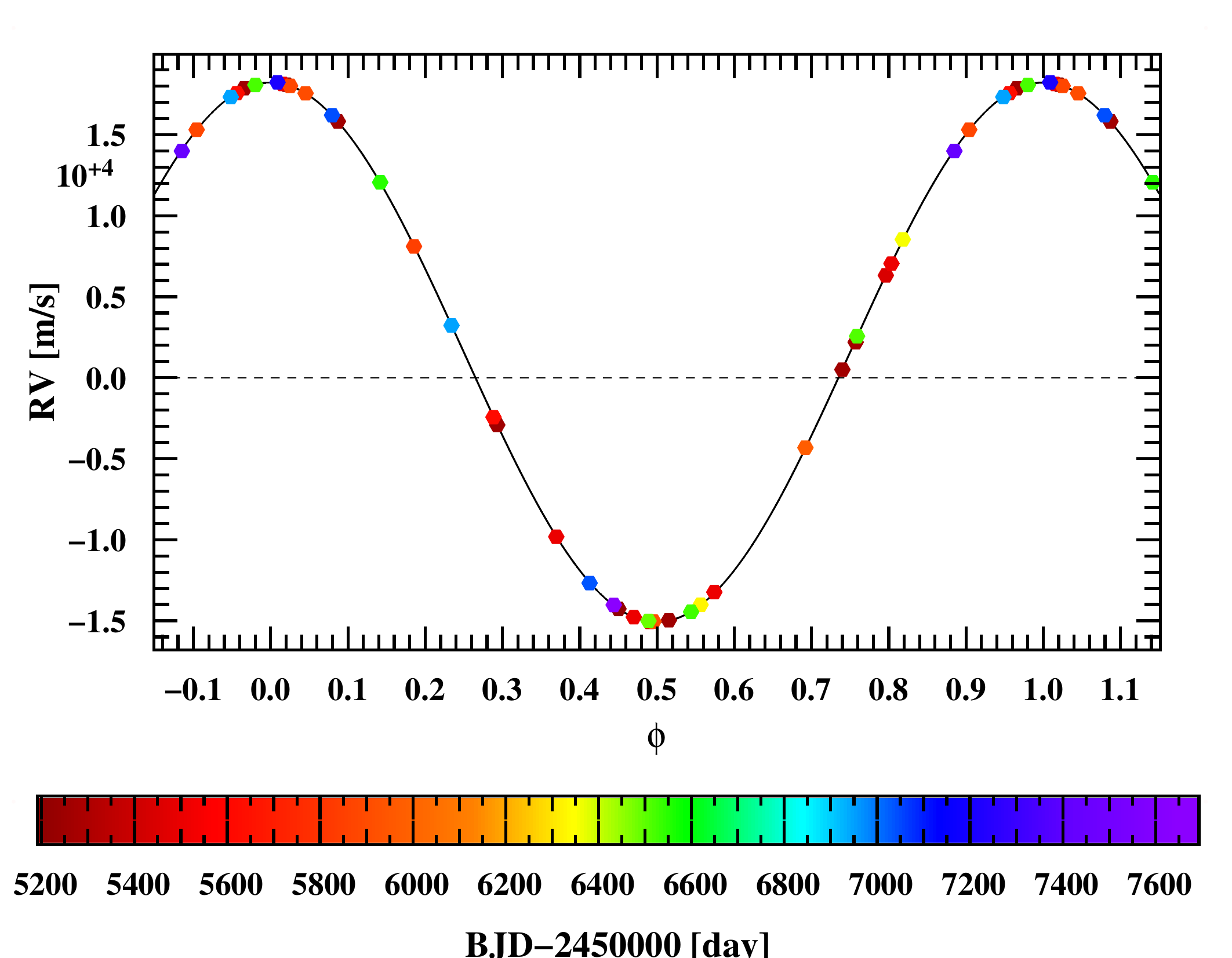}
\end{subfigure}
\begin{subfigure}[b]{0.49\textwidth}
\includegraphics[width=\textwidth,trim={0 0 2cm 0},clip]{orbit_figures/BJD_bar.pdf}
\end{subfigure}
Detection limits
\begin{subfigure}[b]{0.49\textwidth}
\vspace{0.5cm}
\includegraphics[width=\textwidth,trim={0 0 0 0},clip]{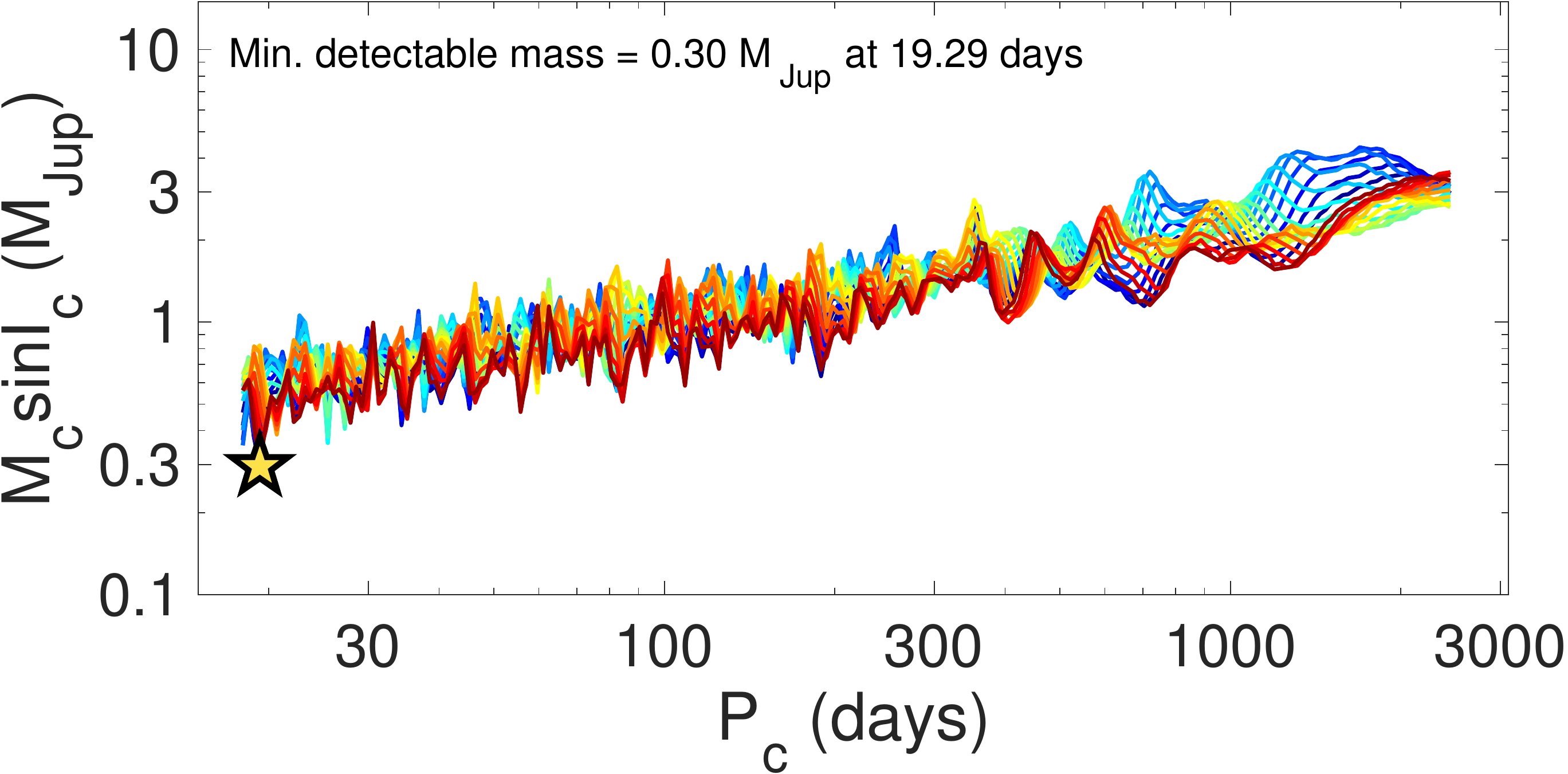}
\end{subfigure}
\end{center}
\end{figure}
\begin{figure}
\begin{center}
\subcaption*{EBLM J0608-59: chosen model = k1 (ecc) \newline \newline $m_{\rm A} = 1.2M_{\odot}$, $m_{\rm B} = 0.325M_{\odot}$, $P = 14.609$ d, $e = 0.156$}
\begin{subfigure}[b]{0.49\textwidth}
\includegraphics[width=\textwidth,trim={0 10cm 0 1.2cm},clip]{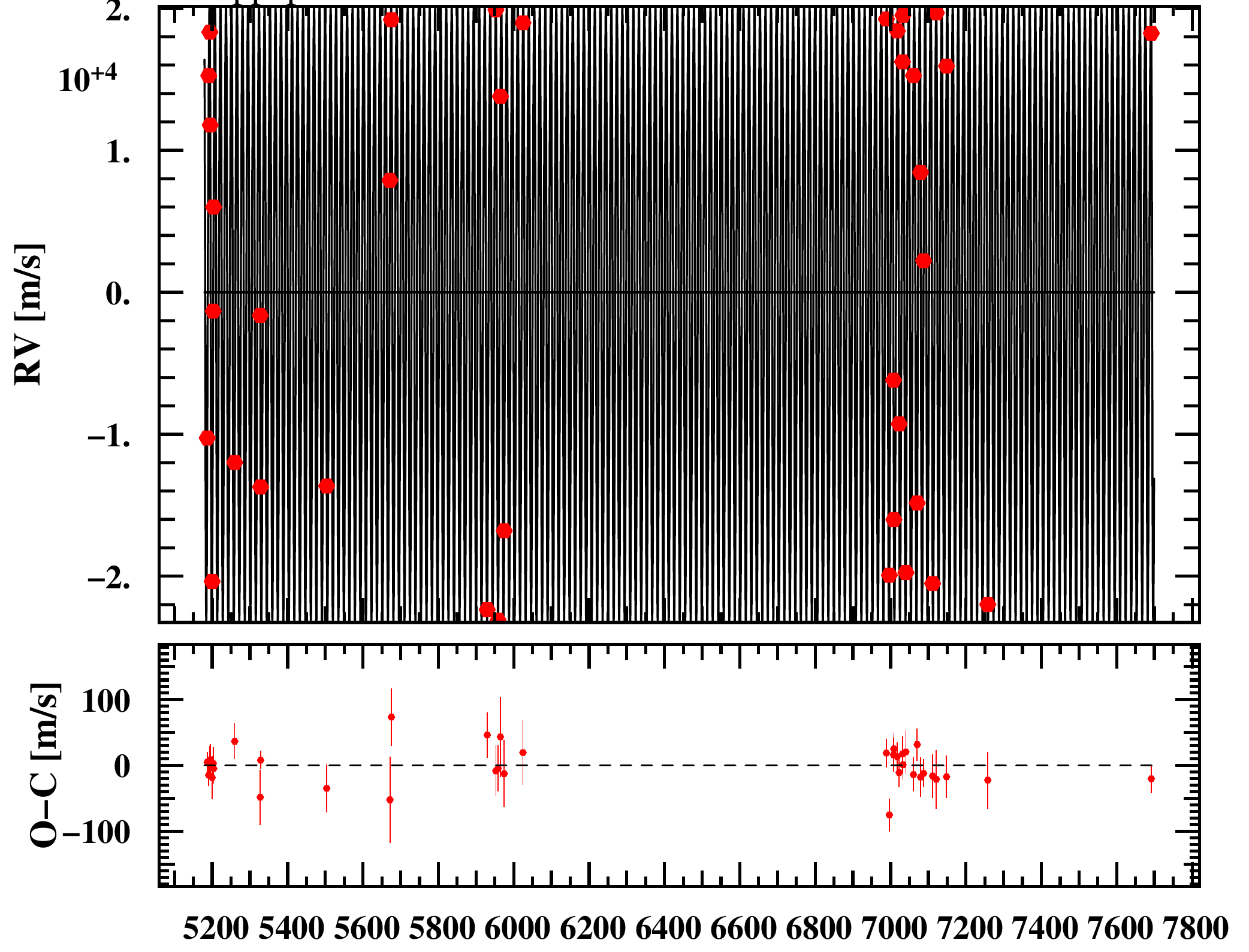}
\end{subfigure}
\begin{subfigure}[b]{0.49\textwidth}
\includegraphics[width=\textwidth,trim={0 0 2cm 0},clip]{orbit_figures/BJD_bar.pdf}
\end{subfigure}
Radial velocities folded on binary phase
\begin{subfigure}[b]{0.49\textwidth}
\includegraphics[width=\textwidth,trim={0 0.5cm 0 0},clip]{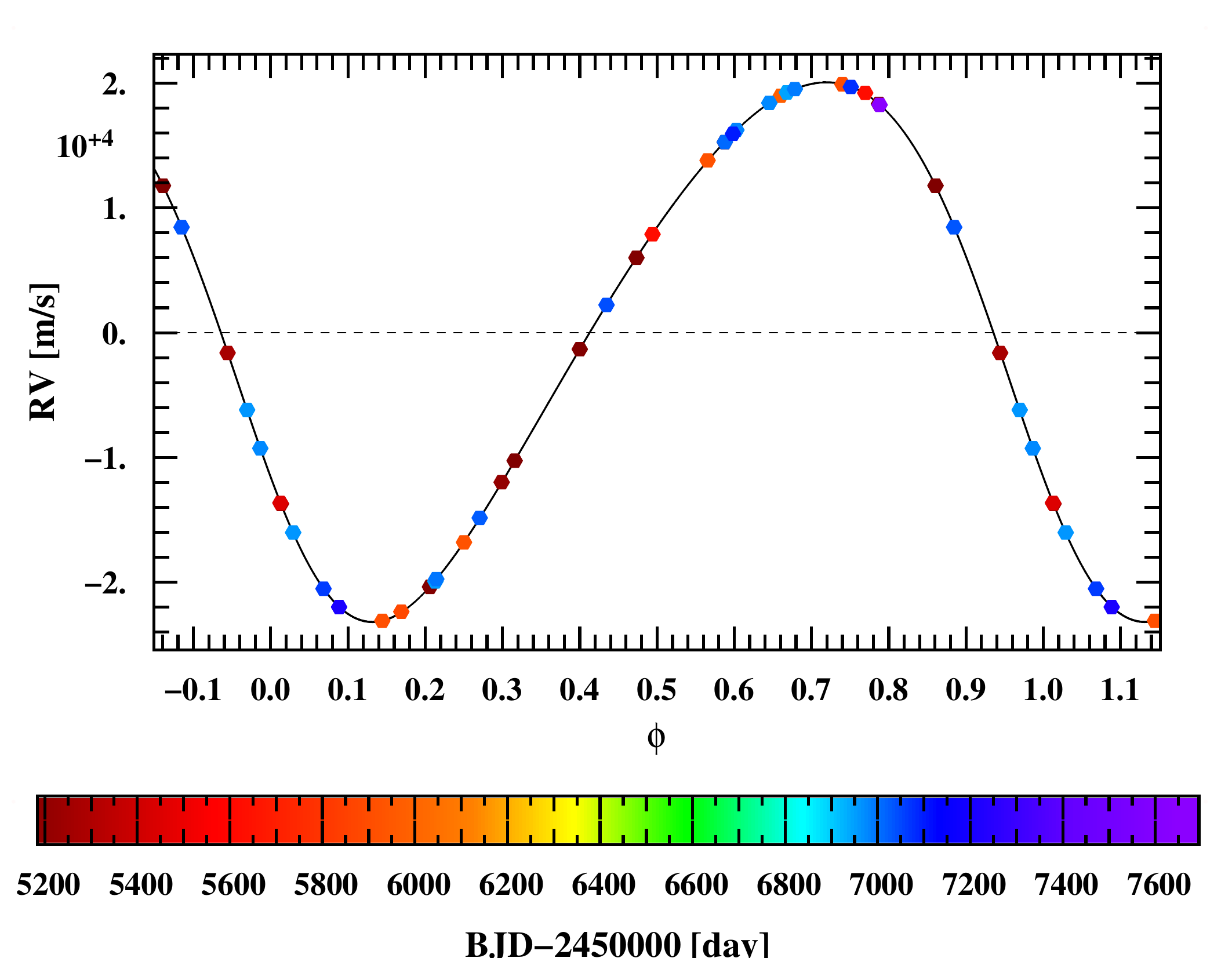}
\end{subfigure}
\begin{subfigure}[b]{0.49\textwidth}
\includegraphics[width=\textwidth,trim={0 0 2cm 0},clip]{orbit_figures/BJD_bar.pdf}
\end{subfigure}
Detection limits
\begin{subfigure}[b]{0.49\textwidth}
\vspace{0.5cm}
\includegraphics[width=\textwidth,trim={0 0 0 0},clip]{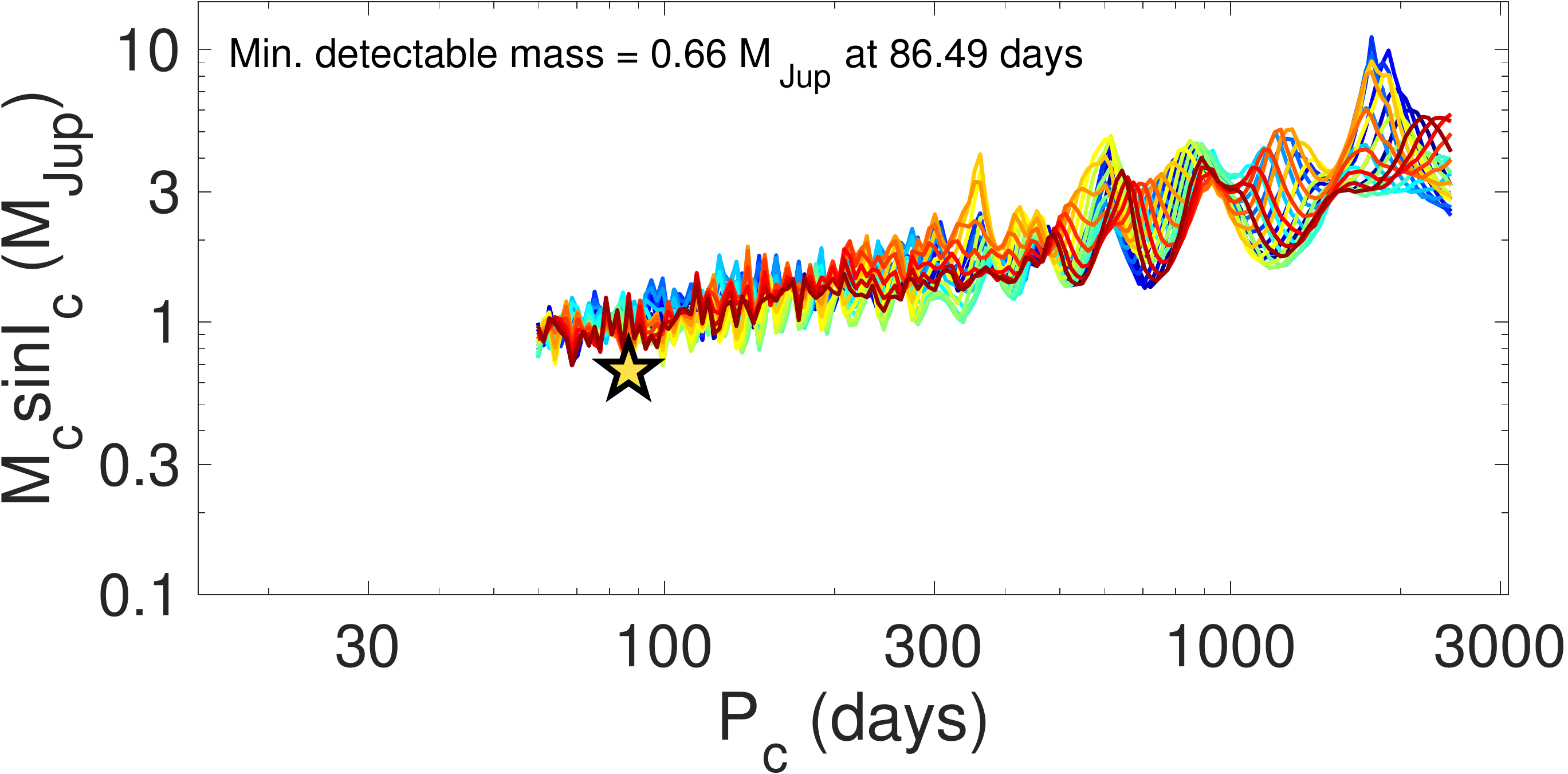}
\end{subfigure}
\end{center}
\end{figure}
\begin{figure}
\begin{center}
\subcaption*{EBLM J0621-50: chosen model = k1 (circ) \newline \newline $m_{\rm A} = 1.23M_{\odot}$, $m_{\rm B} = 0.42M_{\odot}$, $P = 4.964$ d, $e = 0$}
\begin{subfigure}[b]{0.49\textwidth}
\includegraphics[width=\textwidth,trim={0 10cm 0 1.2cm},clip]{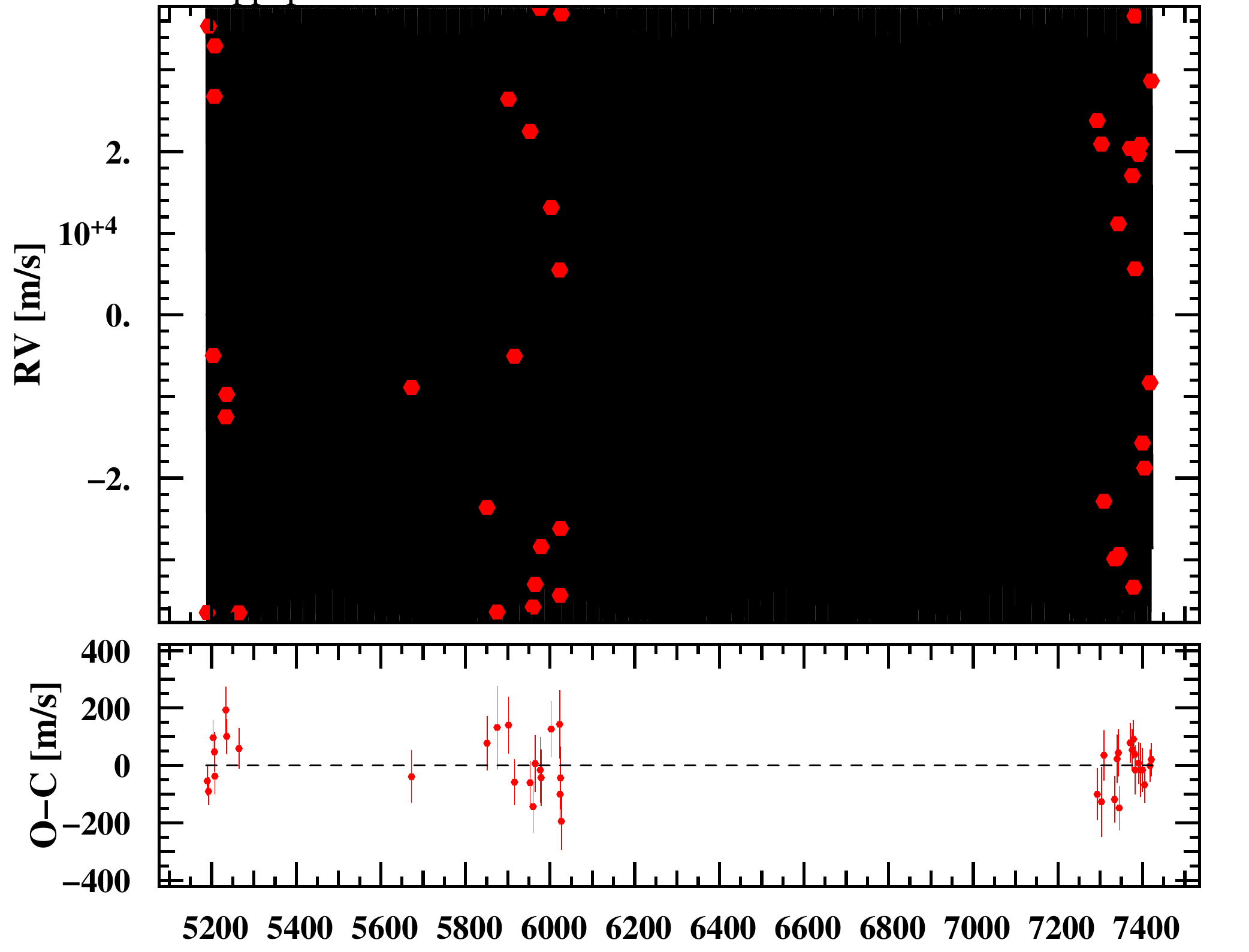}
\end{subfigure}
\begin{subfigure}[b]{0.49\textwidth}
\includegraphics[width=\textwidth,trim={0 0 2cm 0},clip]{orbit_figures/BJD_bar.pdf}
\end{subfigure}
Radial velocities folded on binary phase
\begin{subfigure}[b]{0.49\textwidth}
\includegraphics[width=\textwidth,trim={0 0.5cm 0 0},clip]{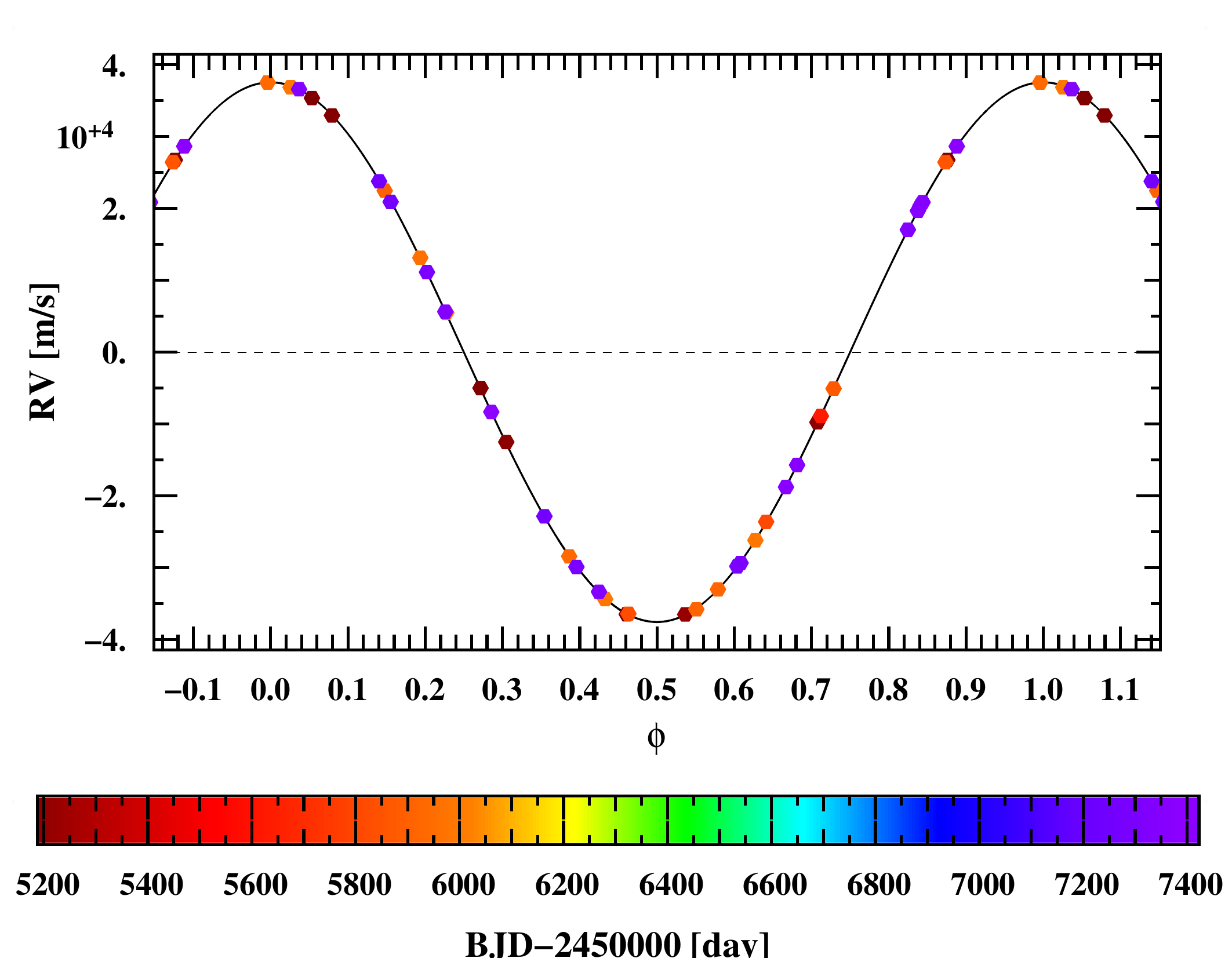}
\end{subfigure}
\begin{subfigure}[b]{0.49\textwidth}
\includegraphics[width=\textwidth,trim={0 0 2cm 0},clip]{orbit_figures/BJD_bar.pdf}
\end{subfigure}
Detection limits
\begin{subfigure}[b]{0.49\textwidth}
\vspace{0.5cm}
\includegraphics[width=\textwidth,trim={0 0 0 0},clip]{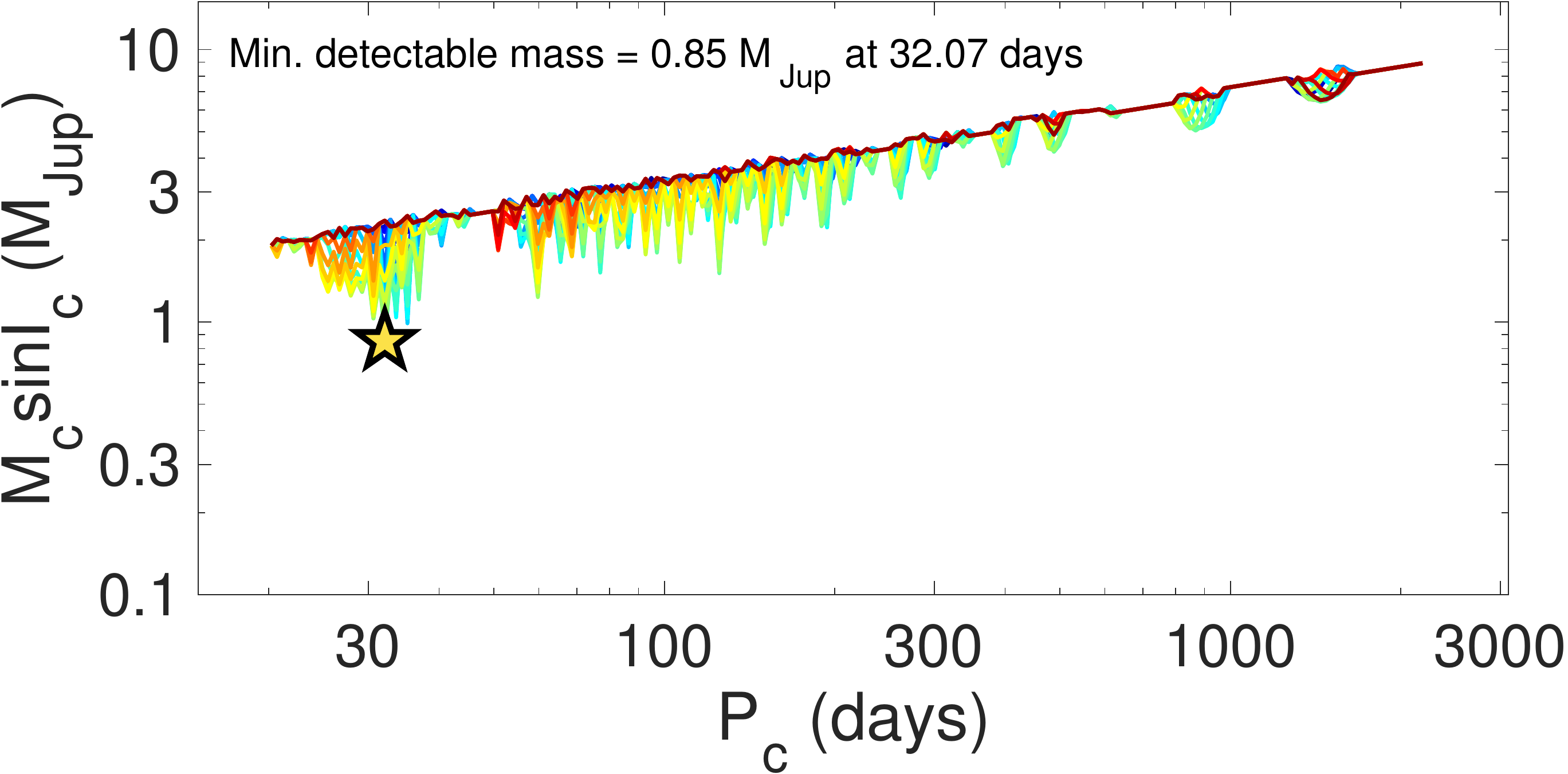}
\end{subfigure}
\end{center}
\end{figure}
\begin{figure}
\begin{center}
\subcaption*{EBLM J0659-61: chosen model = k1d2 (circ) \newline \newline $m_{\rm A} = 1.16M_{\odot}$, $m_{\rm B} = 0.456M_{\odot}$, $P = 4.236$ d, $e = 0$}
\begin{subfigure}[b]{0.49\textwidth}
\includegraphics[width=\textwidth,trim={0 10cm 0 1.2cm},clip]{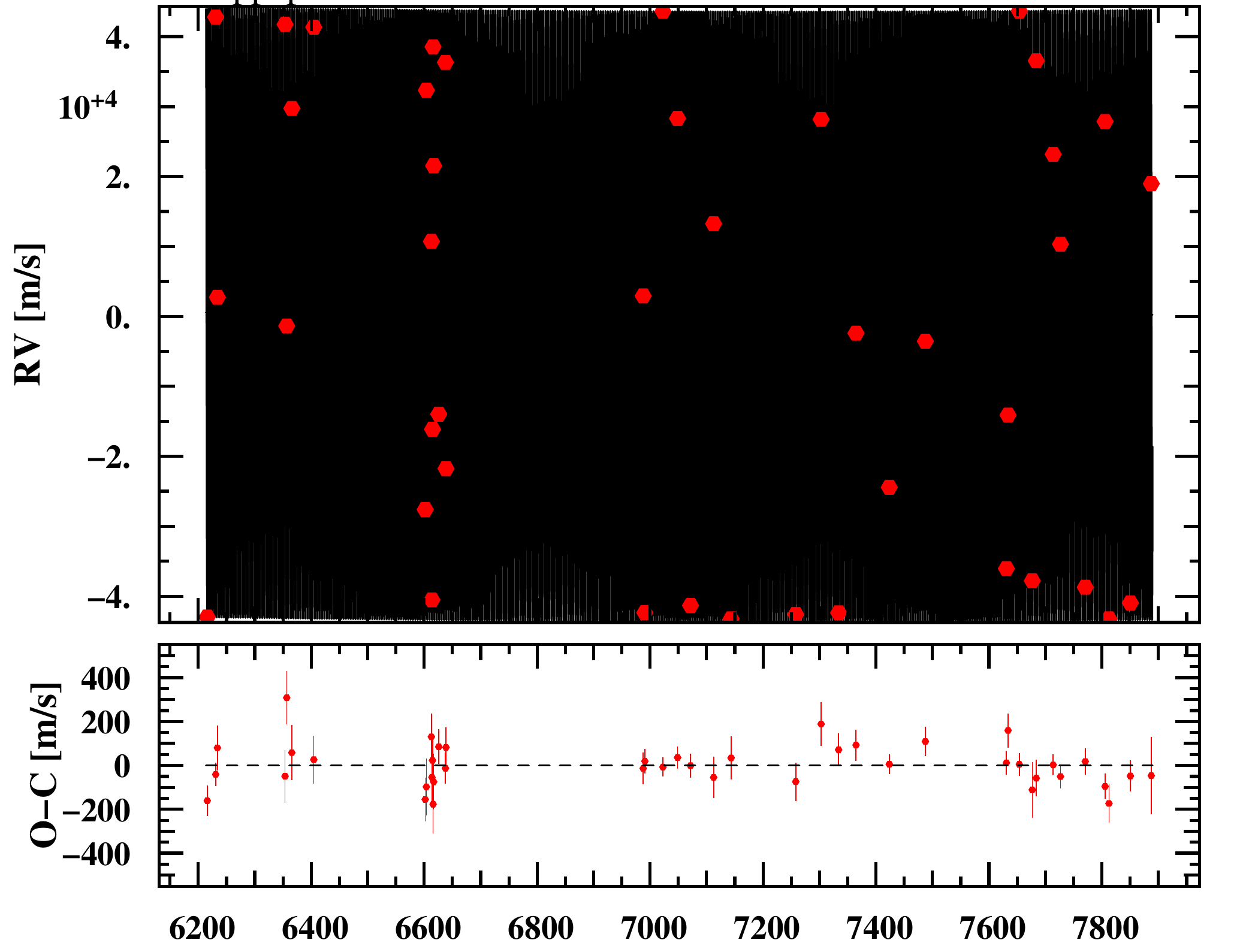}
\end{subfigure}
\begin{subfigure}[b]{0.49\textwidth}
\includegraphics[width=\textwidth,trim={0 0 2cm 0},clip]{orbit_figures/BJD_bar.pdf}
\end{subfigure}
Radial velocities folded on binary phase
\begin{subfigure}[b]{0.49\textwidth}
\includegraphics[width=\textwidth,trim={0 0.5cm 0 0},clip]{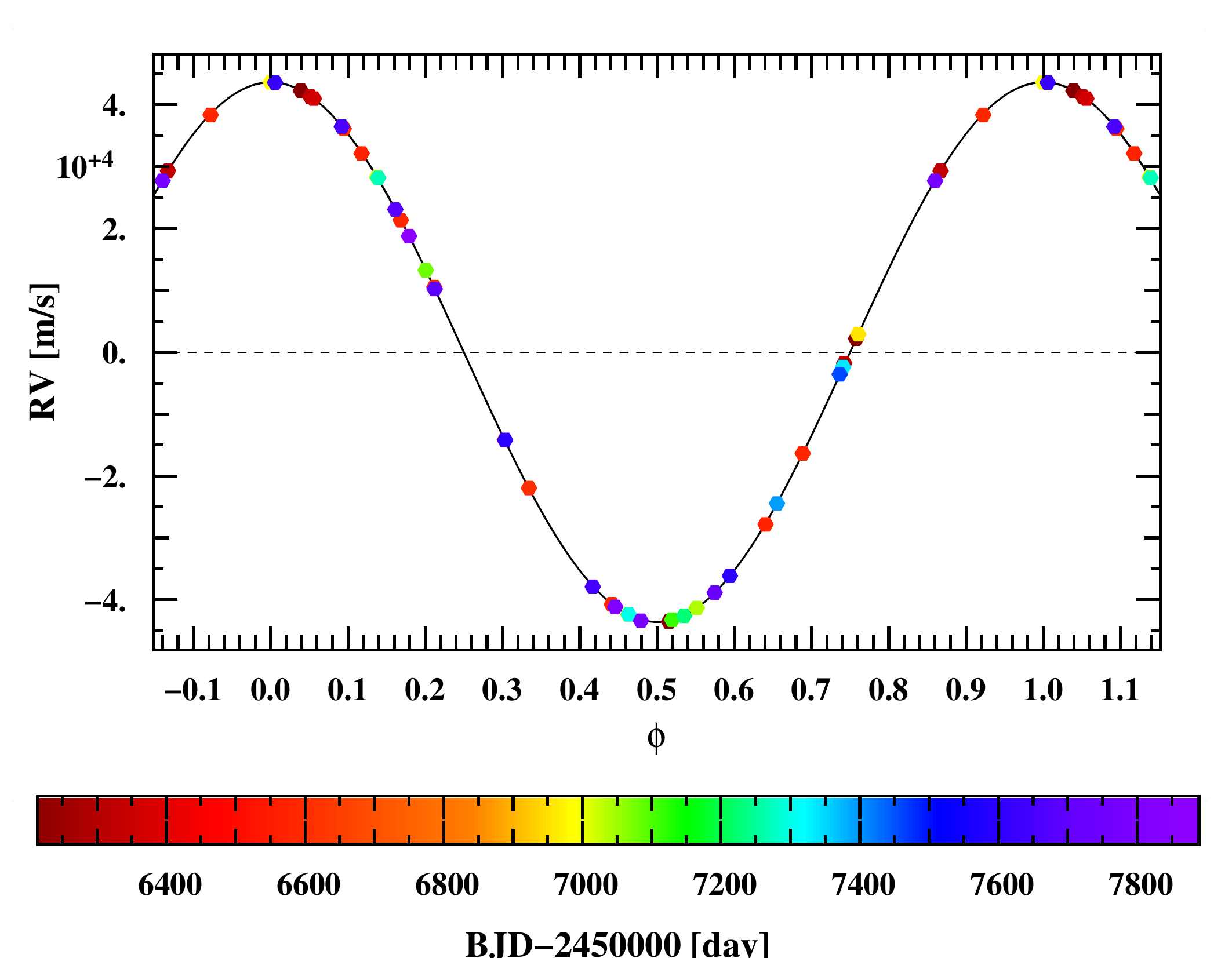}
\end{subfigure}
\begin{subfigure}[b]{0.49\textwidth}
\includegraphics[width=\textwidth,trim={0 0 2cm 0},clip]{orbit_figures/BJD_bar.pdf}
\end{subfigure}
Detection limits
\begin{subfigure}[b]{0.49\textwidth}
\vspace{0.5cm}
\includegraphics[width=\textwidth,trim={0 0 0 0},clip]{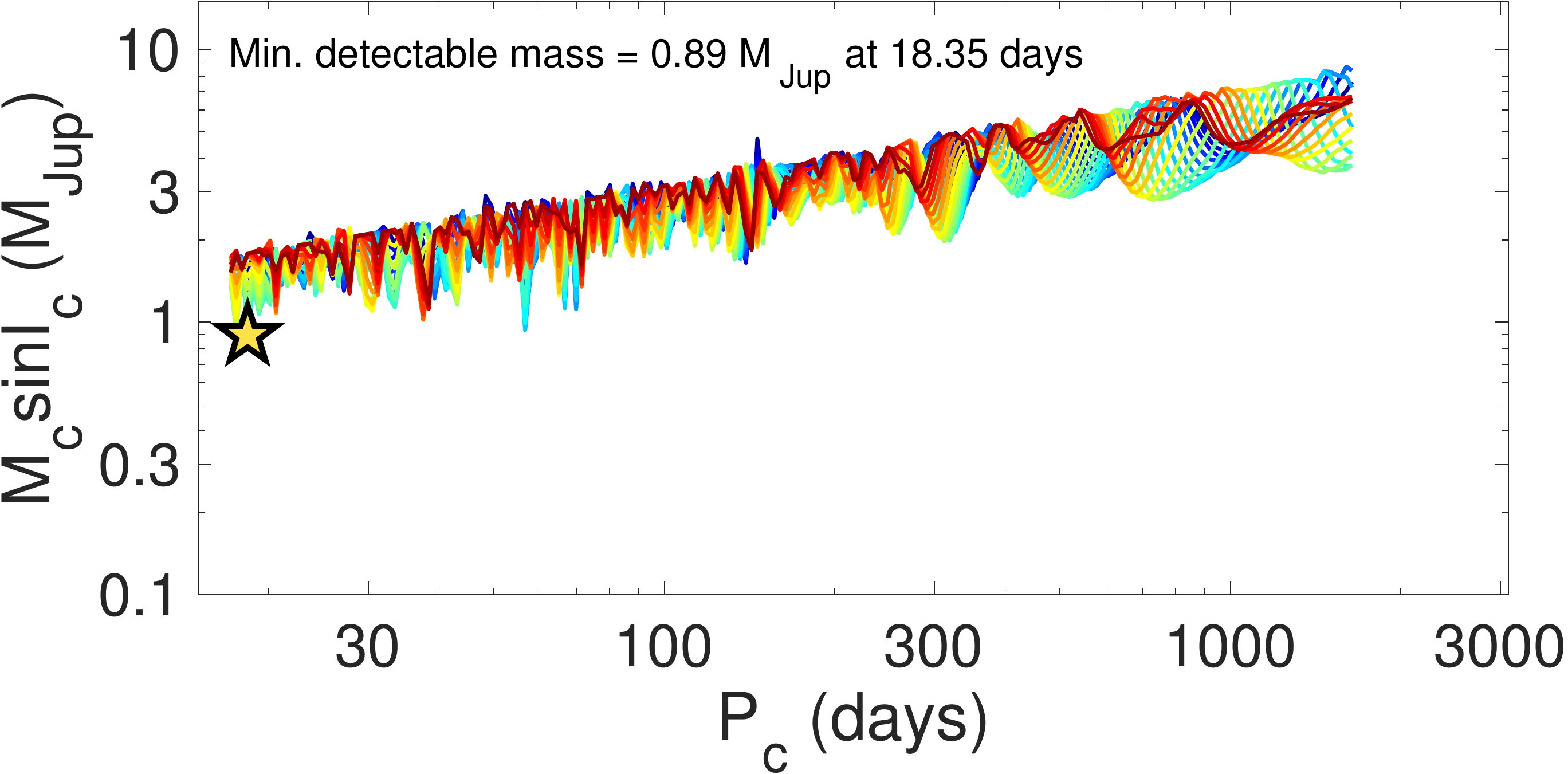}
\end{subfigure}
\end{center}
\end{figure}
\begin{figure}
\begin{center}
\subcaption*{EBLM J0948-08: chosen model = k1d2 (ecc) \newline \newline $m_{\rm A} = 1.41M_{\odot}$, $m_{\rm B} = 0.675M_{\odot}$, $P = 5.38$ d, $e = 0.049$}
\begin{subfigure}[b]{0.49\textwidth}
\includegraphics[width=\textwidth,trim={0 10cm 0 1.2cm},clip]{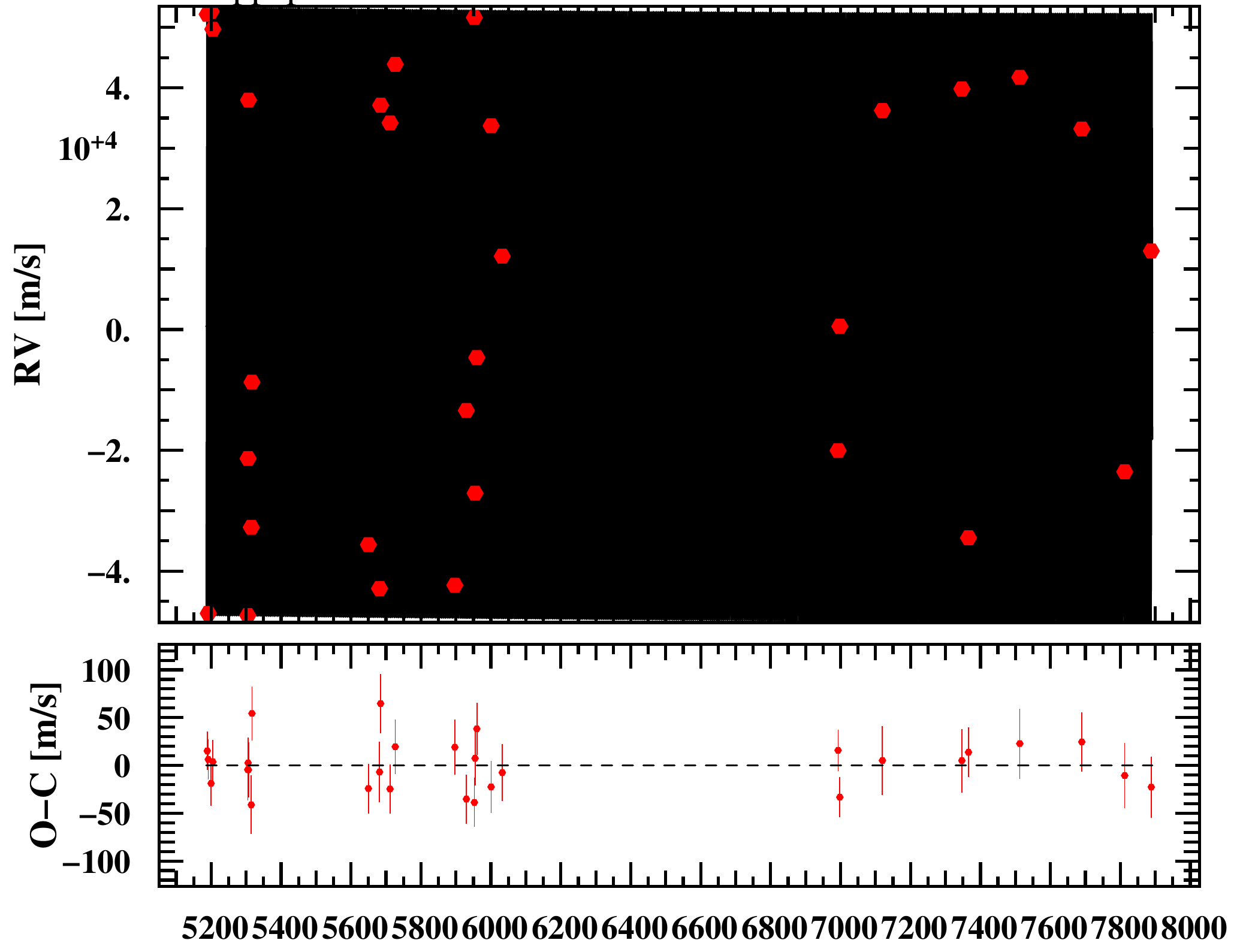}
\end{subfigure}
\begin{subfigure}[b]{0.49\textwidth}
\includegraphics[width=\textwidth,trim={0 0 2cm 0},clip]{orbit_figures/BJD_bar.pdf}
\end{subfigure}
Radial velocities folded on binary phase
\begin{subfigure}[b]{0.49\textwidth}
\includegraphics[width=\textwidth,trim={0 0.5cm 0 0},clip]{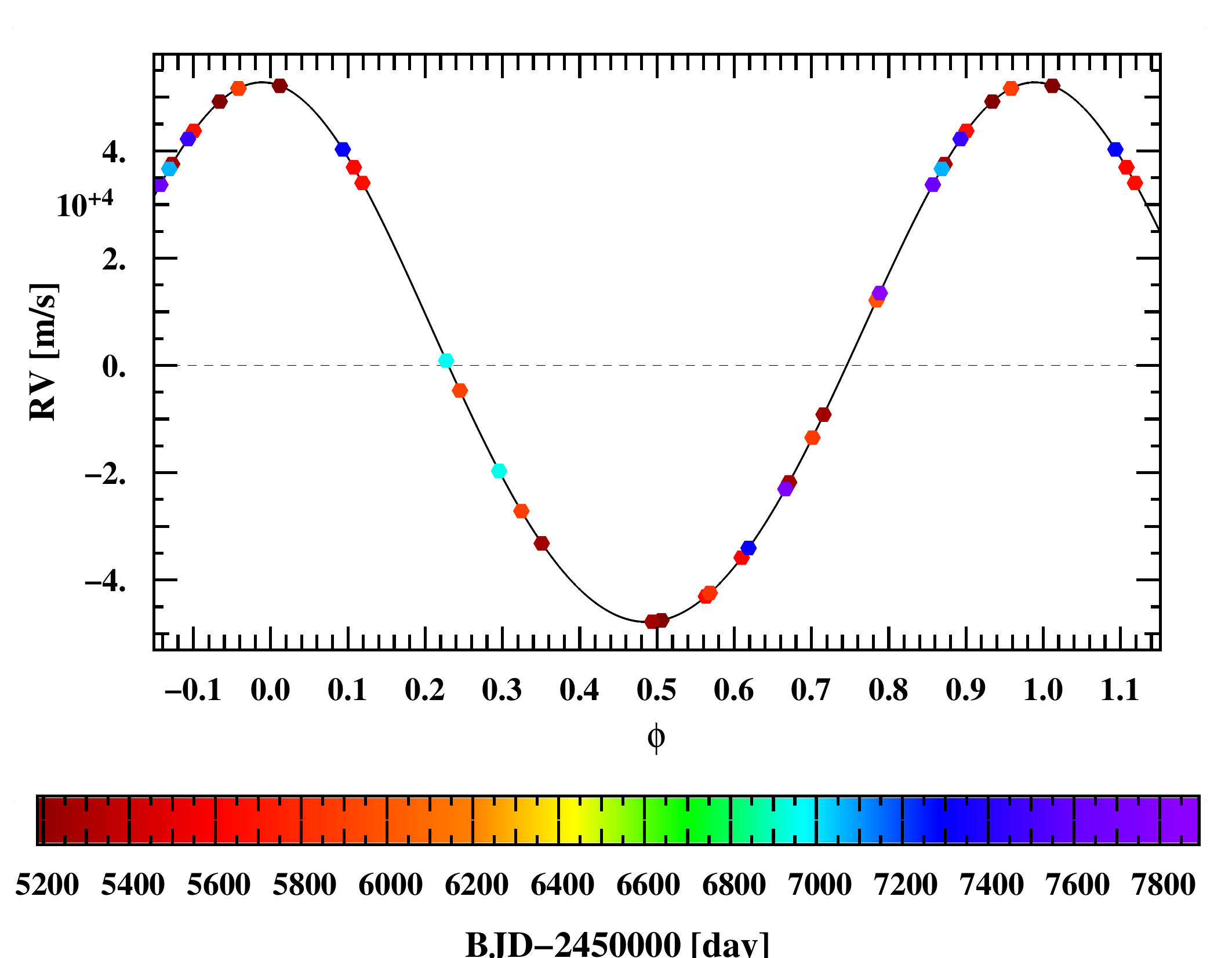}
\end{subfigure}
\begin{subfigure}[b]{0.49\textwidth}
\includegraphics[width=\textwidth,trim={0 0 2cm 0},clip]{orbit_figures/BJD_bar.pdf}
\end{subfigure}
Detection limits
\begin{subfigure}[b]{0.49\textwidth}
\vspace{0.5cm}
\includegraphics[width=\textwidth,trim={0 0 0 0},clip]{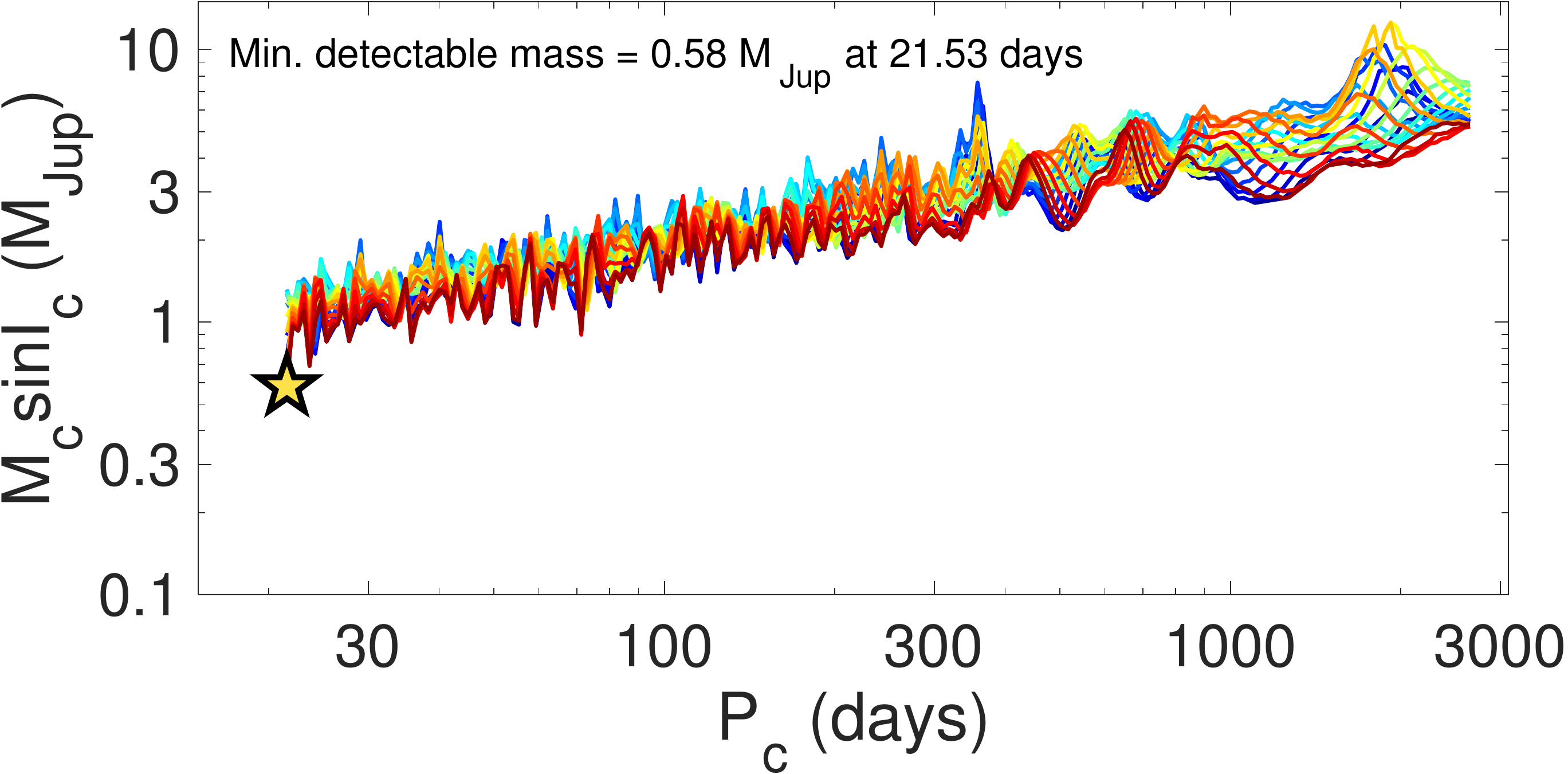}
\end{subfigure}
\end{center}
\end{figure}
\begin{figure}
\begin{center}
\subcaption*{EBLM J0954-23: chosen model = k1 (ecc) \newline \newline $m_{\rm A} = 1.44M_{\odot}$, $m_{\rm B} = 0.107M_{\odot}$, $P = 7.575$ d, $e = 0.042$}
\begin{subfigure}[b]{0.49\textwidth}
\includegraphics[width=\textwidth,trim={0 10cm 0 1.2cm},clip]{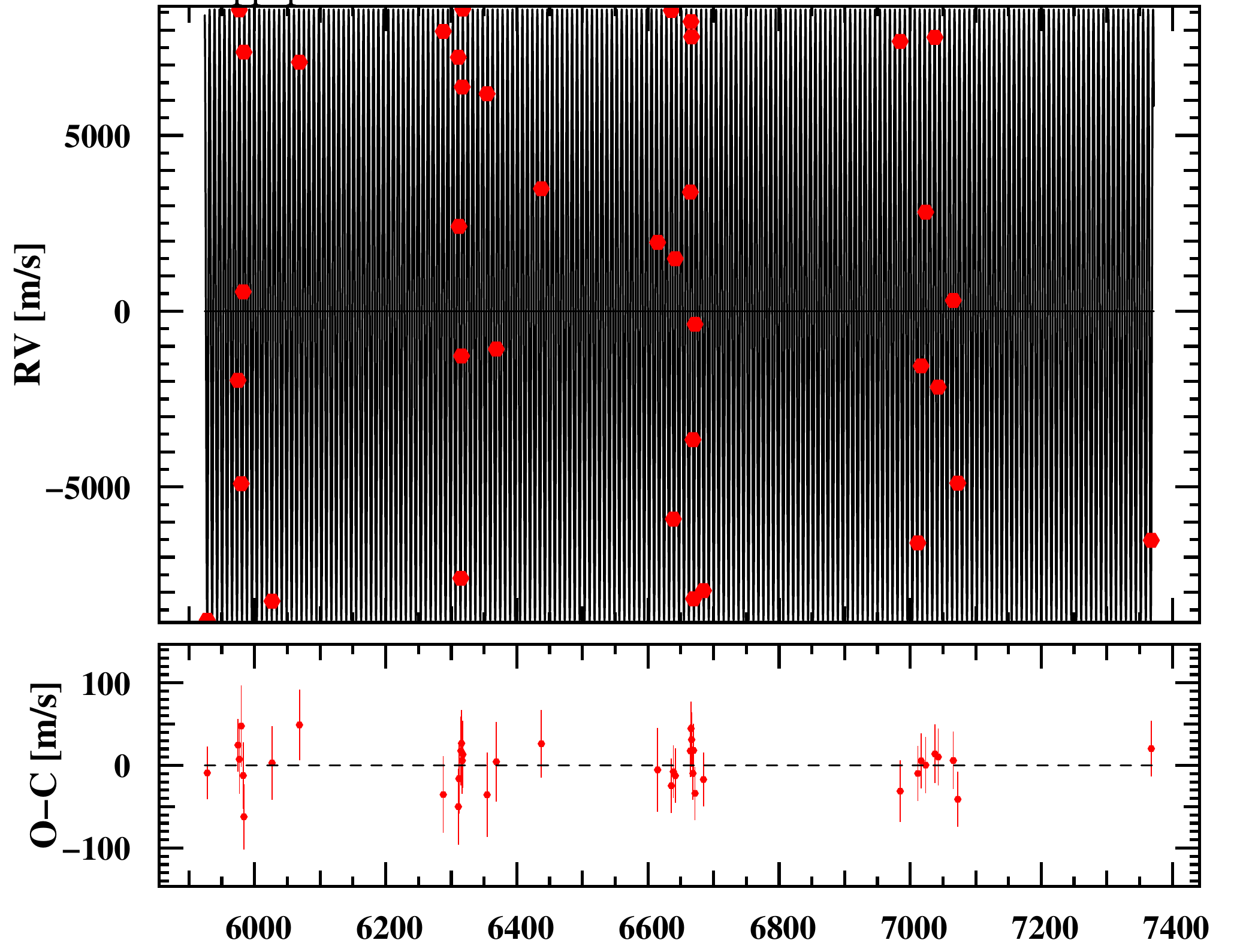}
\end{subfigure}
\begin{subfigure}[b]{0.49\textwidth}
\includegraphics[width=\textwidth,trim={0 0 2cm 0},clip]{orbit_figures/BJD_bar.pdf}
\end{subfigure}
Radial velocities folded on binary phase
\begin{subfigure}[b]{0.49\textwidth}
\includegraphics[width=\textwidth,trim={0 0.5cm 0 0},clip]{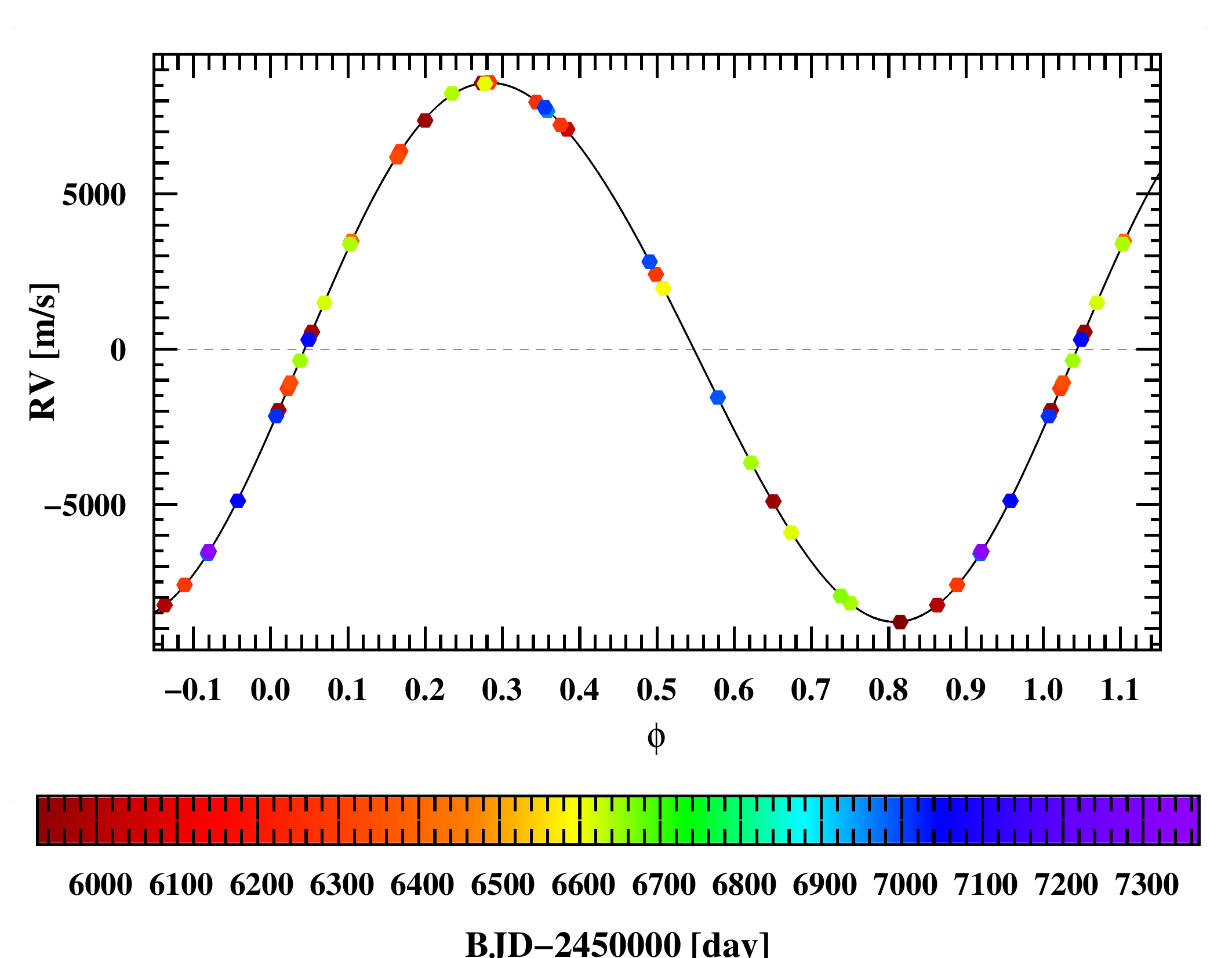}
\end{subfigure}
\begin{subfigure}[b]{0.49\textwidth}
\includegraphics[width=\textwidth,trim={0 0 2cm 0},clip]{orbit_figures/BJD_bar.pdf}
\end{subfigure}
Detection limits
\begin{subfigure}[b]{0.49\textwidth}
\vspace{0.5cm}
\includegraphics[width=\textwidth,trim={0 0 0 0},clip]{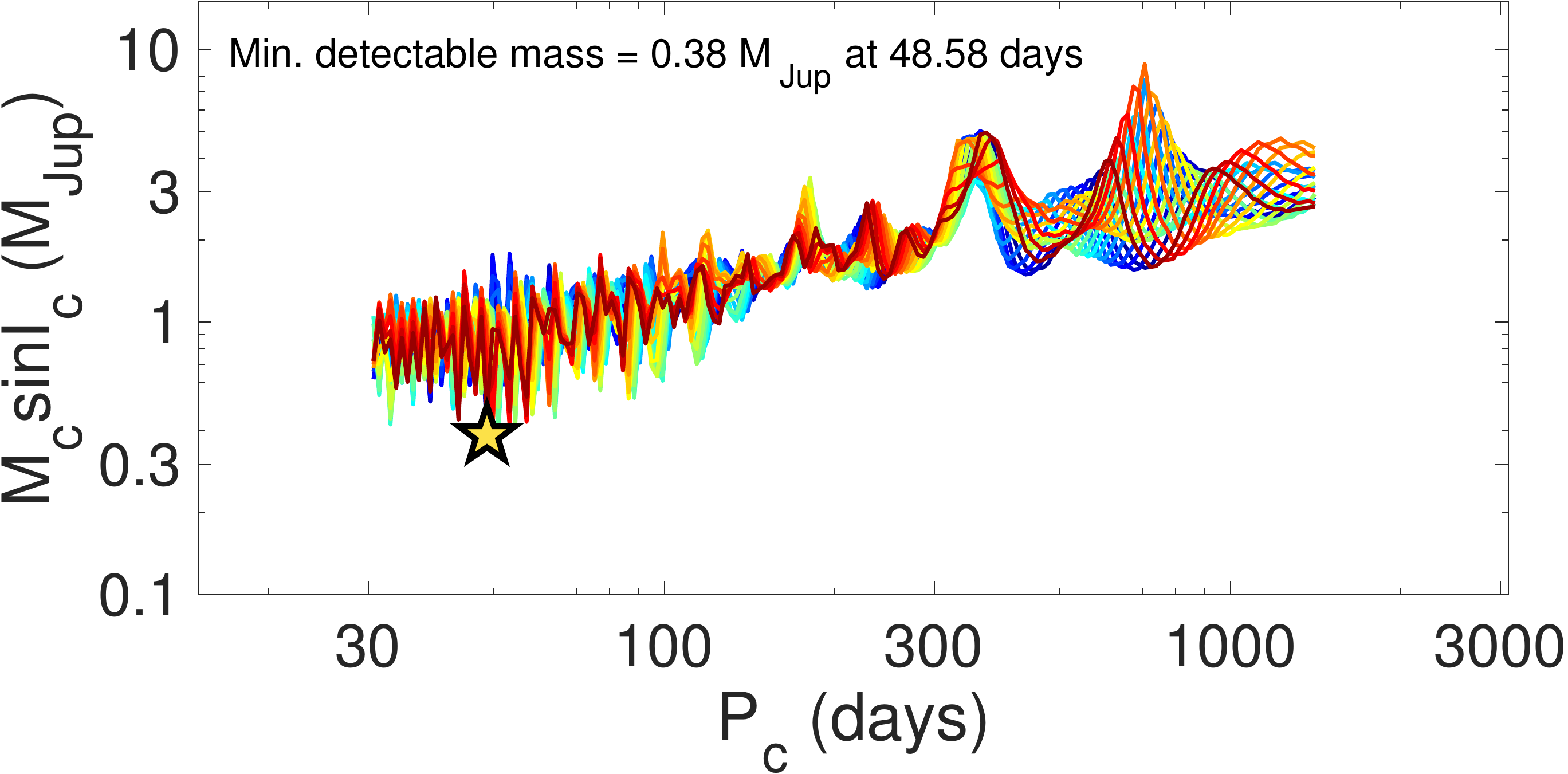}
\end{subfigure}
\end{center}
\end{figure}
\begin{figure}
\begin{center}
\subcaption*{EBLM J0954-45: chosen model = k1 (ecc) \newline \newline $m_{\rm A} = 1.69M_{\odot}$, $m_{\rm B} = 0.412M_{\odot}$, $P = 8.073$ d, $e = 0.295$}
\begin{subfigure}[b]{0.49\textwidth}
\includegraphics[width=\textwidth,trim={0 10cm 0 1.2cm},clip]{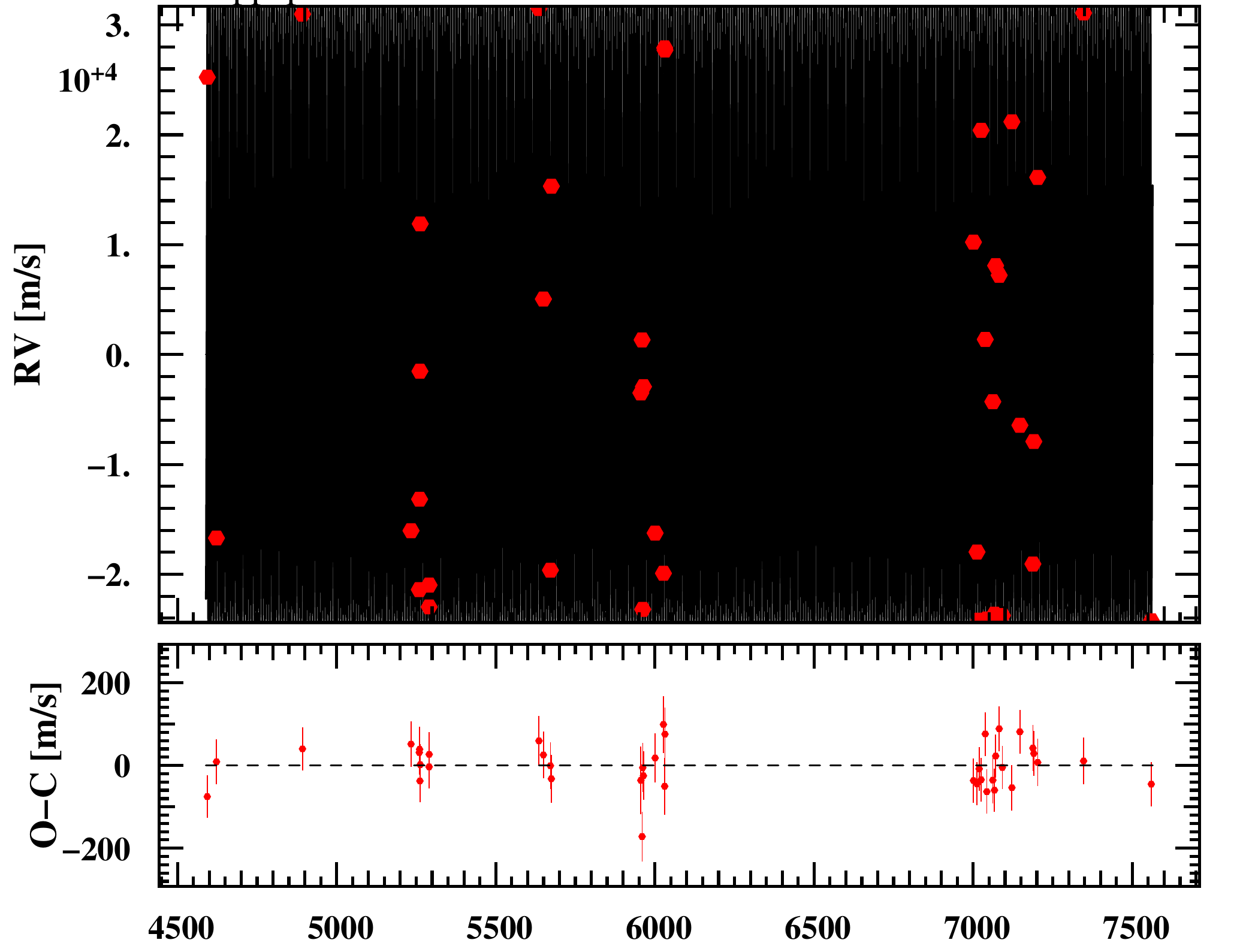}
\end{subfigure}
\begin{subfigure}[b]{0.49\textwidth}
\includegraphics[width=\textwidth,trim={0 0 2cm 0},clip]{orbit_figures/BJD_bar.pdf}
\end{subfigure}
Radial velocities folded on binary phase
\begin{subfigure}[b]{0.49\textwidth}
\includegraphics[width=\textwidth,trim={0 0.5cm 0 0},clip]{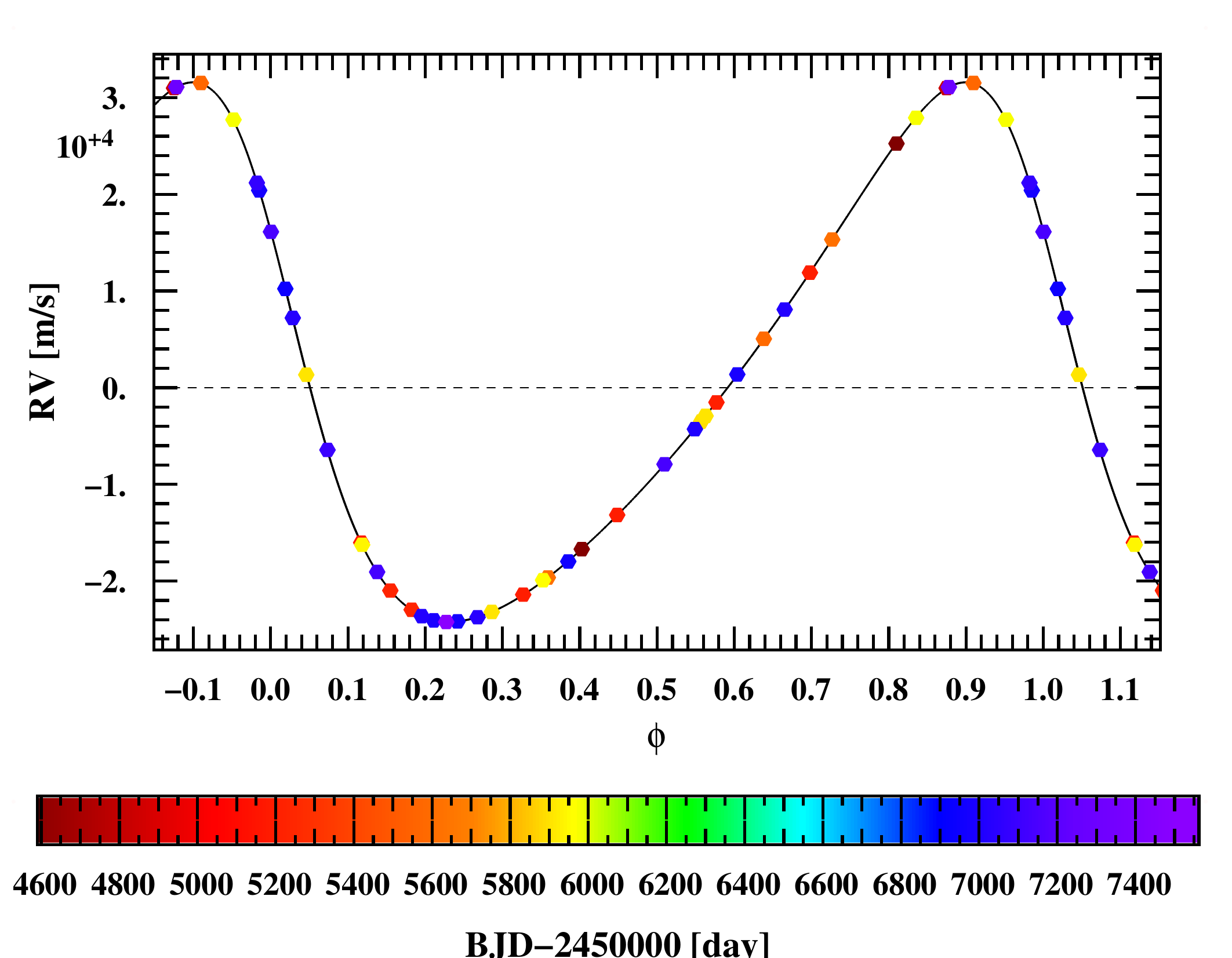}
\end{subfigure}
\begin{subfigure}[b]{0.49\textwidth}
\includegraphics[width=\textwidth,trim={0 0 2cm 0},clip]{orbit_figures/BJD_bar.pdf}
\end{subfigure}
Detection limits
\begin{subfigure}[b]{0.49\textwidth}
\vspace{0.5cm}
\includegraphics[width=\textwidth,trim={0 0 0 0},clip]{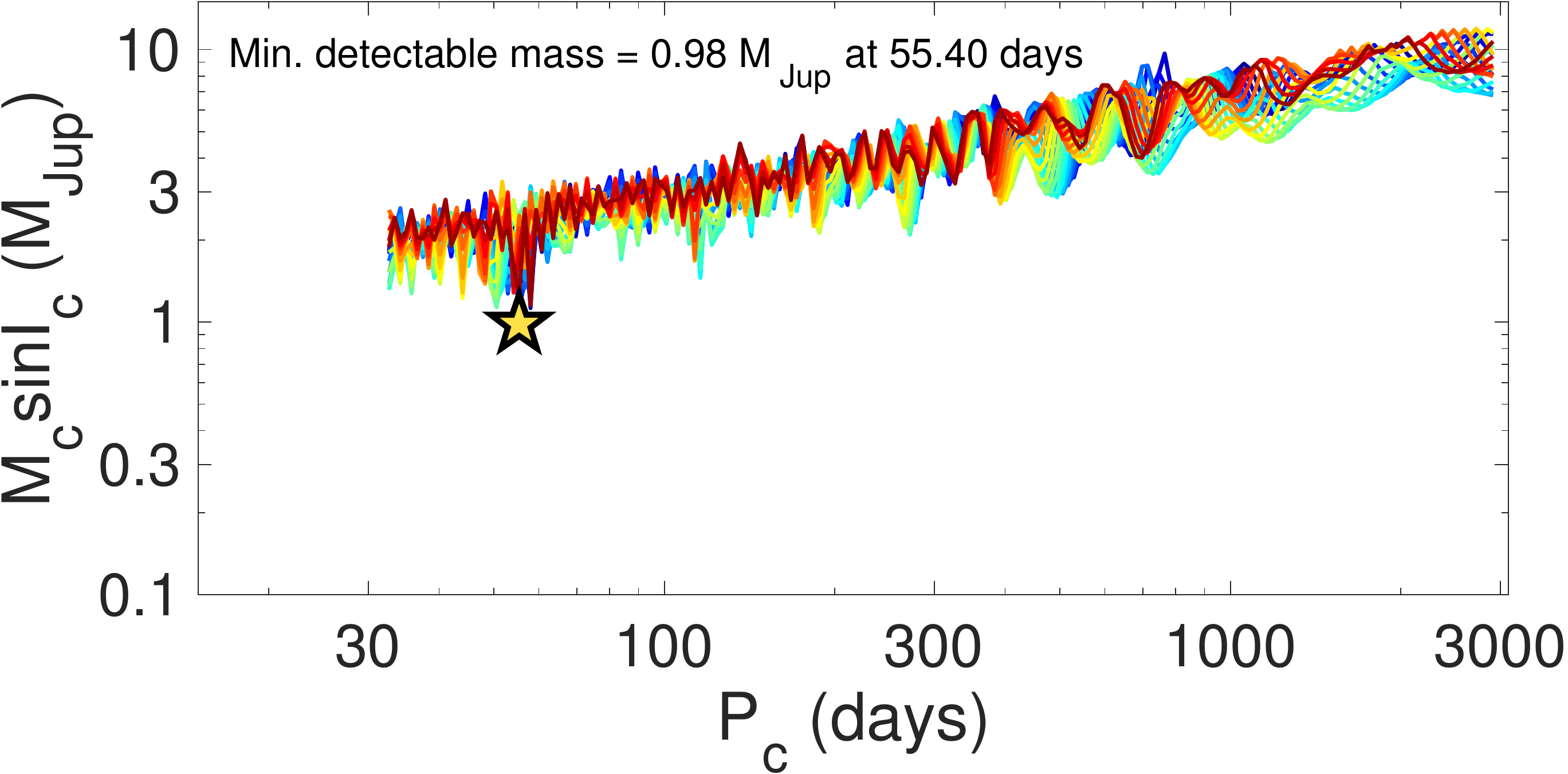}
\end{subfigure}
\end{center}
\end{figure}
\begin{figure}
\begin{center}
\subcaption*{EBLM J1014-07: chosen model = k1d1 (ecc) \newline \newline $m_{\rm A} = 1.3M_{\odot}$, $m_{\rm B} = 0.241M_{\odot}$, $P = 4.557$ d, $e = 0.206$}
\begin{subfigure}[b]{0.49\textwidth}
\includegraphics[width=\textwidth,trim={0 10cm 0 1.2cm},clip]{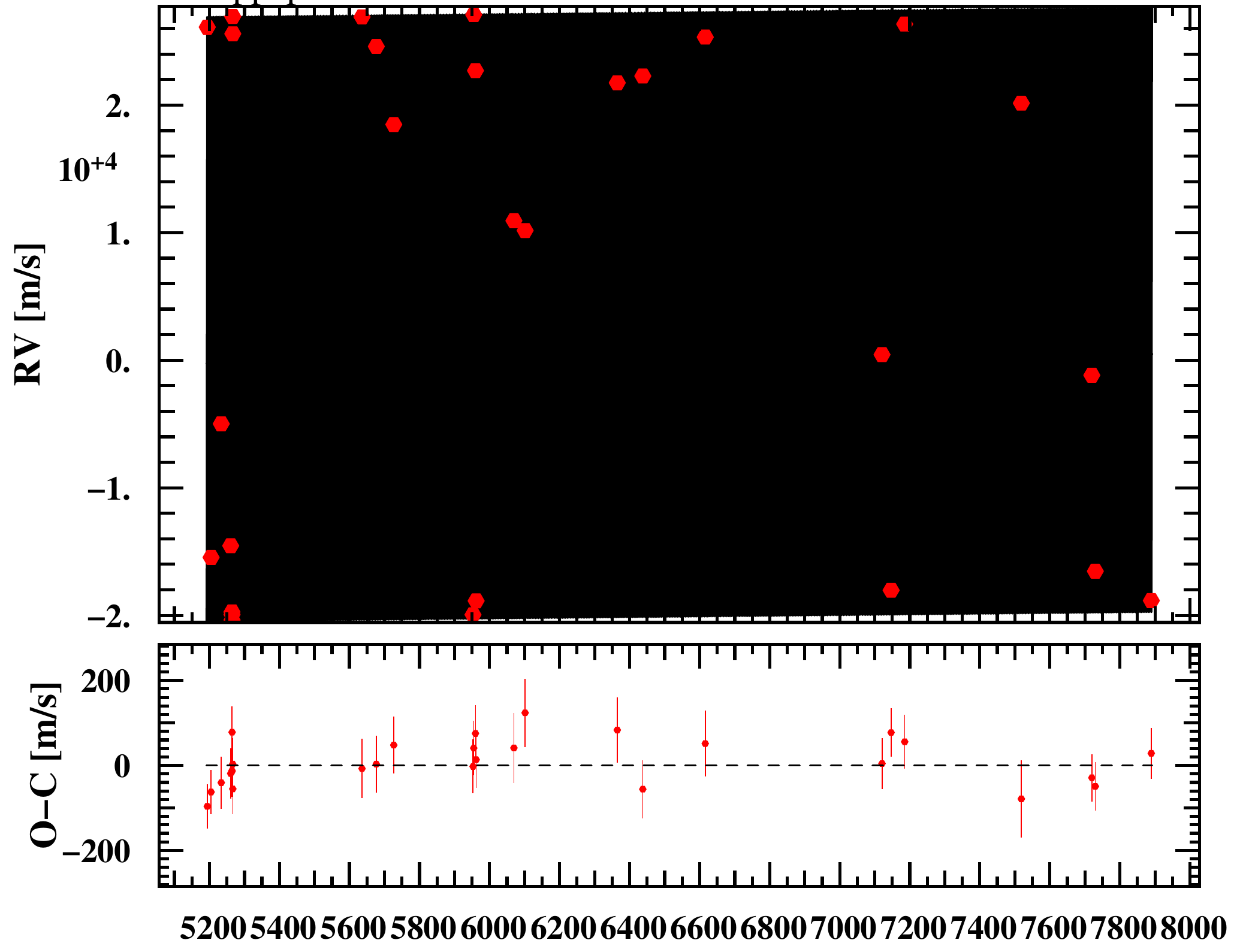}
\end{subfigure}
\begin{subfigure}[b]{0.49\textwidth}
\includegraphics[width=\textwidth,trim={0 0 2cm 0},clip]{orbit_figures/BJD_bar.pdf}
\end{subfigure}
Radial velocities folded on binary phase
\begin{subfigure}[b]{0.49\textwidth}
\includegraphics[width=\textwidth,trim={0 0.5cm 0 0},clip]{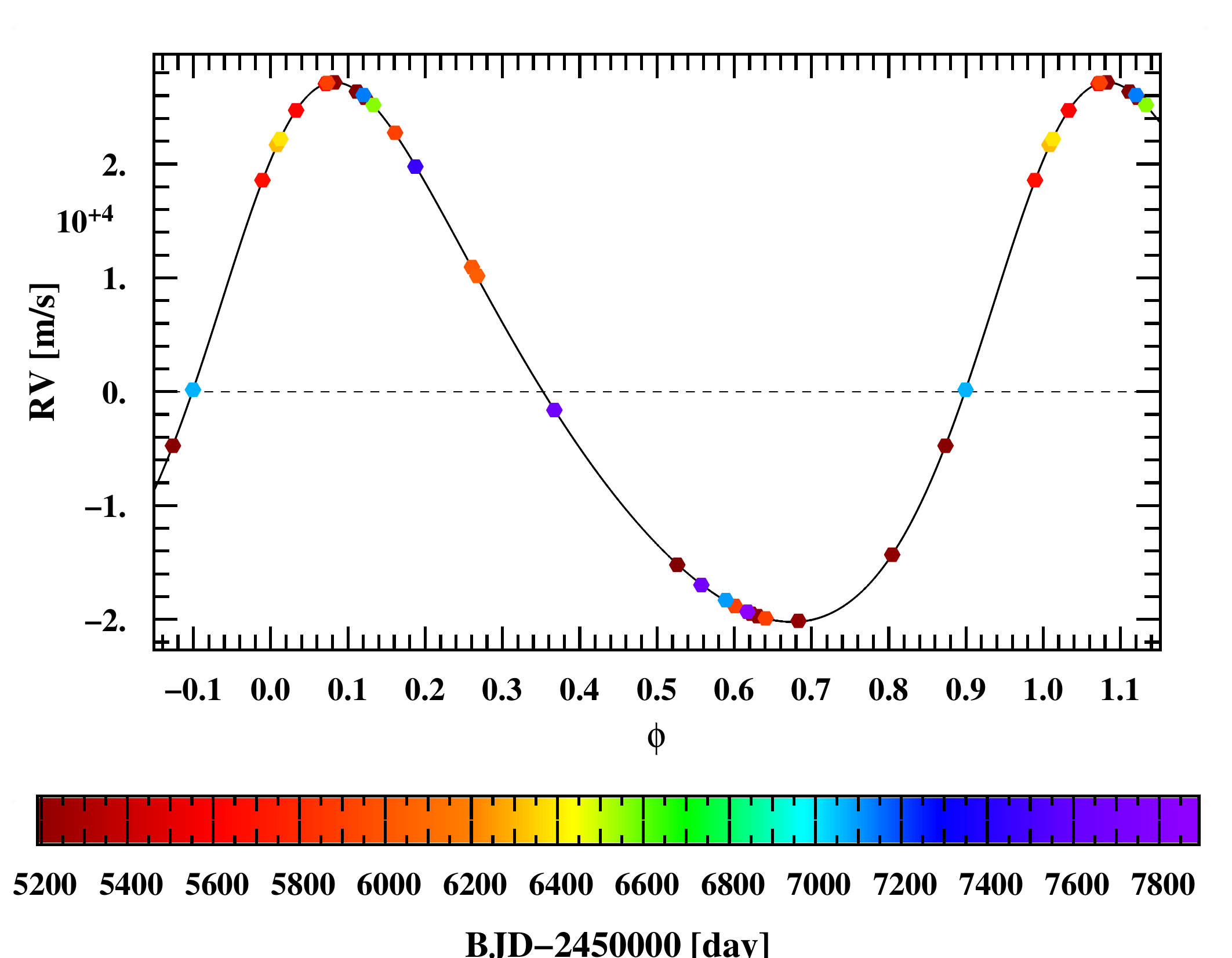}
\end{subfigure}
\begin{subfigure}[b]{0.49\textwidth}
\includegraphics[width=\textwidth,trim={0 0 2cm 0},clip]{orbit_figures/BJD_bar.pdf}
\end{subfigure}
Detection limits
\begin{subfigure}[b]{0.49\textwidth}
\vspace{0.5cm}
\includegraphics[width=\textwidth,trim={0 0 0 0},clip]{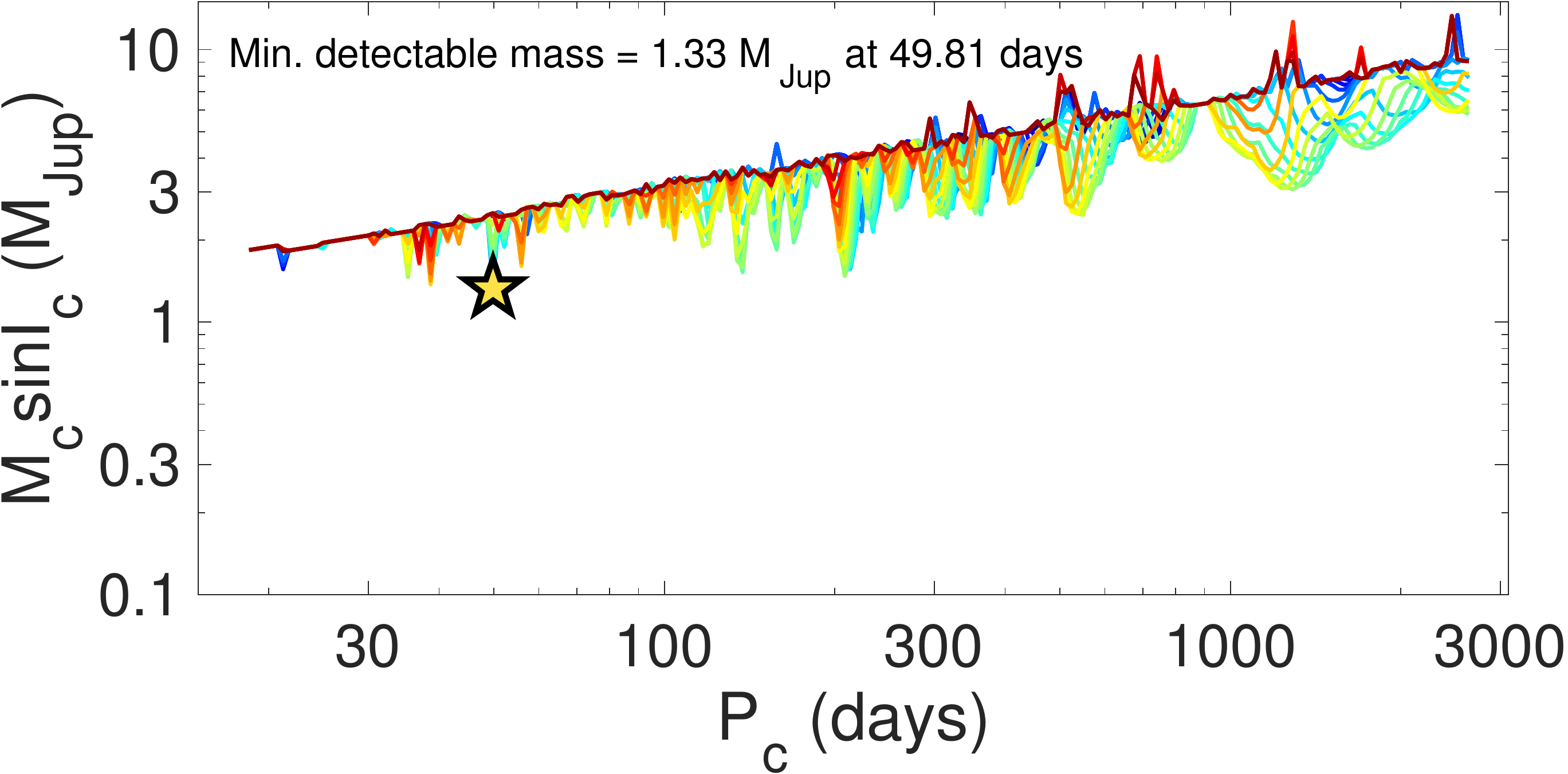}
\end{subfigure}
\end{center}
\end{figure}
\begin{figure}
\begin{center}
\subcaption*{EBLM J1037-25: chosen model = k1 (ecc) \newline \newline $m_{\rm A} = 1.26M_{\odot}$, $m_{\rm B} = 0.26M_{\odot}$, $P = 4.937$ d, $e = 0.121$}
\begin{subfigure}[b]{0.49\textwidth}
\includegraphics[width=\textwidth,trim={0 10cm 0 1.2cm},clip]{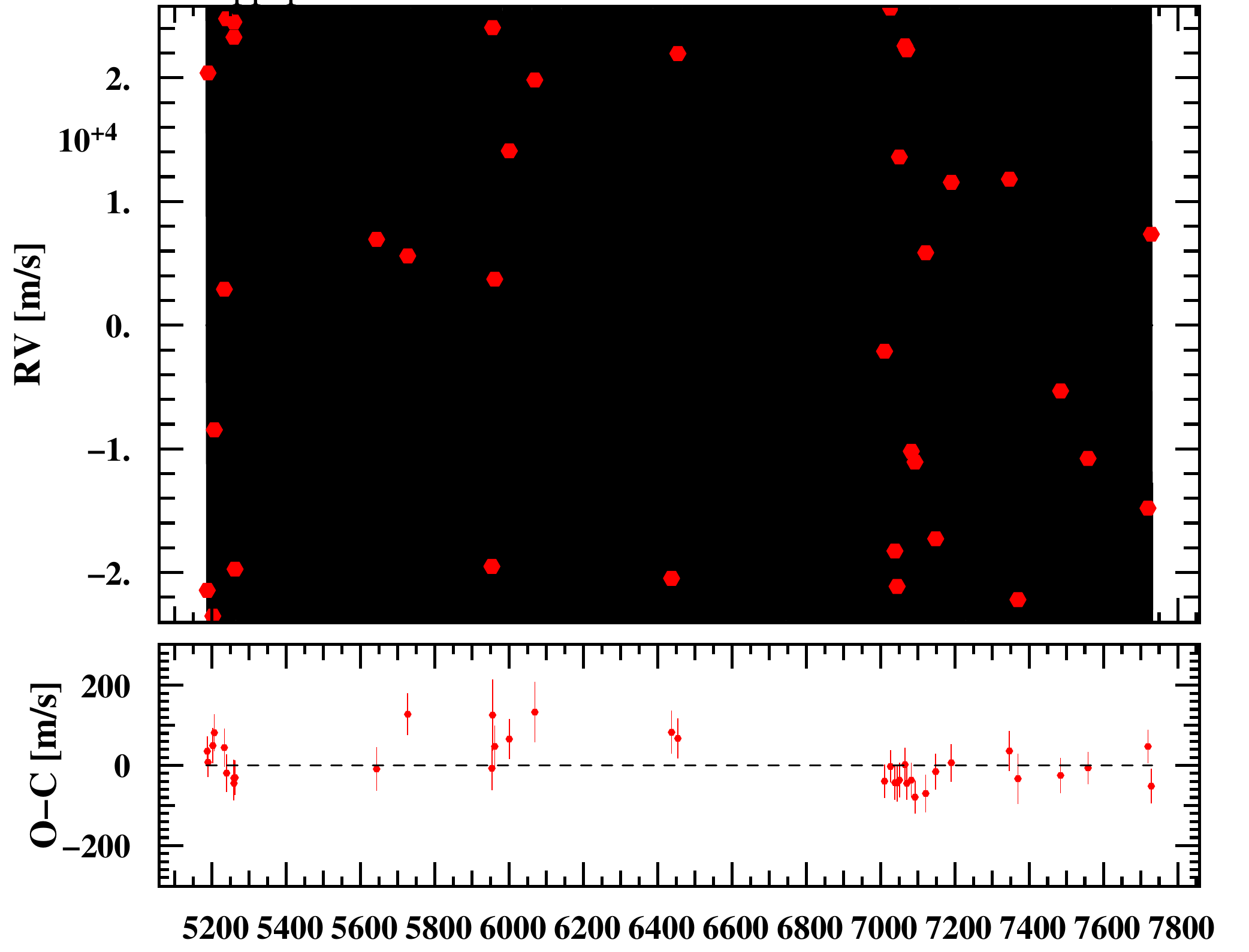}
\end{subfigure}
\begin{subfigure}[b]{0.49\textwidth}
\includegraphics[width=\textwidth,trim={0 0 2cm 0},clip]{orbit_figures/BJD_bar.pdf}
\end{subfigure}
Radial velocities folded on binary phase
\begin{subfigure}[b]{0.49\textwidth}
\includegraphics[width=\textwidth,trim={0 0.5cm 0 0},clip]{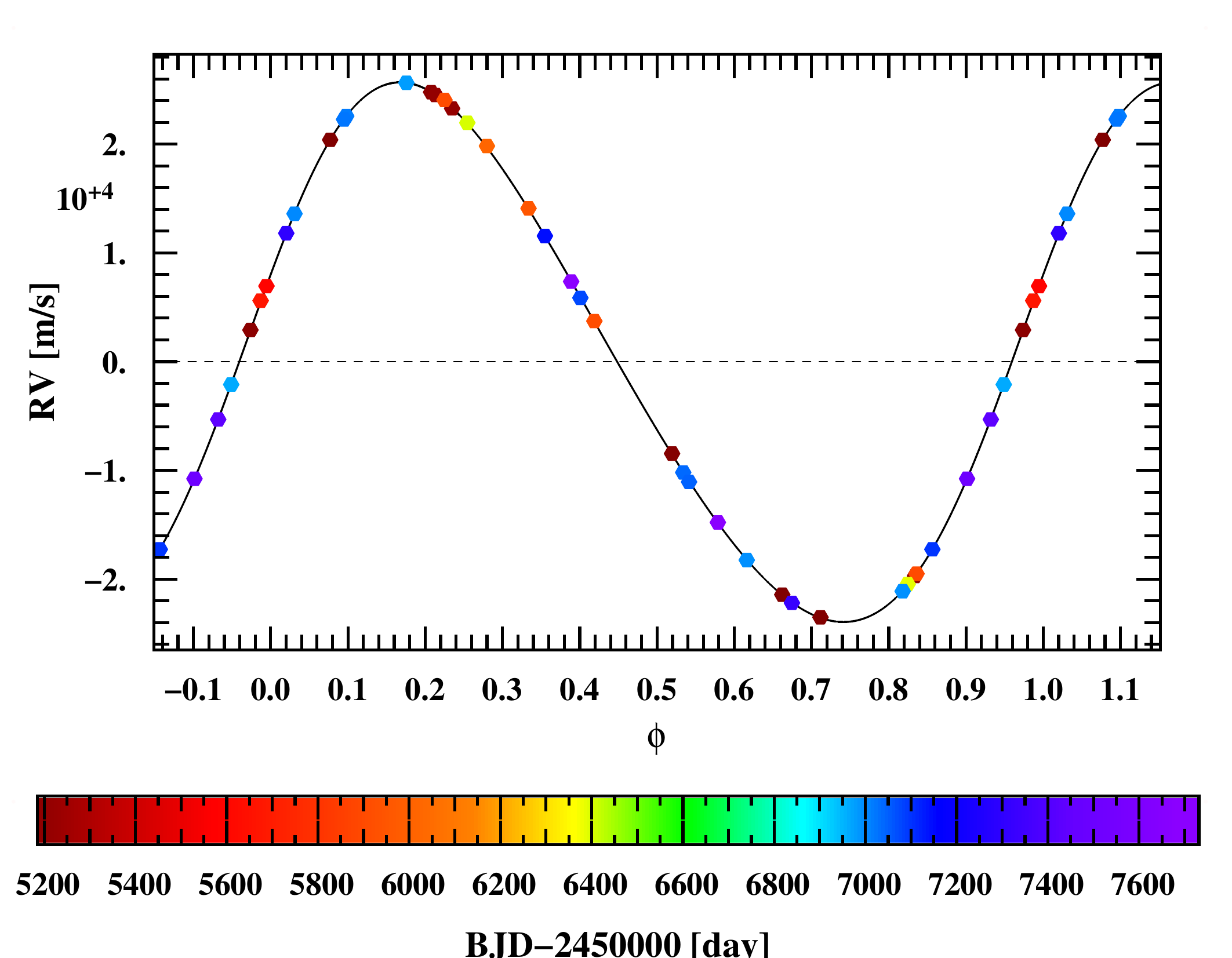}
\end{subfigure}
\begin{subfigure}[b]{0.49\textwidth}
\includegraphics[width=\textwidth,trim={0 0 2cm 0},clip]{orbit_figures/BJD_bar.pdf}
\end{subfigure}
Detection limits
\begin{subfigure}[b]{0.49\textwidth}
\vspace{0.5cm}
\includegraphics[width=\textwidth,trim={0 0 0 0},clip]{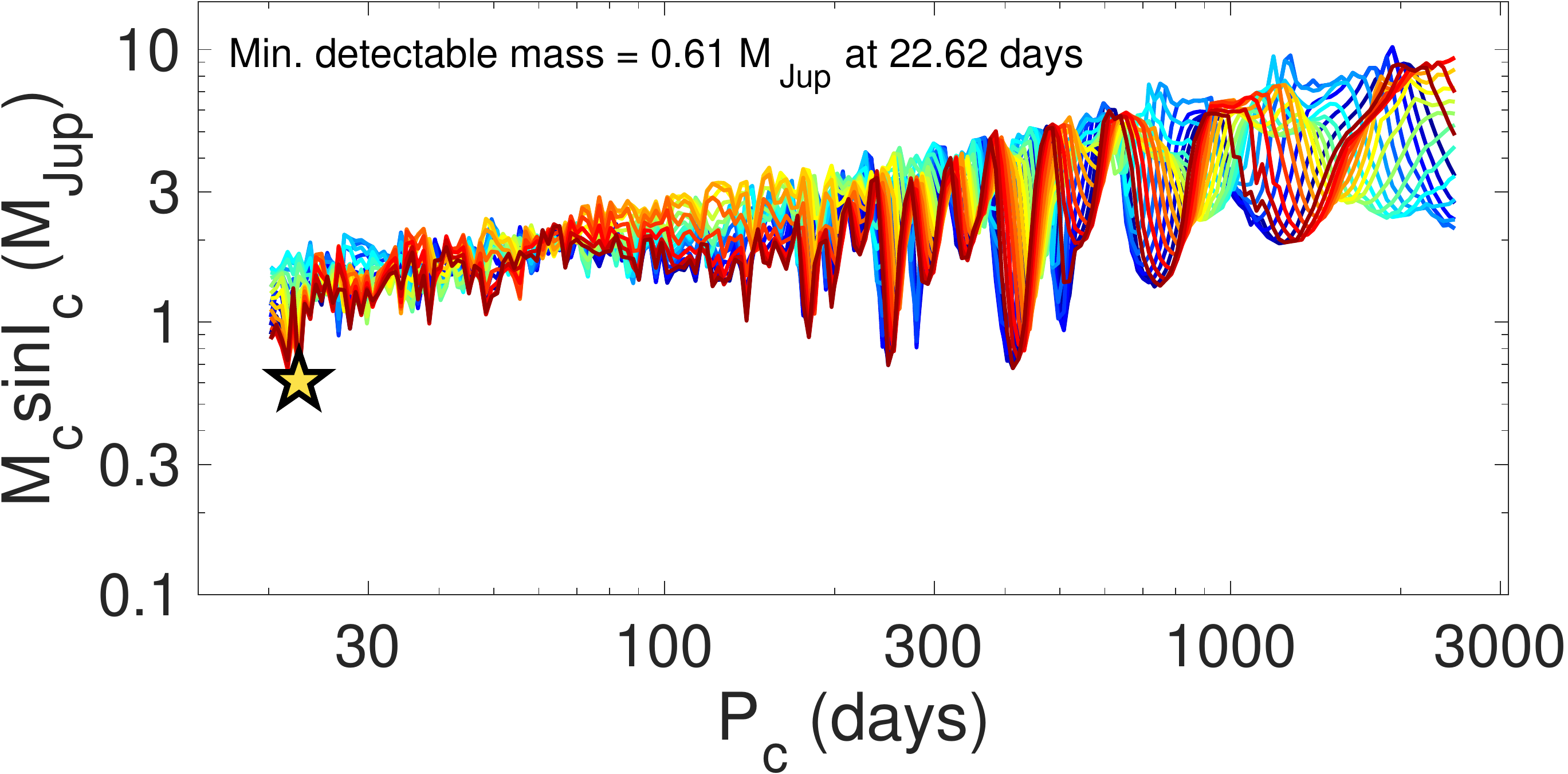}
\end{subfigure}
\end{center}
\end{figure}
\begin{figure}
\begin{center}
\subcaption*{EBLM J1038-37: chosen model = k2 (circ) \newline \newline $m_{\rm A} = 1.17M_{\odot}$, $m_{\rm B} = 0.173M_{\odot}$, $P = 5.022$ d, $e = 0$}
\begin{subfigure}[b]{0.49\textwidth}
\includegraphics[width=\textwidth,trim={0 10cm 0 1.2cm},clip]{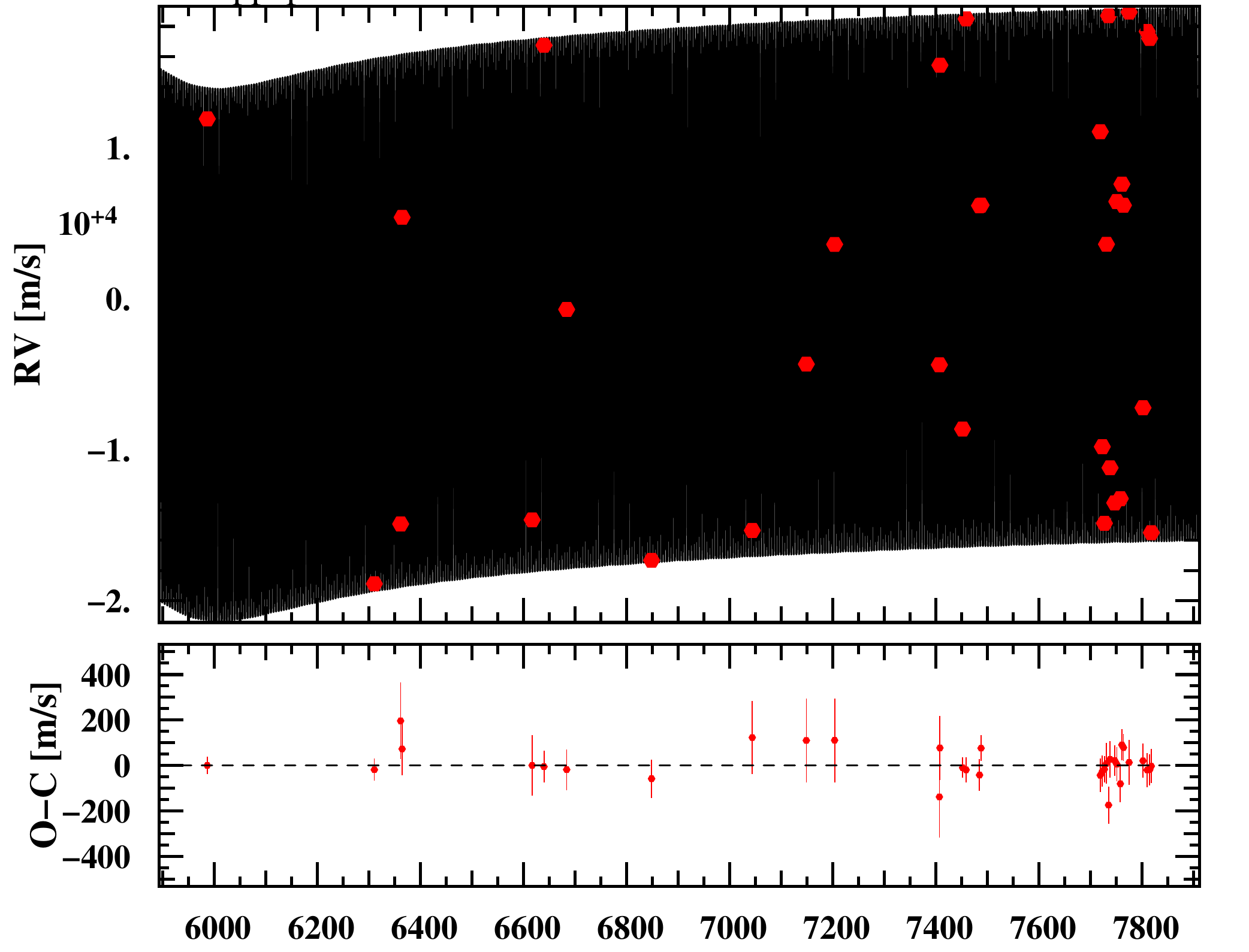}
\end{subfigure}
\begin{subfigure}[b]{0.49\textwidth}
\includegraphics[width=\textwidth,trim={0 0 2cm 0},clip]{orbit_figures/BJD_bar.pdf}
\end{subfigure}
Radial velocities folded on binary phase
\begin{subfigure}[b]{0.49\textwidth}
\includegraphics[width=\textwidth,trim={0 0.5cm 0 0},clip]{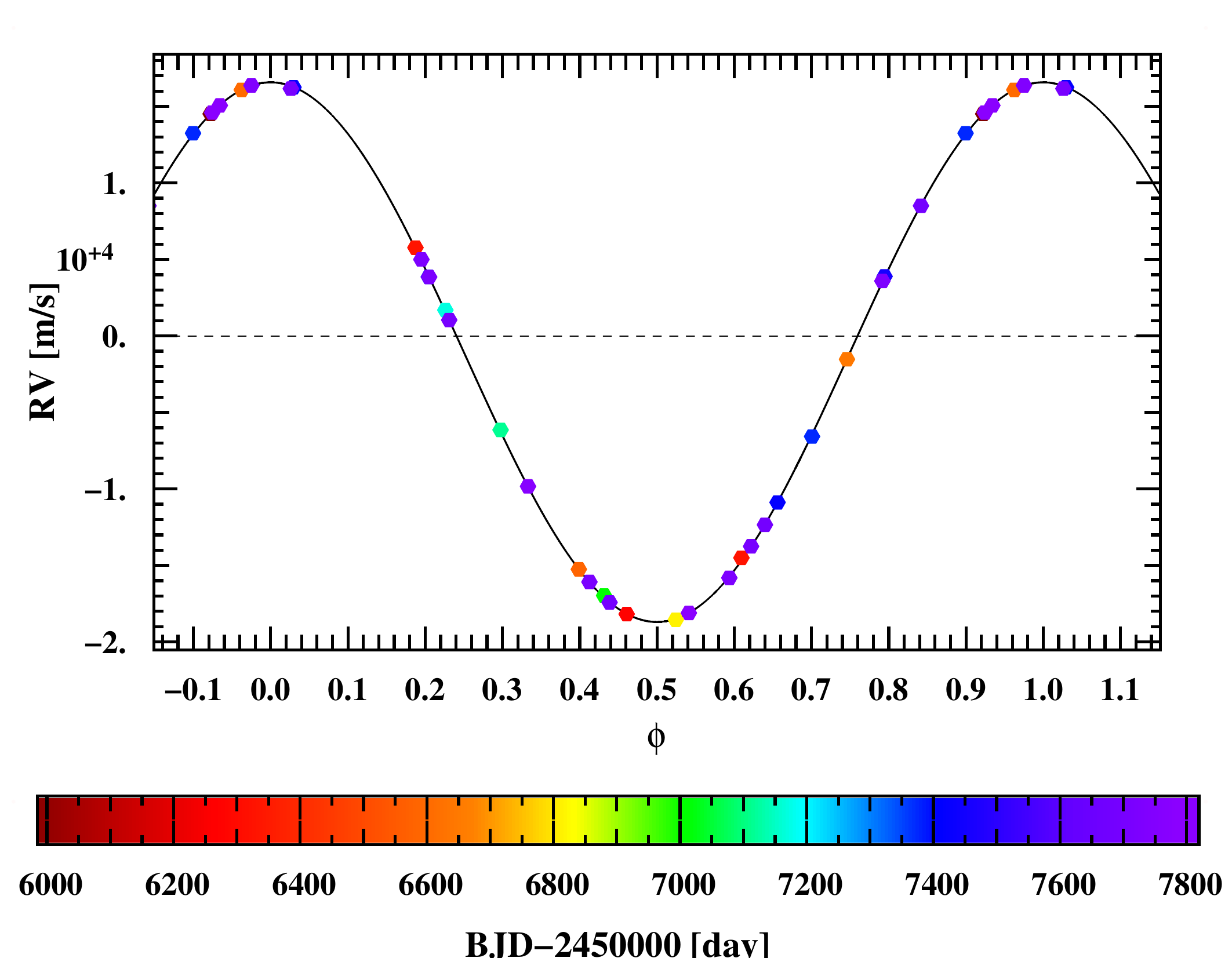}
\end{subfigure}
\begin{subfigure}[b]{0.49\textwidth}
\includegraphics[width=\textwidth,trim={0 0 2cm 0},clip]{orbit_figures/BJD_bar.pdf}
\end{subfigure}
Detection limits
\begin{subfigure}[b]{0.49\textwidth}
\vspace{0.5cm}
\includegraphics[width=\textwidth,trim={0 0 0 0},clip]{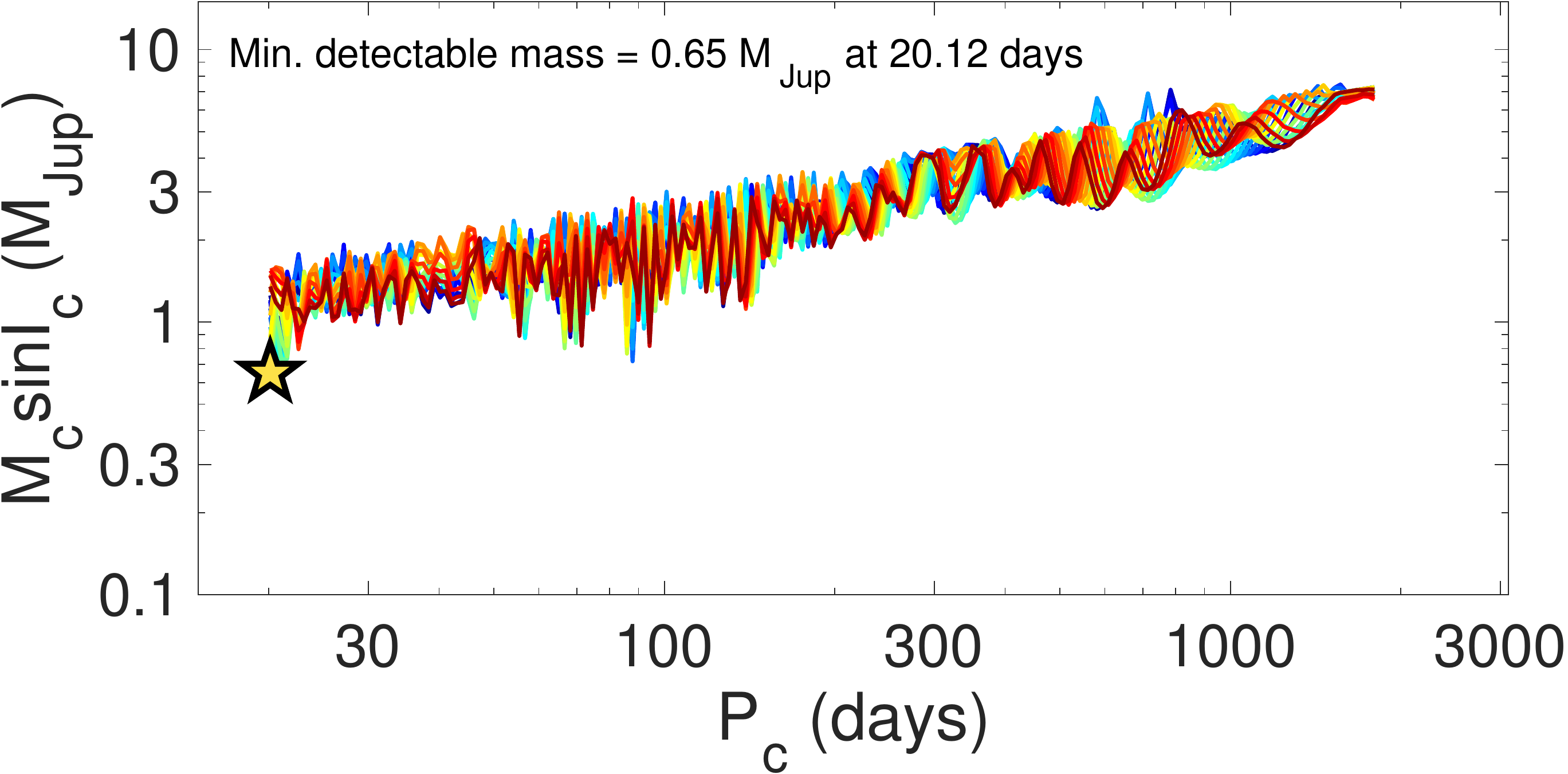}
\end{subfigure}
\end{center}
\end{figure}
\begin{figure}
\begin{center}
\subcaption*{EBLM J1141-37: chosen model = k1 (circ) \newline \newline $m_{\rm A} = 1.22M_{\odot}$, $m_{\rm B} = 0.354M_{\odot}$, $P = 5.148$ d, $e = 0$}
\begin{subfigure}[b]{0.49\textwidth}
\includegraphics[width=\textwidth,trim={0 10cm 0 1.2cm},clip]{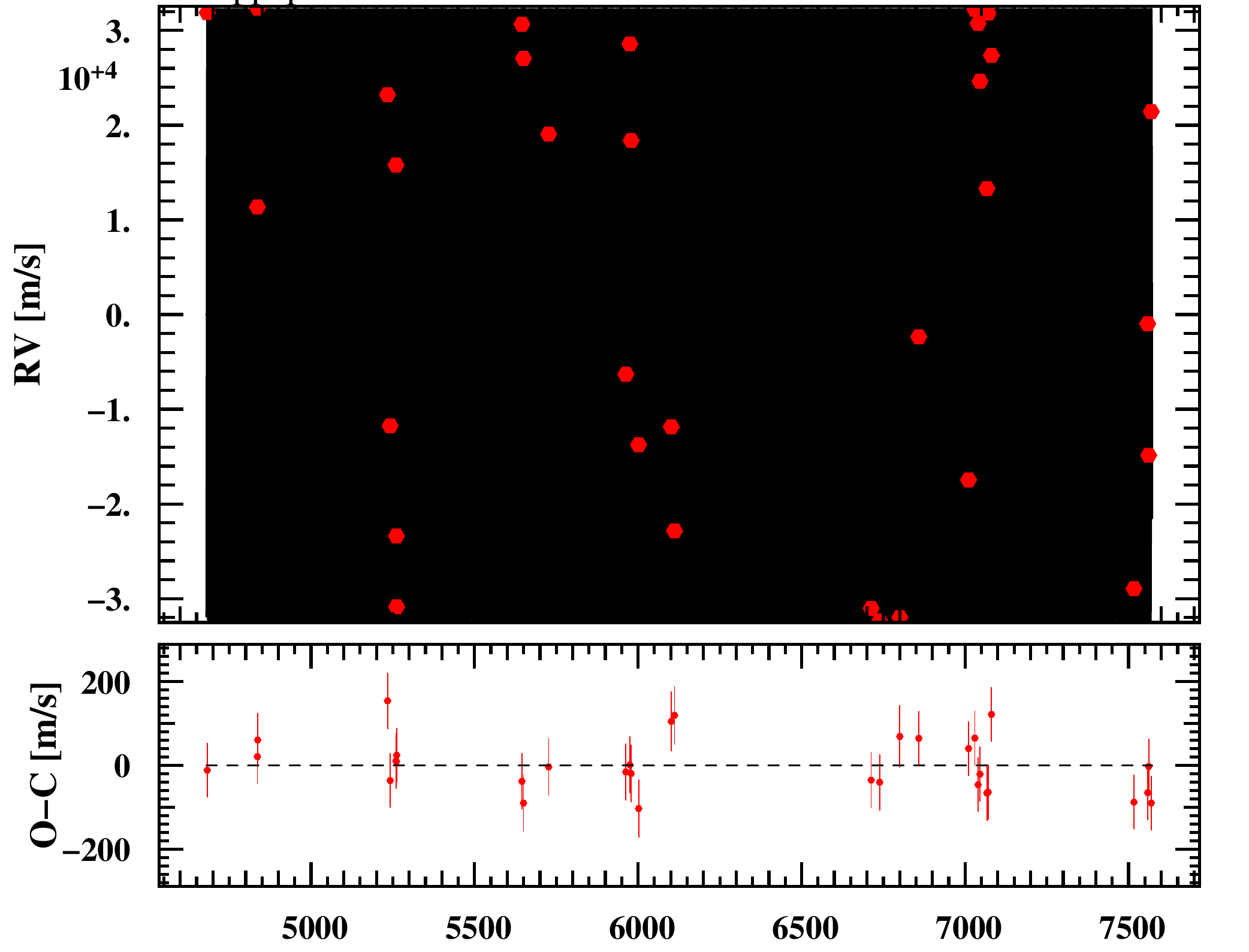}
\end{subfigure}
\begin{subfigure}[b]{0.49\textwidth}
\includegraphics[width=\textwidth,trim={0 0 2cm 0},clip]{orbit_figures/BJD_bar.pdf}
\end{subfigure}
Radial velocities folded on binary phase
\begin{subfigure}[b]{0.49\textwidth}
\includegraphics[width=\textwidth,trim={0 0.5cm 0 0},clip]{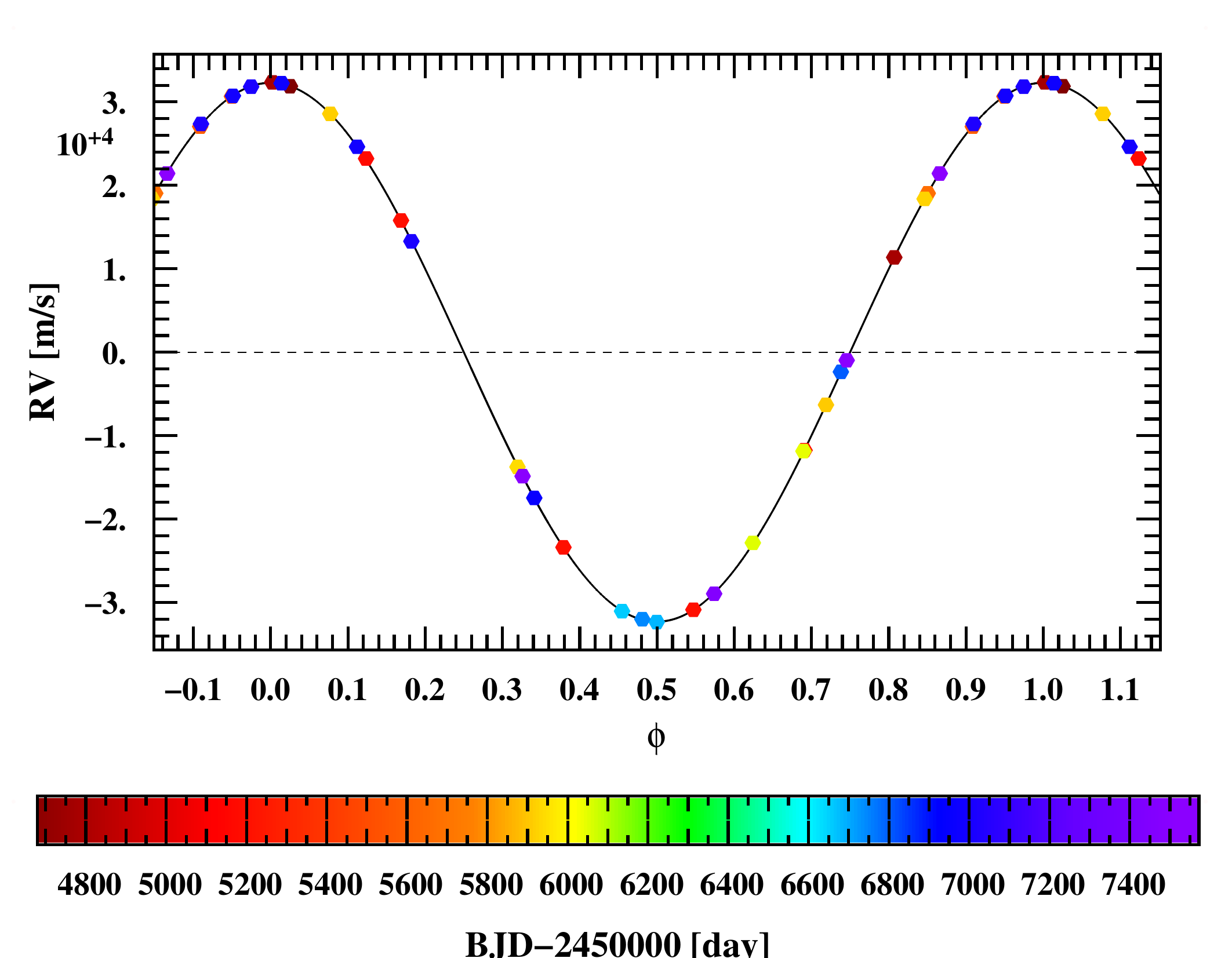}
\end{subfigure}
\begin{subfigure}[b]{0.49\textwidth}
\includegraphics[width=\textwidth,trim={0 0 2cm 0},clip]{orbit_figures/BJD_bar.pdf}
\end{subfigure}
Detection limits
\begin{subfigure}[b]{0.49\textwidth}
\vspace{0.5cm}
\includegraphics[width=\textwidth,trim={0 0 0 0},clip]{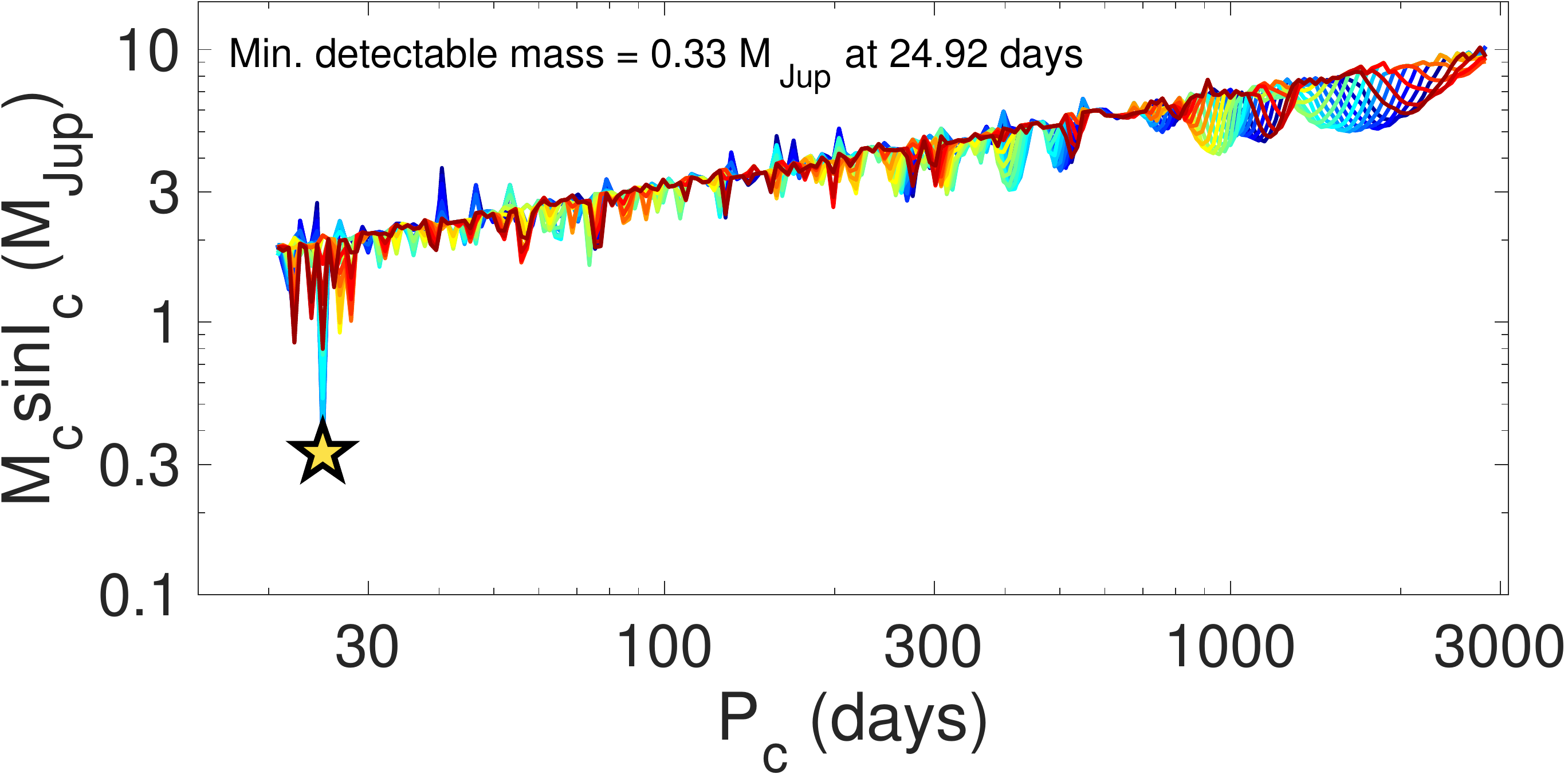}
\end{subfigure}
\end{center}
\end{figure}
\begin{figure}
\begin{center}
\subcaption*{EBLM J1146-42: chosen model = k2 (ecc) \newline \newline $m_{\rm A} = 1.35M_{\odot}$, $m_{\rm B} = 0.536M_{\odot}$, $P = 10.467$ d, $e = 0.052$}
\begin{subfigure}[b]{0.49\textwidth}
\includegraphics[width=\textwidth,trim={0 10cm 0 1.2cm},clip]{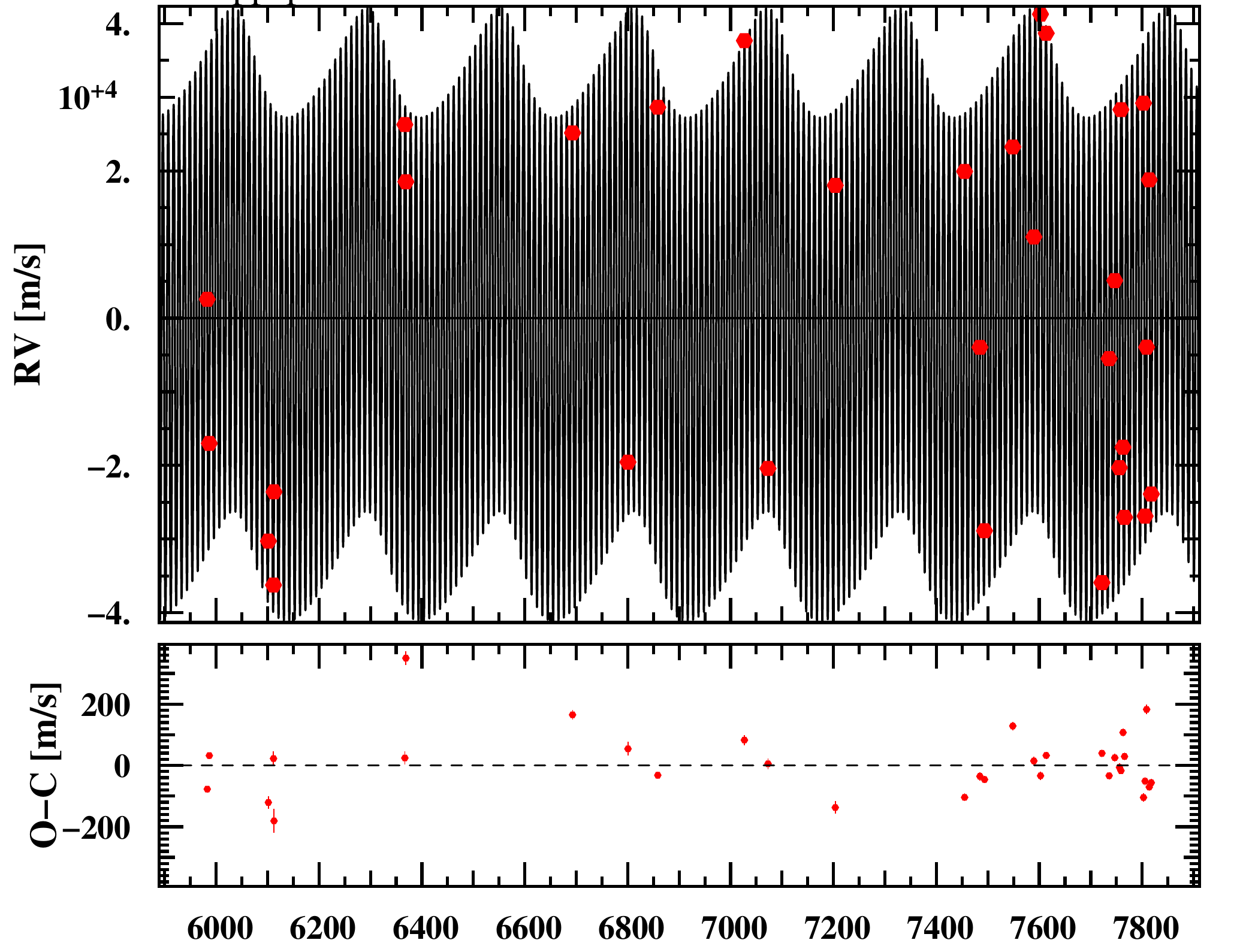}
\end{subfigure}
\begin{subfigure}[b]{0.49\textwidth}
\includegraphics[width=\textwidth,trim={0 0 2cm 0},clip]{orbit_figures/BJD_bar.pdf}
\end{subfigure}
Radial velocities folded on binary phase
\begin{subfigure}[b]{0.49\textwidth}
\includegraphics[width=\textwidth,trim={0 0.5cm 0 0},clip]{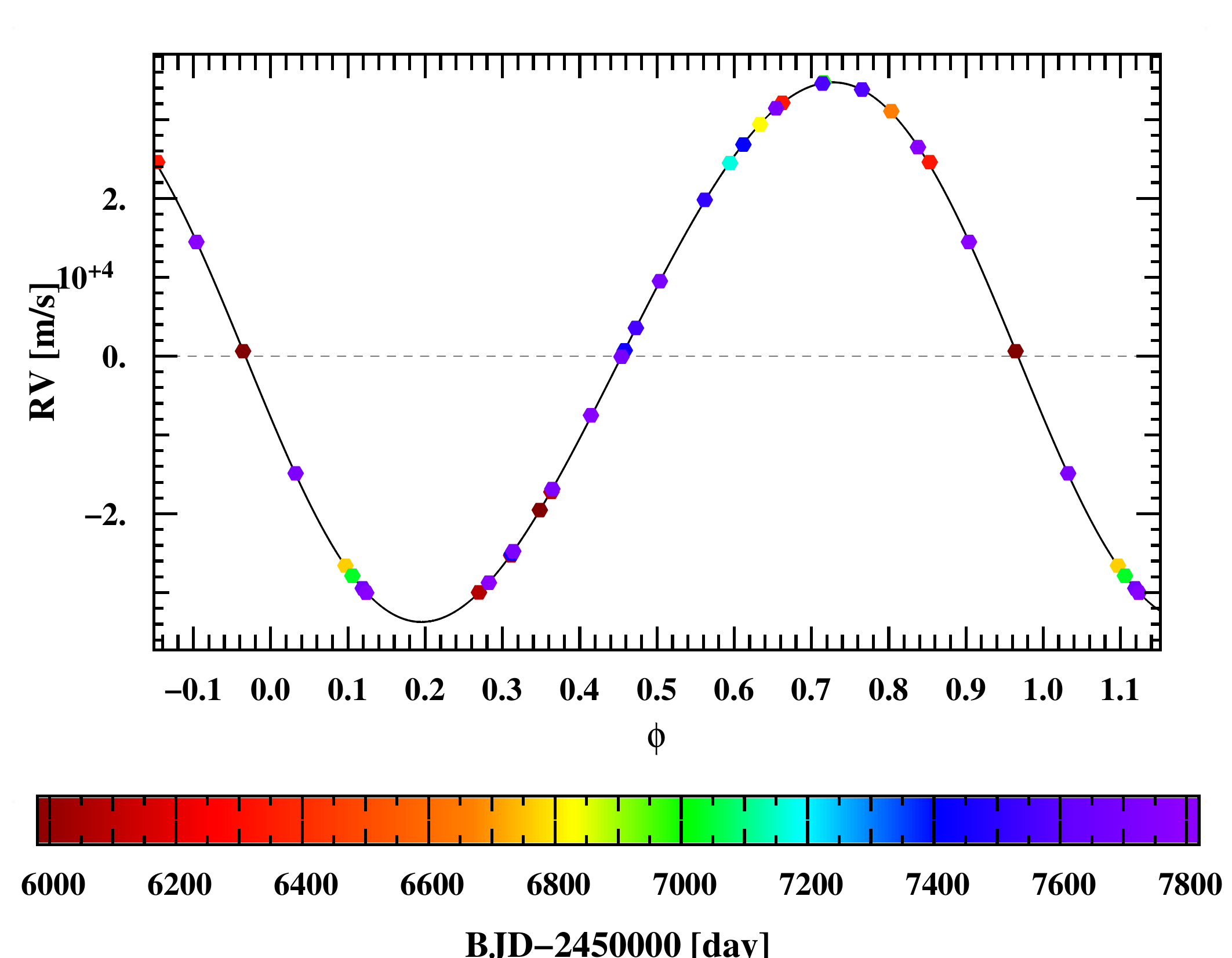}
\end{subfigure}
\begin{subfigure}[b]{0.49\textwidth}
\includegraphics[width=\textwidth,trim={0 0 2cm 0},clip]{orbit_figures/BJD_bar.pdf}
\end{subfigure}
Detection limits
\begin{subfigure}[b]{0.49\textwidth}
\vspace{0.5cm}
\includegraphics[width=\textwidth,trim={0 0 0 0},clip]{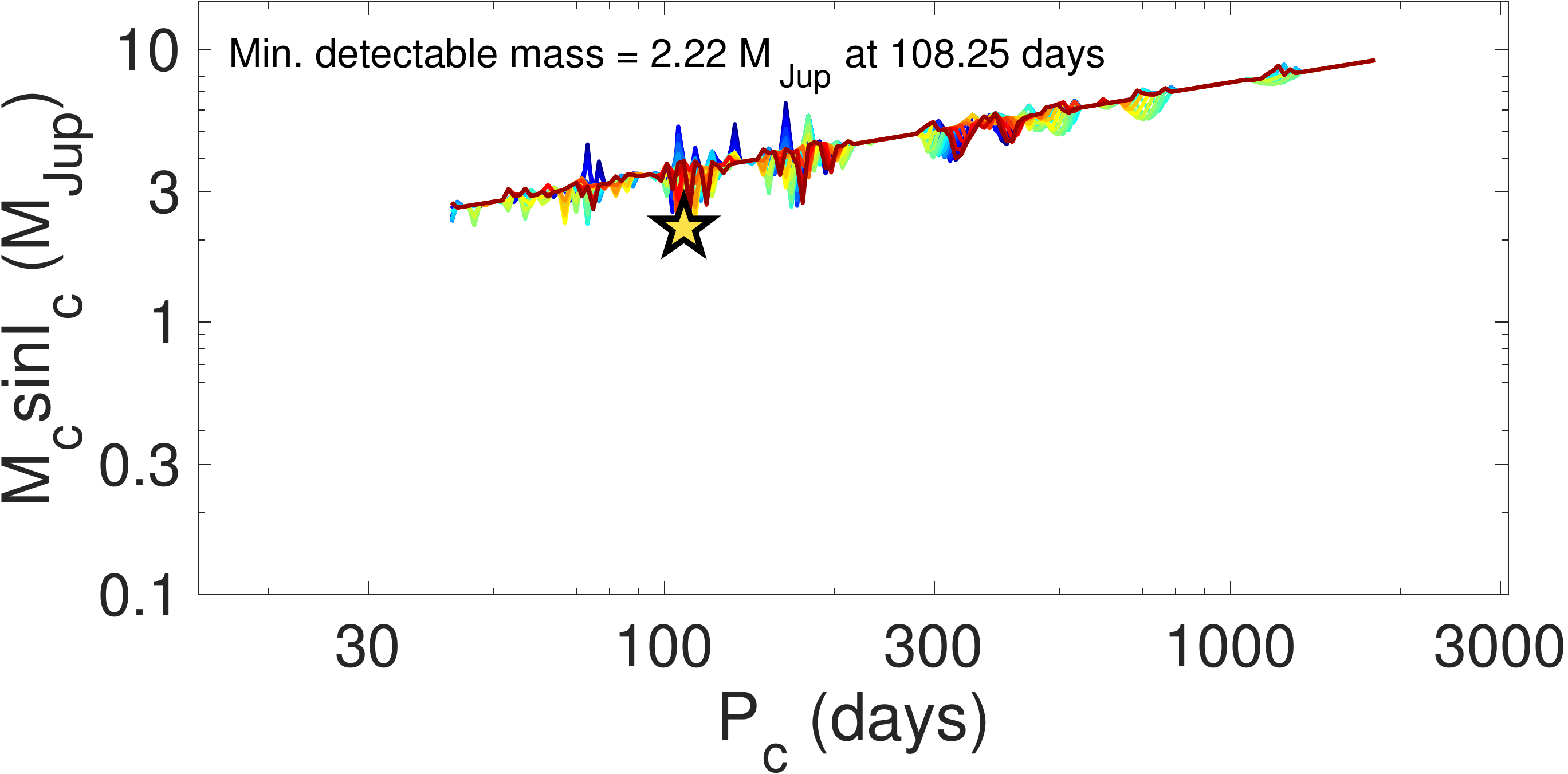}
\end{subfigure}
\end{center}
\end{figure}
\begin{figure}
\begin{center}
\subcaption*{EBLM J1201-36: chosen model = k1 (ecc) \newline \newline $m_{\rm A} = 1.19M_{\odot}$, $m_{\rm B} = 0.101M_{\odot}$, $P = 9.113$ d, $e = 0.152$}
\begin{subfigure}[b]{0.49\textwidth}
\includegraphics[width=\textwidth,trim={0 10cm 0 1.2cm},clip]{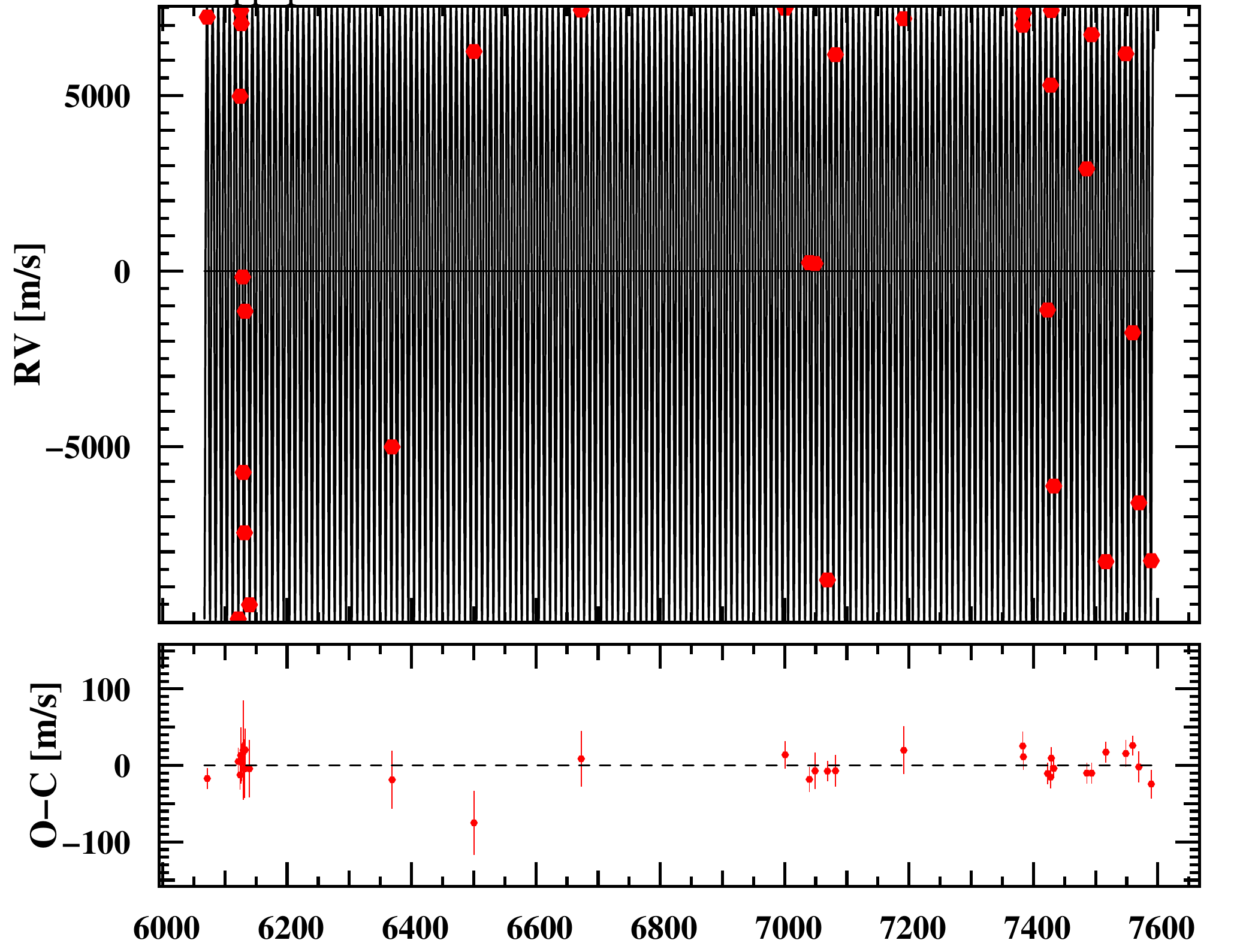}
\end{subfigure}
\begin{subfigure}[b]{0.49\textwidth}
\includegraphics[width=\textwidth,trim={0 0 2cm 0},clip]{orbit_figures/BJD_bar.pdf}
\end{subfigure}
Radial velocities folded on binary phase
\begin{subfigure}[b]{0.49\textwidth}
\includegraphics[width=\textwidth,trim={0 0.5cm 0 0},clip]{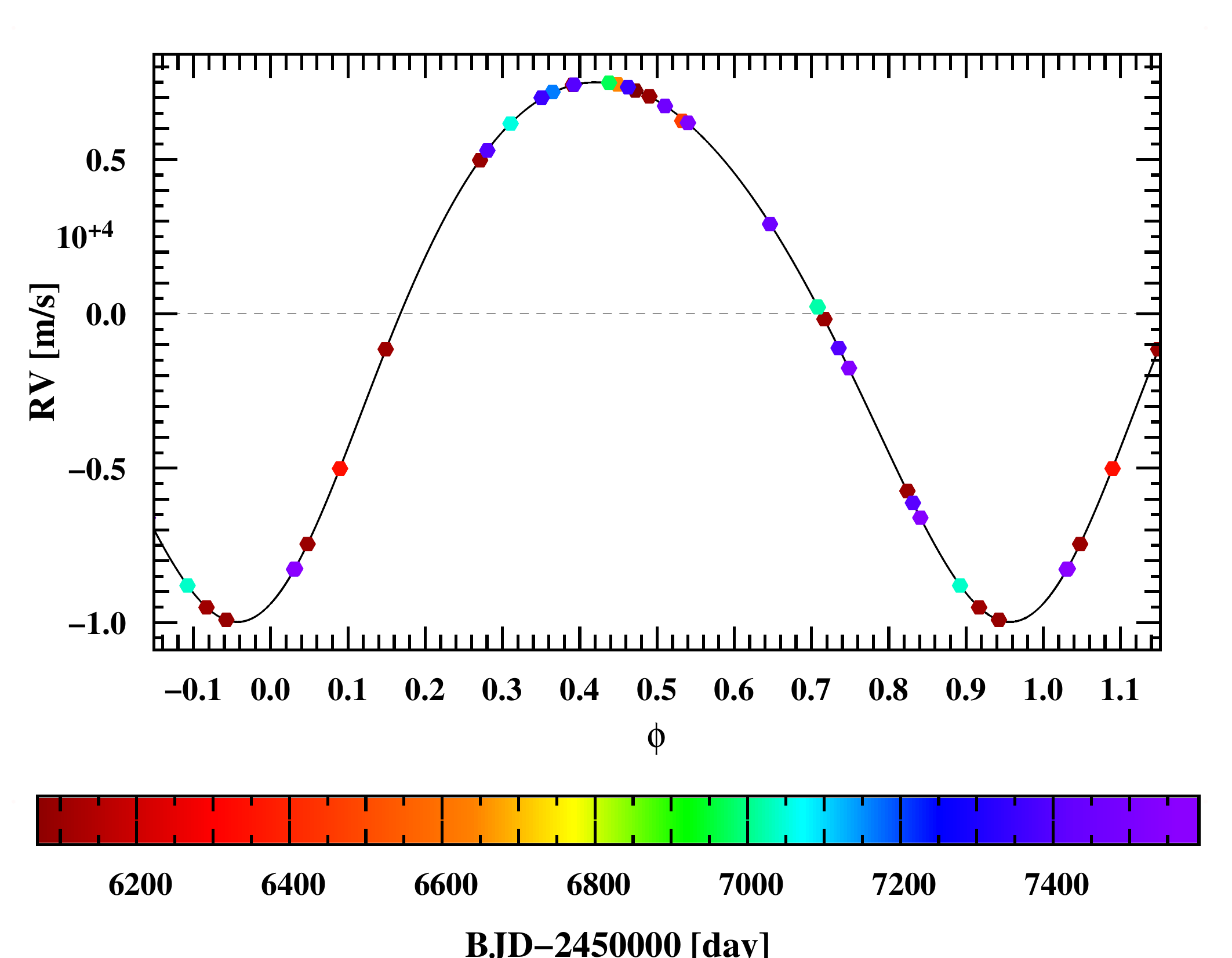}
\end{subfigure}
\begin{subfigure}[b]{0.49\textwidth}
\includegraphics[width=\textwidth,trim={0 0 2cm 0},clip]{orbit_figures/BJD_bar.pdf}
\end{subfigure}
Detection limits
\begin{subfigure}[b]{0.49\textwidth}
\vspace{0.5cm}
\includegraphics[width=\textwidth,trim={0 0 0 0},clip]{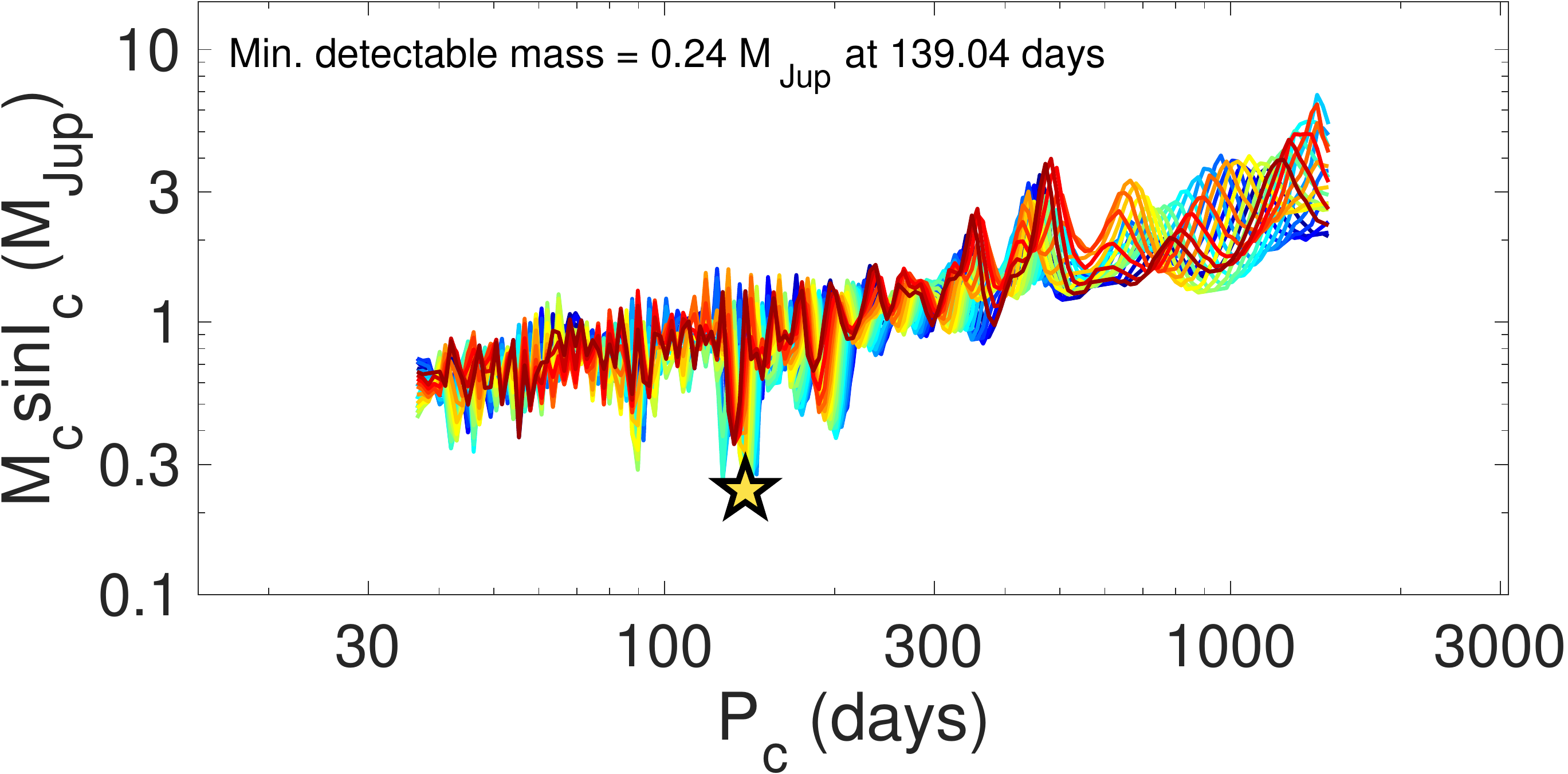}
\end{subfigure}
\end{center}
\end{figure}
\begin{figure}
\begin{center}
\subcaption*{EBLM J1219-39: chosen model = k1 (ecc) \newline \newline $m_{\rm A} = 0.95M_{\odot}$, $m_{\rm B} = 0.1M_{\odot}$, $P = 6.76$ d, $e = 0.056$}
\begin{subfigure}[b]{0.49\textwidth}
\includegraphics[width=\textwidth,trim={0 10cm 0 1.2cm},clip]{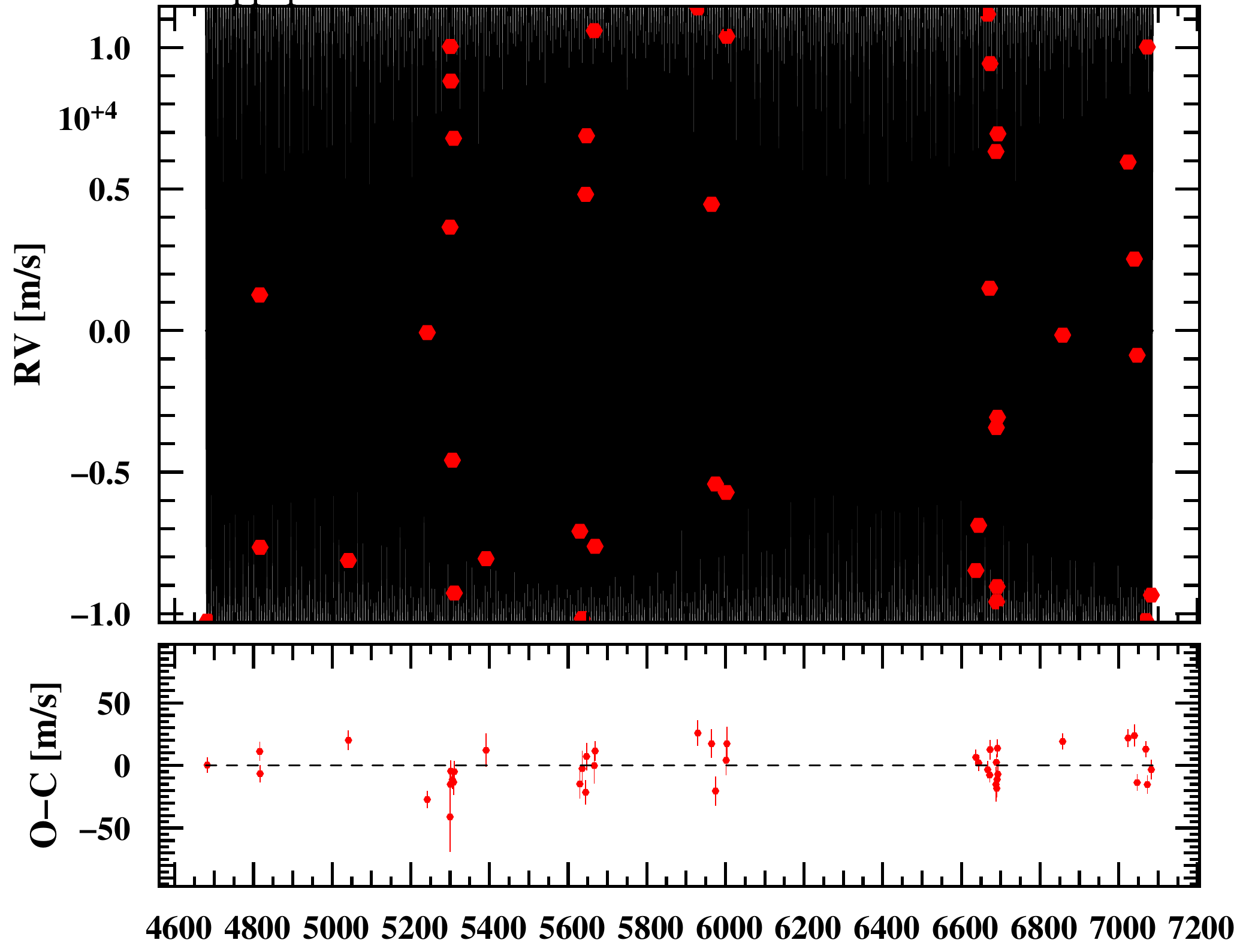}
\end{subfigure}
\begin{subfigure}[b]{0.49\textwidth}
\includegraphics[width=\textwidth,trim={0 0 2cm 0},clip]{orbit_figures/BJD_bar.pdf}
\end{subfigure}
Radial velocities folded on binary phase
\begin{subfigure}[b]{0.49\textwidth}
\includegraphics[width=\textwidth,trim={0 0.5cm 0 0},clip]{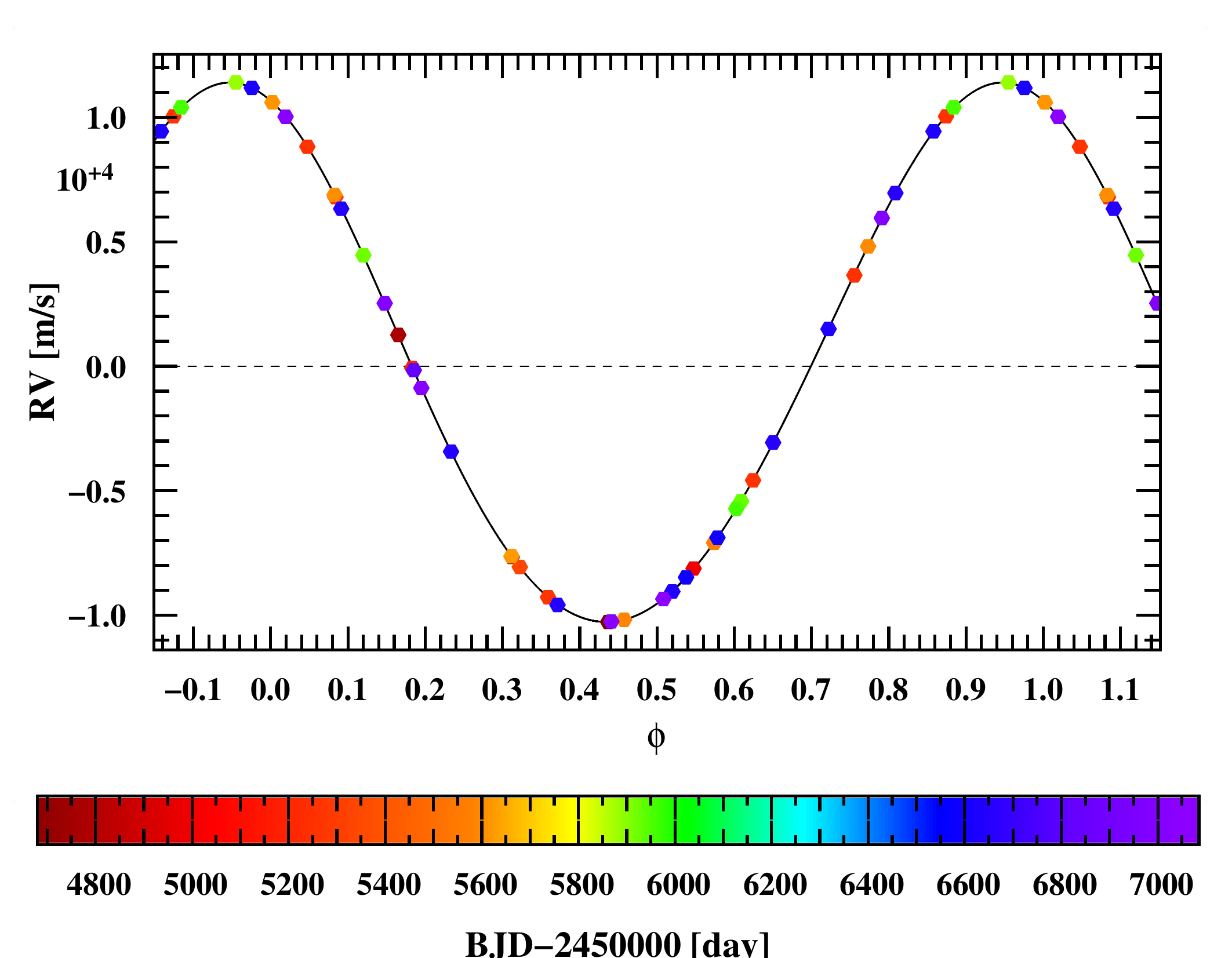}
\end{subfigure}
\begin{subfigure}[b]{0.49\textwidth}
\includegraphics[width=\textwidth,trim={0 0 2cm 0},clip]{orbit_figures/BJD_bar.pdf}
\end{subfigure}
Detection limits
\begin{subfigure}[b]{0.49\textwidth}
\vspace{0.5cm}
\includegraphics[width=\textwidth,trim={0 0 0 0},clip]{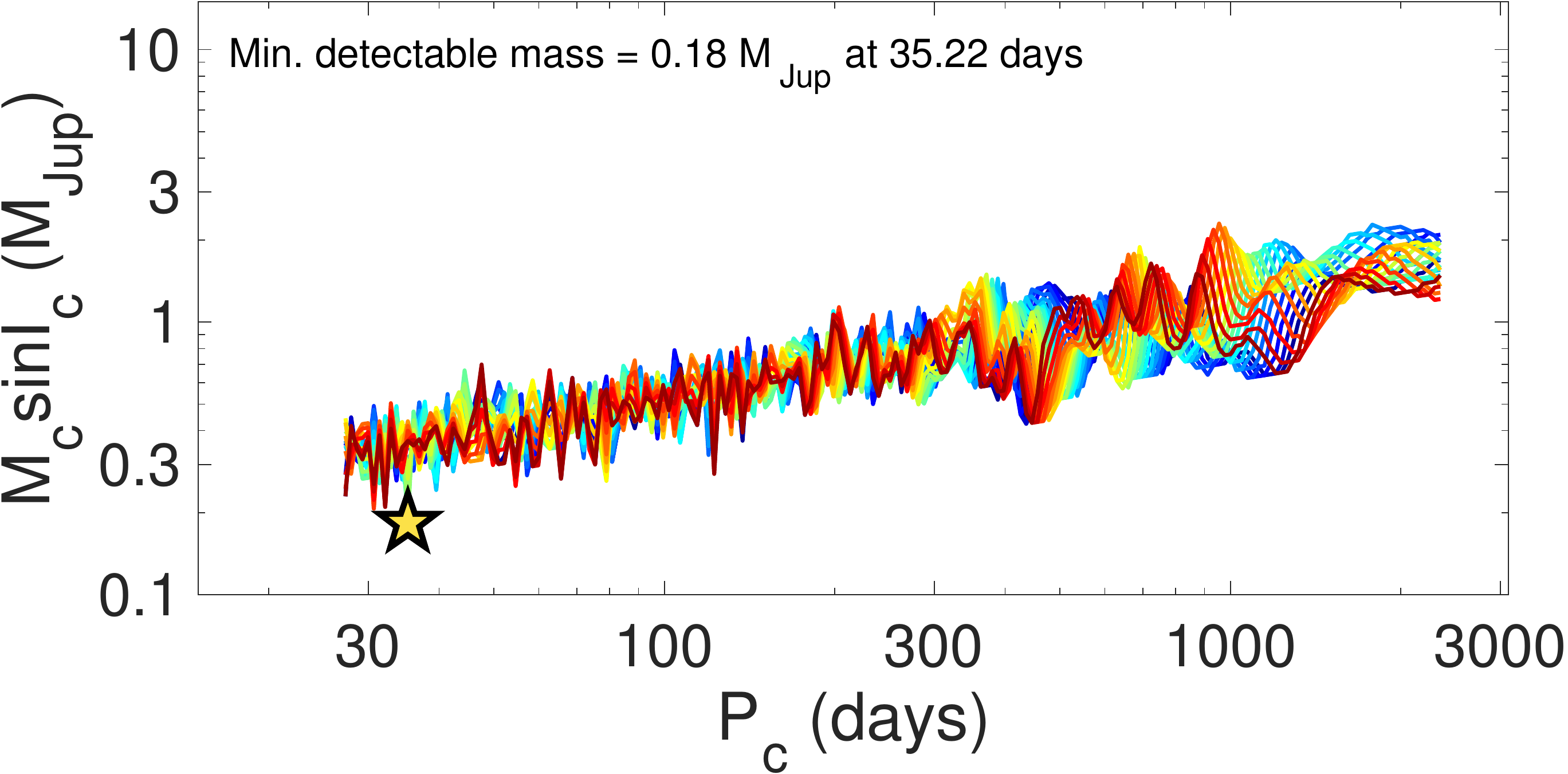}
\end{subfigure}
\end{center}
\end{figure}
\begin{figure}
\begin{center}
\subcaption*{EBLM J1305-31: chosen model = k1 (ecc) \newline \newline $m_{\rm A} = 1.1M_{\odot}$, $m_{\rm B} = 0.288M_{\odot}$, $P = 10.619$ d, $e = 0.037$}
\begin{subfigure}[b]{0.49\textwidth}
\includegraphics[width=\textwidth,trim={0 10cm 0 1.2cm},clip]{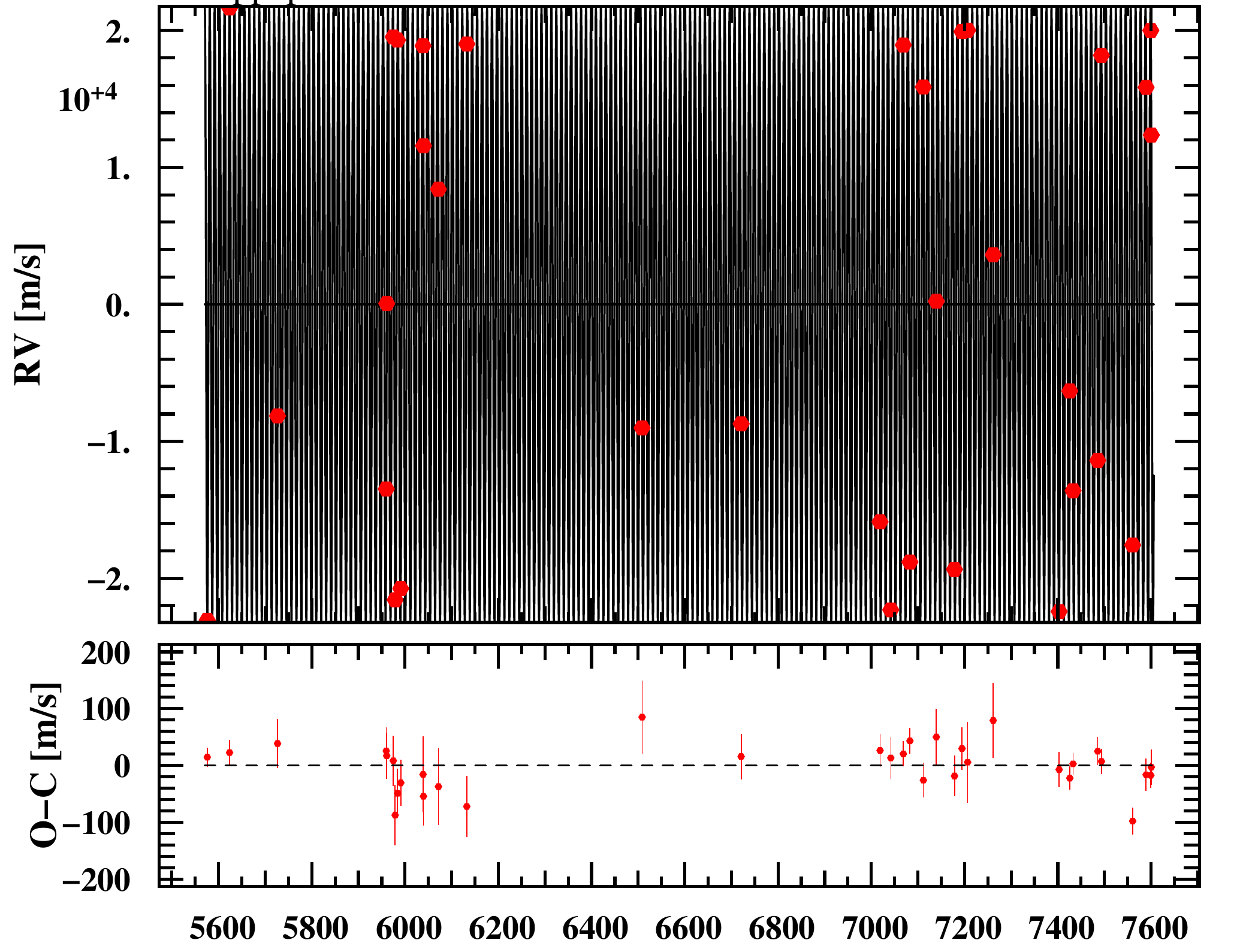}
\end{subfigure}
\begin{subfigure}[b]{0.49\textwidth}
\includegraphics[width=\textwidth,trim={0 0 2cm 0},clip]{orbit_figures/BJD_bar.pdf}
\end{subfigure}
Radial velocities folded on binary phase
\begin{subfigure}[b]{0.49\textwidth}
\includegraphics[width=\textwidth,trim={0 0.5cm 0 0},clip]{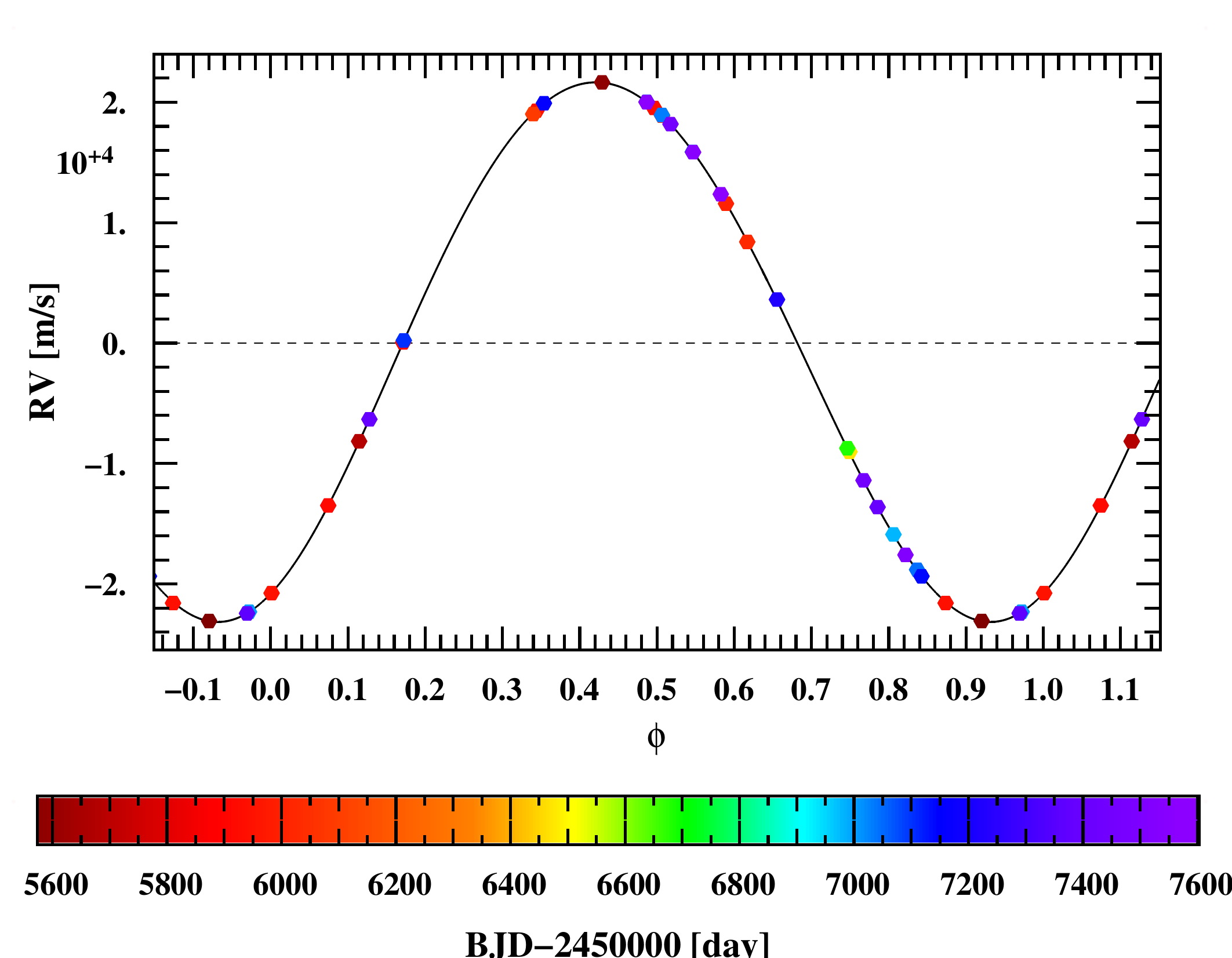}
\end{subfigure}
\begin{subfigure}[b]{0.49\textwidth}
\includegraphics[width=\textwidth,trim={0 0 2cm 0},clip]{orbit_figures/BJD_bar.pdf}
\end{subfigure}
Detection limits
\begin{subfigure}[b]{0.49\textwidth}
\vspace{0.5cm}
\includegraphics[width=\textwidth,trim={0 0 0 0},clip]{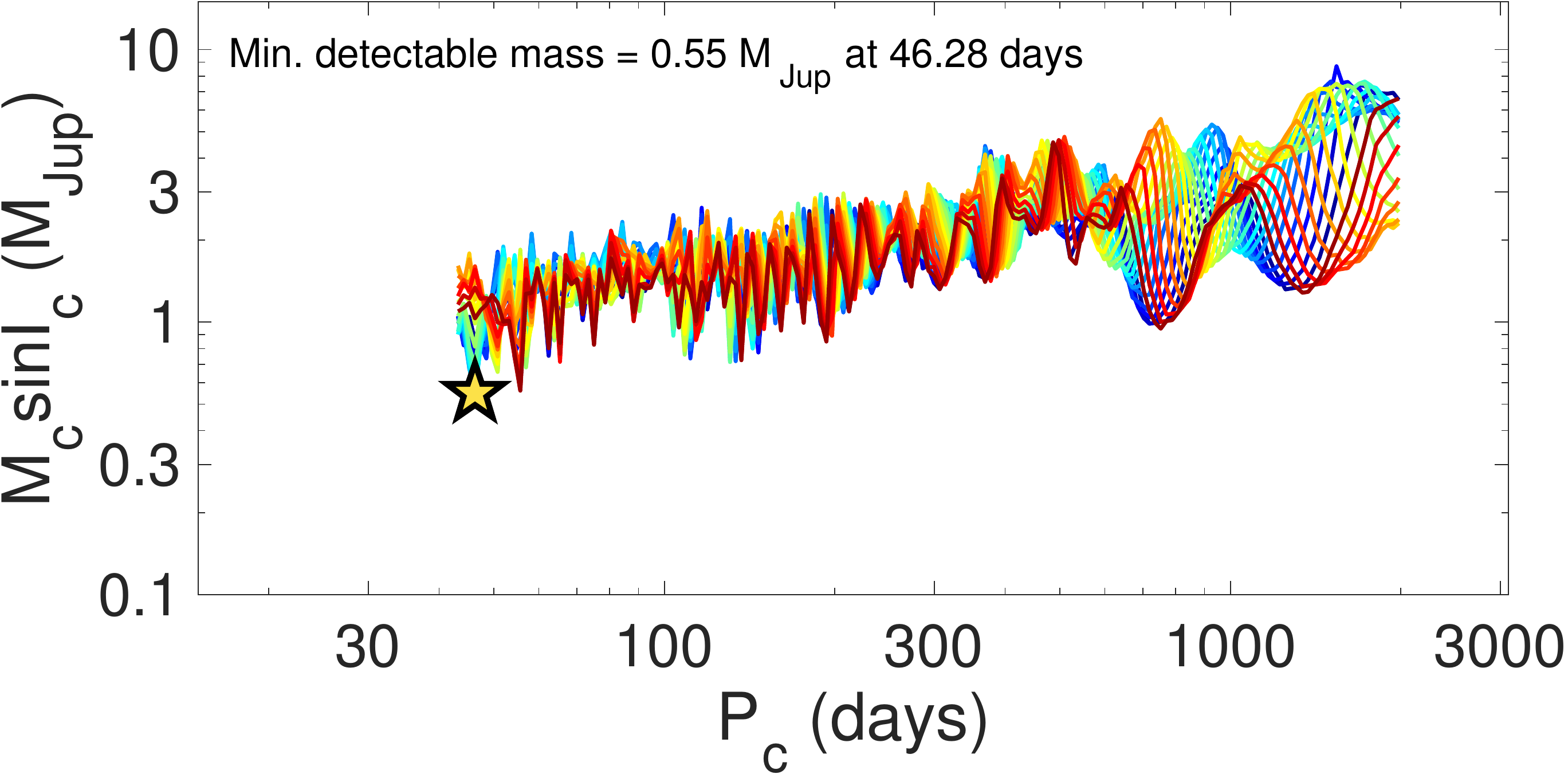}
\end{subfigure}
\end{center}
\end{figure}
\begin{figure}
\begin{center}
\subcaption*{EBLM J1328+05: chosen model = k1 (ecc) \newline \newline $m_{\rm A} = 1.01M_{\odot}$, $m_{\rm B} = 0.341M_{\odot}$, $P = 7.252$ d, $e = 0.004$}
\begin{subfigure}[b]{0.49\textwidth}
\includegraphics[width=\textwidth,trim={0 10cm 0 1.2cm},clip]{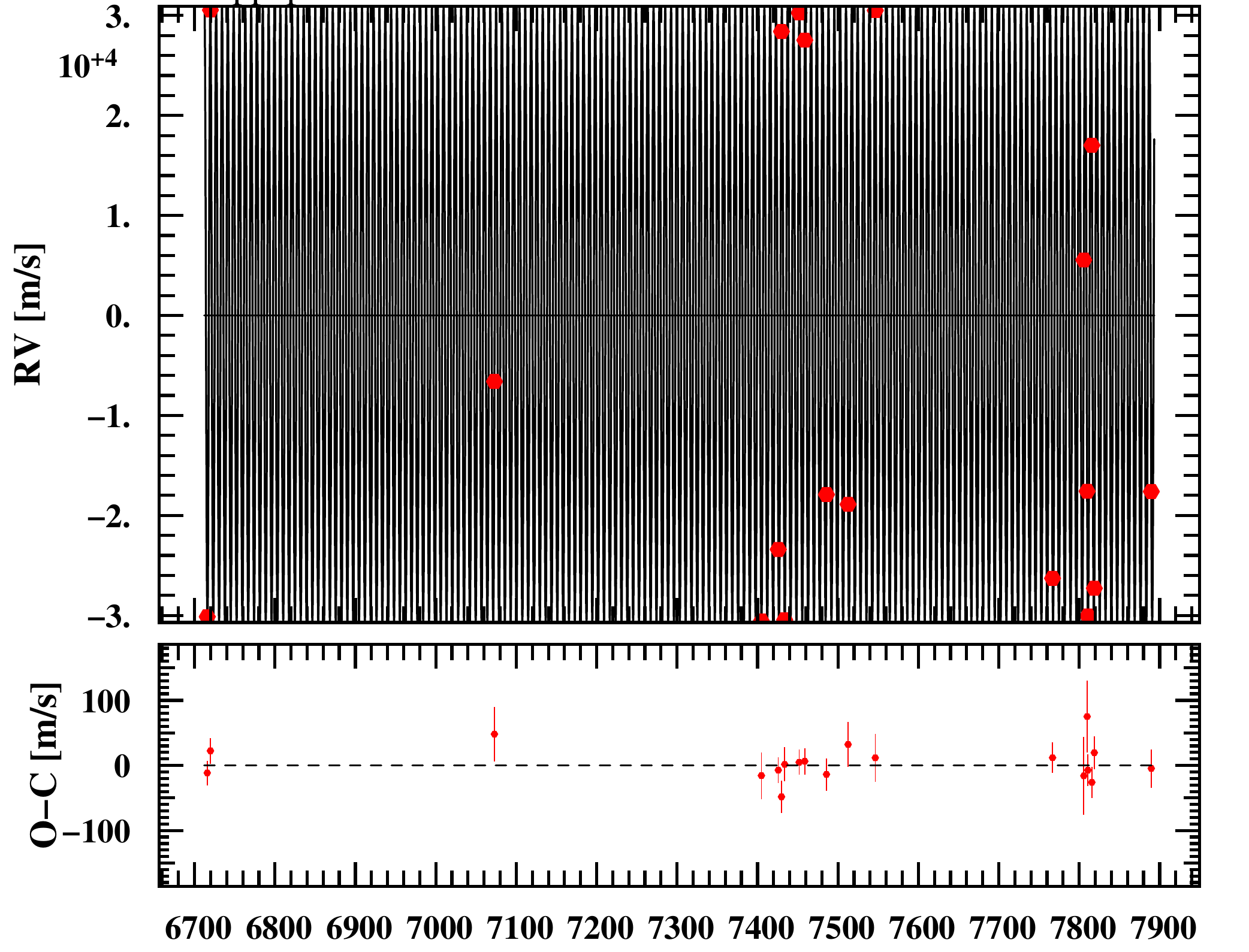}
\end{subfigure}
\begin{subfigure}[b]{0.49\textwidth}
\includegraphics[width=\textwidth,trim={0 0 2cm 0},clip]{orbit_figures/BJD_bar.pdf}
\end{subfigure}
Radial velocities folded on binary phase
\begin{subfigure}[b]{0.49\textwidth}
\includegraphics[width=\textwidth,trim={0 0.5cm 0 0},clip]{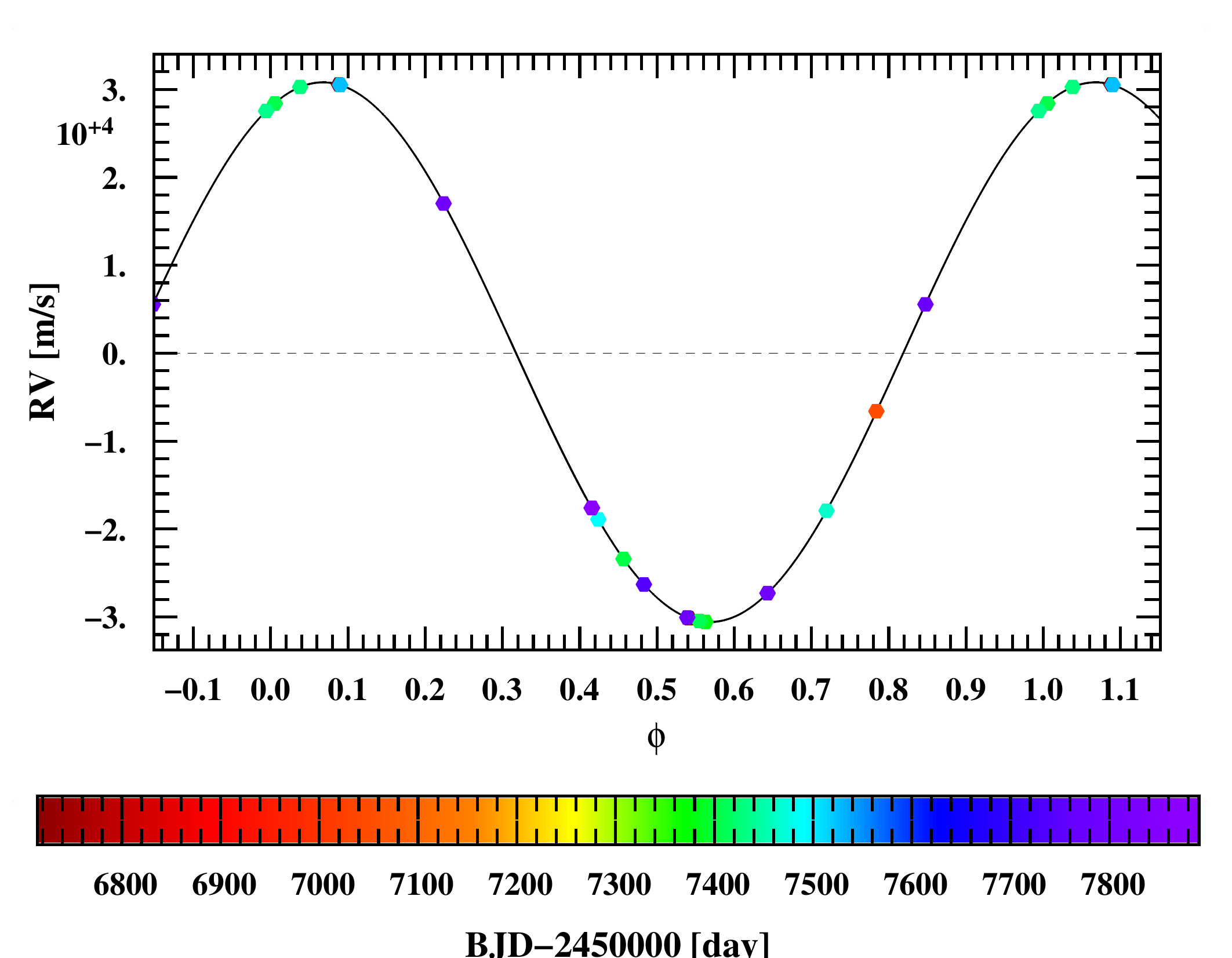}
\end{subfigure}
\begin{subfigure}[b]{0.49\textwidth}
\includegraphics[width=\textwidth,trim={0 0 2cm 0},clip]{orbit_figures/BJD_bar.pdf}
\end{subfigure}
Detection limits
\begin{subfigure}[b]{0.49\textwidth}
\vspace{0.5cm}
\includegraphics[width=\textwidth,trim={0 0 0 0},clip]{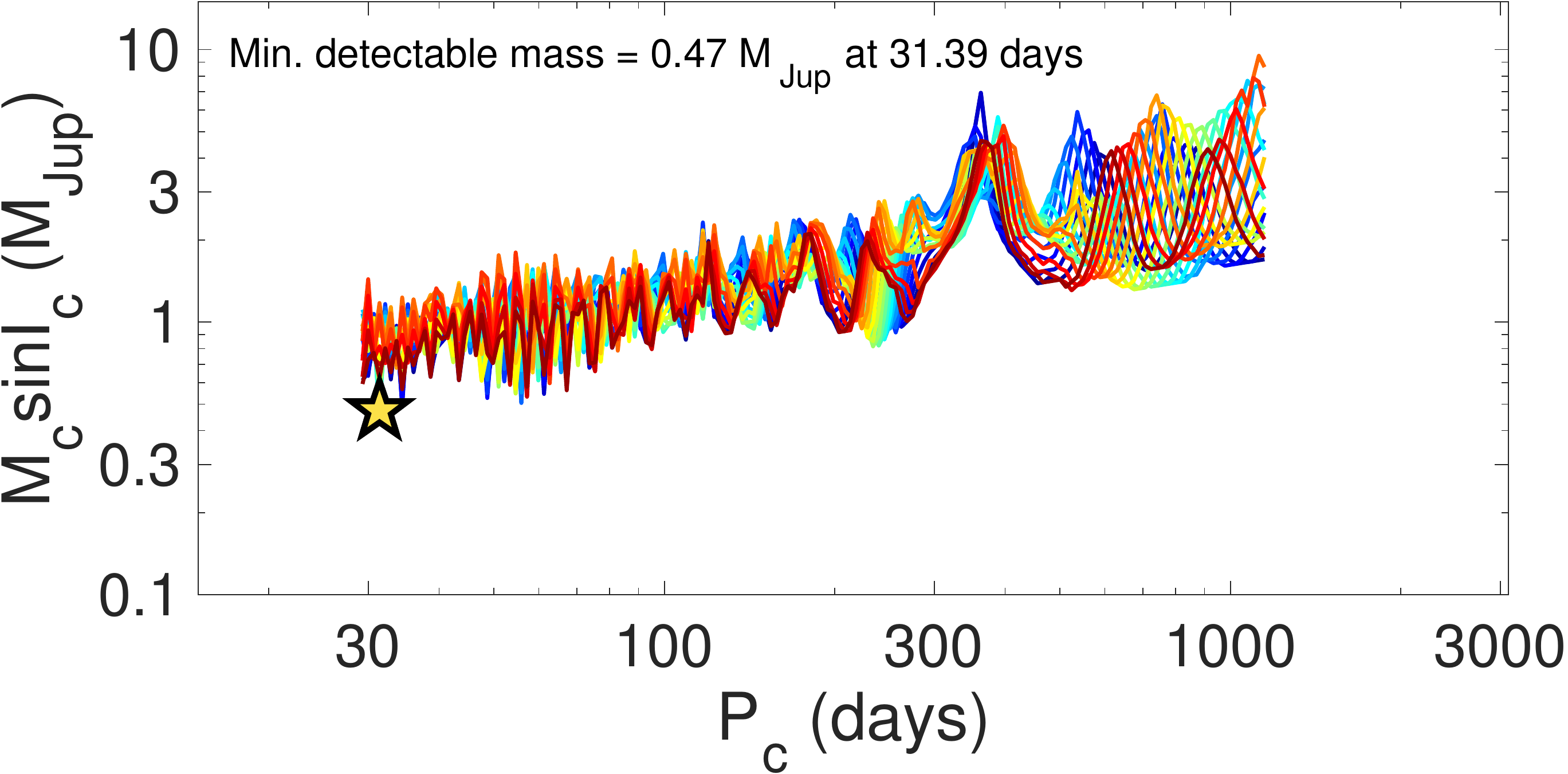}
\end{subfigure}
\end{center}
\end{figure}
\begin{figure}
\begin{center}
\subcaption*{EBLM J1403-32: chosen model = k1 (ecc) \newline \newline $m_{\rm A} = 1.06M_{\odot}$, $m_{\rm B} = 0.27M_{\odot}$, $P = 11.909$ d, $e = 0.109$}
\begin{subfigure}[b]{0.49\textwidth}
\includegraphics[width=\textwidth,trim={0 10cm 0 1.2cm},clip]{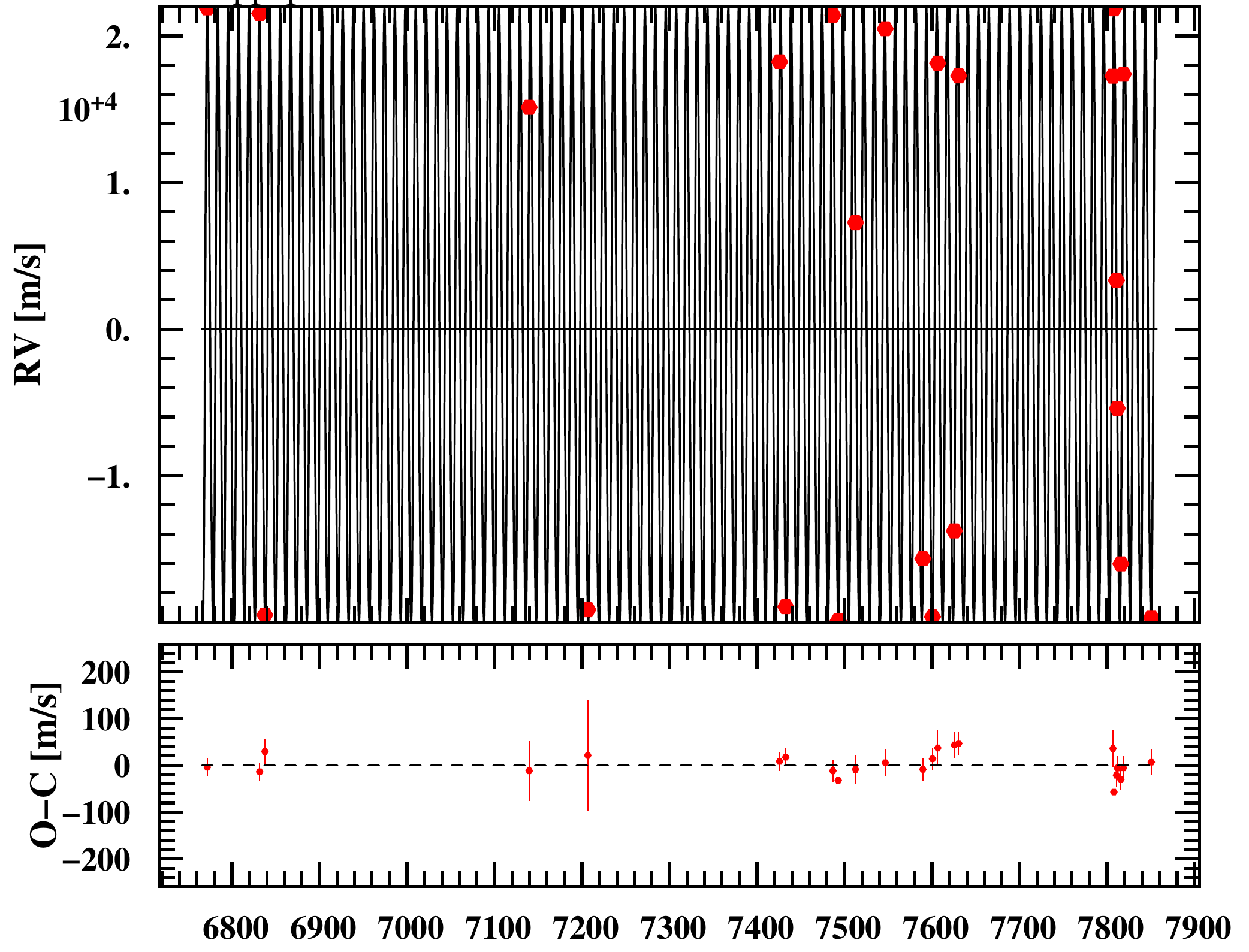}
\end{subfigure}
\begin{subfigure}[b]{0.49\textwidth}
\includegraphics[width=\textwidth,trim={0 0 2cm 0},clip]{orbit_figures/BJD_bar.pdf}
\end{subfigure}
Radial velocities folded on binary phase
\begin{subfigure}[b]{0.49\textwidth}
\includegraphics[width=\textwidth,trim={0 0.5cm 0 0},clip]{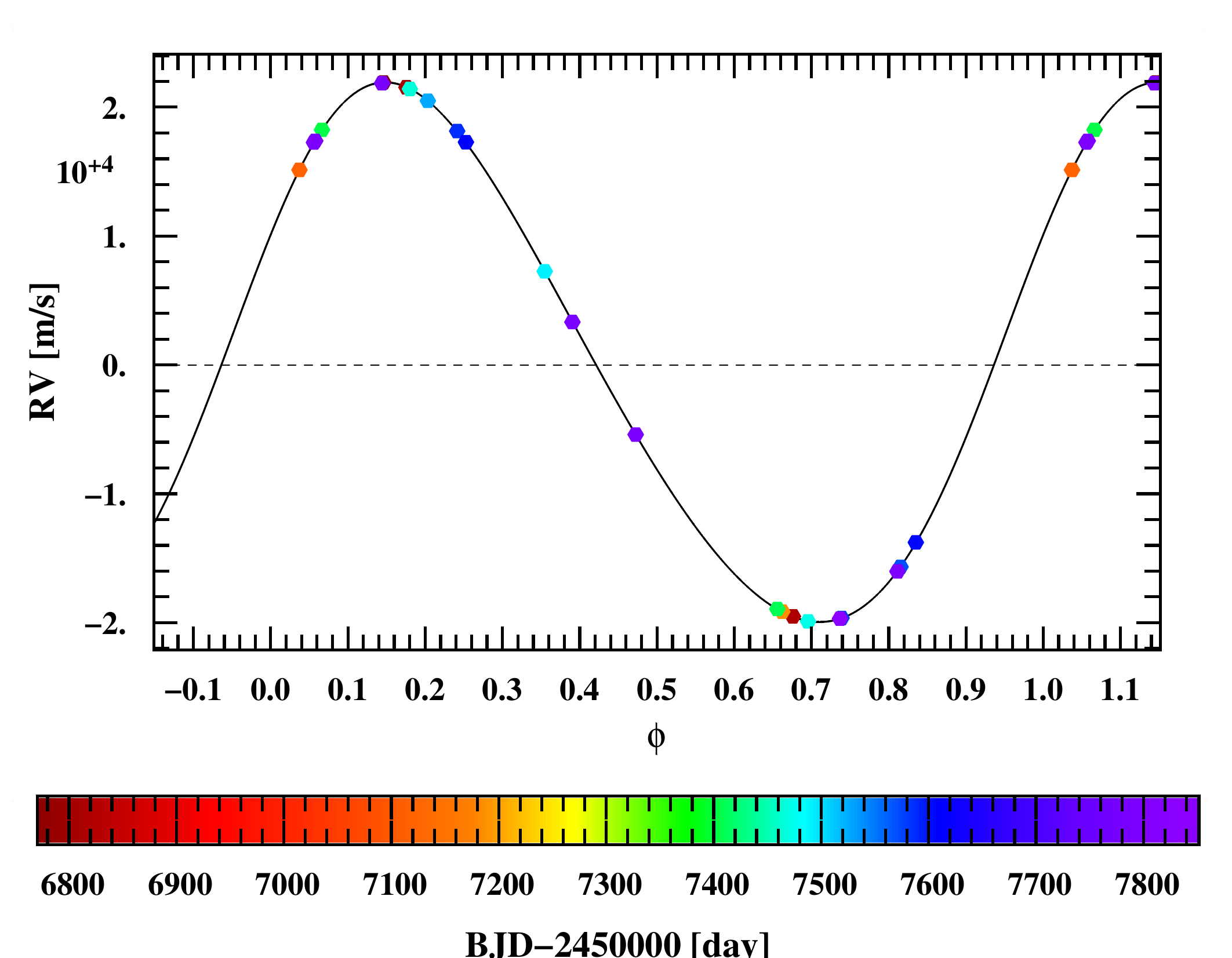}
\end{subfigure}
\begin{subfigure}[b]{0.49\textwidth}
\includegraphics[width=\textwidth,trim={0 0 2cm 0},clip]{orbit_figures/BJD_bar.pdf}
\end{subfigure}
Detection limits
\begin{subfigure}[b]{0.49\textwidth}
\vspace{0.5cm}
\includegraphics[width=\textwidth,trim={0 0 0 0},clip]{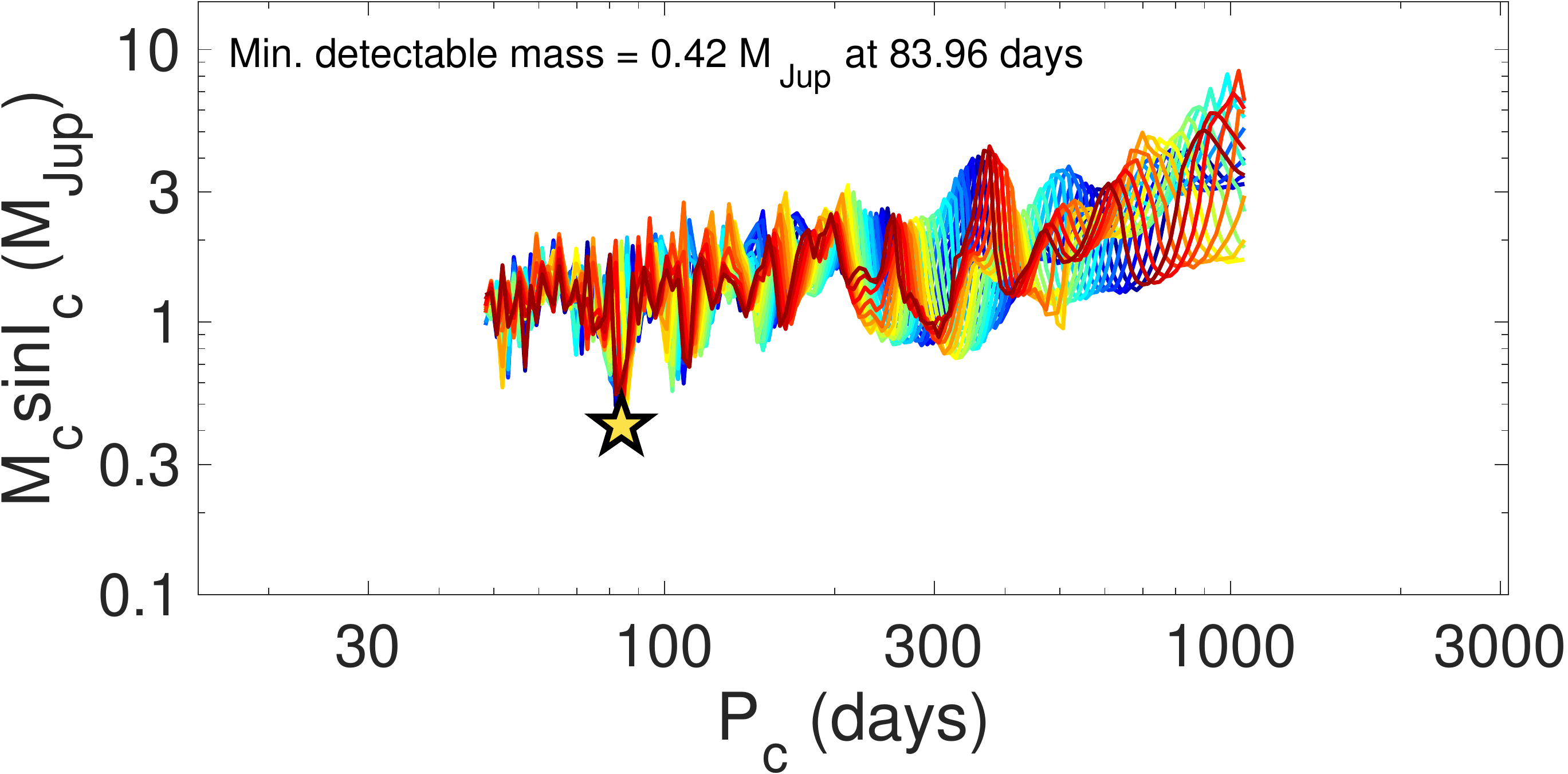}
\end{subfigure}
\end{center}
\end{figure}
\begin{figure}
\begin{center}
\subcaption*{EBLM J1446+05: chosen model = k1d1 (ecc) \newline \newline $m_{\rm A} = 1.04M_{\odot}$, $m_{\rm B} = 0.196M_{\odot}$, $P = 7.763$ d, $e = 0.003$}
\begin{subfigure}[b]{0.49\textwidth}
\includegraphics[width=\textwidth,trim={0 10cm 0 1.2cm},clip]{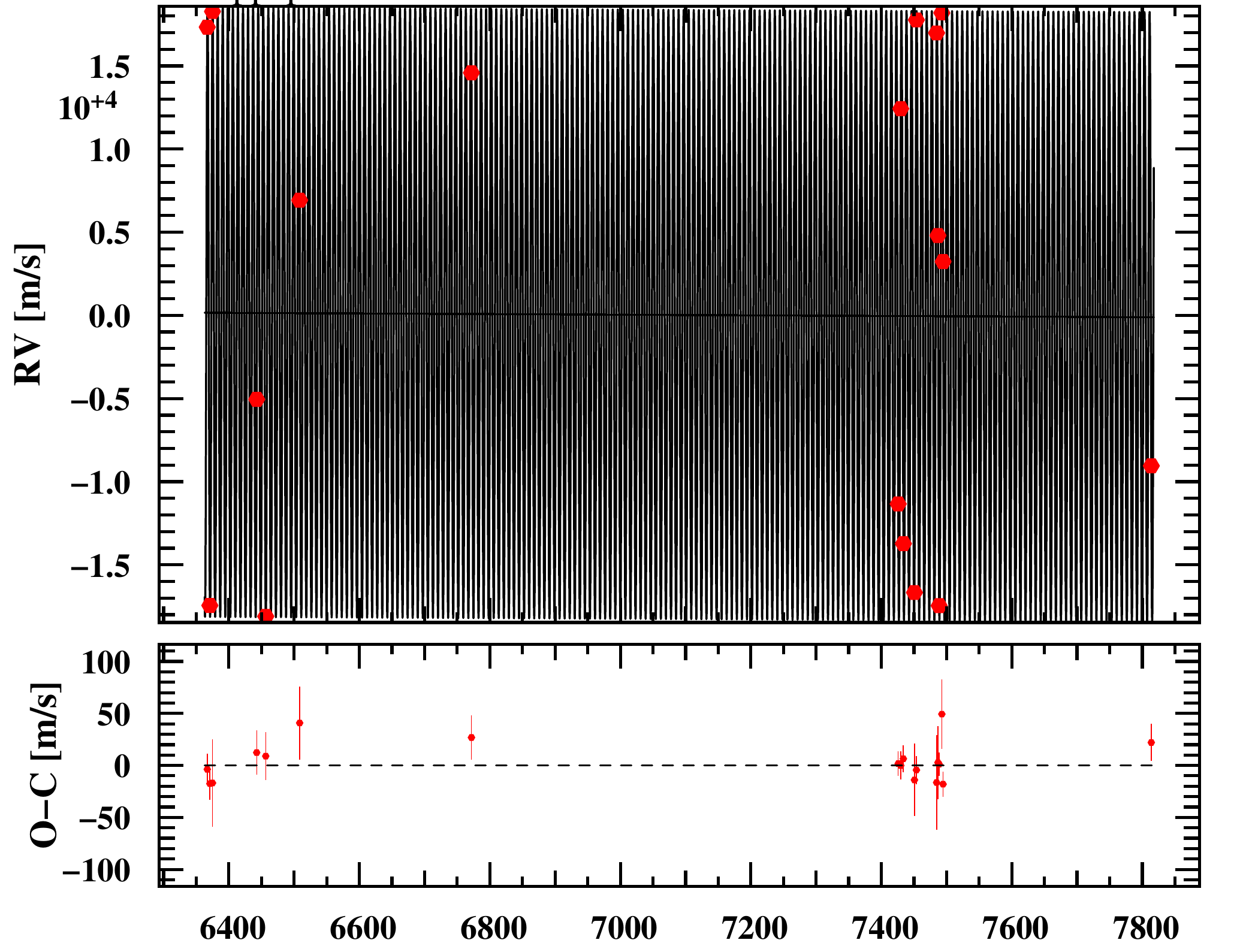}
\end{subfigure}
\begin{subfigure}[b]{0.49\textwidth}
\includegraphics[width=\textwidth,trim={0 0 2cm 0},clip]{orbit_figures/BJD_bar.pdf}
\end{subfigure}
Radial velocities folded on binary phase
\begin{subfigure}[b]{0.49\textwidth}
\includegraphics[width=\textwidth,trim={0 0.5cm 0 0},clip]{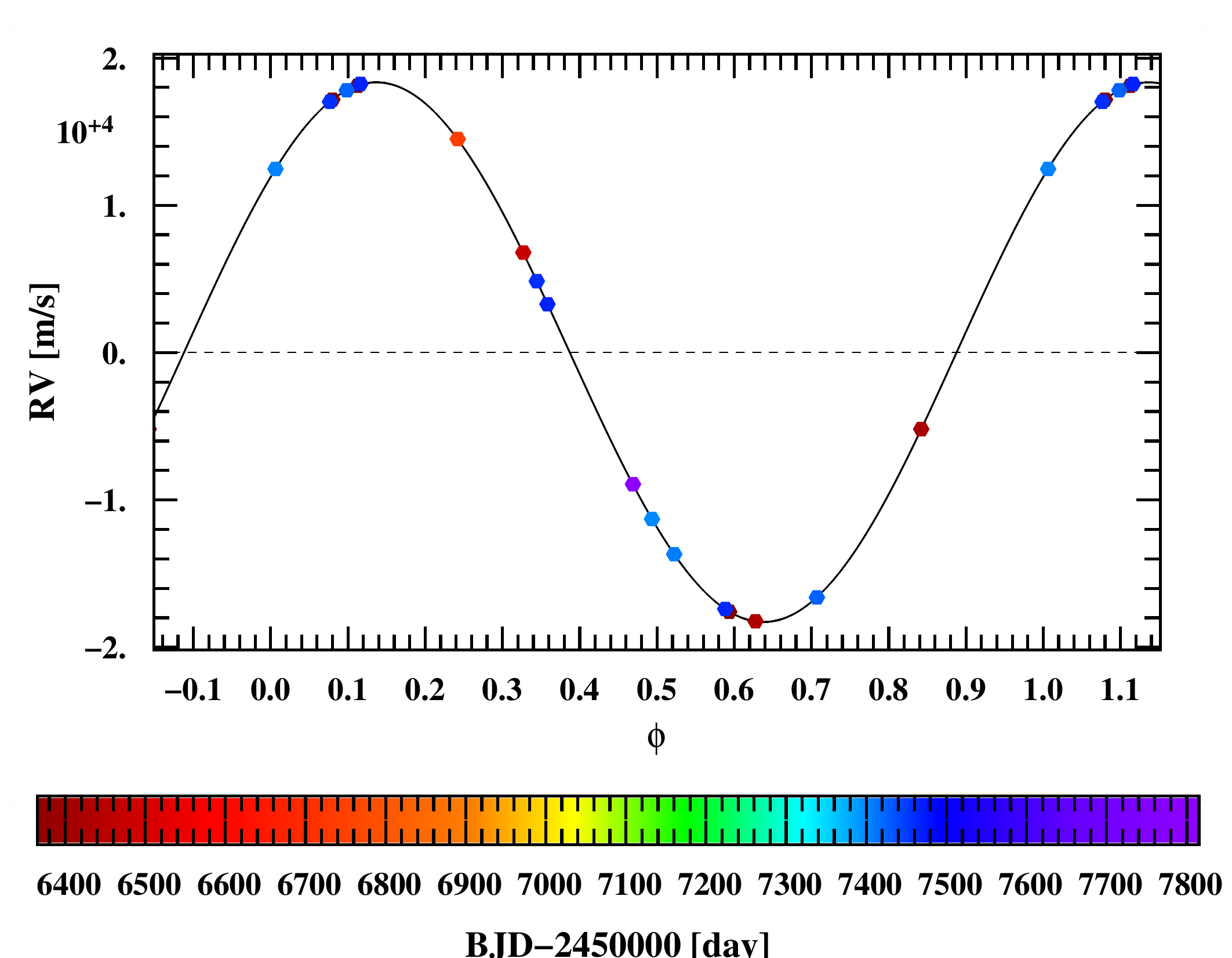}
\end{subfigure}
\begin{subfigure}[b]{0.49\textwidth}
\includegraphics[width=\textwidth,trim={0 0 2cm 0},clip]{orbit_figures/BJD_bar.pdf}
\end{subfigure}
Detection limits
\begin{subfigure}[b]{0.49\textwidth}
\vspace{0.5cm}
\includegraphics[width=\textwidth,trim={0 0 0 0},clip]{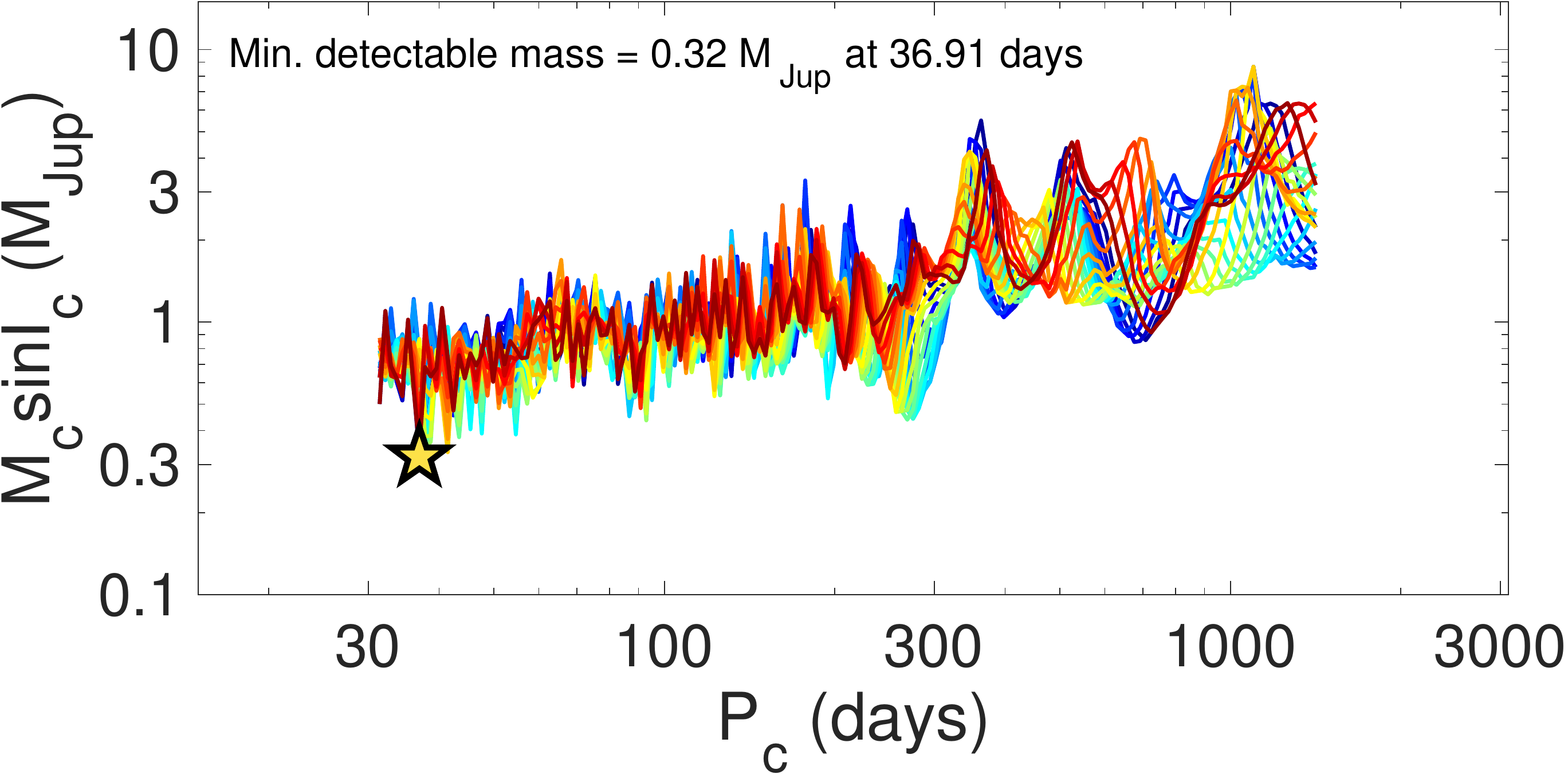}
\end{subfigure}
\end{center}
\end{figure}
\begin{figure}
\begin{center}
\subcaption*{EBLM J1525+03: chosen model = k1 (circ) \newline \newline $m_{\rm A} = 1.23M_{\odot}$, $m_{\rm B} = 0.144M_{\odot}$, $P = 3.822$ d, $e = 0$}
\begin{subfigure}[b]{0.49\textwidth}
\includegraphics[width=\textwidth,trim={0 10cm 0 1.2cm},clip]{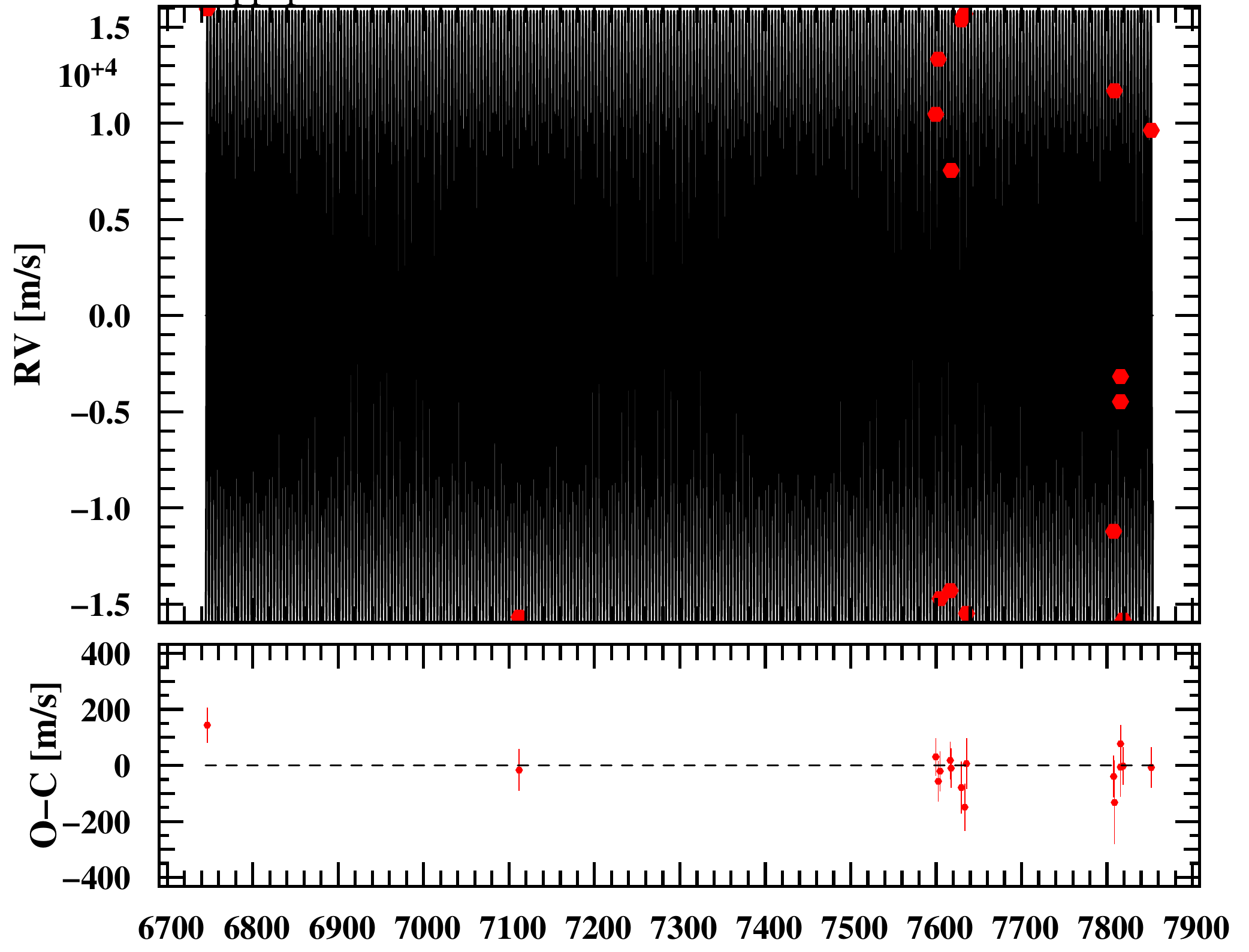}
\end{subfigure}
\begin{subfigure}[b]{0.49\textwidth}
\includegraphics[width=\textwidth,trim={0 0 2cm 0},clip]{orbit_figures/BJD_bar.pdf}
\end{subfigure}
Radial velocities folded on binary phase
\begin{subfigure}[b]{0.49\textwidth}
\includegraphics[width=\textwidth,trim={0 0.5cm 0 0},clip]{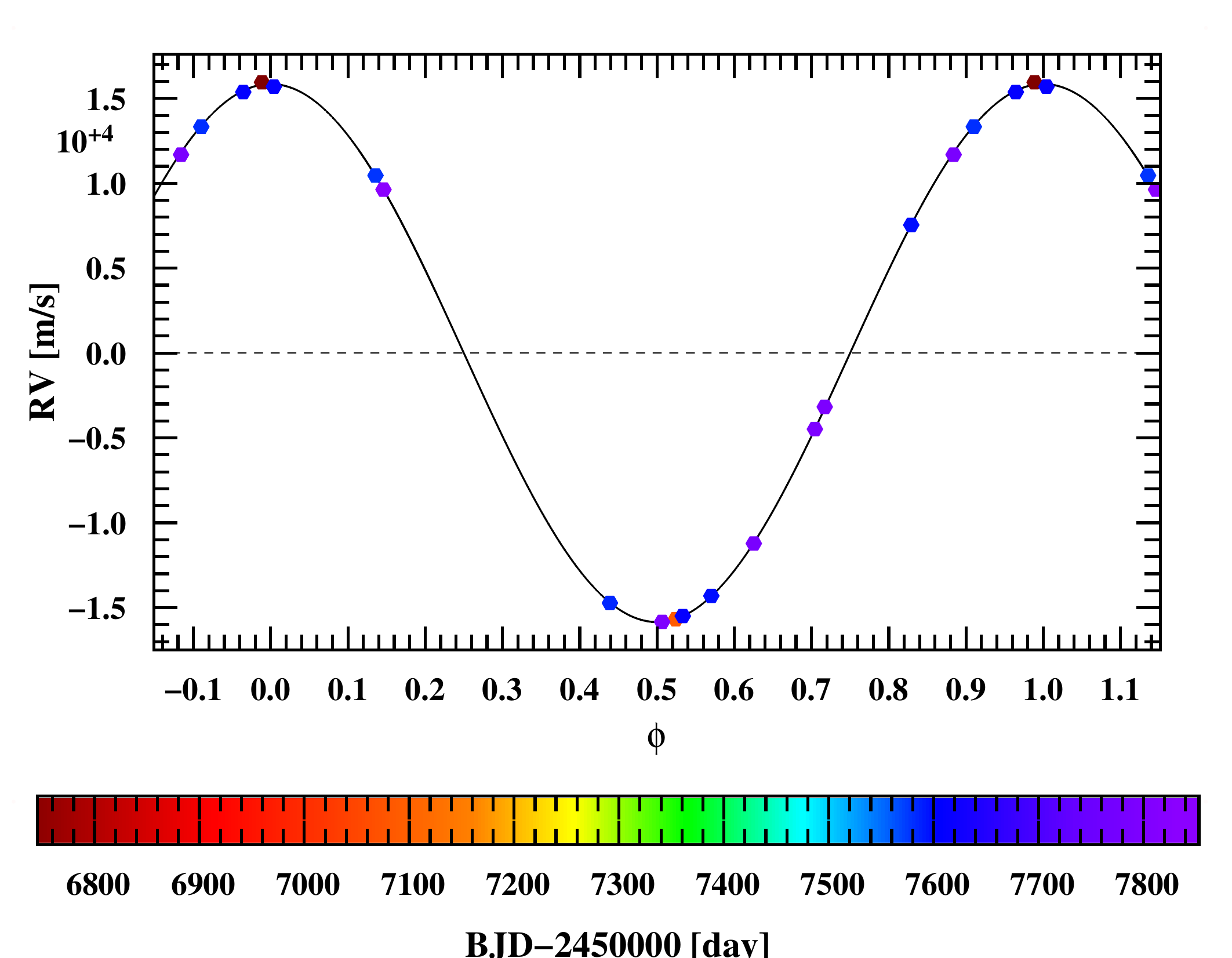}
\end{subfigure}
\begin{subfigure}[b]{0.49\textwidth}
\includegraphics[width=\textwidth,trim={0 0 2cm 0},clip]{orbit_figures/BJD_bar.pdf}
\end{subfigure}
Detection limits
\begin{subfigure}[b]{0.49\textwidth}
\vspace{0.5cm}
\includegraphics[width=\textwidth,trim={0 0 0 0},clip]{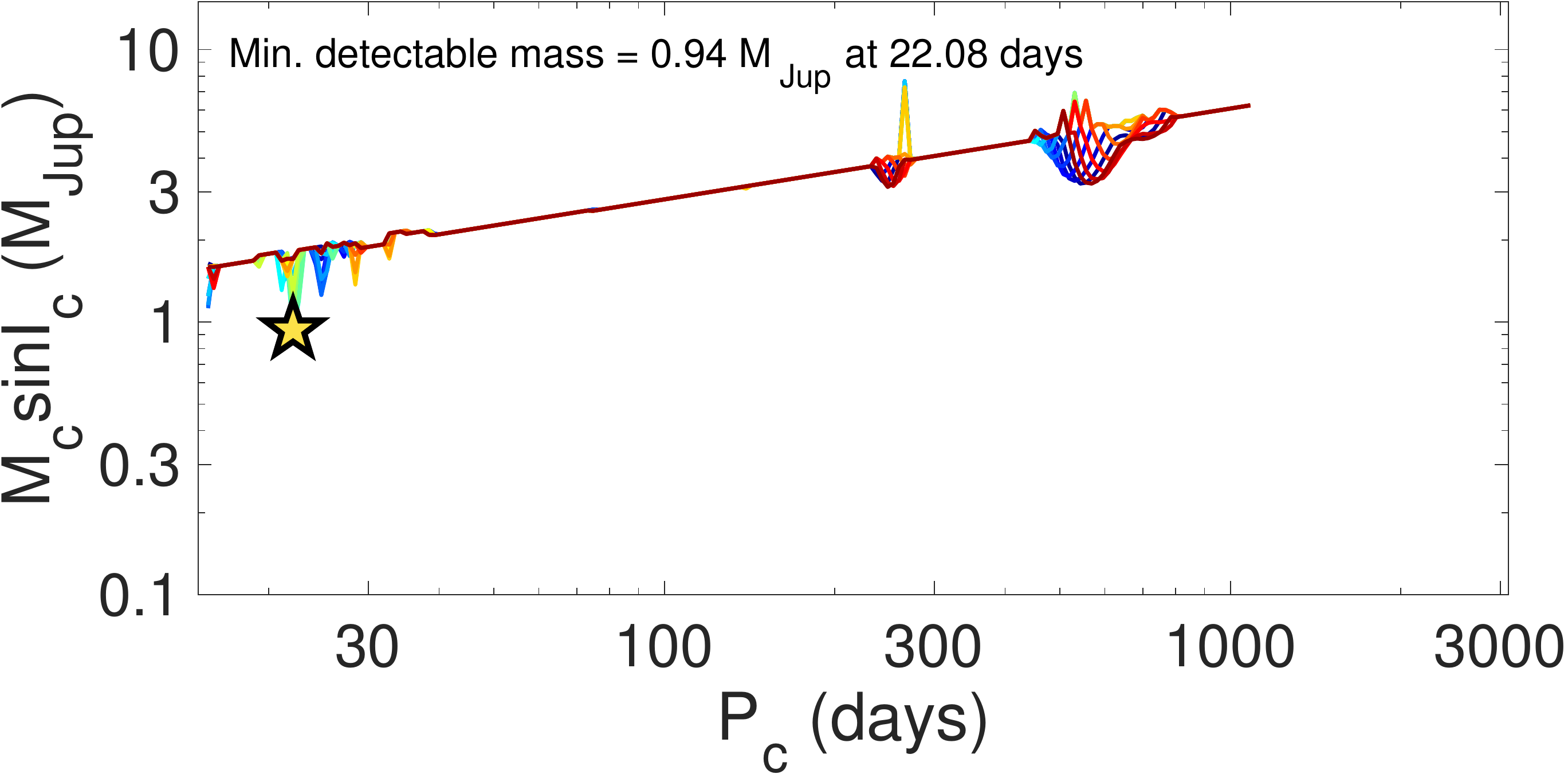}
\end{subfigure}
\end{center}
\end{figure}
\begin{figure}
\begin{center}
\subcaption*{EBLM J1540-09: chosen model = k1 (ecc) \newline \newline $m_{\rm A} = 1.18M_{\odot}$, $m_{\rm B} = 0.444M_{\odot}$, $P = 26.338$ d, $e = 0.12$}
\begin{subfigure}[b]{0.49\textwidth}
\includegraphics[width=\textwidth,trim={0 10cm 0 1.2cm},clip]{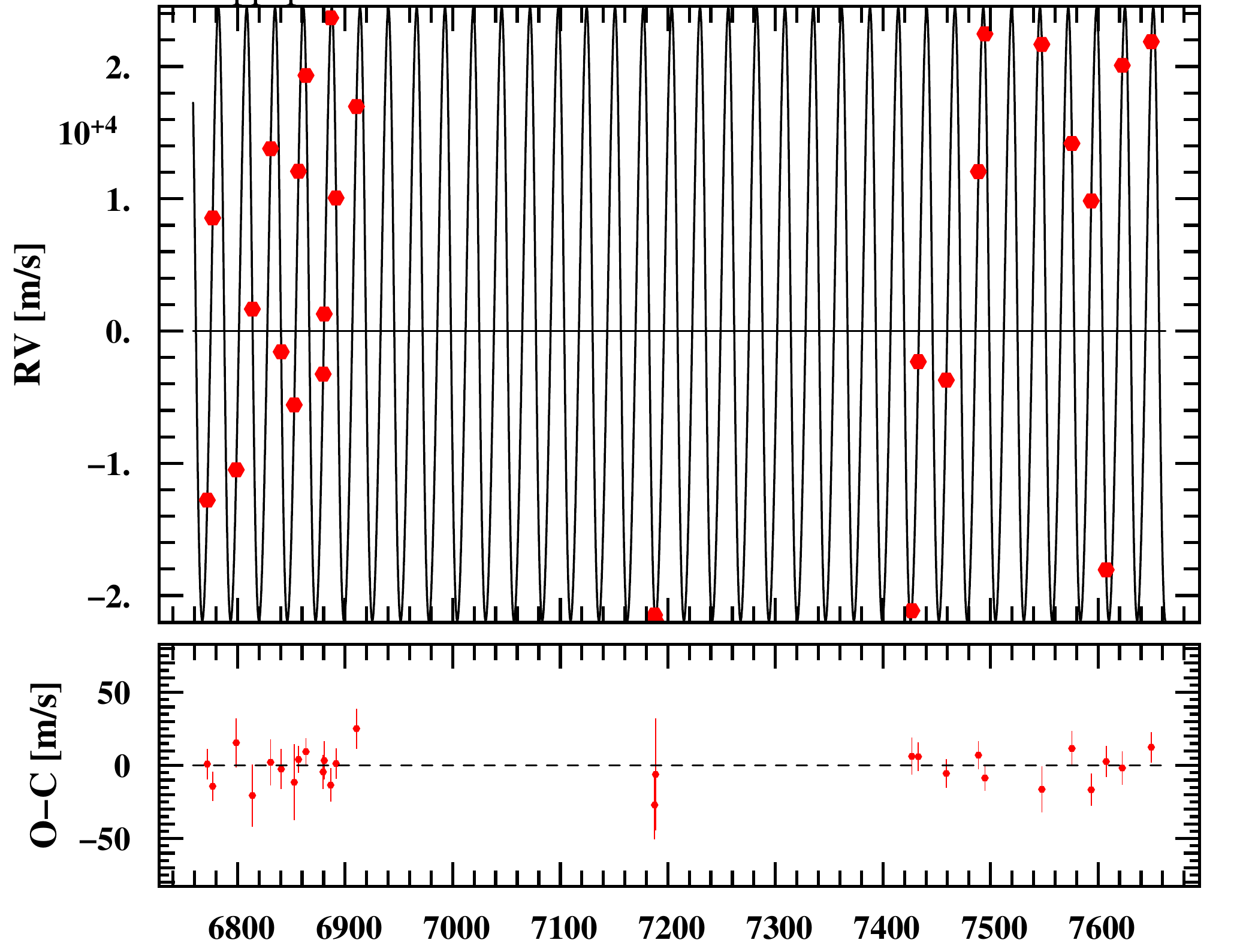}
\end{subfigure}
\begin{subfigure}[b]{0.49\textwidth}
\includegraphics[width=\textwidth,trim={0 0 2cm 0},clip]{orbit_figures/BJD_bar.pdf}
\end{subfigure}
Radial velocities folded on binary phase
\begin{subfigure}[b]{0.49\textwidth}
\includegraphics[width=\textwidth,trim={0 0.5cm 0 0},clip]{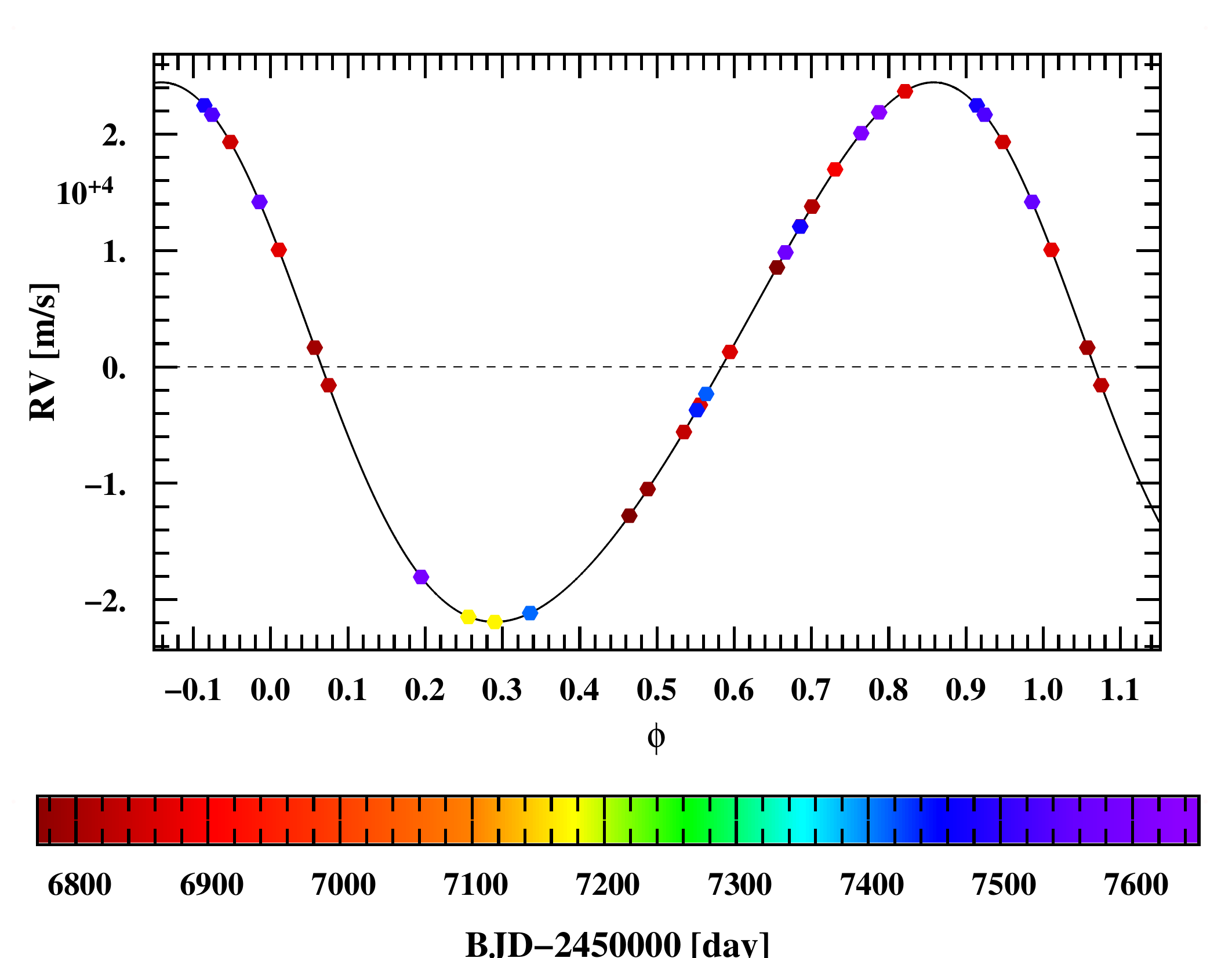}
\end{subfigure}
\begin{subfigure}[b]{0.49\textwidth}
\includegraphics[width=\textwidth,trim={0 0 2cm 0},clip]{orbit_figures/BJD_bar.pdf}
\end{subfigure}
Detection limits
\begin{subfigure}[b]{0.49\textwidth}
\vspace{0.5cm}
\includegraphics[width=\textwidth,trim={0 0 0 0},clip]{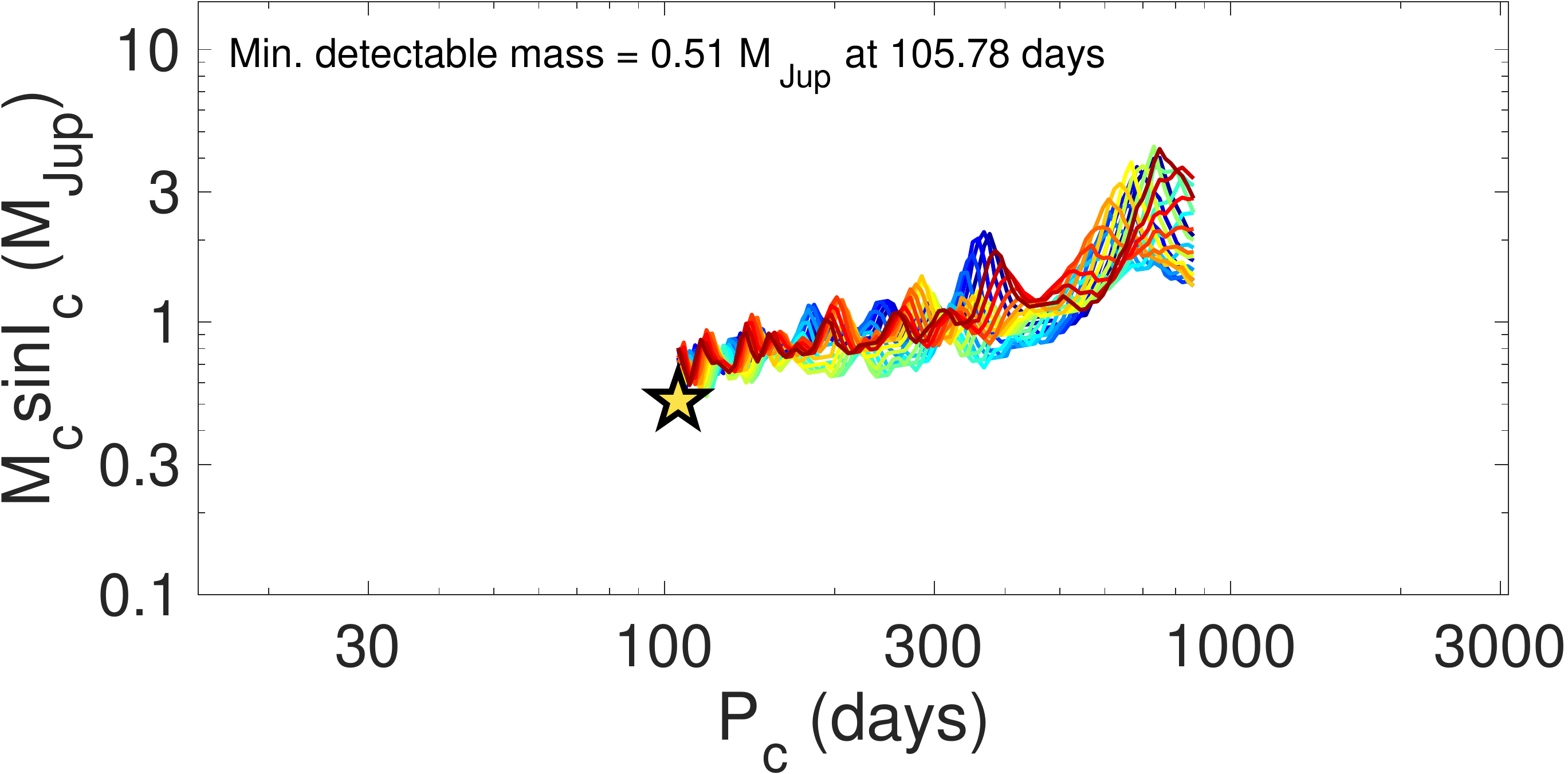}
\end{subfigure}
\end{center}
\end{figure}
\begin{figure}
\begin{center}
\subcaption*{EBLM J1630+10: chosen model = k1d2 (ecc) \newline \newline $m_{\rm A} = 1.07M_{\odot}$, $m_{\rm B} = 0.238M_{\odot}$, $P = 10.964$ d, $e = 0.181$}
\begin{subfigure}[b]{0.49\textwidth}
\includegraphics[width=\textwidth,trim={0 10cm 0 1.2cm},clip]{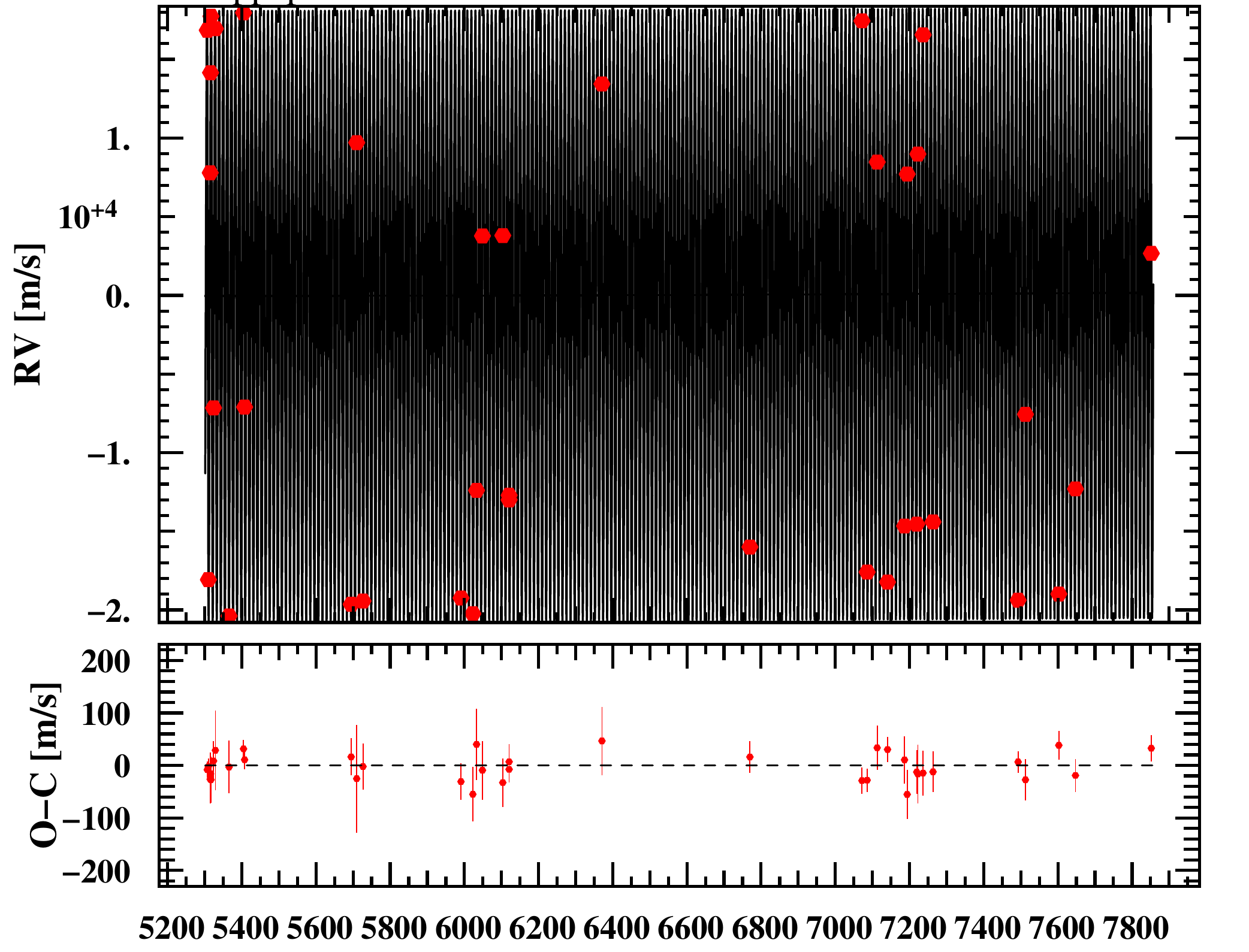}
\end{subfigure}
\begin{subfigure}[b]{0.49\textwidth}
\includegraphics[width=\textwidth,trim={0 0 2cm 0},clip]{orbit_figures/BJD_bar.pdf}
\end{subfigure}
Radial velocities folded on binary phase
\begin{subfigure}[b]{0.49\textwidth}
\includegraphics[width=\textwidth,trim={0 0.5cm 0 0},clip]{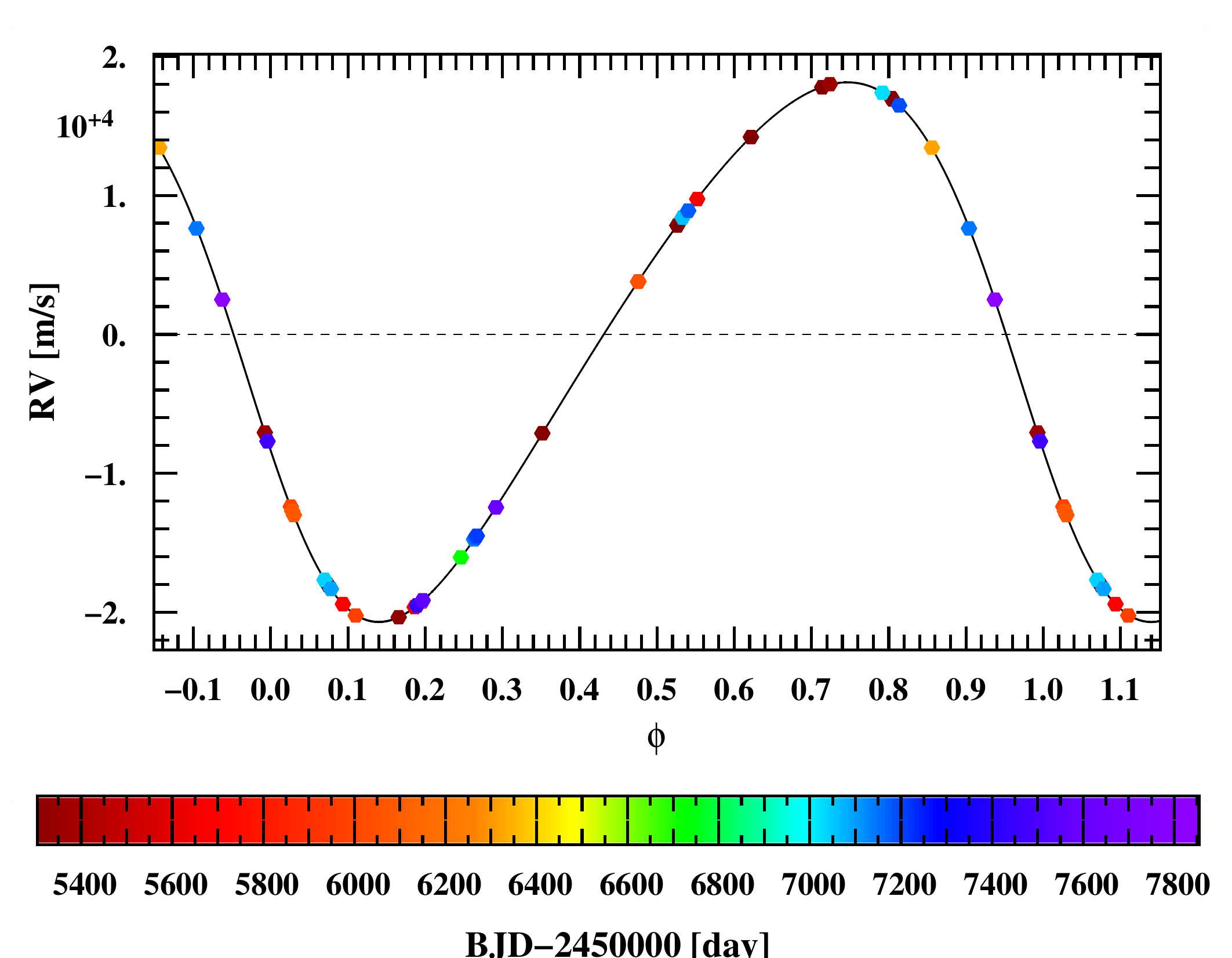}
\end{subfigure}
\begin{subfigure}[b]{0.49\textwidth}
\includegraphics[width=\textwidth,trim={0 0 2cm 0},clip]{orbit_figures/BJD_bar.pdf}
\end{subfigure}
Detection limits
\begin{subfigure}[b]{0.49\textwidth}
\vspace{0.5cm}
\includegraphics[width=\textwidth,trim={0 0 0 0},clip]{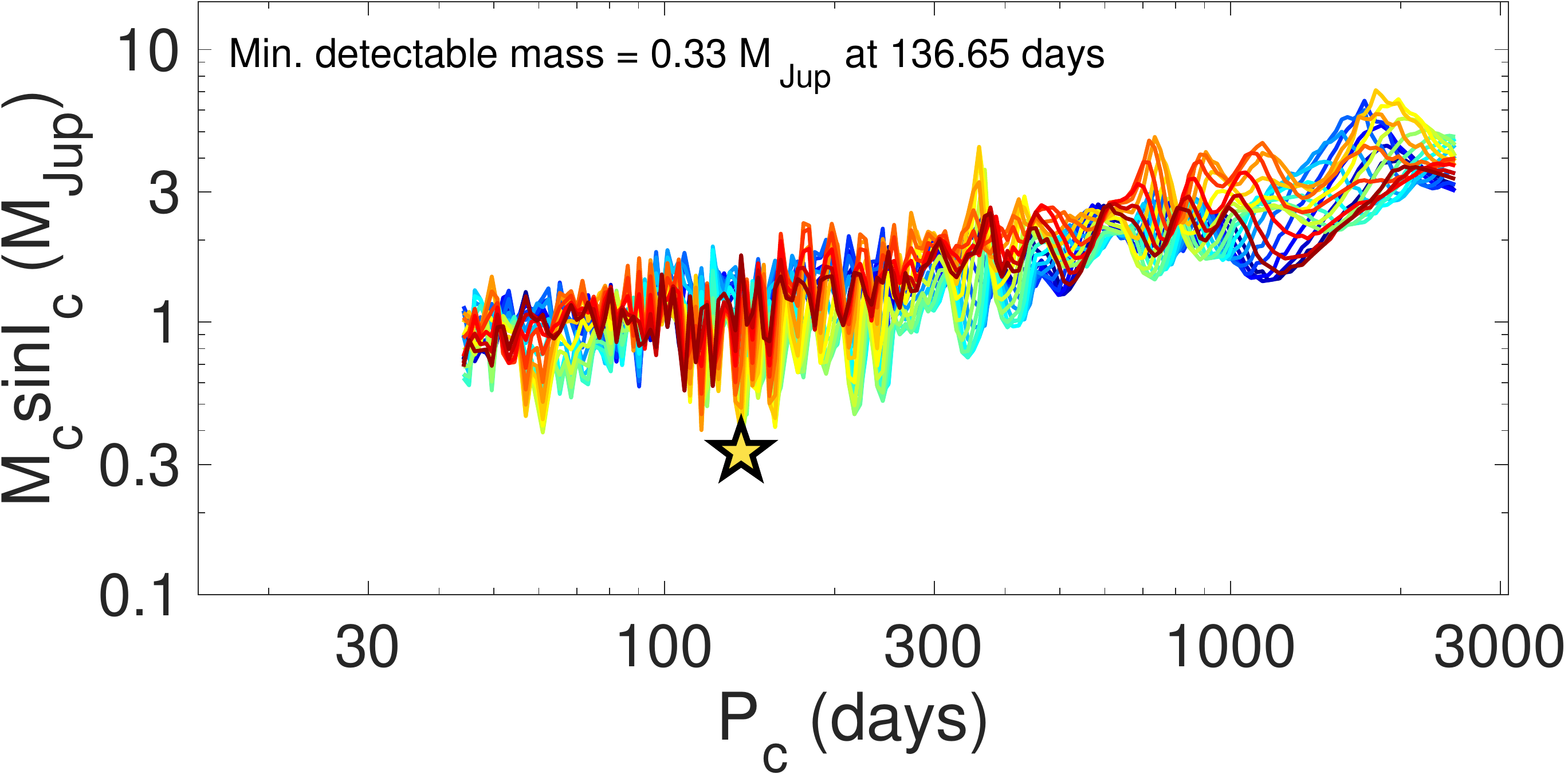}
\end{subfigure}
\end{center}
\end{figure}
\begin{figure}
\begin{center}
\subcaption*{EBLM J1928-38: chosen model = k1 (ecc) \newline \newline $m_{\rm A} = 0.98M_{\odot}$, $m_{\rm B} = 0.268M_{\odot}$, $P = 23.323$ d, $e = 0.074$}
\begin{subfigure}[b]{0.49\textwidth}
\includegraphics[width=\textwidth,trim={0 10cm 0 1.2cm},clip]{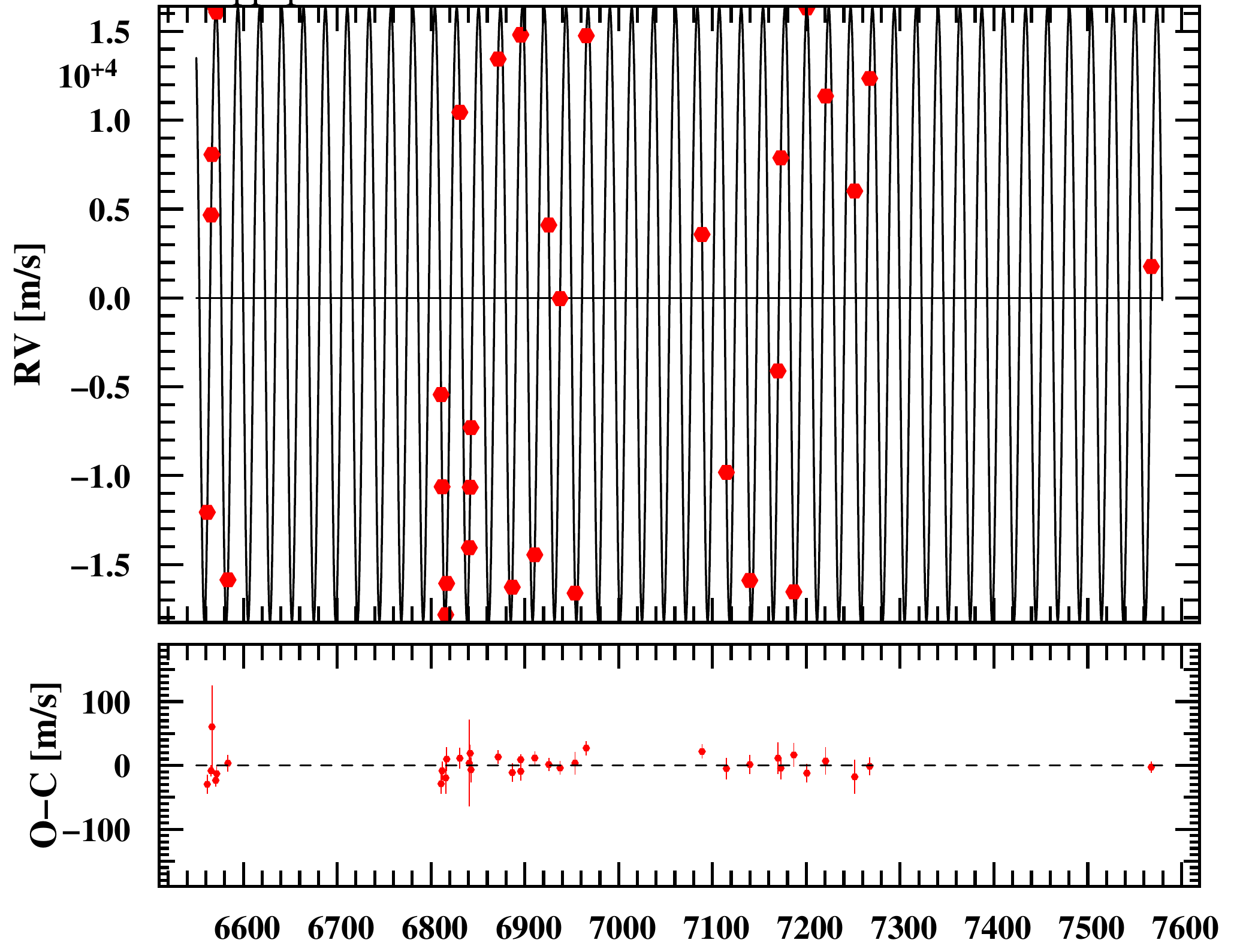}
\end{subfigure}
\begin{subfigure}[b]{0.49\textwidth}
\includegraphics[width=\textwidth,trim={0 0 2cm 0},clip]{orbit_figures/BJD_bar.pdf}
\end{subfigure}
Radial velocities folded on binary phase
\begin{subfigure}[b]{0.49\textwidth}
\includegraphics[width=\textwidth,trim={0 0.5cm 0 0},clip]{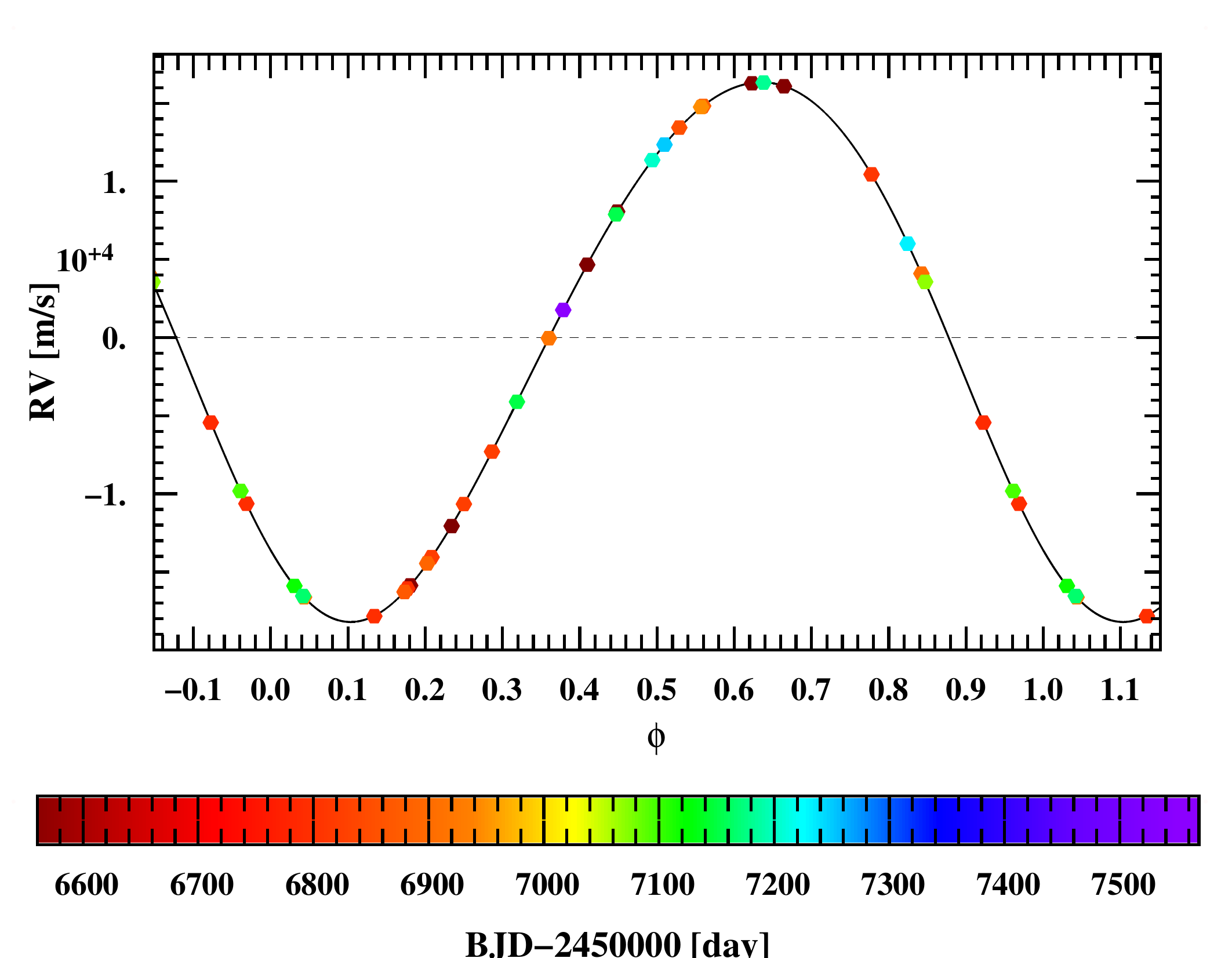}
\end{subfigure}
\begin{subfigure}[b]{0.49\textwidth}
\includegraphics[width=\textwidth,trim={0 0 2cm 0},clip]{orbit_figures/BJD_bar.pdf}
\end{subfigure}
Detection limits
\begin{subfigure}[b]{0.49\textwidth}
\vspace{0.5cm}
\includegraphics[width=\textwidth,trim={0 0 0 0},clip]{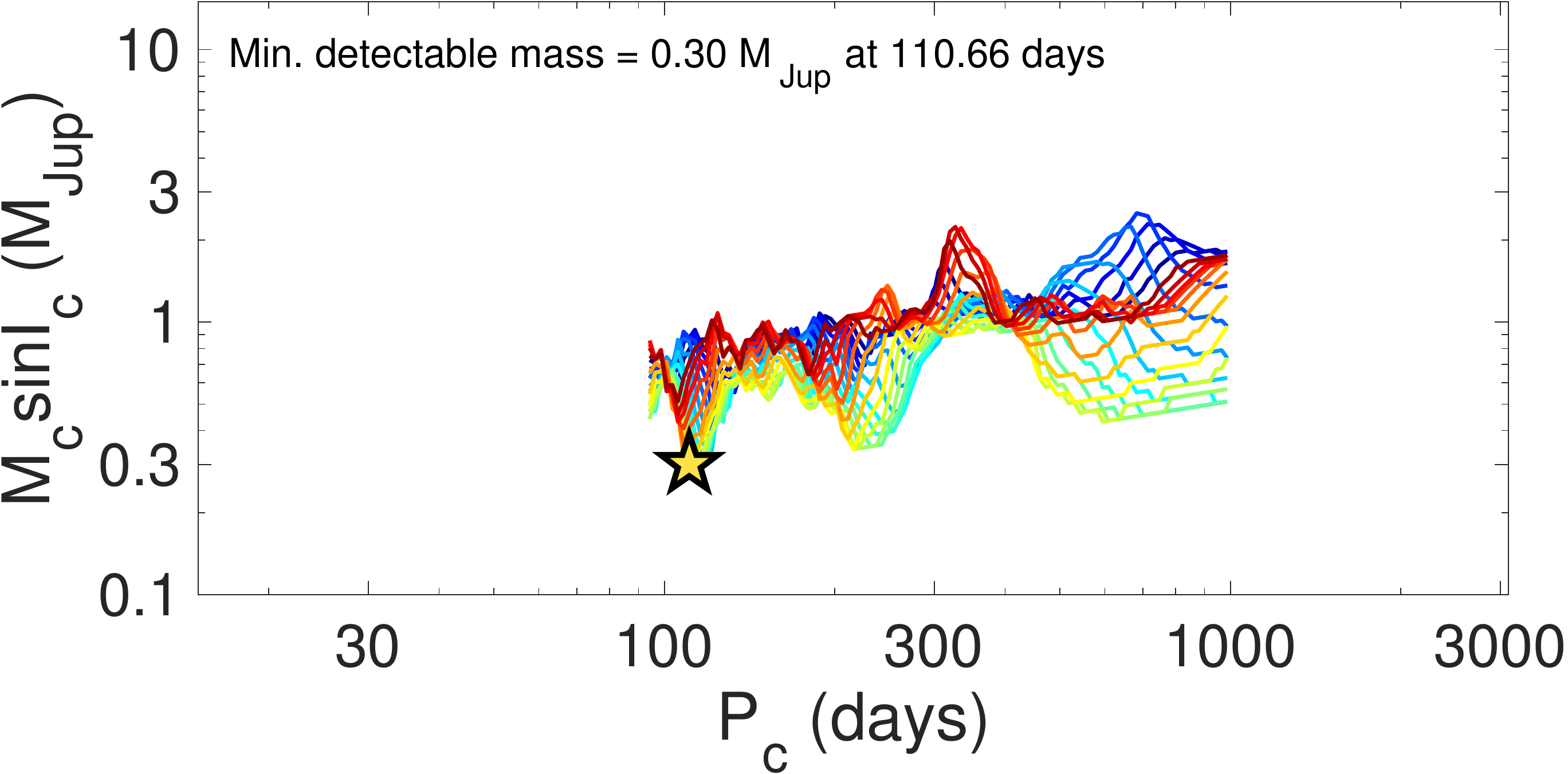}
\end{subfigure}
\end{center}
\end{figure}
\begin{figure}
\begin{center}
\subcaption*{EBLM J1934-42: chosen model = k1 (circ) \newline \newline $m_{\rm A} = 0.97M_{\odot}$, $m_{\rm B} = 0.178M_{\odot}$, $P = 6.353$ d, $e = 0$}
\begin{subfigure}[b]{0.49\textwidth}
\includegraphics[width=\textwidth,trim={0 10cm 0 1.2cm},clip]{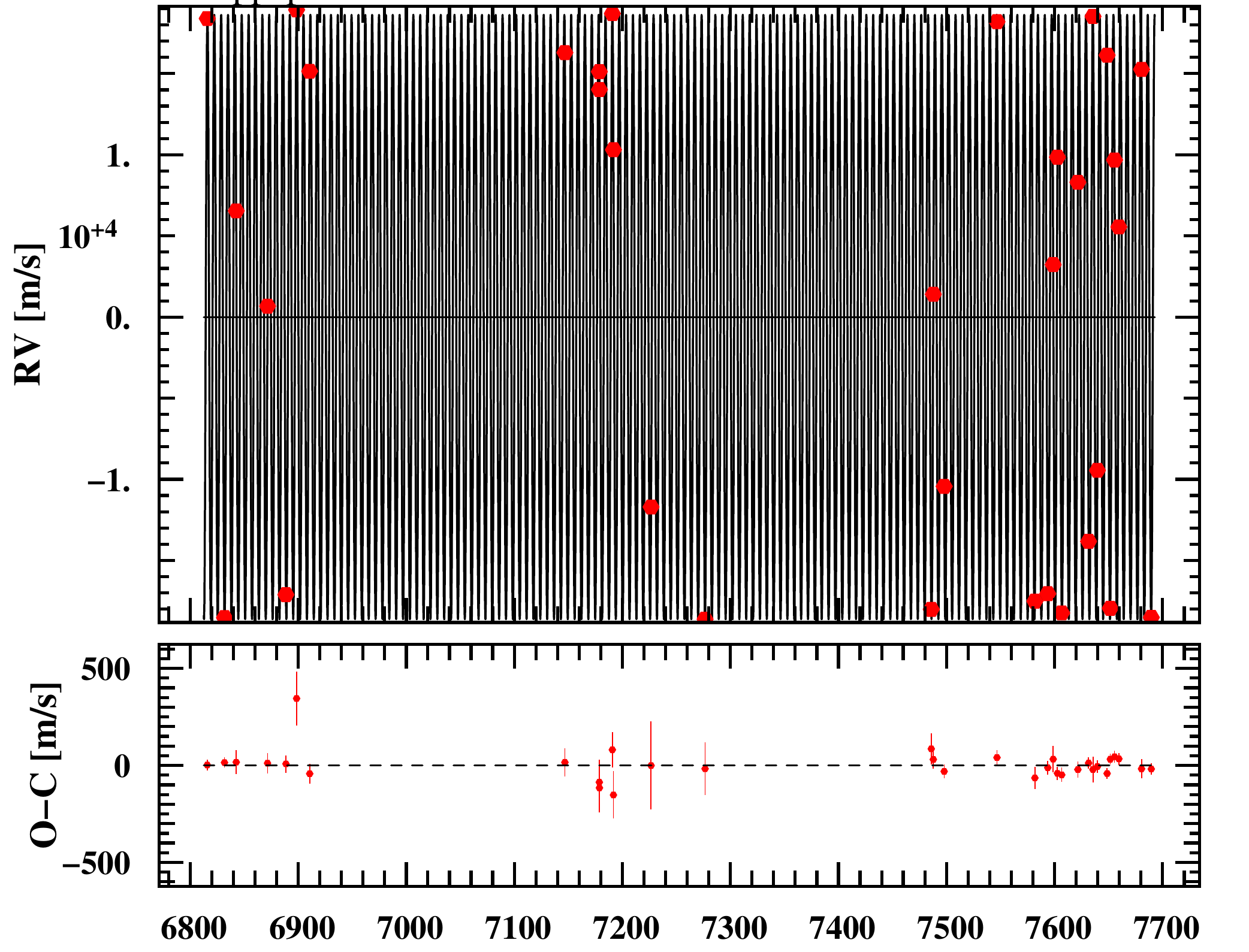}
\end{subfigure}
\begin{subfigure}[b]{0.49\textwidth}
\includegraphics[width=\textwidth,trim={0 0 2cm 0},clip]{orbit_figures/BJD_bar.pdf}
\end{subfigure}
Radial velocities folded on binary phase
\begin{subfigure}[b]{0.49\textwidth}
\includegraphics[width=\textwidth,trim={0 0.5cm 0 0},clip]{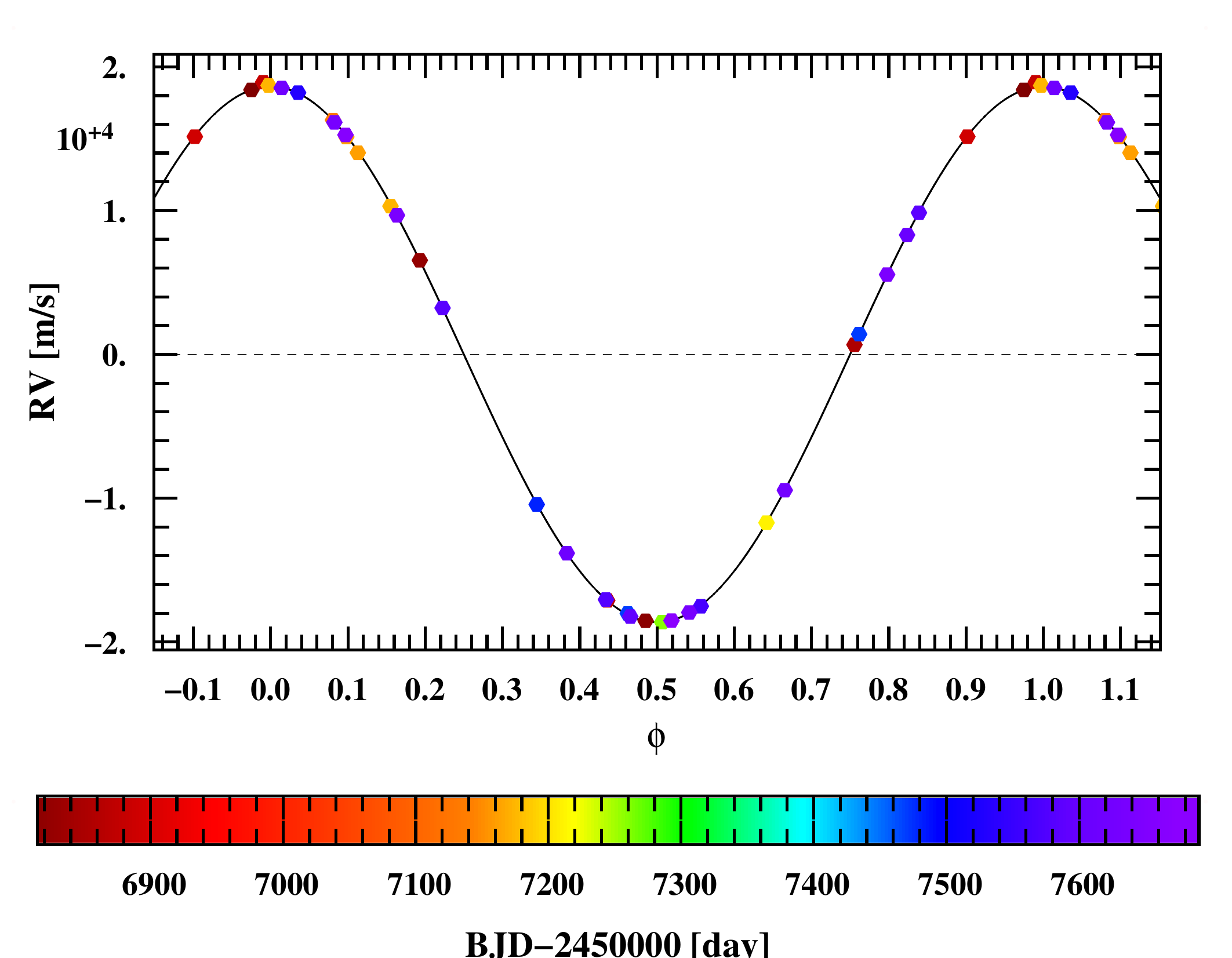}
\end{subfigure}
\begin{subfigure}[b]{0.49\textwidth}
\includegraphics[width=\textwidth,trim={0 0 2cm 0},clip]{orbit_figures/BJD_bar.pdf}
\end{subfigure}
Detection limits
\begin{subfigure}[b]{0.49\textwidth}
\vspace{0.5cm}
\includegraphics[width=\textwidth,trim={0 0 0 0},clip]{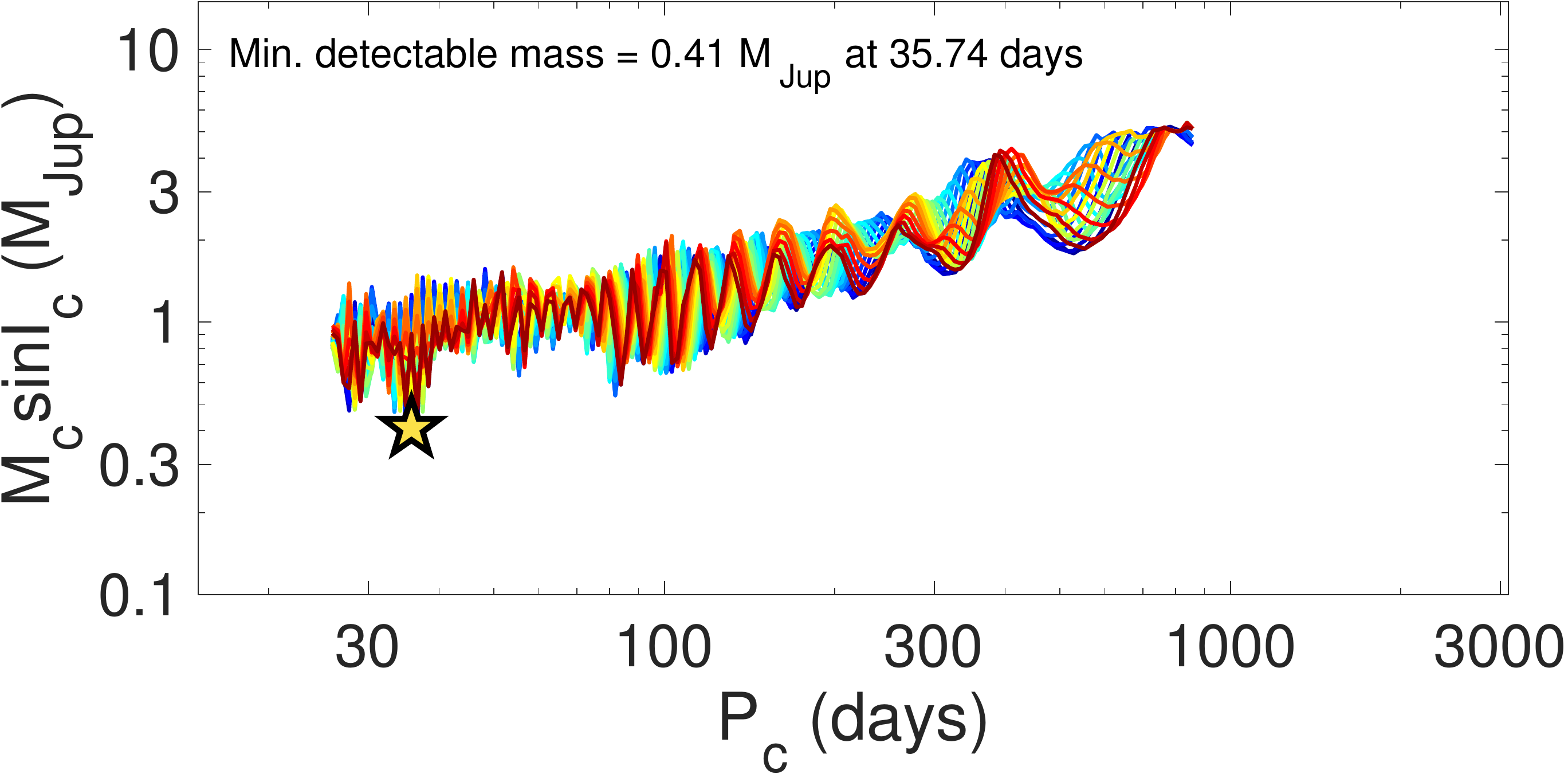}
\end{subfigure}
\end{center}
\end{figure}
\begin{figure}
\begin{center}
\subcaption*{EBLM J2011-71: chosen model = k2 (ecc) \newline \newline $m_{\rm A} = 1.41M_{\odot}$, $m_{\rm B} = 0.285M_{\odot}$, $P = 5.873$ d, $e = 0.031$}
\begin{subfigure}[b]{0.49\textwidth}
\includegraphics[width=\textwidth,trim={0 10cm 0 1.2cm},clip]{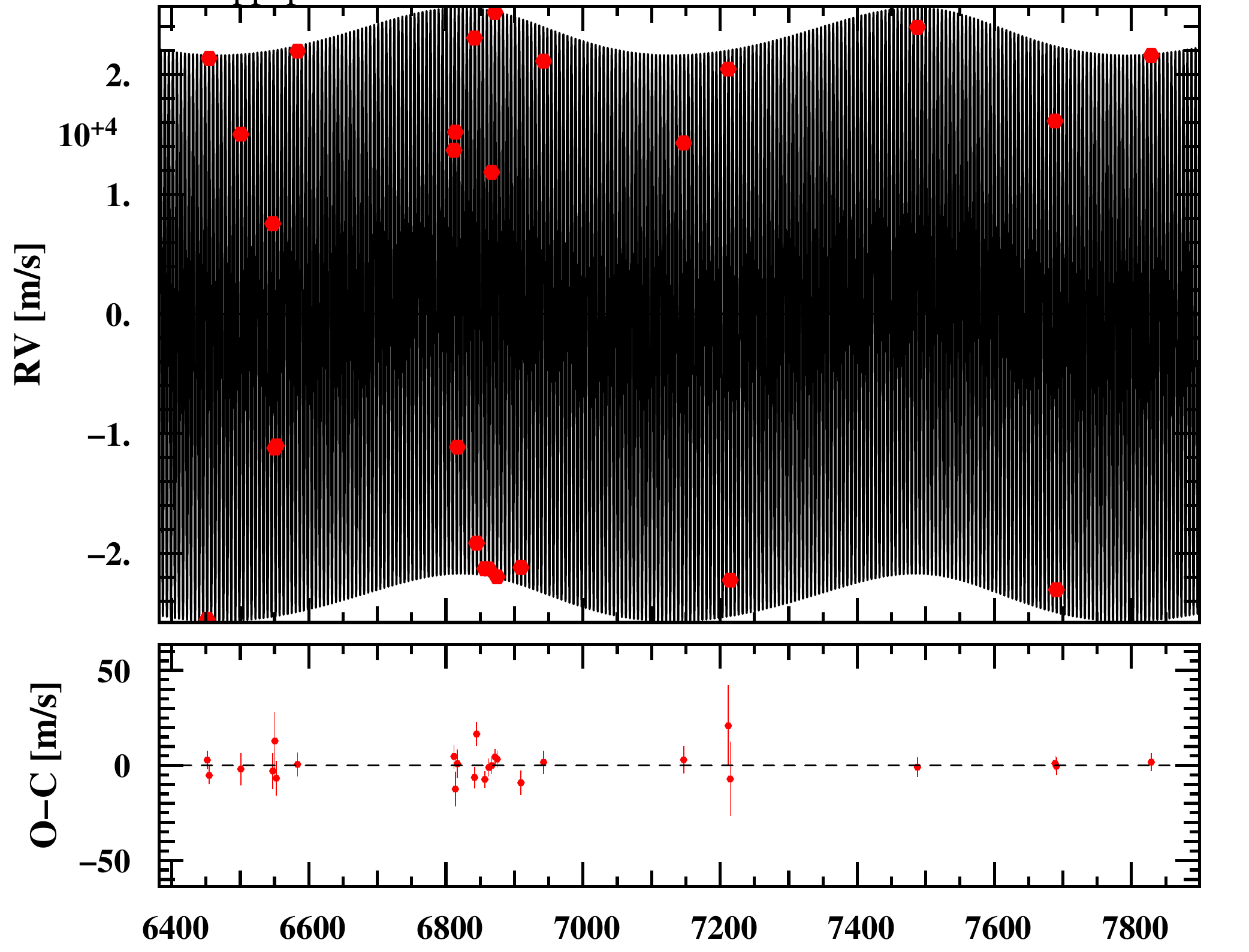}
\end{subfigure}
\begin{subfigure}[b]{0.49\textwidth}
\includegraphics[width=\textwidth,trim={0 0 2cm 0},clip]{orbit_figures/BJD_bar.pdf}
\end{subfigure}
Radial velocities folded on binary phase
\begin{subfigure}[b]{0.49\textwidth}
\includegraphics[width=\textwidth,trim={0 0.5cm 0 0},clip]{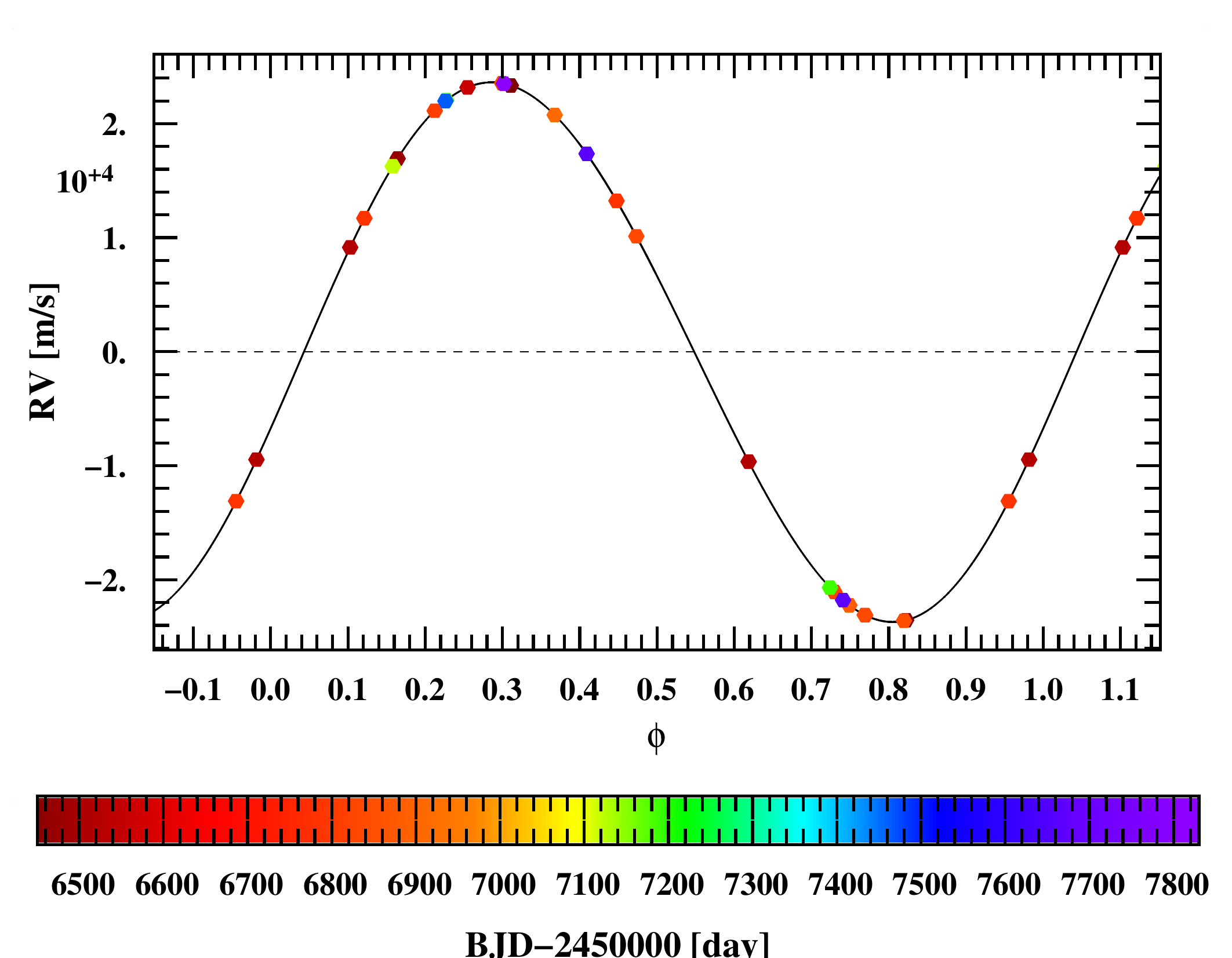}
\end{subfigure}
\begin{subfigure}[b]{0.49\textwidth}
\includegraphics[width=\textwidth,trim={0 0 2cm 0},clip]{orbit_figures/BJD_bar.pdf}
\end{subfigure}
Detection limits
\begin{subfigure}[b]{0.49\textwidth}
\vspace{0.5cm}
\includegraphics[width=\textwidth,trim={0 0 0 0},clip]{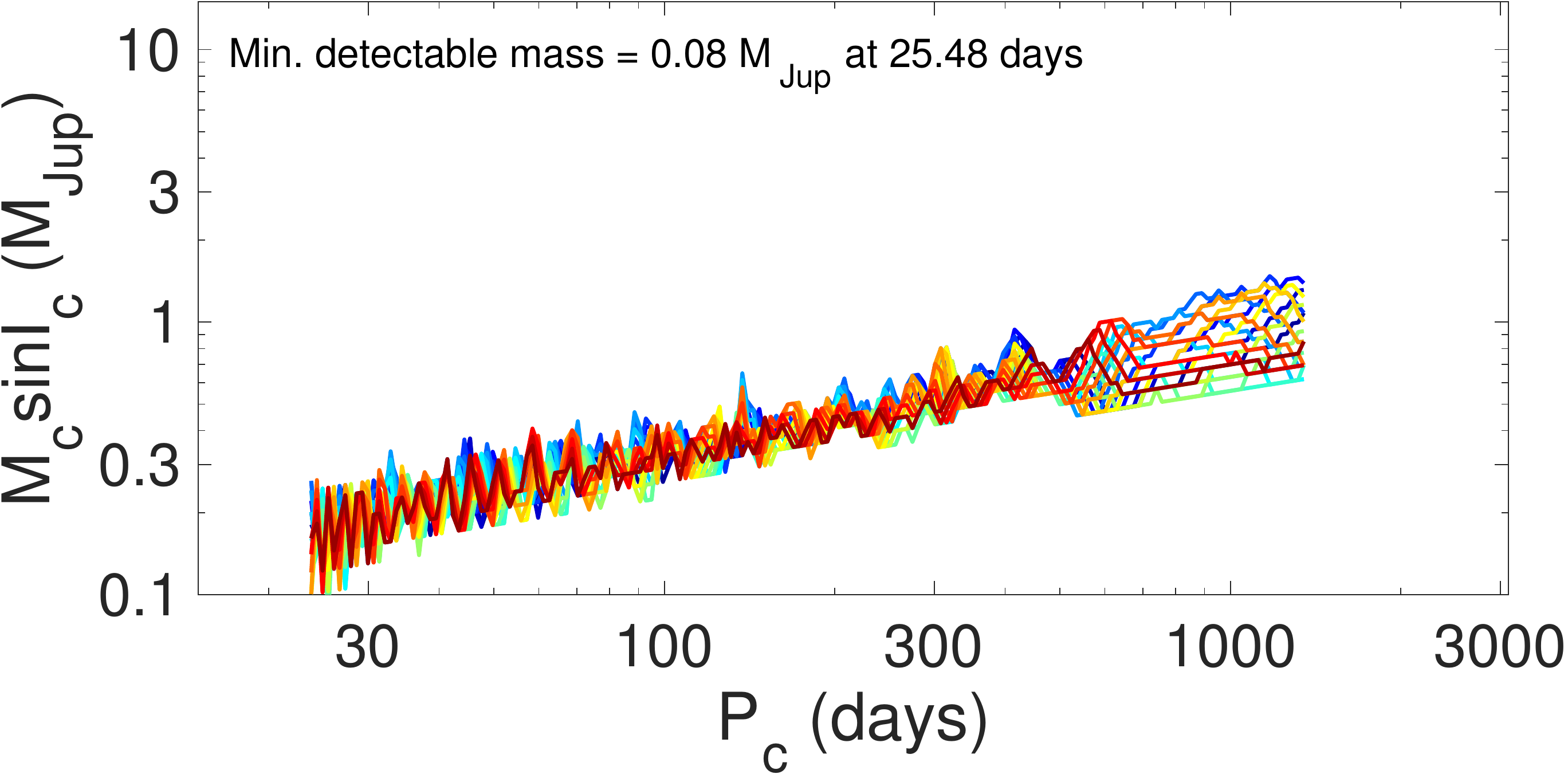}
\end{subfigure}
\end{center}
\end{figure}
\begin{figure}
\begin{center}
\subcaption*{EBLM J2040-41: chosen model = k1 (ecc) \newline \newline $m_{\rm A} = 1.13M_{\odot}$, $m_{\rm B} = 0.165M_{\odot}$, $P = 14.456$ d, $e = 0.226$}
\begin{subfigure}[b]{0.49\textwidth}
\includegraphics[width=\textwidth,trim={0 10cm 0 1.2cm},clip]{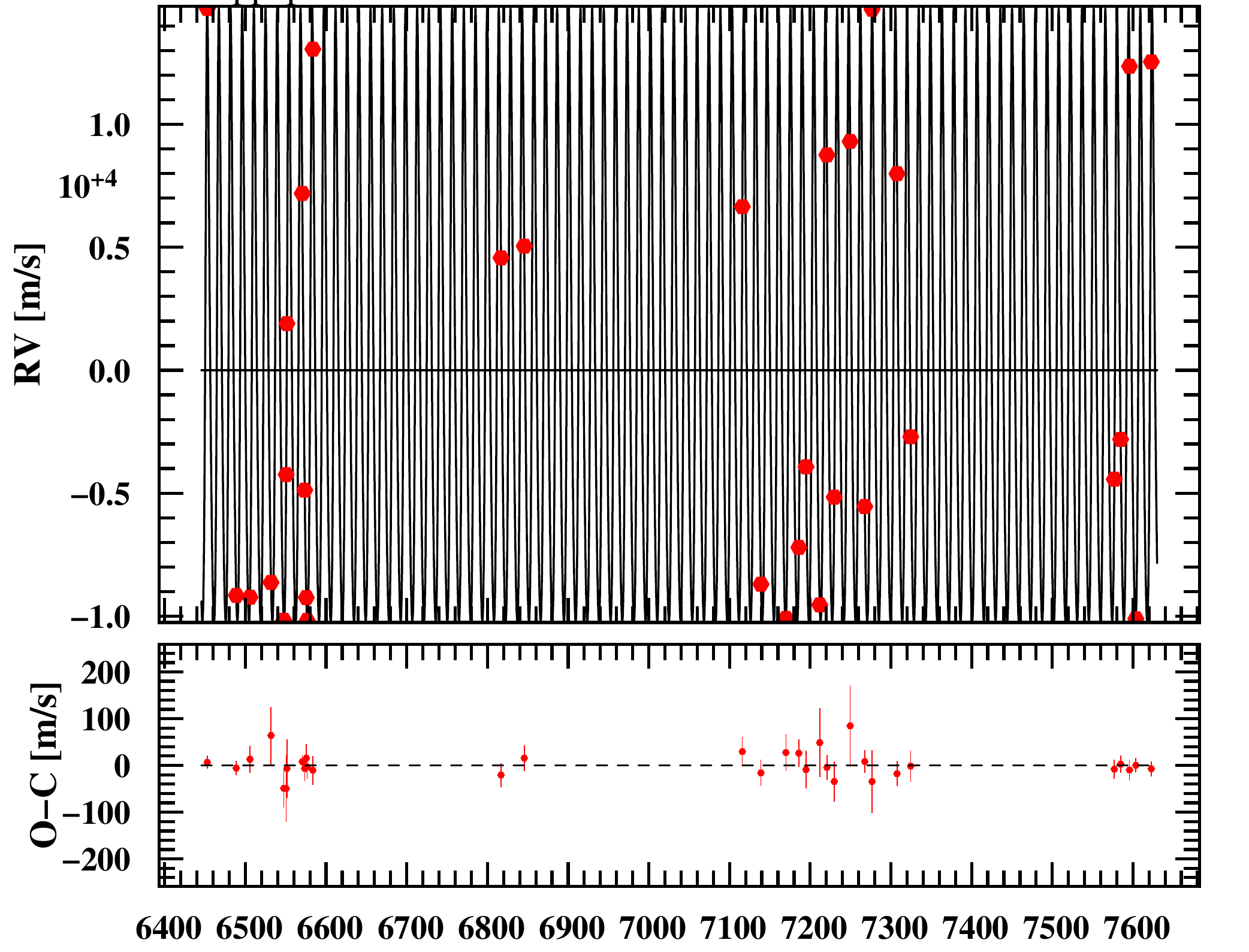}
\end{subfigure}
\begin{subfigure}[b]{0.49\textwidth}
\includegraphics[width=\textwidth,trim={0 0 2cm 0},clip]{orbit_figures/BJD_bar.pdf}
\end{subfigure}
Radial velocities folded on binary phase
\begin{subfigure}[b]{0.49\textwidth}
\includegraphics[width=\textwidth,trim={0 0.5cm 0 0},clip]{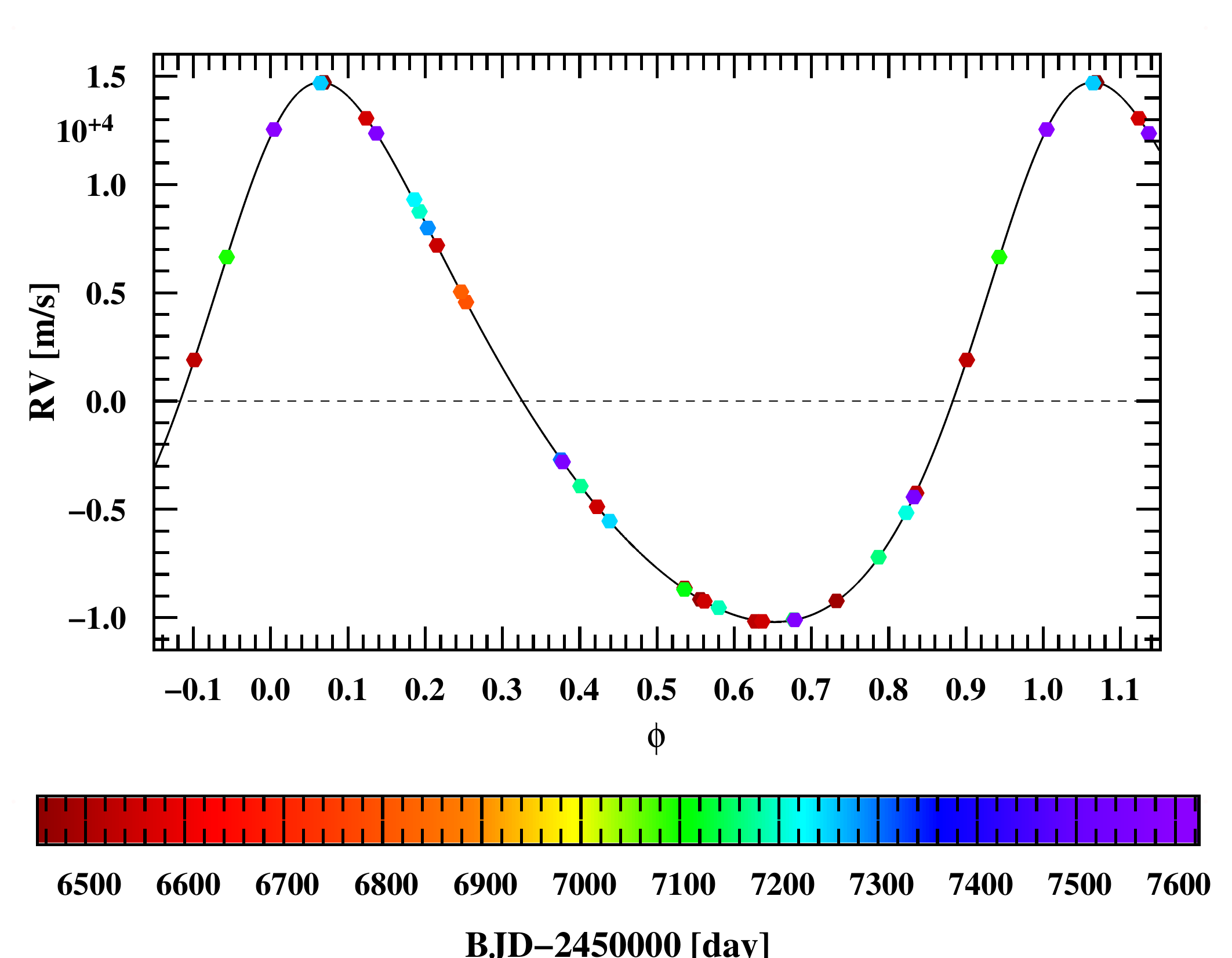}
\end{subfigure}
\begin{subfigure}[b]{0.49\textwidth}
\includegraphics[width=\textwidth,trim={0 0 2cm 0},clip]{orbit_figures/BJD_bar.pdf}
\end{subfigure}
Detection limits
\begin{subfigure}[b]{0.49\textwidth}
\vspace{0.5cm}
\includegraphics[width=\textwidth,trim={0 0 0 0},clip]{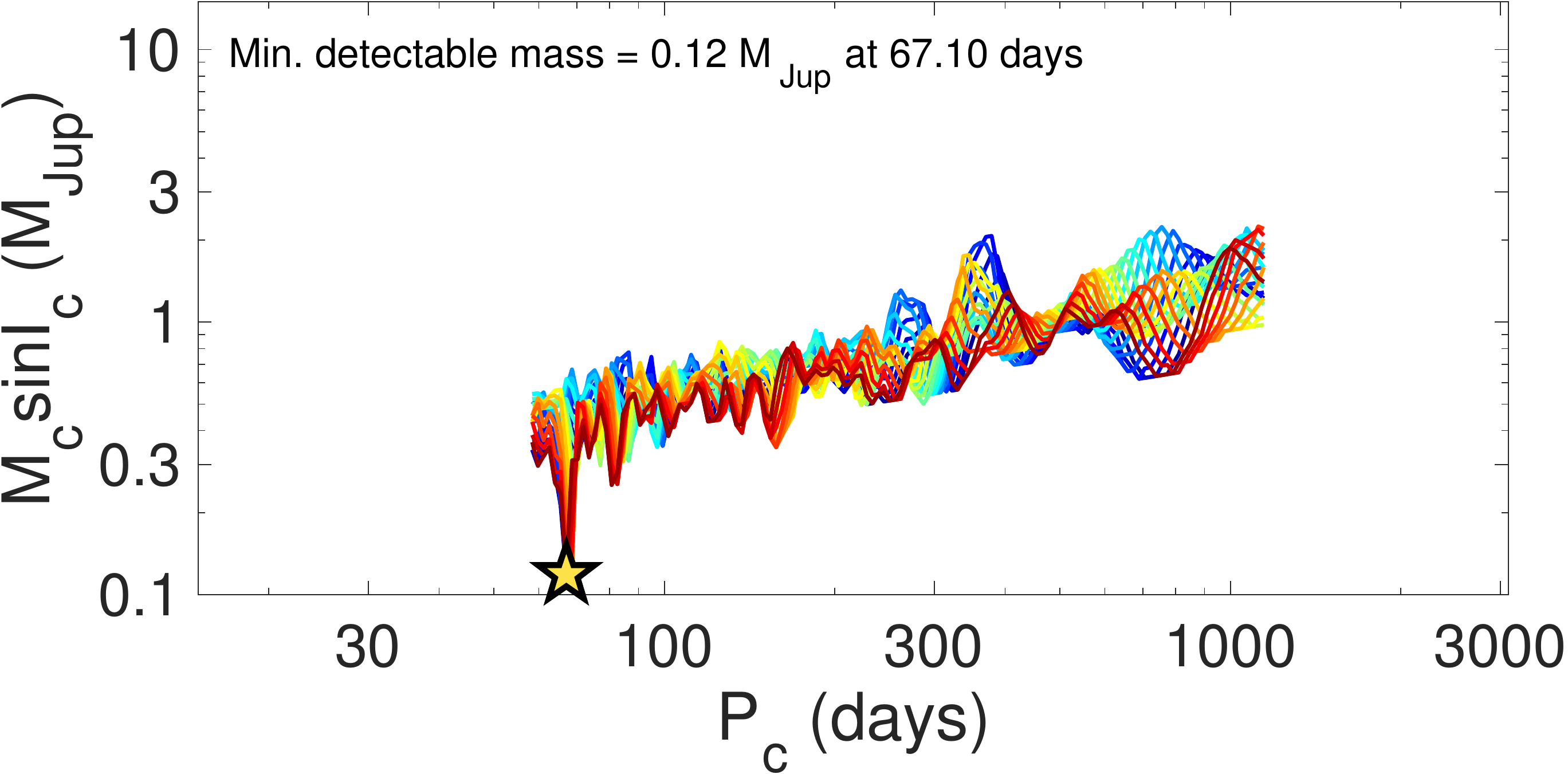}
\end{subfigure}
\end{center}
\end{figure}
\begin{figure}
\begin{center}
\subcaption*{EBLM J2046-40: chosen model = k2 (ecc) \newline \newline $m_{\rm A} = 1.07M_{\odot}$, $m_{\rm B} = 0.193M_{\odot}$, $P = 37.014$ d, $e = 0.473$}
\begin{subfigure}[b]{0.49\textwidth}
\includegraphics[width=\textwidth,trim={0 10cm 0 1.2cm},clip]{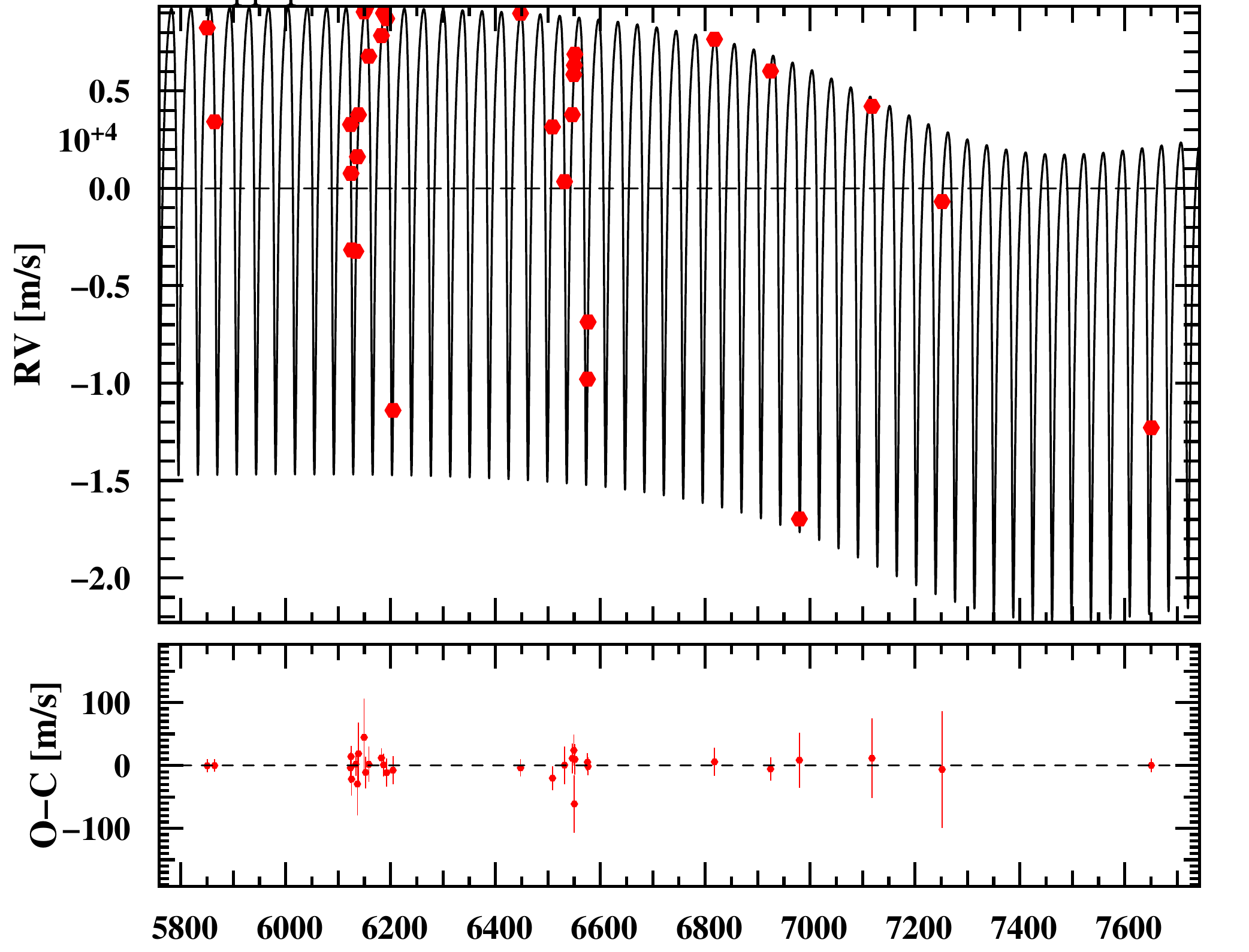}
\end{subfigure}
\begin{subfigure}[b]{0.49\textwidth}
\includegraphics[width=\textwidth,trim={0 0 2cm 0},clip]{orbit_figures/BJD_bar.pdf}
\end{subfigure}
Radial velocities folded on binary phase
\begin{subfigure}[b]{0.49\textwidth}
\includegraphics[width=\textwidth,trim={0 0.5cm 0 0},clip]{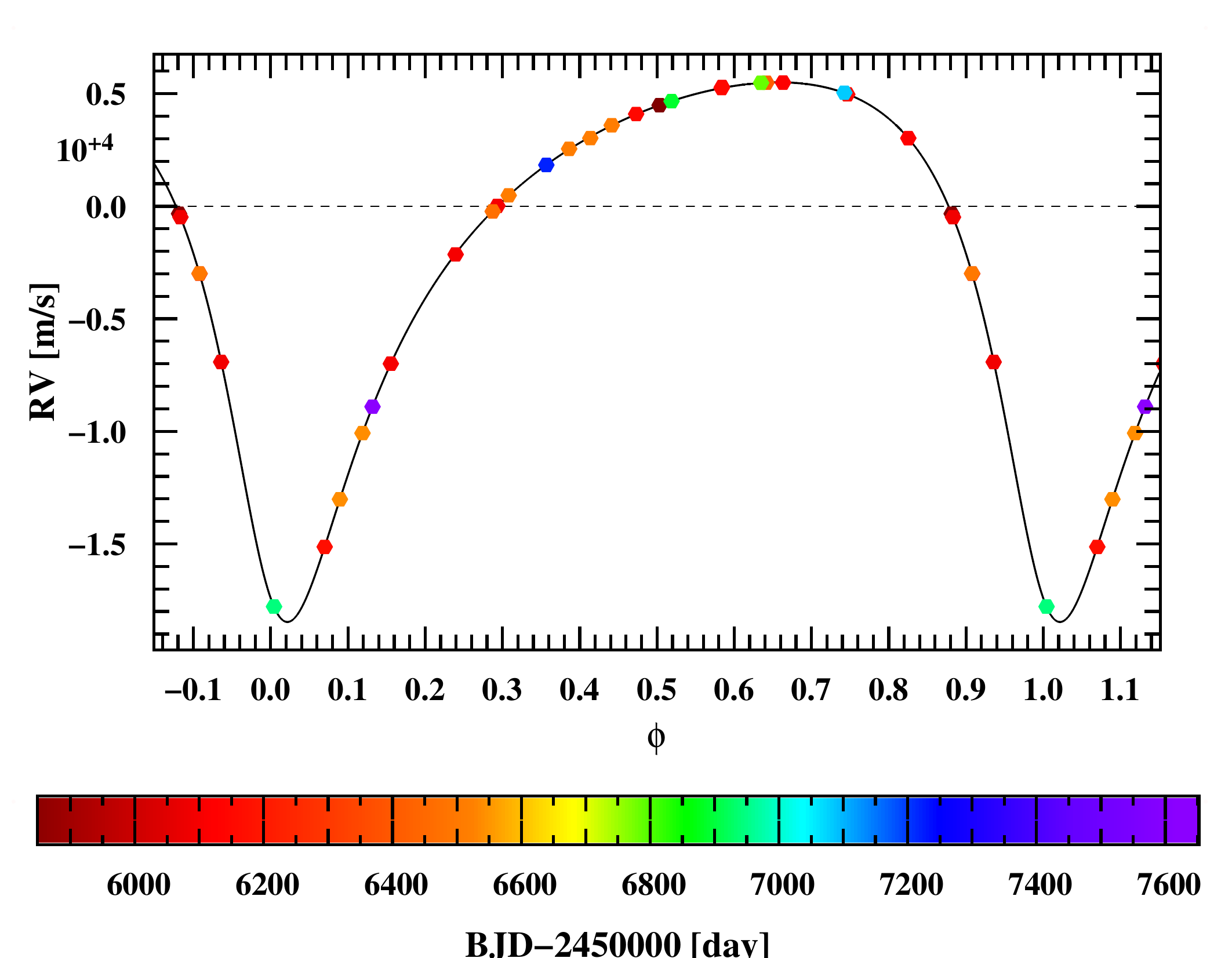}
\end{subfigure}
\begin{subfigure}[b]{0.49\textwidth}
\includegraphics[width=\textwidth,trim={0 0 2cm 0},clip]{orbit_figures/BJD_bar.pdf}
\end{subfigure}
Detection limits
\begin{subfigure}[b]{0.49\textwidth}
\vspace{0.5cm}
\includegraphics[width=\textwidth,trim={0 0 0 0},clip]{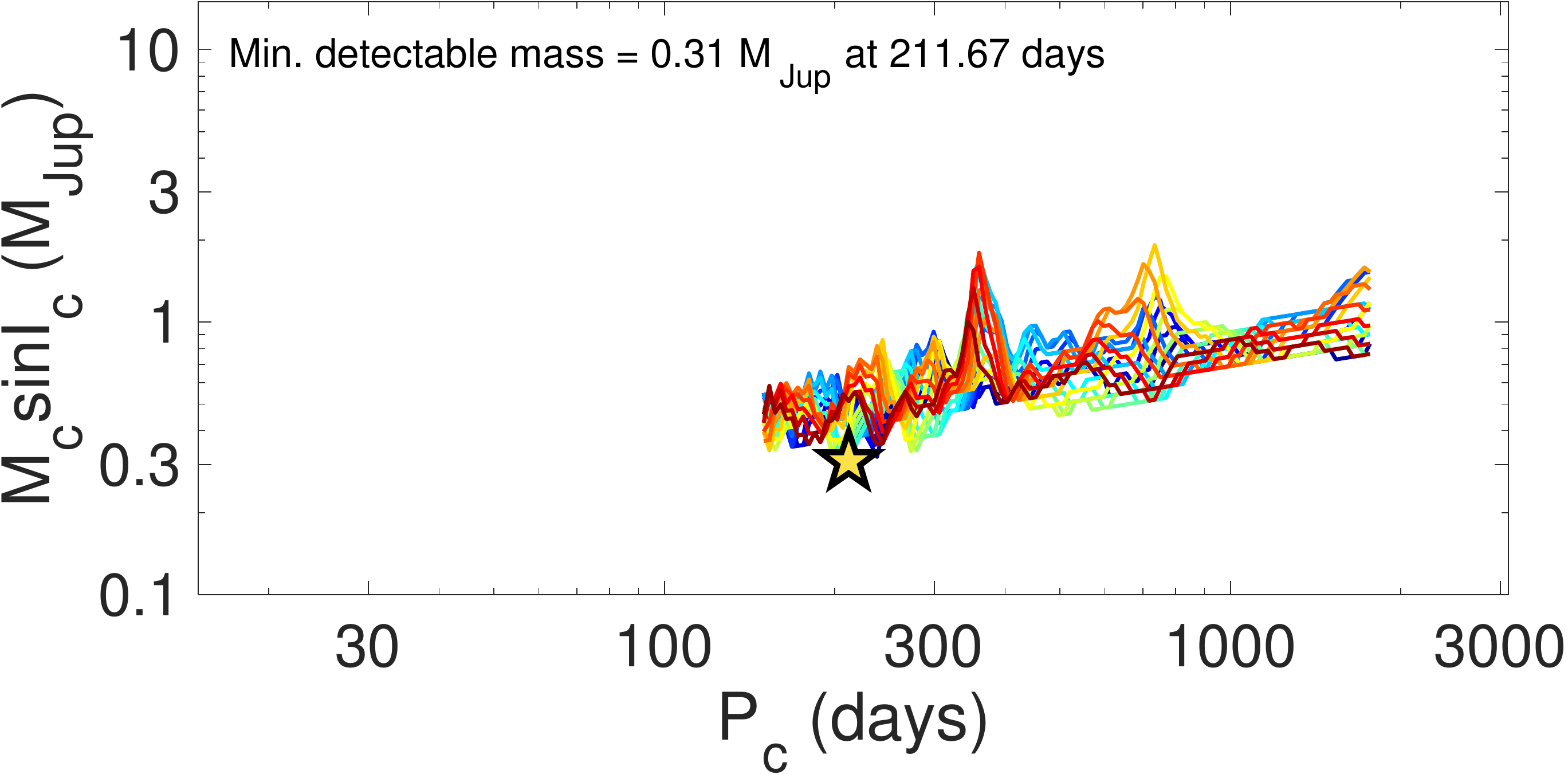}
\end{subfigure}
\end{center}
\end{figure}
\begin{figure}
\begin{center}
\subcaption*{EBLM J2046+06: chosen model = k1 (ecc) \newline \newline $m_{\rm A} = 1.28M_{\odot}$, $m_{\rm B} = 0.192M_{\odot}$, $P = 10.108$ d, $e = 0.344$}
\begin{subfigure}[b]{0.49\textwidth}
\includegraphics[width=\textwidth,trim={0 10cm 0 1.2cm},clip]{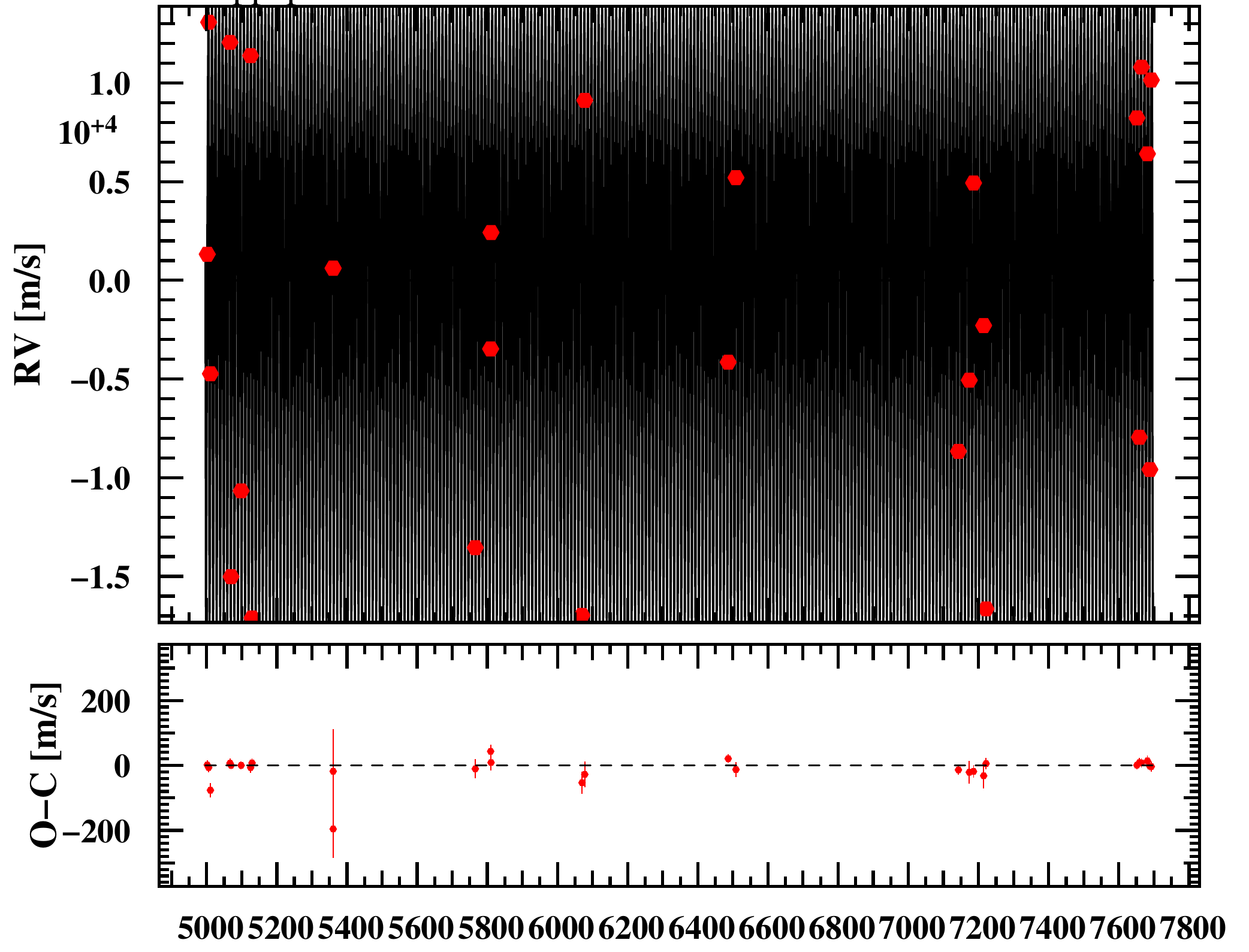}
\end{subfigure}
\begin{subfigure}[b]{0.49\textwidth}
\includegraphics[width=\textwidth,trim={0 0 2cm 0},clip]{orbit_figures/BJD_bar.pdf}
\end{subfigure}
Radial velocities folded on binary phase
\begin{subfigure}[b]{0.49\textwidth}
\includegraphics[width=\textwidth,trim={0 0.5cm 0 0},clip]{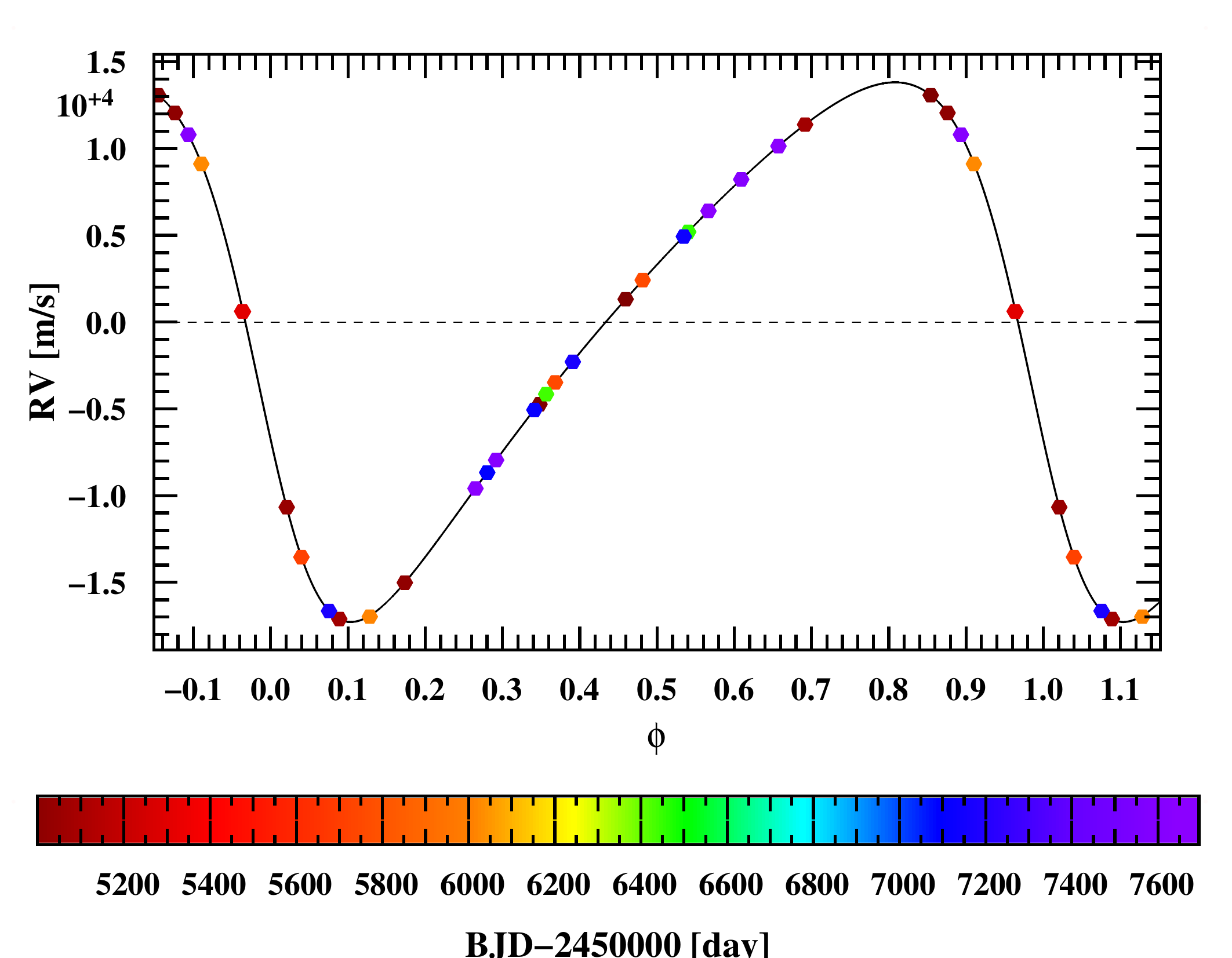}
\end{subfigure}
\begin{subfigure}[b]{0.49\textwidth}
\includegraphics[width=\textwidth,trim={0 0 2cm 0},clip]{orbit_figures/BJD_bar.pdf}
\end{subfigure}
Detection limits
\begin{subfigure}[b]{0.49\textwidth}
\vspace{0.5cm}
\includegraphics[width=\textwidth,trim={0 0 0 0},clip]{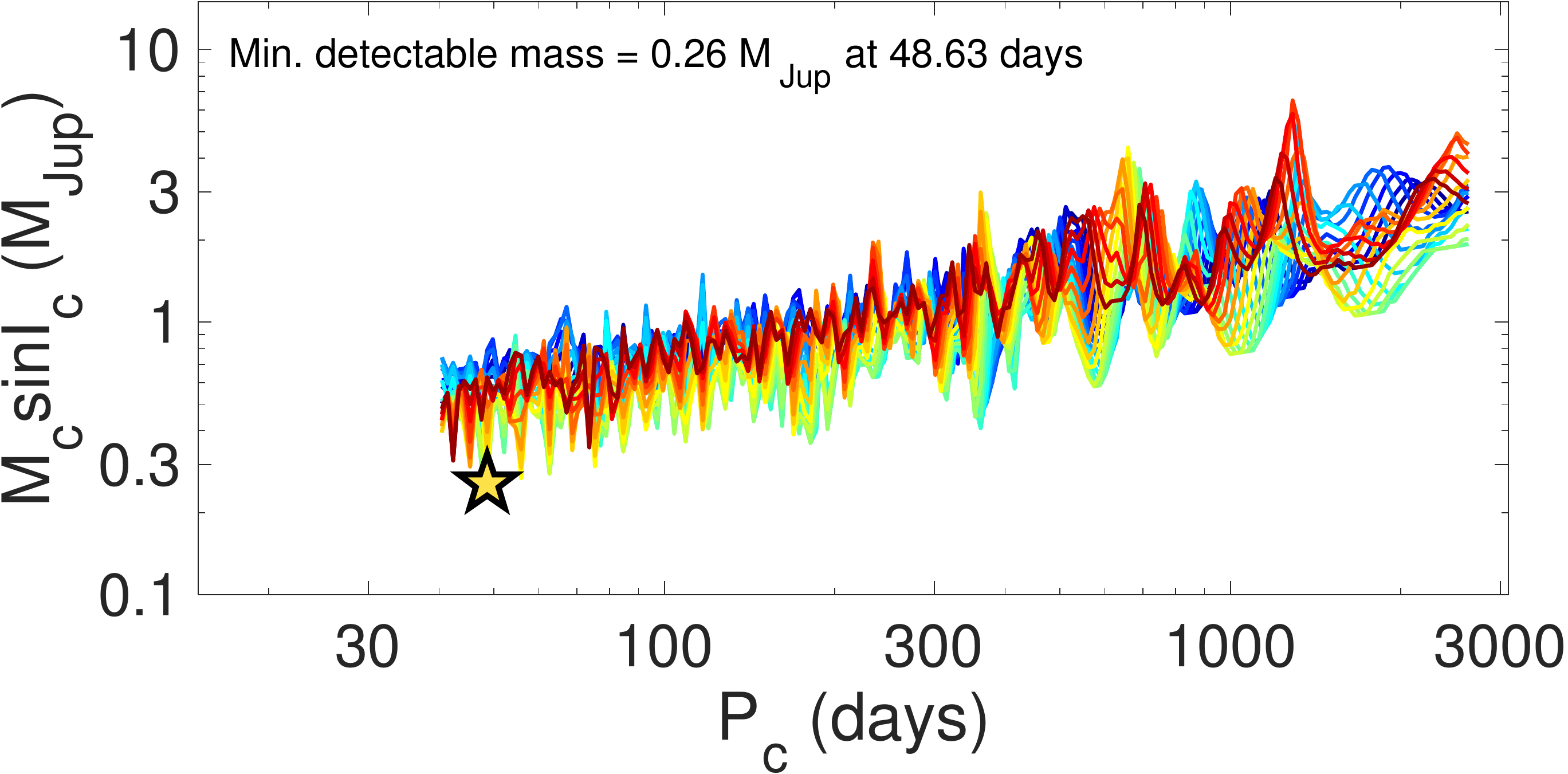}
\end{subfigure}
\end{center}
\end{figure}
\begin{figure}
\begin{center}
\subcaption*{EBLM J2101-45: chosen model = k1 (ecc) \newline \newline $m_{\rm A} = 1.29M_{\odot}$, $m_{\rm B} = 0.523M_{\odot}$, $P = 25.577$ d, $e = 0.091$}
\begin{subfigure}[b]{0.49\textwidth}
\includegraphics[width=\textwidth,trim={0 10cm 0 1.2cm},clip]{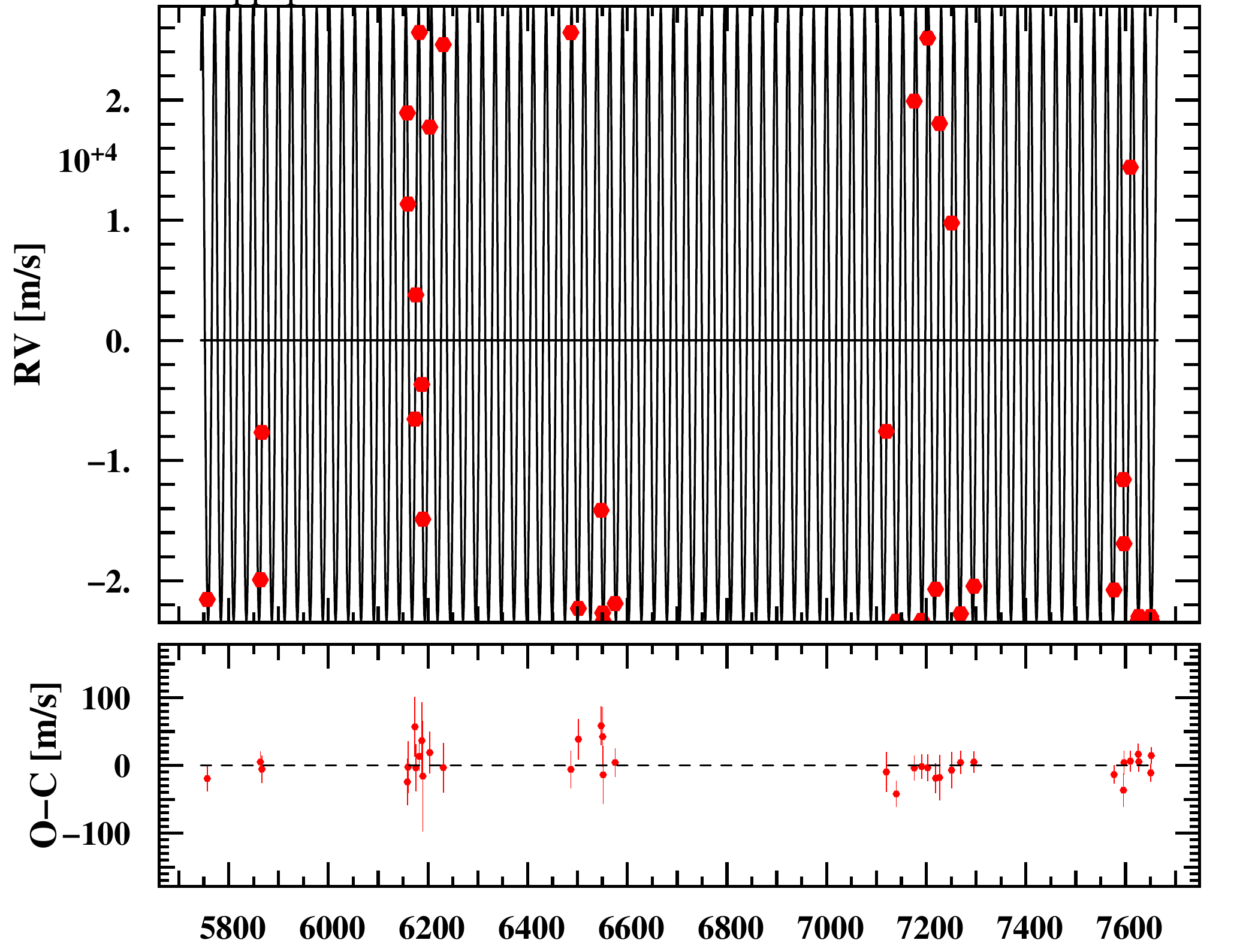}
\end{subfigure}
\begin{subfigure}[b]{0.49\textwidth}
\includegraphics[width=\textwidth,trim={0 0 2cm 0},clip]{orbit_figures/BJD_bar.pdf}
\end{subfigure}
Radial velocities folded on binary phase
\begin{subfigure}[b]{0.49\textwidth}
\includegraphics[width=\textwidth,trim={0 0.5cm 0 0},clip]{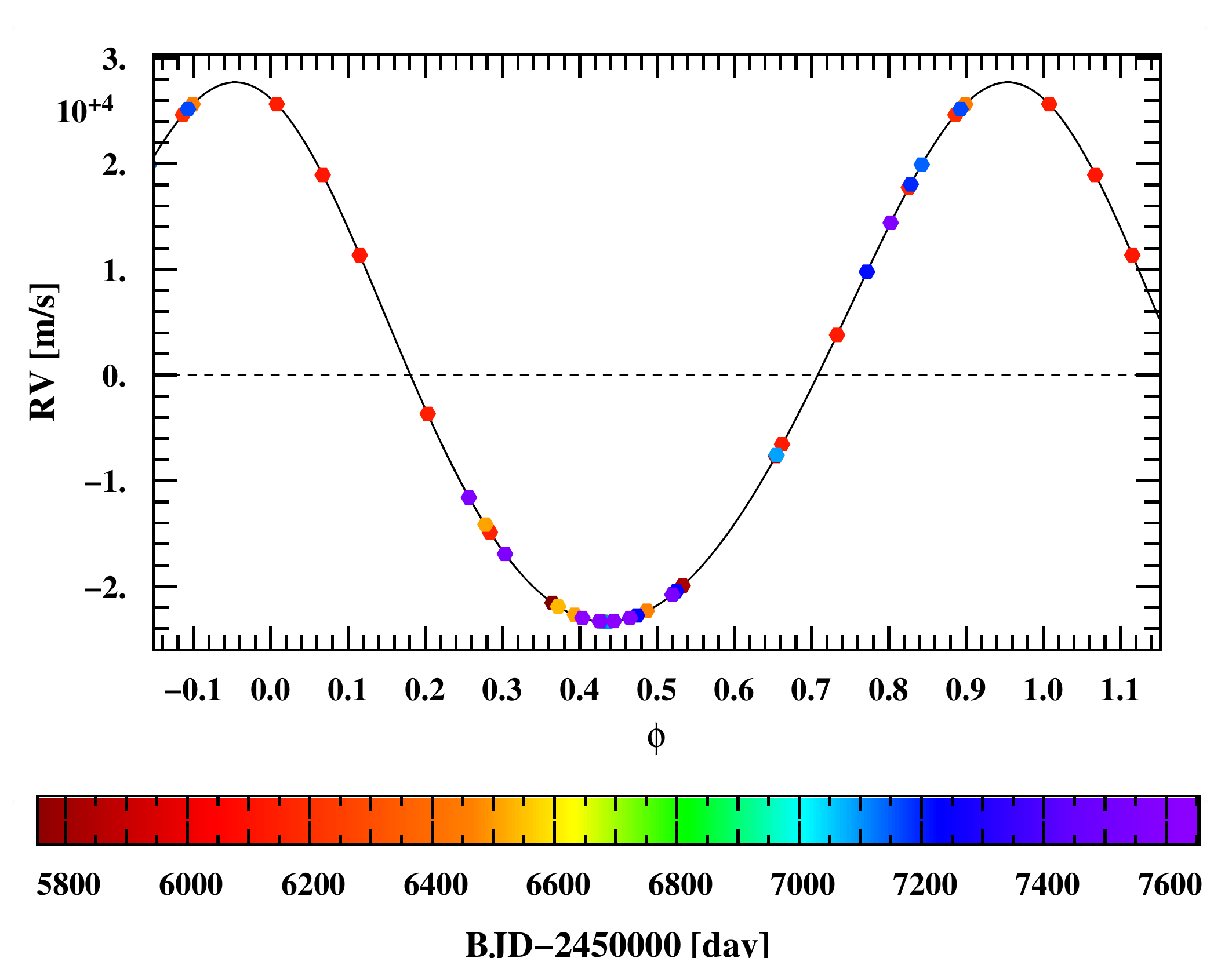}
\end{subfigure}
\begin{subfigure}[b]{0.49\textwidth}
\includegraphics[width=\textwidth,trim={0 0 2cm 0},clip]{orbit_figures/BJD_bar.pdf}
\end{subfigure}
Detection limits
\begin{subfigure}[b]{0.49\textwidth}
\vspace{0.5cm}
\includegraphics[width=\textwidth,trim={0 0 0 0},clip]{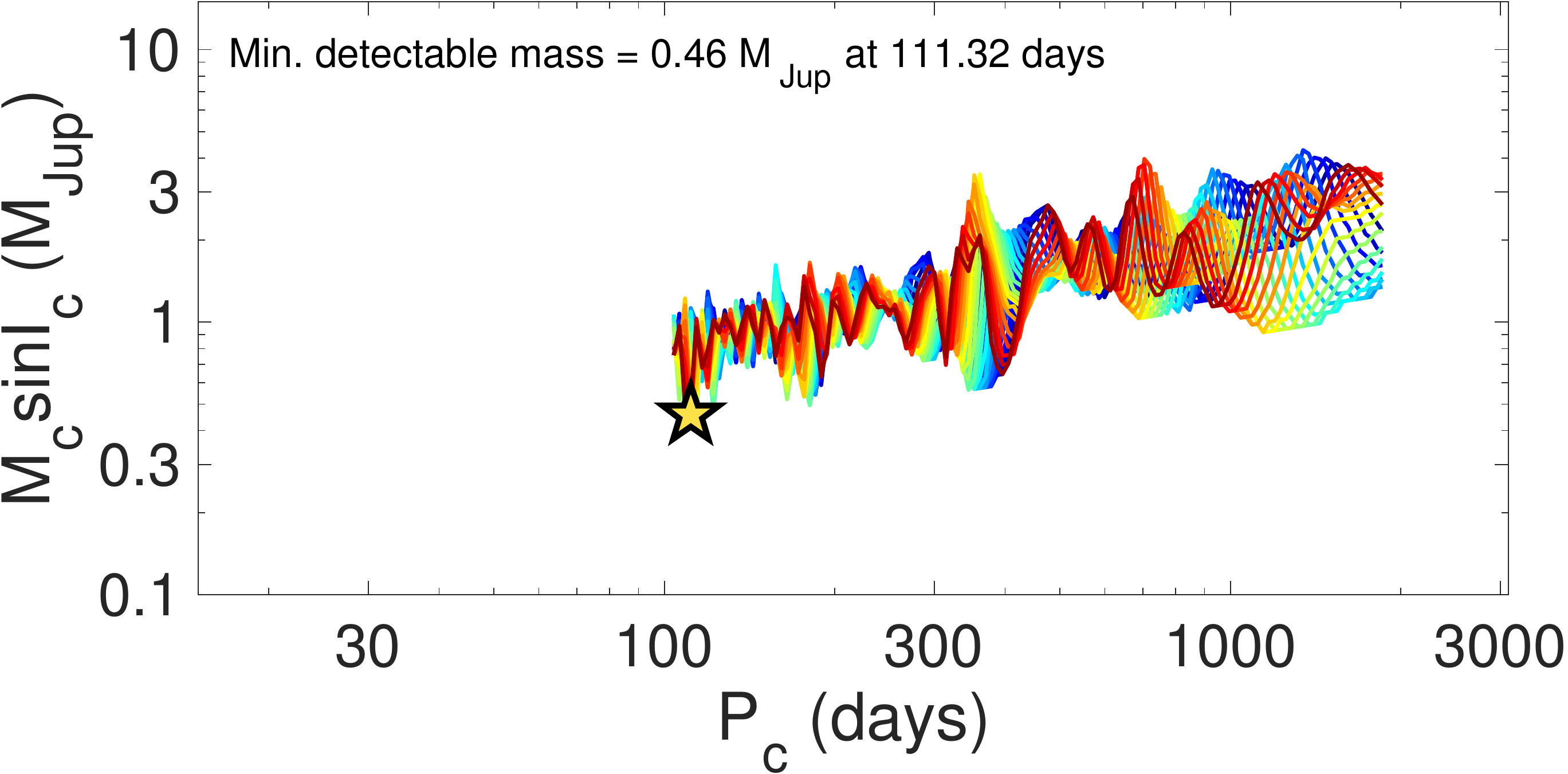}
\end{subfigure}
\end{center}
\end{figure}
\begin{figure}
\begin{center}
\subcaption*{EBLM J2122-32: chosen model = k1 (ecc) \newline \newline $m_{\rm A} = 1.19M_{\odot}$, $m_{\rm B} = 0.592M_{\odot}$, $P = 18.421$ d, $e = 0.405$}
\begin{subfigure}[b]{0.49\textwidth}
\includegraphics[width=\textwidth,trim={0 10cm 0 1.2cm},clip]{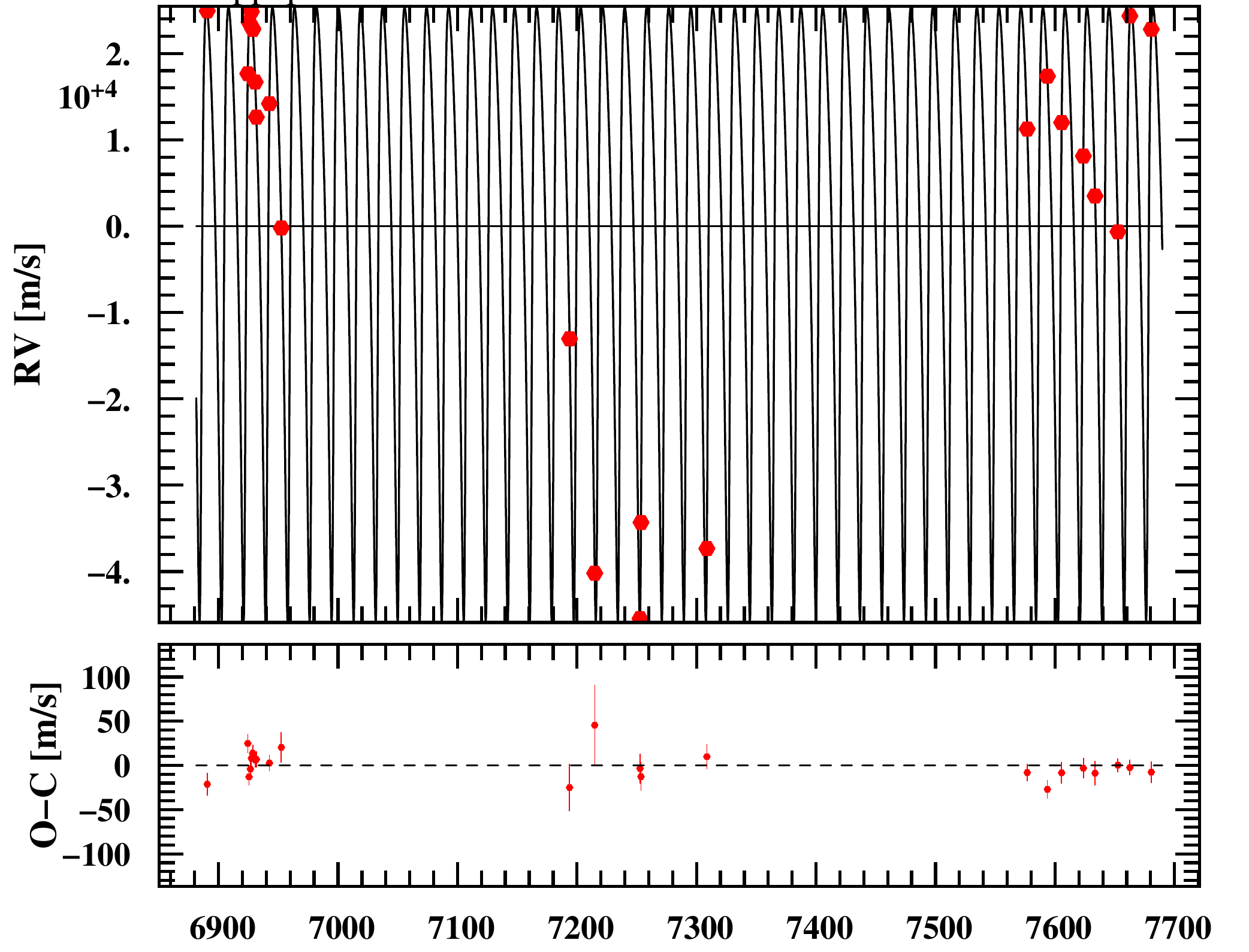}
\end{subfigure}
\begin{subfigure}[b]{0.49\textwidth}
\includegraphics[width=\textwidth,trim={0 0 2cm 0},clip]{orbit_figures/BJD_bar.pdf}
\end{subfigure}
Radial velocities folded on binary phase
\begin{subfigure}[b]{0.49\textwidth}
\includegraphics[width=\textwidth,trim={0 0.5cm 0 0},clip]{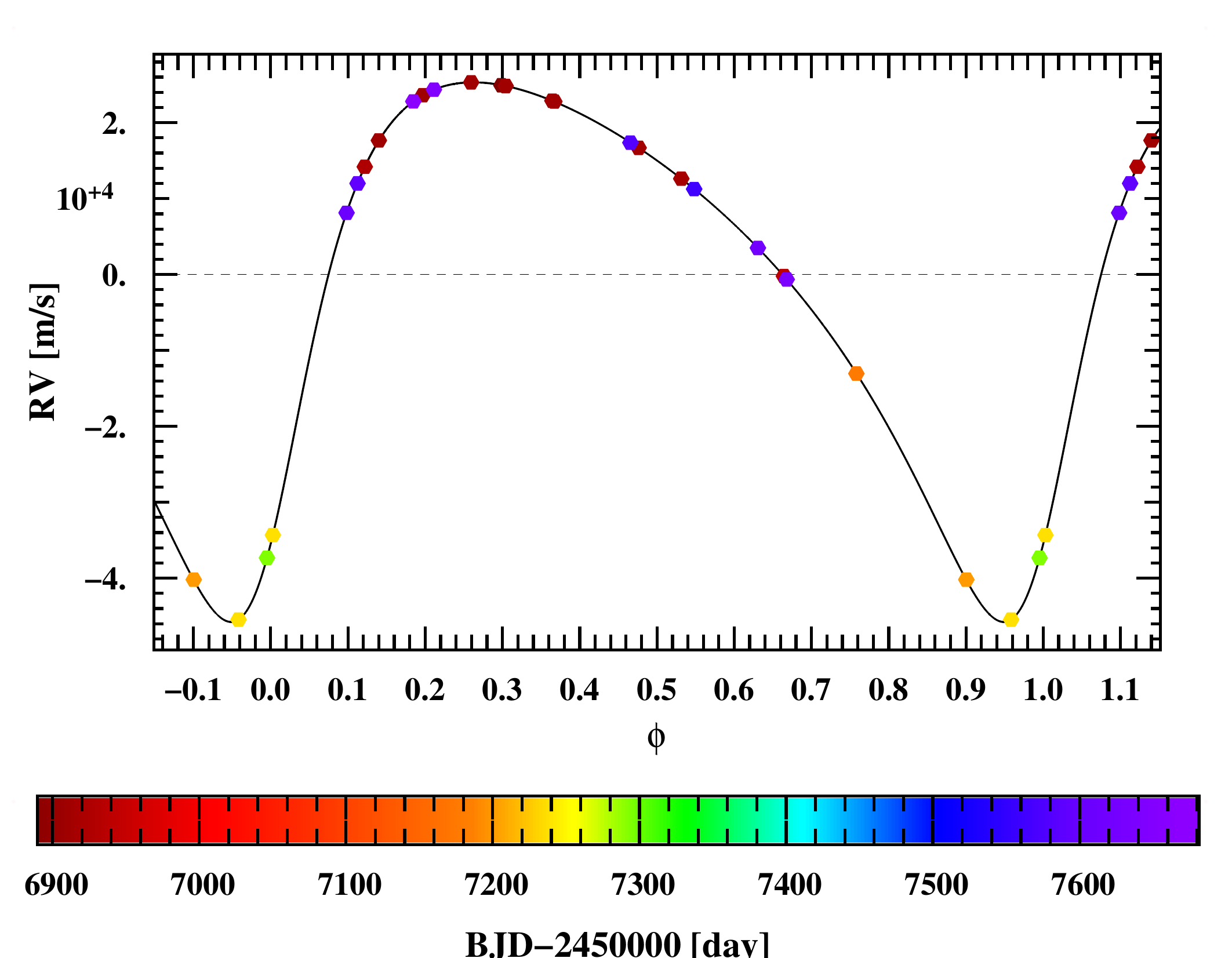}
\end{subfigure}
\begin{subfigure}[b]{0.49\textwidth}
\includegraphics[width=\textwidth,trim={0 0 2cm 0},clip]{orbit_figures/BJD_bar.pdf}
\end{subfigure}
Detection limits
\begin{subfigure}[b]{0.49\textwidth}
\vspace{0.5cm}
\includegraphics[width=\textwidth,trim={0 0 0 0},clip]{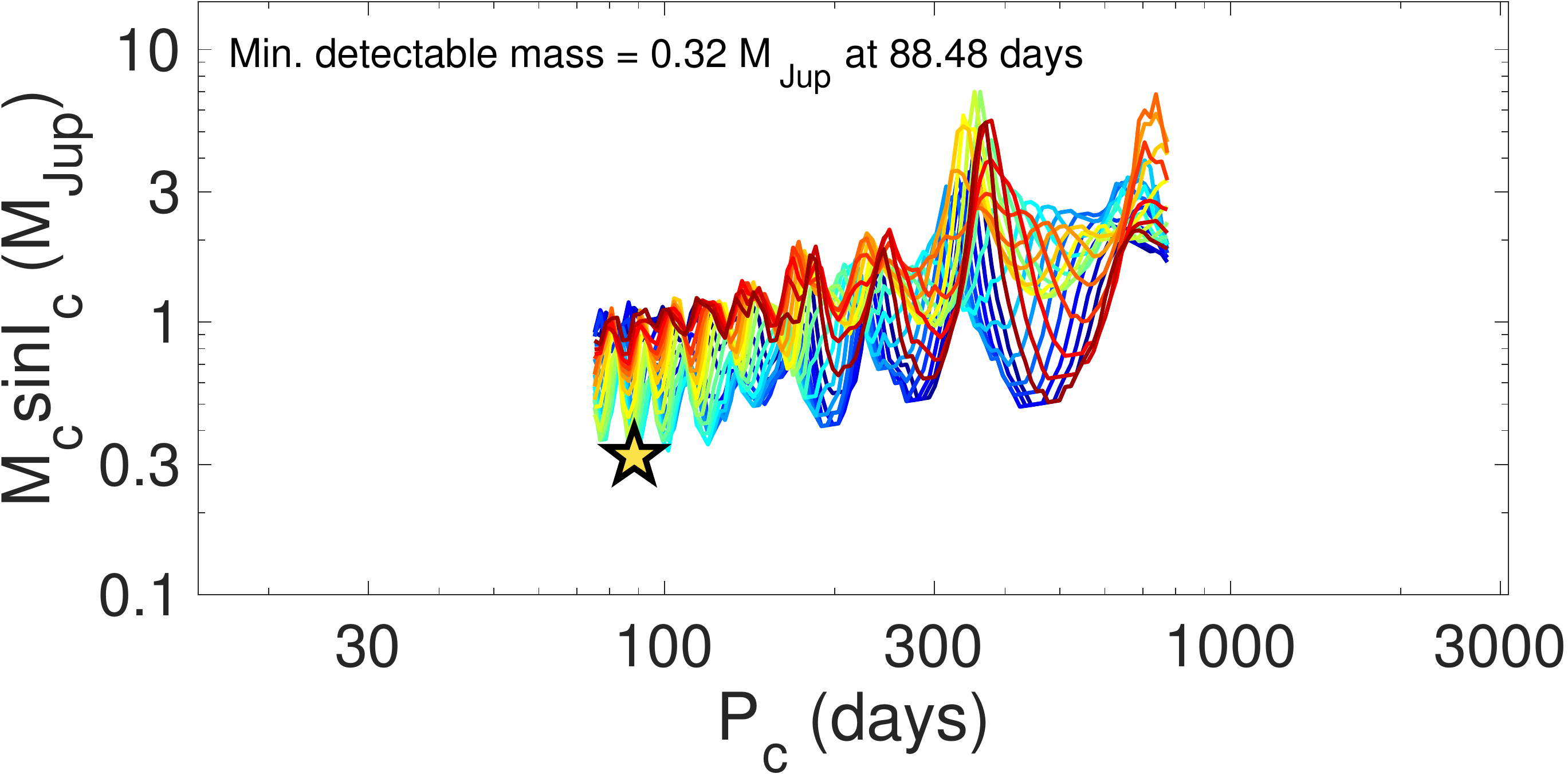}
\end{subfigure}
\end{center}
\end{figure}
\begin{figure}
\begin{center}
\subcaption*{EBLM J2207-41: chosen model = k1 (ecc) \newline \newline $m_{\rm A} = 1.21M_{\odot}$, $m_{\rm B} = 0.121M_{\odot}$, $P = 14.775$ d, $e = 0.068$}
\begin{subfigure}[b]{0.49\textwidth}
\includegraphics[width=\textwidth,trim={0 10cm 0 1.2cm},clip]{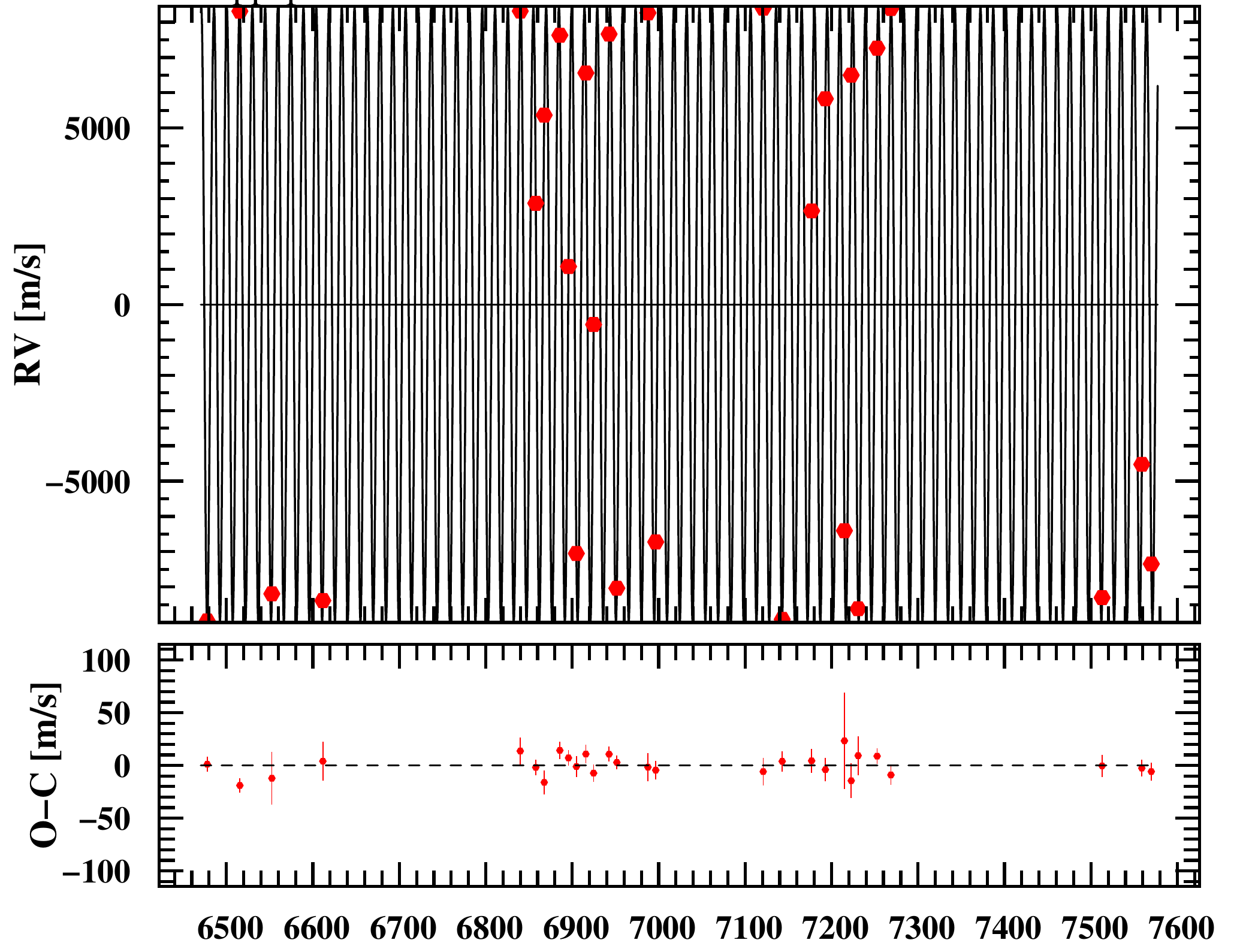}
\end{subfigure}
\begin{subfigure}[b]{0.49\textwidth}
\includegraphics[width=\textwidth,trim={0 0 2cm 0},clip]{orbit_figures/BJD_bar.pdf}
\end{subfigure}
Radial velocities folded on binary phase
\begin{subfigure}[b]{0.49\textwidth}
\includegraphics[width=\textwidth,trim={0 0.5cm 0 0},clip]{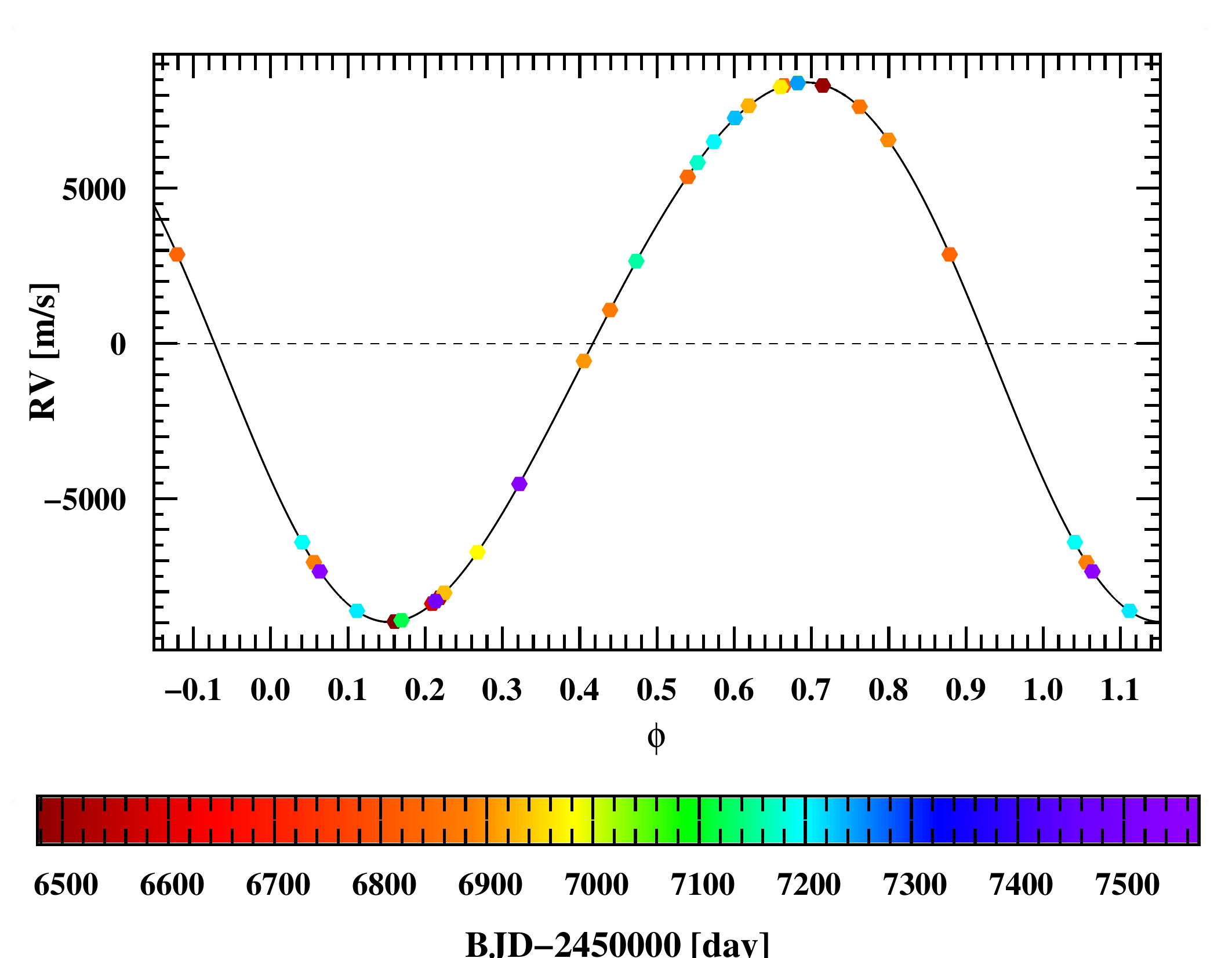}
\end{subfigure}
\begin{subfigure}[b]{0.49\textwidth}
\includegraphics[width=\textwidth,trim={0 0 2cm 0},clip]{orbit_figures/BJD_bar.pdf}
\end{subfigure}
Detection limits
\begin{subfigure}[b]{0.49\textwidth}
\vspace{0.5cm}
\includegraphics[width=\textwidth,trim={0 0 0 0},clip]{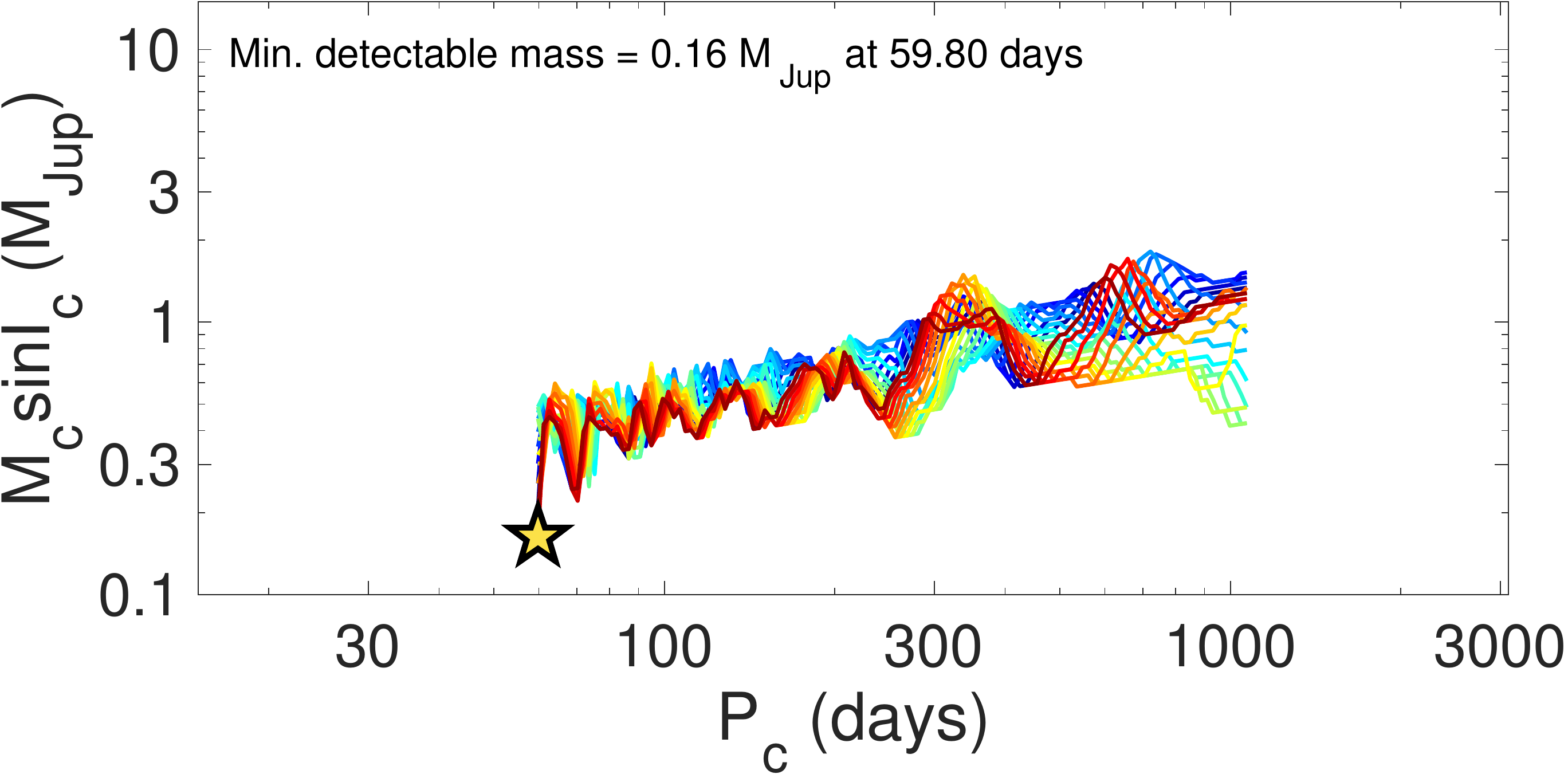}
\end{subfigure}
\end{center}
\end{figure}
\begin{figure}
\begin{center}
\subcaption*{EBLM J2217-04: chosen model = k1d1 (ecc) \newline \newline $m_{\rm A} = 0.95M_{\odot}$, $m_{\rm B} = 0.208M_{\odot}$, $P = 8.155$ d, $e = 0.047$}
\begin{subfigure}[b]{0.49\textwidth}
\includegraphics[width=\textwidth,trim={0 10cm 0 1.2cm},clip]{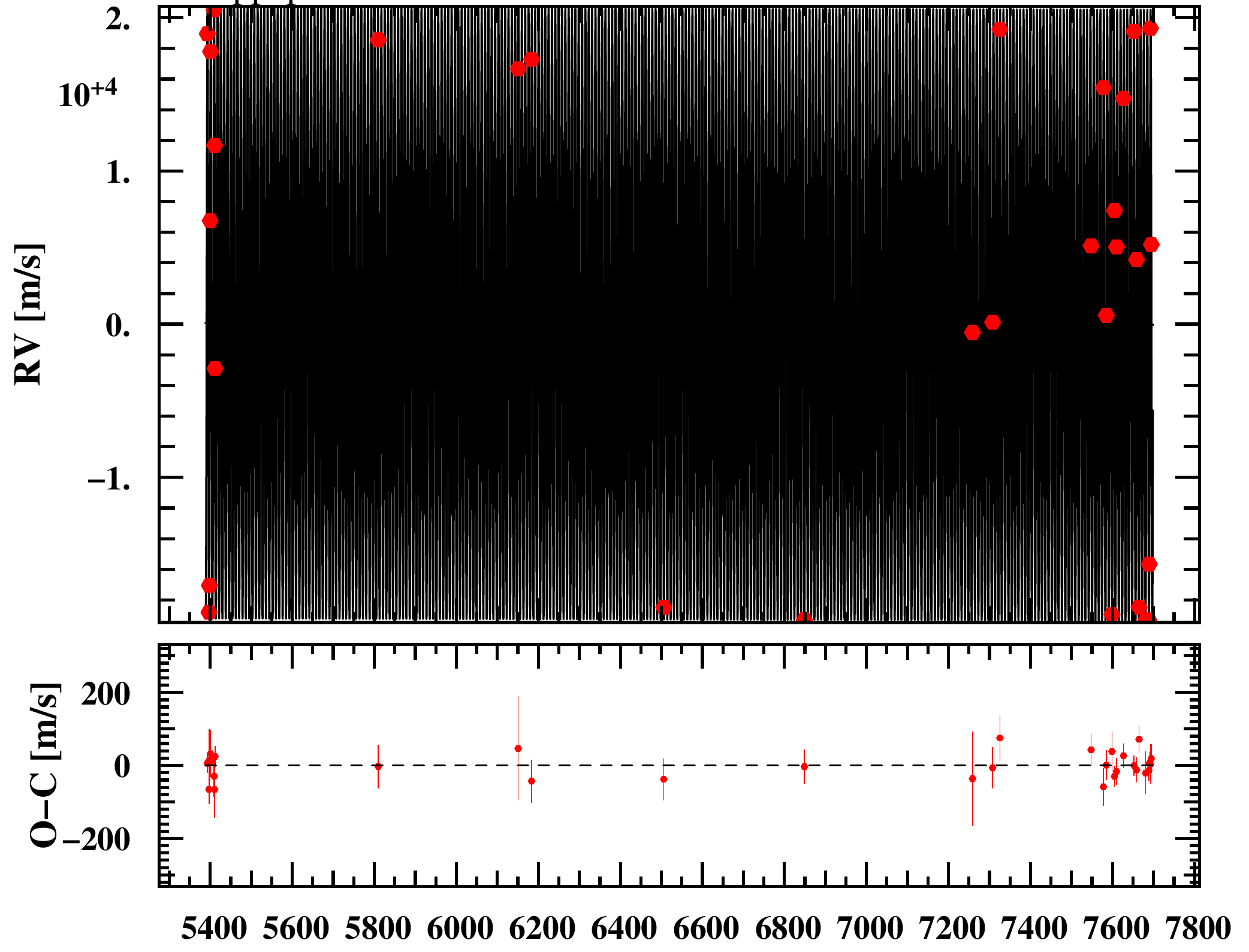}
\end{subfigure}
\begin{subfigure}[b]{0.49\textwidth}
\includegraphics[width=\textwidth,trim={0 0 2cm 0},clip]{orbit_figures/BJD_bar.pdf}
\end{subfigure}
Radial velocities folded on binary phase
\begin{subfigure}[b]{0.49\textwidth}
\includegraphics[width=\textwidth,trim={0 0.5cm 0 0},clip]{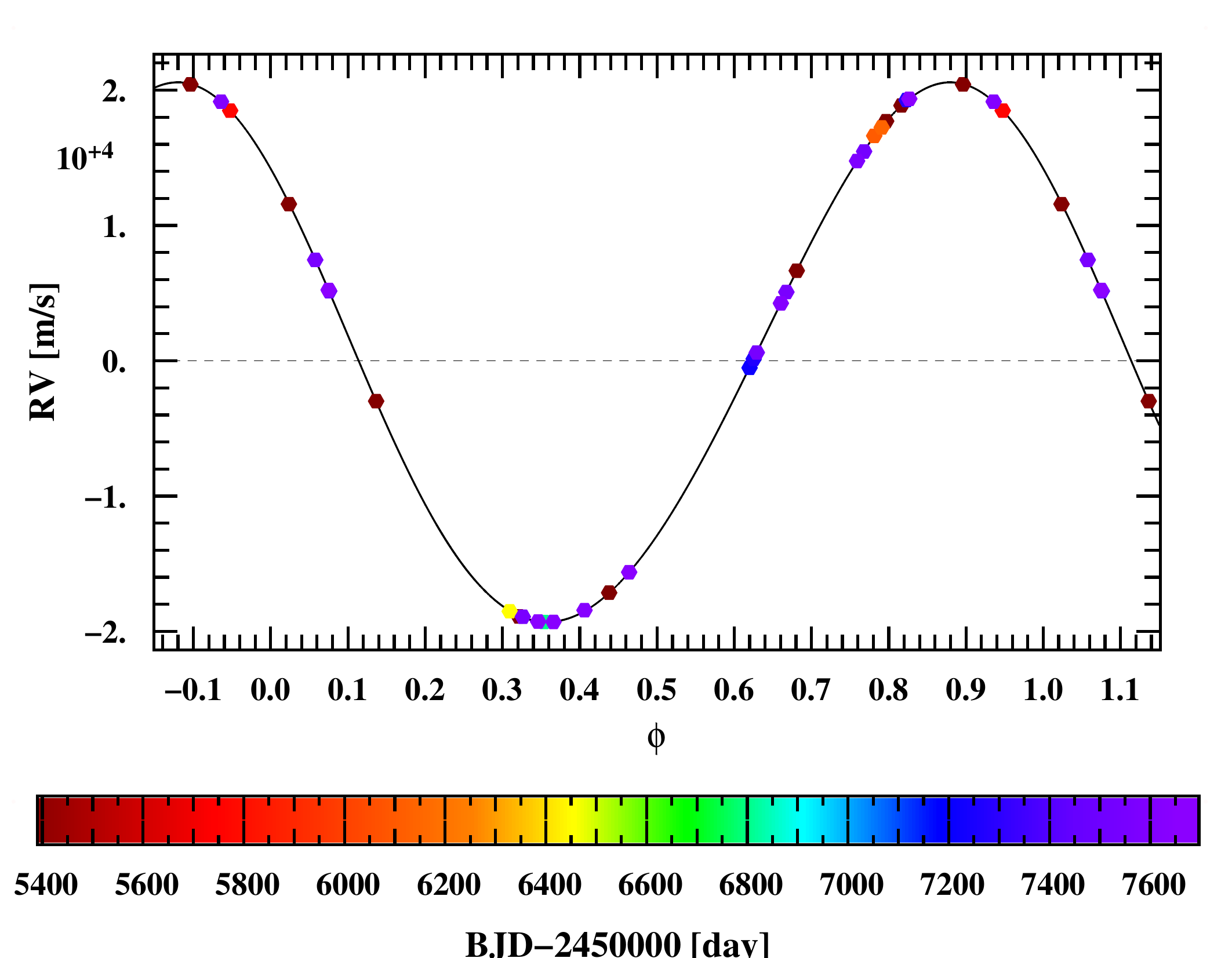}
\end{subfigure}
\begin{subfigure}[b]{0.49\textwidth}
\includegraphics[width=\textwidth,trim={0 0 2cm 0},clip]{orbit_figures/BJD_bar.pdf}
\end{subfigure}
Detection limits
\begin{subfigure}[b]{0.49\textwidth}
\vspace{0.5cm}
\includegraphics[width=\textwidth,trim={0 0 0 0},clip]{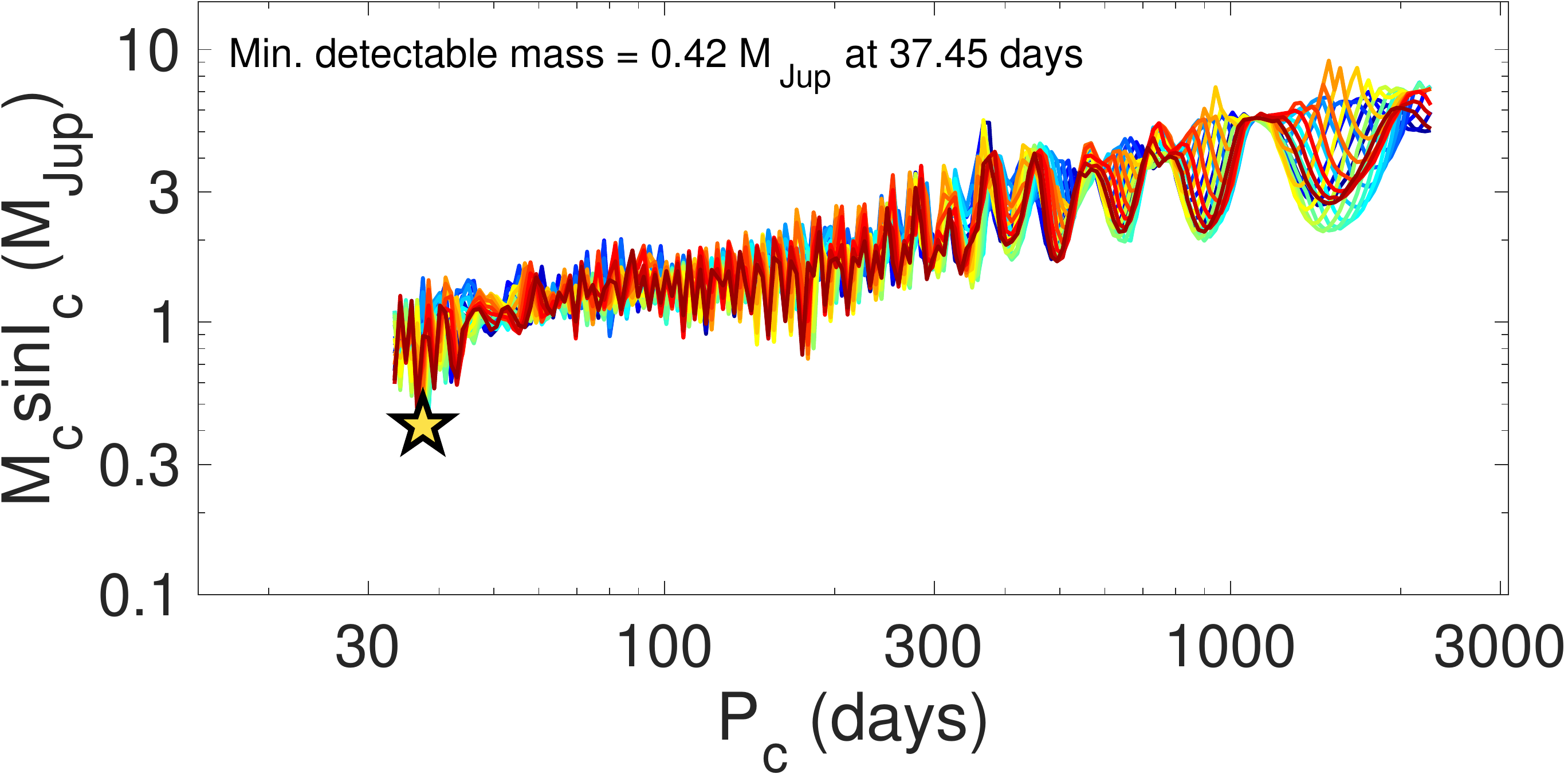}
\end{subfigure}
\end{center}
\end{figure}

\end{document}